\documentclass[aps,floatfix,superscriptaddress,showpacs,showkeys,onecolumn,12pt,vi]{mitthesis}
\usepackage{./STY_files/lgrind}
\usepackage{./STY_files/cmap}
\usepackage[T1]{fontenc}
\usepackage{lmodern}
\usepackage{afterpage}
\newcommand\blankpage{%
    \null
    \thispagestyle{empty}%
    \addtocounter{page}{-1}%
    \newpage}
\usepackage[caption=false]{subfig}
\usepackage{textcomp}
\usepackage{./STY_files/kbordermatrix}
\usepackage{graphicx,bm,color}
\usepackage{mathtools}
\usepackage{blindtext}

\usepackage{braket}

\usepackage{amsfonts,amsmath,amssymb,amsthm}
\usepackage{color}
\usepackage{url}
\usepackage{booktabs}
\usepackage{lscape}
\usepackage{rotating}
\usepackage{array}
\usepackage{./STY_files/algorithm2e}
\usepackage{algpseudocode}
\usepackage{braket}
\usepackage[dvipsnames]{xcolor}
\usepackage{hyperref}
\hypersetup{
    colorlinks=true,
    linkcolor=blue,
    filecolor=blue,      
    urlcolor=blue,
}
\usepackage{epstopdf}
\newtheorem{theorem}{Theorem}
\begin{document}

%
%
%
%
%
%
%
%
%
%
%

\title{\textbf{{\Huge Aspects of Quantum Entanglement and Indistinguishability}}}

\author{\textbf{{\LARGE Soumya Das}}}
\department{\textbf{Cryptology and Security Research Unit, \\ Indian Statistical Institute \\203, B. T. Road, Kolkata, West Bengal 700108}}

\degree{{\Large Doctor of Philosophy in Computer Science}}

\thesisdate{August, 2022}


\supervisor{Goutam Paul}{Associate Professor}

\chairman{Utpal Garain}{Chairman, Department Committee on Graduate Theses}
\maketitle



\setcounter{savepage}{\thepage}

\afterpage{\blankpage}
\newpage
\newpage
\medskip
\medskip
\vspace{2cm}
\begin{center}
\textbf{\textit{{\Large I dedicate this thesis to}
\\
{\LARGE \hspace{2cm} My Parents}}}
\end{center}

\newpage

\cleardoublepage

\section*{Acknowledgements}

\indent First and foremost I am extremely grateful to my supervisor \textbf{Dr. Goutam Paul} for his invaluable advice, continuous support, and patience from the first day when I have decided to start preparation for the entrance of my Ph. D. His immense knowledge and plentiful experience have encouraged me throughout my academic research and in my daily life.

Most importantly, none of this could have happened without my family. 
I am deeply grateful to my parents \textbf{Subrata Das} and \textbf{Swapna Das} and my wife \textbf{Sampurna Das} for providing me with unfailing support and continuous encouragement throughout my years of study and through the process of researching and writing this thesis. I would also like to thank my sister, my other family members and all the relatives, friends and well wishers for their continuous support throughout my life. Without their tremendous understanding and encouragement in the past few years, it would be impossible for me to complete my study.

I would like to thank my collaborators \textbf{Dr. Anindya Banerji} and \textbf{Dr. Ritabrata Sengupta} for  their support and tutelage in my research papers. I would like to thank my junior collaborator \textbf{Adarsh Chandrashekar} for his work.

Getting through my dissertation required more than academic support, and I have many, many people to thank for listening to and, at times, having to tolerate me over the past three years. I would like to thank \textbf{Dr. Manas Mukherjee} and \textbf{Dr. Alexander Ling} for hosting my  internship in Centre for Quantum Technologies, Singapore. I would like to extend my sincere thanks to	\textbf{Dr. Ravindra Pratap Singh} of Physical Research Laboratory  for a visit to his lab of quantum optics. 

I would like to thank all the faculty members, specially \textbf{Prof. Bimal Kumar Roy} and \textbf{Prof. Subhamoy Maitra}, and all the other staffs of our institute who have helped me for all the academic as well as non-academic issues.

I express my gratitude and appreciation to thank my colleagues
\textbf{Mostafizur Rahaman}, \textbf{Probal Banerjee}, \textbf{Avishek Majumder}, \textbf{Samir Kundu}, \textbf{Laltu Sardar}, \textbf{Diptendu Chatterjee}, \textbf{Pritam Chattopadhyay}, \textbf{Amit Jana}, \textbf{Snehalika Lall}, \textbf{Nayana Das}, \textbf{Bikash Santra} and others for their support, company and healthy discussions.

Finally, I want to thank  my Master's guide \textbf{Dr. Tamaghna Acharya} and my colleagues  \textbf{Dr. Surajit Basak}, \textbf{Dr. Aritra Roy}, \textbf{Dr. Amartya Banerjee} and \textbf{Dr. Avijit Dutta} and others who help be during the entrance exam and interview of this institute. Also I would like to thanks my school tuition teacher \textbf{Dr. Indraniv Ray} for his constant support throughout this journey.

 \begin{abstractpage}
%
%
%

\indent Entanglement of distinguishable and indistinguishable particles under different scenarios and related properties and results constitute the core component of this thesis. 
 We propose a new error-modeling for Hardy's test and also perform experimental verification of it in superconducting qubits. Further, we point out the difficulties associated with the practical implementation of quantum protocols based on Hardy's test and propose possible remedies. We also propose two performance measures for any two qubits of any quantum computer based on superconducting qubits.

Next, we prove that if quantum particles (either distinguishable or indistinguishable) can simultaneously produce and perform hyper-hybrid entangled state and unit fidelity quantum teleportation respectively then using that cloning of any arbitrary quantum state is possible. This theorem results two no-go theorems: (1) hyper-hybrid entangled state is not possible for distinguishable particles and (2) unit fidelity quantum teleportation is not possible for indistinguishable particles. These theorems establish that there exists some quantum correlation or application unique to indistinguishable particles only and yet some unique to distinguishable particles only, giving a separation between the two domains.
We also establish that the hyper-hybrid entangled state is possible using two indistinguishable fermions and we generalize it for bosons and fermions.

 We establish a generalized degree of freedom trace-out rule that covers single or multiple degree of freedom scenarios for both distinguishable and indistinguishable systems. Using this,  we propose  generalized expressions for teleportation fidelity and  singlet fraction and derive their relations, applicable for both distinguishable and indistinguishable particles with single or multiple degrees of freedom. We also derive an upper bound for the generalized singlet fraction for distinguishable and indistinguishable particles.  We further show how our  relation helps to characterize different types of composite states in terms of their distinguishability, separability, presence of maximally entangled structure and the number of degrees of freedom. Finally, we demonstrate an optical circuit to generate entanglement for distinguishable particles each having two degrees of freedom and characterize it using our relation. 

Further, using generalized degree of freedom trace-out rule, we show that, for two indistinguishable particles each having more than one degree of freedom, the monogamy of entanglement can be violated maximally using the  measures that are monogamous for distinguishable particles. This results the following theorem that ``In qubit systems, indistinguishability is a necessary criterion for maximum violation of monogamy of entanglement by the same measures that are monogamous for distinguishable particles". 

 For three indistinguishable particles each having multiple degree of freedom, we show that monogamy of entanglement is obeyed using squared concurrence as an entanglement measure.
 We  also establish that the monogamy inequality becomes equality for all pure indistinguishable states, but the inequality remains for mixed indistinguishable states. This can be used as a one-sided test of distinguishability for particles in pure states

 We show that the cost of adding an ancilla particle can be bypassed by using an additional degrees of freedom and creating multi-degree of freedom entanglement. 
Next, we show that entangled indistinguishable particles may alter certain important parameters in cryptographic protocols, in particular, we demonstrate how indistinguishability can change Hardy's probability 
Finally we propose a novel entanglement swapping protocol without Bell state measurement using only two indistinguishable particles that will be very useful in quantum networks specially in quantum repeaters.

 \end{abstractpage}
\pagestyle{plain}
\chapter*{List of publications/preprints}
\begin{itemize}
\item As a part of this thesis
\begin{enumerate}
\item \textbf{A New Error-Modeling of Hardy's Paradox for Superconducting Qubits and Its Experimental Verification} \\
\textit{\textbf{Soumya Das}} and \textit{Goutam Paul} \\
ACM Transactions on Quantum Computing, Volume 1, Issue 1, December 2020, Article No. 4, pp 1-24, Published : 2nd October 2020 \\
DOI: \href{https://doi.org/10.1145/3396239}{10.1145/3396239}\\
Also available at  [arXiv:\href{https://arxiv.org/abs/1712.04925}{1712.04925}].

\item \textbf{Hyper-hybrid entanglement, indistinguishability, and two-particle entanglement swapping} \\
\textit{\textbf{Soumya Das}, Goutam Paul,} and \textit{Anindya Banerji} \\
Physical Review A 102, 052401,  Published 2 November 2020 \\
DOI: \href{https://doi.org/10.1103/PhysRevA.102.052401}{10.1103/PhysRevA.102.052401} \\
Presented at \href{https://youtu.be/J8oaq3V_z0c}{Quantum Physics and Logic (QPL) 2021}\\
Also available at [arXiv:\href{https://arxiv.org/abs/2101.10089}{2101.10089}]. \\

\item \textbf{Maximum violation of monogamy of entanglement for indistinguishable particles by measures that are monogamous for distinguishable particles} [arXiv:\href{https://arxiv.org/abs/2102.00780}{2102.00780}] \\
\textit{Goutam Paul}, \textbf{\textit{Soumya Das}}, and \textit{Anindya Banerji} \\
Physical Review A 104, L010402 (as a \textbf{Letter}) - Published 20 July 2021 \\
DOI: \href{https://doi.org/10.1103/PhysRevA.104.L010402}{10.1103/PhysRevA.104.L010402}\\
Also available at [arXiv:\href{https://arxiv.org/abs/2102.00780}{2102.00780}].

\item \textbf{Monogamy of Entanglement for Indistinguishable Particles} \\
\textit{\textbf{Soumya Das}}, \textit{Goutam Paul}, and \textit{Ritabrata Sengupta}\\
Status: Communicated on 7th April, 2022 to Physical Review A with manuscript id: LJ17132AR.	  \\

\item \textbf{Generalized Relation between Teleportation Fidelity and Singlet Fraction for
(In)-distinguishable Particles} \\
\textit{\textbf{Soumya Das}}, \textit{Goutam Paul}, and \textit{Anindya Banerji} \\
Status: Communicated on 29th September, 2021 to Physical Review A with manuscript id: LJ17778AR.
\end{enumerate}
\item Not included in this thesis
\begin{enumerate}
\item \textbf{Experimental Characterization of Noise using variants of Unitarity Randomized Benchmarking} \\
\textit{Adarsh Chandrashekar}, \textbf{\textit{Soumya Das}}, and \textit{Goutam Paul} \\
Status: Communicated on 8th October, 2021 to Physical Review A with manuscript id: AK12142.
\end{enumerate}
\end{itemize}

\tableofcontents
\newpage
\listoffigures
\newpage
\listoftables

\chapter*{Notation and list of symbols}
\begin{center}
 \begin{tabular}{c | c } 
 \hline
 Symbol & Meaning  \\ [0.5ex] 
 \hline
  $q$ & Value of Hardy's probability \\
 $q_{max}$ & maximum value of Hardy's probability  $= \frac{5\sqrt{5}-11}{2} \approx 0.09017  $ for two qubits \\
 $q_{lb}$ & lower bound on Hardy's probability \\ 
 $d$ & Dimension of the system \\ 
 $\mathbb{E}$ & Entanglement measure\\
 $\mathcal{C}$ & Concurrence \\
 $\mathcal{N}$ & Negativity \\
 $E_{\mathcal{N}}$ & Log-negativity\\
 $\mathbb{N}_n$ & $\lbrace 1, 2, \ldots, n \rbrace$ \\
 $\eta$  & +1 for bosons and -1 for fermions \\
  $f$ & Teleportation fidelity \\
 $f_g$ & Generalized teleportation fidelity \\
  $F$ & Singlet fraction \\
 $F^{(n)}_g$ & Generalized singlet fraction \\
 $\mu$ & mean \\
 $\sigma$ & Standard deviation \\
 $\mathbb{S}^P$ & $\lbrace s^1, s^2, \ldots, s^P \rbrace$
\end{tabular}
\end{center}
\chapter*{Abbreviations}
\begin{center}
 \begin{tabular}{c | c } 
 \hline
Short form  &  Full form  \\ [0.5ex] 
 \hline
 BSM & Bell State Measurement \\
 CHSH & Clauser-Horne-Shimony-Holt \\
 CI & Confidence interval \\
 DOF & Degrees of Freedom  \\ 
 HHES & Hyper Hybrid Entangled State \\
 HUP & Heisenberg's uncertainty principle \\ 
 LOCC & Local Operations and Classical Communications\\
 LHV & Local hidden variable \\
 MOE & Monogamy of Entanglement\\
 MES & Maximally entangled state \\
 NMES & Non maximally entangled state  \\
NMR & Nuclear magnetic resonance \\
 PS & Product state \\
 QPQ & Quantum private query\\ 
 QBA & Quantum Byzantine agreement \\
 SSR & Super selection rule\\
 UFQT & Unit Fidelity Quantum Teleportation \\ 
\end{tabular}
\end{center}
\chapter{Introduction} \label{Chap1}
Max Planck, the father of quantum physics, was advised by his supervisor Philipp von Jolly \textit{not to study theoretical physics as it was probably not a great idea, since there was not much left to do} when he was about to start his journey with physics in 1874. A similar quotation is also found in 1900 when William Thomson, a British mathematician, physicist and engineer, (better known as Lord Kelvin)  addressed the British Association for the Advancement of Science that ``\textit{There is nothing new to be discovered in physics now. All that remains is more and more precise measurement.}'' Coincidentally in that same year, quantum physics was born when Max Planck explained the black body radiation using the quantized formulation of electromagnetic energy. 

\section{Quantum physics}
At the beginning of the 20th century, there were a number of experimental observations that cannot be explained by the existing theories, often considered classical physics. This includes black-body radiation~\cite{Planck01,planck1972original}, photo-electric effect~\cite{einstein1905generations}, Stern-Gerlach experiment~\cite{gerlach1922experimentelle}, Mach-Zehnder interferometer~\cite{zehnder1891neuer,mach1892ueber}, etc.  
To solve these shortcomings of physics, a new theory came out with the hands of Max Planck, Albert Einstein, etc., and later formulated by Max Born, Werner Heisenberg, Wolfgang Pauli, Erwin Schr$\"o$dinger, etc. They have revolutionized the history of physics by their theories which are later verified by experimental results. 

Quantum theory is a mathematical tool to model and understand how physical phenomenons are happening and to predict what will happen next. It basically describes the behavior of the atomic and the subatomic particles using the wave function. The evolution of this wave function is governed by the Schr$\"o$dinger equation~\cite{PhysRev.28.1049}. When a measurement is performed, how this wave function collapses is the most important problem in quantum theory and is known as the measurement problem~\cite{RevModPhys.85.471}. Different solutions are proposed to solve this problem known as the interpretations of quantum theory~\cite{jarmmer1974philosophy}. These theories mainly concerned about whether quantum mechanics is deterministic or stochastic, the description about the nature of reality, the process of measurement, etc. Most popular interpretation of quantum mechanics is Copenhagen interpretation~\cite{faye2002copenhagen}. Other interpretations includes many world interpretation~\cite{dewitt2015many}, QBism~\cite{TIMPSON2008579}, consistent histories~\cite{griffiths1984consistent}, Bohm theory~\cite{PhysRev.85.166,durr1996bohmian}, Transactional interpretation~\cite{RevModPhys.58.647,cramer1988overview,cramer2009transactional}, etc. This thesis is based on the Copenhagen interpretation which can be explained by the four \textit{postulates of quantum mechanics}~\cite{Nielsenbook}. 

\textit{Postulate 1} describes the basic building block of quantum physics, i.e., the state of an isolated quantum system. 

\textit{Postulate 2} concerned about the dynamics of closed quantum systems. This is governed by the Schr$\"o$dinger equation, i.e., by the unitary evolution. 

\textit{Postulate 3} explains the most interesting phenomenon of quantum mechanics, i.e., measurement, which is a process of extracting information from quantum systems. 

\textit{Postulate 4} deals with the description of the composite system, i.e., a combination of different quantum systems. The rest of quantum mechanics is just the derivation and applications of these postulates.

Till now, quantum theory can be used to mathematically model all the natural events (that cannot be explained using classical theory as said above)  other than gravity. Most of the present-day technologies are successful applications of this quantum theory, for example in the electronics industry, chemical industry, communication industry, and most important in the cryptographic industry.

\section{Quantum information}
Quantum information theory deals with how quantum messages can be sent over quantum communication channels.
 More specifically how information can be compressed as a quantum message and transmitted reliably from the sender to receiver in the presence of noise. The measure of quantum information is von-Neumann entropy~\cite{von2018mathematical}, like Shannon entropy~\cite{shannon2001mathematical} in classical communication. Since  the aim of quantum communication is to compress the information as much as possible by the sender keeping in mind that it should be decoded at the receiving end. The amount of compression is given by the Schumacher's quantum noiseless coding theorem~\cite{PhysRevA.51.2738,jozsa1994new} that states that von-Neumann entropy is the optimal compression factor for quantum information. The basic fact for quantum information transfer is that the communicated quantum messages are not generally orthogonal in nature and thus cannot be decoded properly. The limits to which information can be accessible to the receiver is given by the Holevo bound~\cite{holevo1973bounds}.
 
The basic unit of quantum information is qubit, a two-state system that can be realized using any physical two-level device. Normally, quantum information are encoded in quantum particles degree of freedom (DoF). For example, spin DoF of an electron where up-spin and down-spin is taken as the two level system. Similarly, polarization of a photon can be used as a qubit with horizontal polarization and vertical polarization are used as two levels. Unlike classical bit, qubit can be present as a superposition of the two levels.
 
Mathematically, qubits are described by a two-dimensional complex Hilbert space. The two levels are represented as a normalized and mutually orthogonal vectors in that space. The notations are used for this representation was given by Paul Dirac which is known as the Dirac-notation~\cite{dirac1939new}. In this notation, two levels are commonly represented as $\ket{0}$ and $\ket{1}$ which is called \textit{ket} vector. These ket vectors are commonly corresponds to column vectors such as $\ket{0}=(\begin{smallmatrix} 1 \\ 0 \end{smallmatrix}) $ and $\ket{1}=(\begin{smallmatrix} 0 \\ 1 \end{smallmatrix}) $. Another representation of qubits are as a point in a sphere of unit radius, known as the block sphere. The azimuth and elevation angles are the two parameters of the sphere that define the quantum state. Block sphere representation is very much useful to visualize the qubits in the three-dimensional space.

The state vector representations are useful for a single quantum systems. The sub-system of a a composite quantum systems cannot be represented by the state vector. Thus the density matrix formalisms are introduced by von Neumann and Lev Landau.   
In the block sphere representation, the pure states are those which are represented at the circumference of the sphere. Mixed states are those who reside inside the sphere.

The journey of quantum mechanics was not like a dream one. Some of the scientists were doubtful about it due to its counter-intuitive nature. One of them was famous Albert Einstein, who in 1935, with Boris Podolsky and Nathan Rosen, using a thought experiment~\cite{epr} questioned the \textit{completeness} of quantum theory which is known as \textit{Einstein-Podolsky-Rosen (EPR)  paradox}. It states that any physical theory should be \textit{complete}, i.e., they should respect two conditions: \textit{locality} and \textit{reality}. Since the quantum theory does not satisfy these conditions, it should no longer be a complete theory. The founders and followers of quantum theory argued and debated with this theory for almost two decades but none came up with a complete answer.

In 1964, John Bell came up with an elegant and experimentally testable solution to the EPR paradox. He showed that in quantum theory there exist some correlations which are not compatible with the two assumptions of the EPR paradox which is known as \textit{Bell's Theorem}~\cite{Bell}. The legacy of Bell's work is that he showed he has designed an experiment to test this theorem in lab by using an inequality. If some correlation violates this inequality, then that correlation would not be compatible with locality and reality. This theorem proves that quantum theory is the ultimate theory, however, if some ultimate theory exists, it would not be the complete theory as defined by EPR paradox.

In 1969, four scientists John Clauser, Michael Horne, Abner Shimony, and Richard Holt came up with the most famous Bell inequality known as \textit{CHSH inequality}~\cite{CHSH}. This experiment consists of two people, Alice and Bob who have shared a set of pair of particles where each pair has some specific correlation. Now Alice and Bob performed some measurements in their pairs and note down the results. Now after a series of calculations, they check whether their result is greater than some specific value or not? if so then it violated the CHSH inequality else not. This experiment has been performed a lot of time since its mathematical discovery and it supports that quantum correlation indeed violated the CHSH inequality.

Rather than using the statistical inequalities, the contradiction between quantum theory and any Local hidden variable theory (LHV) theory can also be demonstrated by a simple and elegant way, which is known as all-versus-nothing (AVN) proof of Bell's nonlocality~\cite{Mermin_poly90}. In this method, a logical paradox is formed in such a way that while doing experiments, in principle, only a single event can be used to reveal the non-locality. Among several AVN proofs, in~\cite{GHZ}, the authors have demonstrated non-locality without using inequalities which is known as the Greenberger-Horne-Zeilinger (GHZ) paradox. It has been verified experimentally in~\cite{Pan2000}. However, this paradox applies to three~\cite{GHZ} or more qubits~\cite{Brunner14}, but not for two qubits. In 1992, through a thought experiment, Hardy constructed the test of local realism without using inequalities for two qubits, which is called Hardy's test~\cite{hardy92,hardy93}. It is known as the ``Best version of Bell's theorem" as indicated by Mermin~\cite{mermin}. This test provides a direct contradiction between the predictions of quantum theory and any LHV theory for two qubits~\cite{hardy92,hardy93} and also for multi-qubits~\cite{Cereceda04}. The applications of Hardy's paradox includes device-independent randomness~\cite{ramij_random}, device-independent quantum key distribution~\cite{DIQKD}, quantum Byzantine agreement (QBA)~\cite{QBA}, etc. 

\subsection{Quantum entanglement} \label{Quant_Ent}
One of the most interesting features of quantum physics is quantum entanglement~\cite{HHHH08}. This counterintuitive property suggest the presence of global states of a composite system which cannot be represented as a product of the states of separate subsystems. Although quantum entanglement itself does not carry useful information, but using it some of the classically impossible tasks can be performed like dense coding~\cite{DC92}, teleportation~\cite{QT93}, Entanglement swapping~\cite{ES93}, quantum key distribution~\cite{BB84},  etc. 

One of the non-intuitive features of quantum physics is the presence of certain types of correlation in the composite systems which are not possible using classical particles. The strongest of quantum correlation is Bell non-locality~\cite{Brunner14}. One of the important applications of it in device-independent quantum cryptography~\cite{mayers1998quantum}. The next one is quantum steering~\cite{Uola20} which is mainly used for subchannel discrimination~\cite{Piani15}. The most popular quantum correlation is quantum entanglement~\cite{HHHH08} which is used for quantum teleportation~\cite{QT93}, entanglement swapping~\cite{ES93}, quantum key distribution etc~\cite{BB84}.
\begin{figure}[h!]
\centering
\includegraphics[width=\columnwidth]{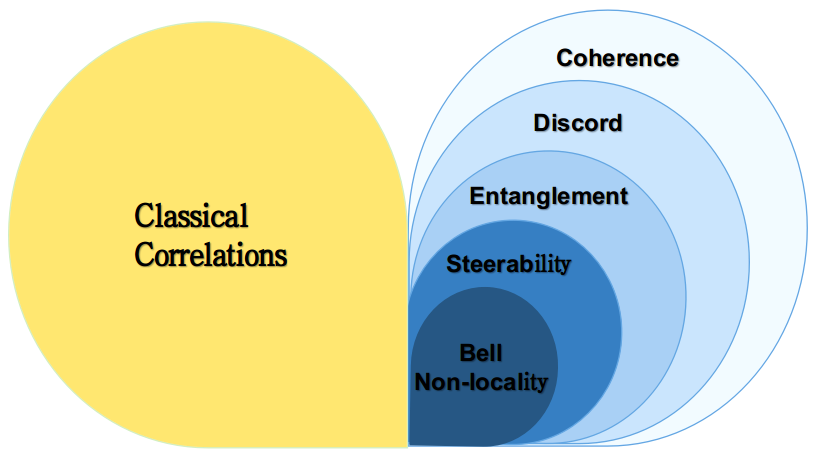} 
\caption{Schematic description of classical and all the quantum correlations. Here coherence $\supseteq$ Discord $\supseteq$ Entanglement $\supseteq$ Steerable $\supseteq$ Bell non-local where $A\supseteq B$ represents $A$ is the superset of $B$.}
\label{Quantum_correl}
\end{figure}
The properties of mixed states can be better explained by quantum discord~\cite{PhysRevLett.88.017901} when entanglement fails to capture the necessary information. Although quantum coherence~\cite{Streltsov17} is the weakest of all quantum correlations, still it is useful for quantum thermodynamics, quantum metrology, quantum phase transitions, etc. All these correlations are shown by set-theoretic notions in Fig.~\ref{Quantum_correl}.

There are some unique properties and applications of quantum entanglement which are absent in other correlations. These applications are the main reason why modern industries are also exploring this unique feature.

\subsubsection{Quantum teleportation} 
Quantum teleportation is a process to transfer quantum information from one location to another distant location. In this method, the sender may not know the location of the receiver and does know about the information being transferred. This process was first proposed in the seminal paper by Bennett \textit{et al.} in 1993~\cite{QT93} and first experimentally realized in~\cite{Bouwmeester97} in 1997. The quality of teleportation is measured by teleportation fidelity which measures the overlap between the initial quantum state before teleporation at the sender and the final teleported state at the receiver. Quantum teleportation has applications in entanglement swapping, quantum communication network, etc. 

\subsubsection{Singlet fraction}
Singlet fraction~\cite{Horodecki99} of any quantum state signifies the amount of overlap between a maximally entangled state with it. Singlet fraction is very useful to measure the quantum teleportation fidelity.

\subsubsection{Entanglement swapping}
Entanglement swapping~\cite{ES93} is the process of transferring quantum entanglement from one place to another distant place. In this method, four particles are required as resource and Bell state measurements (BSM) and local operations and classical communications (LOCC) are required as tools.  Better versions with only three distinguishable particles were proposed in two subsequent works, one~\cite{Pan10} with BSM and another~\cite{Pan19} without BSM. Entanglement swapping is widely applied in quantum repeaters.

\subsubsection{Trace-out operation}
When two particles are entangled then it is not possible to know the state of the individual particles, rather the information about the global state of the two particles is available. Now the mathematical representation of the partial state of the individual particles can be known by the trace-out operation~\cite{Nielsenbook}. If suppose two particles $A$ and $B$ are entangled, then if trace-out operation is performed over $A$, then the partial information about the state of the particle $B$ is known. Trace-out operation is very useful to detect, measure and quantify entanglement.

 \subsubsection{Entanglement measures}
For various quantum information processing protocols, it is required to know how much  entanglement is contained in a quantum state. The basic properties of an entanglement measure~\cite{Plenio07,Guine09} that it should vanish for non-entangled state, should give a positive value for entangled state and should give maximum value for maximally entangled states. There are other properties of entanglement measures like it should be invariant under local unitary transformations, it should not increase under LOCC, etc. Some commonly used entanglement measures are the entanglement of formation~\cite{Bennett96}, log-negativity~\cite{Zyczkowski98,Vidal02}, Tsallis-q entropy~\cite{Kim10,Luo16}, R\'{e}nyi-$\alpha$ entanglement~\cite{Kim_Sanders10,Song16}, Unified-(q, s) entropy~\cite{Kim11,Khan19}, etc.
 
\subsubsection{Monogamy of entanglement}
The most interesting feature of quantum entanglement is it's restriction upon the shareability among composite systems which is known as \emph{monogamy}~\cite{CKW00}. 
Qualitatively, it states that if two particles share a maximally entangled state, then they cannot share entanglement or even classical correlations with any other particles. Intuitively,  it may seem that monogamy feature reduces the usefulness and the possible applications of quantum entanglement in quantum information processing tasks, but surprisingly it has applications on the security of quantum key distribution~\cite{Pawlowski10}, quantum games~\cite{Tomamichel13,Johnston16}, quantum state classification~\cite{Dur00}, interconvertibility between asymptotic quantum cloning and state estimation~\cite{Bae06,Chiribella06}, condensed-matter physics~\cite{Ma11,Garcia13}, quantum-to-classical transition~\cite{Brandao15}, frustrated spin systems~\cite{Rao13}, black-hole physics~\cite{Lloyd14}, etc.

\subsubsection{No Cloning theorem}
One of most counter-intuitive feature of quantum physics that it forbids the creation of independent and identical copy of any arbitrary unknown quantum state known as no-cloning theorem~\cite{Wootters82,QC05,QC14}. It is derived from the linearity of quantum mechanics. Fundamentally, this theorem preserves the Heisenberg's uncertainty principle (HUP)~\cite{Heisenberg27} in quantum physics. If one could make copies of any unknown quantum states, then using those many copies, one can measure position and momentum precisely. That will violate the Heisenberg's uncertainty principle . No cloning theorem has applications in the security of quantum cryptographic protocols~\cite{Pawlowski10},  quantum error correcting codes~\cite{PhysRevA.55.900}, etc.

\subsection{Quantum indistinguishability}
\begin{figure}[h!]
\centering
\includegraphics[width=8.6cm]{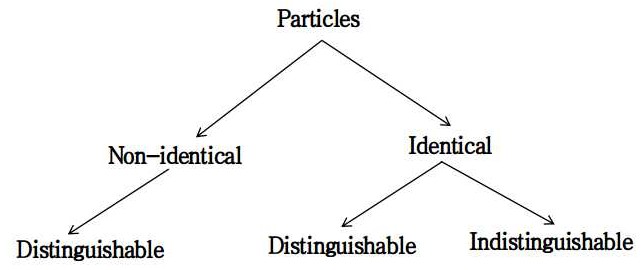} 
\caption{Schematic description of all the particles based on their identity and distinguishability.}
\label{fig:Particles}
\end{figure}

All particles in the world can be divided into two categories based on their identity: (a) identical particles and (b) non-identical particles. If two particles cannot be distinguished by all of their intrinsic properties like mass, color, shape, charge etc. is known as non-identical particles. All the electrons, positrons, photons, protons, neutrons, up quarks, neutrino, hydrogen atom etc have the same intrinsic properties and they behave the same ways. Classical particles can also be identified if two classical particles have same properties like same shape, color, weight, smell etc. Non-identical particles are always distinguishable in nature but identical particles may become distinguishable by some specific method~\cite{Feynman94,Sakurai94}. This division is shown schematically in Fig.~\ref{fig:Particles}.

Two identical particles can be made indistinguishable using spatial measurements. When two identical particles are spatially separated and measured by their individual degrees of freedom, then they are distinguishable. However, when two identical particles have all degrees of freedom equal and their wave functions are spatially overlapped (partial or full), then they are indistinguishable. The concept of indistinguishability is related to the measurement process and to the eye of a detector~\cite{Compagno2018,Nosrati2020}. 
Some of these aspects are better described in Sec.~\ref{IDP_dis_back}. 

\section{Quantum computation}
Classical computers perform computation using the laws of classical mechanics. However, according to Moore's law~\cite{Moore65}, the device fabrication technology will saturate to reduce its size in the first two decades of the twenty-first century. Thus as an alternate solution, quantum computation comes into the picture. Quantum computation is the method to perform computation using quantum mechanical phenomenons like superpositions, entanglement, etc. The theory of quantum computation suggests that there is a huge advantage of performing a task in quantum computers rather than currently available best classical computers in terms of computing time. It can be shown theoretically that any task that the classical computers take years to complete, quantum computers can do it within a few days.

The abstract notion of computation was developed by the famous mathematician Alan Turing~\cite{Turing36}. He proposed the first prototype of modern day computer known as Turing machine. He showed that any algorithmic task cam be emulated by the Turing machine
The equivalence between physical concept of computation and the practical implementation in to some physical device was asserted in the famous \textit{Church-Turing thesis}.  It states that If any task can be performed in any physical device, then there is an equivalent algorithm that can be performed in a Turing machine.
In 1980, famous physicist Richard. P. Feynman proposed that quantum systems can be effectively simulated using a quantum computer, rather than a classical one~\cite{Feynman18}.

\subsection{Quantum computing technologies}
The most important question to make a physical quantum computer is how to make the hardware. There are different kind of technologies are available such as photons, electrons, trapped ions, semiconductors, etc. But there are no consensus among scientists that which technology will be most useful for future quantum computers.
However, there are some properties that must be present in any quantum computer.

(i) \textit{Isolated system}: The quantum computers must be isolated from the rest of the universe, otherwise a small amount of noise will interfere the internal operations and may alter the output.

(ii) \textit{Small decoherence time}: An ideal noise-free quantum computer is not practically possible. A negligible amount of noise will destroy the quantum mechanical arrangements over time. The time taken this phenomenon is known as decoherence time. An effective quantum computer should have large decoherence time.

(iii) \textit{Fault-tolerence}: How much error in the quantum computer can be corrected is known as its fault-tolerance capability. The requirement for a fault-tolerant quantum computer was proposed by DiVincenzo, known as DiVincenzo's criteria~\cite{divincenzo2000physical}. All the quantum computers much obey this criteria.

(iv) \textit{Scalability}: The quantum computer much be scalable enough so that when the dimension of the Hilbert space grows, the cost of operation much not be exponentially increasing. 

(v) \textit{Universal set of logic}: The quantum computer much operate by a finite set of logical control operations. 

\begin{figure}[h!]
\centering
\includegraphics[width=\columnwidth]{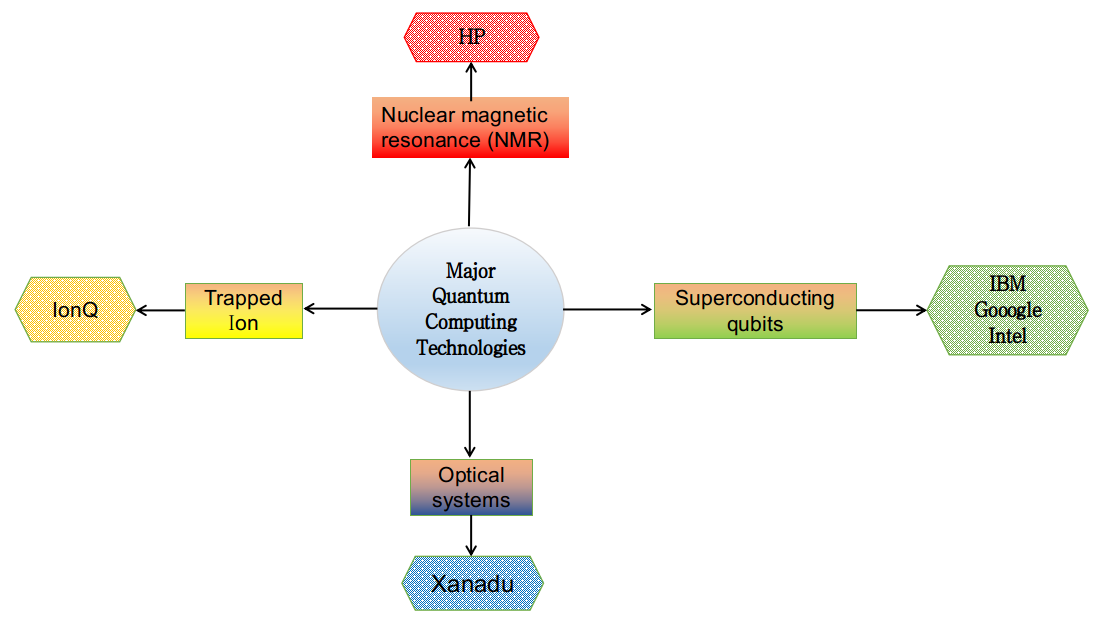} 
\caption{A schematic of major quantum computing technologies with corresponding quantum processors available.}
\label{Quantum_tech}
\end{figure}

 Here, we briefly review some of the major quantum computing platforms and corresponding quantum processors based on the above mentioned properties as shown in Fig.~\ref{Quantum_tech}.
 
\subsubsection{Nuclear magnetic resonance (NMR)}
The spin of the nuclei of the molecules can be used as a qubit for quantum computing. The spin states can be directly controlled using the radio frequency known as nuclear magnetic resonance~\cite{cory1997ensemble,oliveira2011nmr,lu2016nmr} and acts as a single qubit. Two-qubit interactions can be performed by the indirect coupling of the molecular electrons. The measurement of a qubit can be done by observing the current induced in a coil. The major drawback of the NMR technology is the preparation of the pure states. As large number of molecules should be gathers to output a reasonable amount of signal, thus a single qubit is represented by an ensemble of the molecules. 

\subsubsection{Trapped ion}
The spin states of atoms and nuclei can be use as a qubit by trapping them in an electromagnetic cavity known as Ion trapped quantum computers~\cite{Haffner08,Benhelm08}. Once these atoms are trapped, then they are cooled until their spin energy is more than their kinetic energy. Then these atoms are controlled effectively using monochromatic light source. Ion trapped systems have very large decoherence time than other technologies. However controlling large number of atoms and performing joint state operations such as CNOT gates are difficult using ion trap quantum computer.

\subsubsection{Superconducting qubits}
The benefit of  using a superconducting circuit for quantum computing is to control a large number of qubits~\cite{krantz19,kjaergaard20}. Josephson junctions are normally used as a qubit because of its longer coherence time. In a superconductor, a single super-fluid is formed by electron condensate in the cooper pairs that moves without any resistance. There are two types of superconducting qubits, the charge qubit and the flux qubit that are related to the amplitude and the phase of the circuit respectively. However, superconducting qubits have relatively shorter decoherence time than the other quantum technologies. Also, very low temperature  have to be created to use it effectively.  

\subsubsection{Optical systems}
For optical systems~\cite{o07,kok07}, photons can be used as a qubit due to negligible amount of decoherence time which is a major advantage compare to other technologies. Different degrees of freedom of photons can be used for encoding the quantum information, such as polarization, orbital angular momentum, path, etc. However, the major drawback for optical quantum computer is efficient control of multi-qubit systems due to non-availability of the required optical non-linearities. Although, with the invention of Knill-Laflamme-Milburn scheme~\cite{knill01}, optical quantum computing becomes a possibility with required scalability. 

\subsubsection{Other technologies}
Many other quantum computing technologies have been proposed based on the different technologies. Semiconductor based quantum computer known as quantum dots~\cite{loss98,kloeffel13,lent97} are one of them where single atom in vacuum is used a qubit after cooling and trapping them. Also, single ballistic electrons~\cite{ionicioiu01} in very low temperature can be used as qubit instead of photons. Another example is of using rare-earth ions in crystalline hosts~\cite{ohlsson02} which has long coherence time.

\subsection{Quantum processors and simulators} 
Several quantum computing platforms with different archetypes as discussed above are available now, i.e., the way qubits are represented and manipulated.  Here, we briefly review some of the major quantum processors available today.

\subsubsection{IBM}
IBM has given access to its quantum computer that uses superconducting qubits in the cloud and this opens a new door for testing quantum phenomena to the researchers~\cite{IBM}. Till now, there are over 20 devices are made my IBM, from five qubit to 53 qubits, six of which are available online for free for the students and the researchers. Various algorithms and experiments can be performed via cloud access using these simulators.

As there are researchers accessing these real processors, then there can be a long queue to perform experiments. So, IBM came up with quantum simulators that can provide an output like a real quantum computer with used defined noise models for the quantum gates.

The CHSH inequality and the GHZ paradox are already performed in the IBM quantum computer~\cite{IBM}.
In~\cite{devitt16}, the author has implemented some protocols in quantum error correction, quantum arithmetic, quantum graph theory, and fault-tolerant quantum computation  in the IBM 
 quantum computer. In~\cite{berta2016}, the authors have tested the theoretical predictions of entropic uncertainty relation with quantum side information (EUR-QSI) in the IBM quantum computer. Leggett-Garg test~\cite{huffman2017}, compressed quantum computation~\cite{alsina2017}, 
fault-tolerant state preparation~\cite{Takita17}, fault-tolerant logical gates~\cite{Harper19}, quantum cheque~\cite{Qcheque}, quantum permutation algorithm~\cite{Yalçınkaya17}, Deutsch-Jozsa-like algorithm~\cite{DJ_ALGO_Pani_18}, Shor's factoring algorithm~\cite{Amico19}, hybrid quantum linear equation algorithm~\cite{Lee19} are also recently performed in the IBM quantum computer.

\subsubsection{Gooogle}
Google has build a 54 transmon qbit quantum computer named Sycamore based on nonlinear superconducting resonators which was released in 2019~\cite{Sycamore}. It is programmable quantum computer capable of performing various quantum algorithms. This is not available for public. Only Google has access for this quantum processor  and the are currently research topics are near-term applications such as quantum physics simulation, quantum chemistry and generative machine learning.

\subsubsection{Intel}
Intel has manufactured a 17 qubits superconducting quantum processor, named Tangle Lake which can incorporate upto 49 qubits~\cite{IntelQ}. Their research interests are to simulate and analyze natural phenomena, quick answers to phenomenological questions that would take excessive amounts of time on available computers, etc, They are also interests in medicine, astrophysics and weather predictions.

\subsubsection{Xanadu}
Xanadu is a quantum computer company which provides hardware and softwares for various quantum experiments based on photonic quantum computers~\cite{Xanadu} which can be accessed via cloud. Their hardware technology uses programmable Gaussian Boson Sampling (GBS) devices. They also provide some open-source quantum software for simulation of quantum algorithms. 

\subsubsection{Rigetti}
Rigetti builds multi-chip quantum processors using superconducting systems which can be accessible via cloud~\cite{Rigetti}. Their processors incorporates with existing computing infrastructure which aim to solve scaling challenges of fault-tolerant quantum computers. Their mission is to build a 80-qubit quantum computing chip which can be accessible via cloud service. 

\subsubsection{D-Wave}
D-Wave has build a 2000 qubit quantum computer based on quantum annealing technology~\cite{DWave}. It may be noted that although the number of qubits in D-Wave is more than the above quantum computer, that does not mean it is the best among them. D-Wave search solutions to a problem using large number of quits. Thus D-Wave's qubits are can be affected by noise and their  quantum states are also more fragile, and their manipulation  are less precise. 

\subsubsection{Other processors and simulators}
Other than the above mentioned quantum processors, there are also other quantum computers such as IonQ~\cite{IonQ}, Microsoft~\cite{Microsoft}, Toshiba~\cite{Tosiba}, Alibaba~\cite{Alibaba}, Amazon web services~\cite{Amazon}, 1QBit~\cite{1Qbit}, etc.

\section{Motivation of current work} 
Now we will briefly describe the open questions associated with the above topics and the motivation of the current works.

There are several tests of non-locality which has been performed in various quantum technologies. For example, Mermin inequalities~\cite{Mermin_poly90} have been tested experimentally using photons and ion traps~\cite{Zhao03,Lanyon14}, subsequently the authors of~\cite{mermin2016} have tested Mermin polynomial of three, four and five qubits in the IBM quantum computer based on superconducting qubits. Hardy's experiment is the only method to test non-locality for two qubits without using inequalities. Several experiments have been performed to demonstrate Hardy's paradox using polarization, energy-time and orbital angular momentum of photons, entangled qubits, classical light, and two-level quantum states~\cite{Giuseppe1997,irvine2005,Lundeen2009,Yokota2009,vallone2011,chen2012,karimi2014,zhang2016,fan2017,ladder2017}, but none in superconducting qubits. However, none of the experimental verifications of Hardy's non-locality have used superconducting qubits. \textit{This motivates us to test Hardy's paradox for two qubits in a quantum computer using superconducting qubits.}  This is the motivation behind Chapter~\ref{Chap3}.

In the last century, physicists were puzzled about whether ``the characteristic trait of Quantum Mechanics"~\cite{Schrodinger}, i.e., entanglement~\cite{epr}, is real and, if so, whether it can show some nontrivial advantages over classical information processing tasks. The answers to both are positive. In the current century, entanglement of indistinguishable particles and its similarity with as well as difference from that of distinguishable ones have been extensively studied~\cite{Li01,You01,John01,Zanardi02,Ghirardhi02,Wiseman03,Ghirardi04,Vedral03,Barnum04,Barnum05,Zanardi04,Omar05,Eckert02,Grabowski11,Sasaki11,Tichy13,Kiloran14,Benatti17,LFC16,Braun18,LFC18}. 
However, the entanglement between distinguishable and indistinguishable particles raises several non-trivial open questions: \\
  {\em (i) Hyper-hybrid entangled state (HHES) was created by two indistinguishable bosons~\cite{HHNL}. Is it also  possible for two indistinguishable fermions}? \\
{\em (ii) Is the scheme for HHES, as proposed by  Li \textit{et al.}~\cite{HHNL}, applicable for two distinguishable particles}? \\
 {\em (iii) If the scheme for HHES, as proposed by  Li \textit{et al.}~\cite{HHNL}, is not possible applicable for two distinguishable particles, the can distinguishable particles exhibit HHES through some other scheme}? \\
 {\em (iv) Is there exist some quantum correlations and applications which unique in the case of distinguishable particles and indistinguishable particles}? \\
  These open problems are the motivations behind Chapter~\ref{Chap4}.
  
  \textit{Partial trace-out operation}~\cite{HHHH08,Nielsenbook} is a typical method of finding the reduced density matrix of a subsystem which can be either one whole particle or a single DoF for distinguishable systems. However, for indistinguishable systems, applying the above method results in a contradiction in identifying entanglement~\cite{Ghirardi04,Tichy11}. Experimental works on such systems~\cite{Bloch08,Hayes07,Anderlini07,Petta05,Lundskog14,Tan15,Veldhorst15} existed earlier, but a common mathematical framework for a consistent theoretical interpretation was first attempted in~\cite{LFC16,LFC18}, by providing a method of partial trace-out for a whole indistinguishable particle. One may be tempted to think that the same rule can trace out a single DoF also. However, this is not so straightforward. When particles become indistinguishable, performing the partial trace-out of a particular DoF is challenging, because a DoF cannot be associated with a specific particle. This motivates to search for a DoF trace-out rule which can be applicable to both distinguishable and indistinguishable particles when each particle have multiple DoFs. This is the basic motivation for Chapter~\ref{Chap5}.
  
  Thus quantum teleportation fidelity~\cite{Jozsa94} plays an important role in quantum information both for distinguishable and indistinguishable particles. One way to measure the teleportation fidelity is by using the singlet fraction~\cite{Horodecki99} of the quantum channel used for teleportation. Singlet fraction for any state is defined as the maximum overlap of that state with the maximally entangled state. The relation between teleportation fidelity and singlet fraction was proposed by Horodecki \textit{et al.}~\cite{Horodecki99}. This relation is only applicable for distinguishable particle with a single degrees of freedom (DoF). However, for indistinguishable particles with multiple DoF, this relation no longer holds. This motivated us to find a new relation between teleportation fidelity and singlet fraction applicable both for distinguishable and indistinguishable particles where each particle have multiple DoFs. This is the basic motivation behind Chapter~\ref{Chap6}
 
One important feature of quantum entanglement of distinguishable particles~\cite{HHHH08} is its restriction upon the shareability among composite systems (consisting of particles or degrees of freedom (DoFs)), known as~\emph{monogamy of entanglement} (MoE). Monogamy of entanglement is widely regarded as one of the basic principles of quantum physics~\cite{Terhal04}. Qualitatively, it is always expected to hold, as a maximal violation will have consequences for the no-cloning theorem. However, it was an open problem that monogamy of entanglement can be violated maximally using indistinguishable particles and using the DoFs of the indistinguishable particles? The problem of monogamy of entanglement using DoFs of indistinguishable particles for two-qubits and more that two qubits are discussed in Chapter~\ref{Chap7} and Chapter~\ref{Chap8} respectively. 

The main motivation of Chapter~\ref{Chap9} is to find some new applications of indistinguishable particles. Security and efficiency are two major criteria of a cryptographic protocols. If two cryptographic systems with different resource requirements provide the same level of security, the one with less resources becomes the natural choice. The motivation behind the first application is to find how to reduce resources in the cryptographic protocols without compromising the security. There are several types of attacks such as intercept and resend attack, man-in-the-middle attack, Trojan-horse attack, etc. The motivation of our second application is to find a new attack for quantum cryptographic protocols.
 The standard Entanglement swapping (ES)~\cite{ES93} required four distinguishable particles as a resource along with Bell state measurement (BSM)~\cite{BSM99} and local operations and classical communications (LOCC)~\cite{LOCC} as tools. Better versions with only three distinguishable particles were proposed in two subsequent works, one~\cite{Pan10} with BSM and another~\cite{Pan19} without BSM. Recently, Castellini \textit{et al.}~\cite{LFCES19} have shown that ES for the indistinguishable case is also possible with four particles (with BSM for bosons and without BSM for fermions). The motivation of our final application is reduce the number of particles further in entanglement swapping protocol using distinguishable or indistinguishable particles? 

\section{Thesis organization}
This thesis is organized as follow:

\begin{itemize}
\item In Chapter~\ref{Chap2}, we present the necessary background studies, mathematical formulation, and applications needed to understand our proposed works in rest of the thesis.  

\item In Chapter~\ref{Chap3}, We experimentally verify Hardy's paradox for two qubits on a quantum computer based on superconducting circuits. We argue that for practical verification of Hardy's test, the error-modeling used for optical circuits cannot be used for superconducting qubits in~\ref{Pro_error}. We propose a new error-modeling and a new method to estimate the lower bound on Hardy's probability for superconducting qubits. We also point out that the earlier tests performed in optical circuits and in the IBM quantum computer have not analyzed the test results in a statistically correct and coherent way in~\ref{Experimental Results and Discussion}. We analyze our data using Student's t-distribution~\cite{feller} which is the statistically correct way to represent the test results. Our statistical analysis leads to the conclusion that any two-qubit non-maximally entangled state (NMES) gives a nonzero value of Hardy's probability, whereas any two-qubit maximally entangled state (MES) as well as any product state (PS) yields a zero value of Hardy's probability.  We identify the difficulties associated with the practical implementation  of quantum protocols based on Hardy's paradox  and discuss how to overcome them in~\ref{App_shift}. We propose two performance measures for any two qubits of any quantum computer based on superconducting qubits. Finally, we discuss benchmarking of superconducting quantum devices using Hardy's paradox in~\ref{Ben_hardy}.

\item  In Chapter~\ref{Chap4}, some of the open questions related to distinguishable and indistinguishable particles and their properties and applications. Using systematic calculations we establish that hyper-hybrid entangled state is possible using two indistinguishable fermions in Chapter~\ref{HHESfermions} and also presented a generalized version for hyper-hybrid entangled state applicable for bosons and fermions in Chapter~\ref{Gen_boson_fermion}. Next, we show that the scheme of Li \textit{et al.}~\cite{HHNL} for producing hyper-hybrid entanglement using two indistinguishable particles does not work for distinguishable particles in Chapter~\ref{DisHHES}. After that, in Chapter~\ref{Sig_UFQT_HHES}, we prove that \textit{If quantum particles (either distinguishable or indistinguishable) can simultaneously produce and perform hyper hybrid entangled state and unit fidelity quantum teleportation respectively then using that cloning of any arbitrary quantum state is possible.}. Next, 
in Chapter~\ref{NoHHES}, we establish the first \textit{no-go} result: HHES is not possible for distinguishable particles;	 otherwise, exploiting it, signaling can be achieved. In Chapter~\ref{NoUFQT}, we prove our second \textit{no-go} result that  unit fidelity quantum teleportation is not possible for indistinguishable particles. Using the above two no-go theorem, we establish a separation result using quantum properties and applications between distinguishable and indistinguishable particles in Chapter~\ref{Sepresult}. 

\item In Chapter~\ref{Chap5}, we have proposed a new degrees of freedom trace-out rule applicable for both distinguishable and indistinguishable particles, each having multiple degrees of freedom. First we represent two indistinguishable particles each having two DoFs in~\ref{2P2DoF} and $n$ DoFs in~\ref{2PnDoF}. Then we generalize it for $p$ indistinguishable particles, each having $n$ DoFs in~\ref{pPnDoF}. Next, We establish a generalized DoF trace-out rule for two indistinguishable particles each having two DoFs in~\ref{Tr2p2DoF} and $n$ DoFs in~\ref{Tr2PnDoF}. This rule is generalized for $p$ indistinguishable particles, each having $n$ DoFs in~\ref{TrpPnDoF}. Finally we present the physical significance of our DoF trace-out rule in~\ref{DoF_Tr_Phy_Sig}.

\item In Chapter~\ref{Chap6}, we establish a generalized relation between teleportation fidelity~\cite{Jozsa94} and singlet fraction~\cite{Horodecki99} for both distinguishable and indistinguishable particles with where each particle has multiple DoFs. First we have defined the generalized teleportation fidelity in~\ref{Gen_TF} and generalized singlet fraction in~\ref{Gen_SF}. After generalizing the relation between above two in~\ref{TF_ST_Rel}, we prove an upper bound for generalized singlet fraction for distinguishable and indistinguishable particles in~\ref{up_bound_Fn}. Finally, we state the physical significance of our new generalized relation in~\ref{Phy_Sig_TF_SF}.

\item In Chapter~\ref{Chap7}, using this generalized DoF trace-out, we show that monogamy of entanglement can be violated maximally by indistinguishable particles in qubit systems for measures (such as squared concurrence, log-negativity, etc.) that are monogamous for distinguishable particles. First, we describe the condition for violation of no-cloning theorem using the maximum violation of monogamy of entanglement in~\ref{Con_noclong}. Then we give an example of an apparent violation of particle-based monogamy
of entanglement in~\ref{App_vio}. To remove this ambiguity, we generalized 
the monogamy relation from particle view to degree of freedom view in~\ref{Inter_DoF_MoE}. 
Then we prove the following theorem that 
\textit{In qubit systems, indistinguishability is a necessary criterion for maximum violation of monogamy of  entanglement by the same measures that are monogamous for distinguishable particles} in~\ref{MoE_Vio}. Finally, we discuss the physical significance of the maximum violation
of monogamy in~\ref{Phy_sig_moe_vio}.

\item In Chapter~\ref{Chap8}, we have shown that monogamy of entanglement using three indistinguishable particles is obeyed using squared concurrence as an entanglement measure. First, we show how to calculate the concurrence between two spatial regions between any DoFs in~\ref{Cal_con}. Then we present the proof of monogamy inequality for three or more indistinguishable particles becomes equality for pure states in~\ref{Pure_MoE} and for mixed states it remains an inequality in~\ref{MOE_mixed}. At last, we discuss the physical significance of our results in~\ref{MoE_phy_sig}.

\item In Chapter~\ref{Chap9}, we present some applications of indistinguishable particles. First, we show We show  that for device independent test such as quantum pseudo-telepathy game~\cite{Brassard05,Jyoti18}, the cost of adding an ancilla particle can be bypassed by using an additional degrees of freedom and creating multi-DoF entanglement in quantum private query protocol where the success probability remains the same but the generalized singlet
fraction changes in~\ref{QPQ_app}. Next we show how indistinguishability can change Hardy's probability which can be used as an attack in quantum cryptographic protocols in~\ref{Hardy_appli}. Finally, in Section~\ref{2PES}, we have proposed an Entanglement Swapping protocol using only two indistinguishable particles without using Bell state measurement. 

\item In Chapter~\ref{conclusion}, we summarize each of the contributing chapters, discuss the open problems related to those chapter, and our future plans. 
\end{itemize}

\chapter{Background} \label{Chap2}
In this section, all the technical backgrounds to understand the rest of the thesis  is presented.

\section{Statistical interpretation of experimental results} \label{Stat}
Here, we discuss how any experimental analysis should be performed using some basics of statistics.

\subsection{Standard normal distribution vs. Students t-distribution} \label{How_stat}
It is known from the central limit theorem~\cite{feller}  that if $\lbrace X_{1},X_{2},\ldots, X_{n}\rbrace $ are independent and identically distributed random samples drawn from any population with mean $\mu$ and variance $\sigma^{2}$ and if $n$ is large, then the sample mean 
\begin{equation}
\bar{X}=\dfrac{1}{n} \sum^{n}_{i} X_{i}
\end{equation}
 follows a normal distribution with mean $\mu$ and variance $\sigma^{2}/n$, i.e., $ \bar{X} \backsim N (\mu, \sigma^{2}/n) $. It follows that the variable
\begin{equation*}
Z=\dfrac{\bar{X}-\mu}{\sqrt{\sigma^{2}/n}}, 
\end{equation*}
follows the standard normal distribution, i.e., $Z \backsim N(0,1)$.
If the population variance $\sigma^{2}$ is unknown, it is replaced by its closest approximation, i.e., the sample variance $S^{2}$. Then the quantity follows a Students t-distribution~\cite{feller} described as
\begin{equation*}
T=\dfrac{\bar{X}-\mu}{\sqrt{S^{2}/n}}, \hspace{0.3cm} \text{where} \hspace{0.3cm} S^2=\dfrac{1}{n-1} \sum_{i=1}^{n}(X_{i}-\bar{X})^{2}.
\end{equation*} 
Moreover, while normal approximation works only for very large $n$ (ideally infinite), the Students t-distribution holds for small $n$ as well. This distribution is a function of the degrees of freedom, which is one less than the number of times the experiment is repeated. As the number of degrees of freedom tends to infinite, the Student's t-distribution converges to the normal distribution.

\subsection{Hypothesis testing and confidence interval} \label{Hypo_CI}
We know that it is impossible to conduct any experiment without any error. So, when the conclusion is drawn from an experimental result, it is not appropriate to claim that the results are  $100\%$ correct. Statistically speaking, we can only test a hypothesis and conclude the test based on our experimental results with some percentage of confidence in our conclusion.
In hypothesis testing~\cite{feller}, we turn a question of interest into a hypothesis about the value of a parameter or a set of parameters. In our case, suppose we want to test whether a given state is NMES or not. Then, according to the notion of hypothesis testing, we can have the following formulation from Equation~(\ref{q_bound}).

Test the null hypothesis 
$$\mathcal{H}_{0}: q = 0  \hspace{0.5cm} \text{(The unknown state is not an NMES)}$$
against the alternative hypothesis
$$\mathcal{H}_{1}: q > 0  \hspace{0.5cm} \text{(The unknown state is an NMES)}.$$

So it is evident that our test will be a one-sided test\footnote[3]{Two-sided test typically applies to $\mathcal{H}_{0}: X =p$ vs $\mathcal{H}_{1}: X \neq p$ and one-sided test $\mathcal{H}_{0}: X = p$ vs $\mathcal{H}_{1}: X > p$ or $\mathcal{H}_{1}: X < p$ where $X$ is the chosen parameter under test and $p$ is a specific value.} as $q$ is a non-negative quantity. The significance level $\alpha$ is equal to the false-positive error or Type I error which is defined by the probability $\Pr$(reject $\mathcal{H}_{0} \mid \mathcal{H}_{0}$ is correct) where $ 0 \leq \alpha \leq 1 $. The level of confidence is defined by $\left( 1-\alpha\right)$ or $100(1-\alpha)\%$. 
Let us assume that an experiment is repeated $n$ number of times, where $\mu$, $ \bar{X}$, $\sigma$,  and $S$  are the population mean, sample mean, population standard deviation and sample standard deviation respectively. 
Again, let $100(1-\alpha)\%$  confidence interval (CI) of $\bar{X}$  be the interval $\left[  X_{lb} ,  X_{ub} \right] $. 
This means that we have $100(1-\alpha)\%$ confidence that $\mu$ will lie in between $ X_{lb}$ and $ X_{ub}$.

For the standard normal distribution, the expression for $100(1-\alpha)\%$ confidence interval for the mean in the above situation is $\left(  \bar{X} \pm z_{\frac{\alpha}{2}}\frac{\sigma}{\sqrt{n}} \right) $, where $z_{\frac{\alpha}{2}}$ is the value of the standard normal variable $Z$ such that
\begin{equation*}
\text{Pr}(-z_{\frac{\alpha}{2}} < Z < z_{\frac{\alpha}{2}})=(1-\alpha).
\end{equation*}
For this case, $ X_{lb} =\left(  \bar{X} - z_{\frac{\alpha}{2}}\frac{\sigma}{\sqrt{n}} \right)  $ and  $ X_{ub} =\left(  \bar{X} + z_{\frac{\alpha}{2}}\frac{\sigma}{\sqrt{n}} \right) $. Here, $z_{\frac{\alpha}{2}}$ depends only on the value of $\alpha$.
For the Student's t-distribution, the expression for $100(1-\alpha)\%$ 	 confidence interval is given by $\bar{X} \pm t_{\frac{\alpha}{2}}\frac{S}{\sqrt{n}} $ where $t_{\frac{\alpha}{2}}$ is given by the following expression
\begin{equation*}
\text{Pr}(-t_{\frac{\alpha}{2}} < T< t_{\frac{\alpha}{2}})=(1-\alpha).
\end{equation*}
For this case, $ X_{lb} =\left(  \bar{X} - t_{\frac{\alpha}{2}}\frac{S}{\sqrt{n}} \right)  $ and  $ X_{ub} =\left(  \bar{X} + t_{\frac{\alpha}{2}}\frac{S}{\sqrt{n}} \right) $. Here, $t_{\frac{\alpha}{2}}$ depends on the value of $\alpha$ and the degrees of freedom $\nu = \left( n -1\right)$, where $n$ is the number of samples used for the experiment. As $\nu \to\infty $, we have $t_{\frac{\alpha}{2}} \to  z_{\frac{\alpha}{2}}$.

\section{Quantum non-locality}
Quantum non-locality as shown in Fig.~\ref{Quantum_correl} is the strongest form of quantum correlations. The existence of this type of correlations was first proved by the violations of Clauser-Horne-Shimony-Holt (CHSH) inequality. Later, non-locality was proved without using inequality. For, two-qubits, Hardy's test is used to verify non-locality. 
\subsection{Clauser-Horne-Shimony-Holt (CHSH) inequality}
The verifications of Bell's theorem can be experimentally performed using CHSH inequality. Let us assume, two parties, Alice and Bob are sharing a physical system which they can measure and see the outcome. They us denote the the measurements of Alice are $A_1$ and $A_2$. Similarly, the measurements of Bob are denoted by $B_1$ and $B_2$. The measurements outputs are denoted by $+1$ or $-1$. Now let us calculate the quantity
\begin{equation} \label{CHSH_ini}
\text{CHSH}= A_1B_1+A_1B_2+A_2B_1-A_2B_2.
\end{equation}
Clearly, from the basic algebra, it can be shown that the maximum and the minimum value of Eq.~\eqref{CHSH_ini} is $+2$ and $-2$ respectively. Let, $E(\bullet)$ denote the exception value of a quantity then we can write 
\begin{equation} \label{CHSH_f}
E(\text{CHSH}) = E(A_1B_1)+E(A_1B_2)+E(A_2B_1)-E(A_2B_2) \leq 2
\end{equation}
This is known as the CHSH inequality. It can be shown that for some quantum systems this inequality can be violated which supports the existence of non-locality.

\subsection{Hardy's test of non-locality} \label{Hardy_gen}
Hardy's test of non-locality for two qubits involves two non-communicating distant parties, Alice and Bob. A physical system consisting of two subsystems is shared between them. Alice and Bob can freely measure and observe the measurement results of their own subsystems. Alice can perform the measurement on her own subsystem by choosing freely one of the two $\lbrace+1,-1\rbrace$-valued random variables $A_{1}$ and $A_{2}$. Similarly, Bob can also choose freely one of the two $\lbrace+1,-1\rbrace$-valued random variables $B_{1}$ and $B_{2}$ for measuring the subsystem in his possession. 

Hardy's test of non-locality starts with the following set of joint probability equations.
\begin{align}
P(+1,+1|A_{1},B_{1})=&0,\label{eq:1}\\
P(+1,-1|A_{2},B_{1})=&0,\label{eq:2}\\
P(-1,+1|A_{1},B_{2})=&0,\label{eq:3}\\
P(+1,+1|A_{2},B_{2})=&q, \hspace{0.2cm} \text{where} \hspace{0.2cm} 
  \begin{cases}
       q=0 & \text{for LHV theory,} \label{eq:4}\\
       q>0& \text{for non-locality.} 
  \end{cases}
\end{align}

\noindent Here $P(x,y|A_{i},B_{j})$ denotes the joint probability of obtaining outcomes $x , y \in \left\lbrace +1,-1 \right\rbrace $ given that $ A_{i} $ and $ B_{j} $ were the experimental choices made where $i , j \in \left\lbrace 1 , 2 \right\rbrace $. If an experiment is designed in such a way that Equations~(\ref{eq:1}),~(\ref{eq:2}), and~(\ref{eq:3}) are satisfied, then for any LHV theory, the right-hand side of Equation~(\ref{eq:4}) becomes zero. But if this value is found to be greater than zero for some values of $q$, then non-locality is established. The set of Equations~(\ref{eq:1})-(\ref{eq:4}) are called \textit{Hardy's equations} and $q$ is called \textit{Hardy's probability}. 

It can be easily shown that a greater than zero value of Equation~(\ref{eq:4}) implies non-locality under the assumptions of Equations~(\ref{eq:1})-(\ref{eq:3}). Let us assume that, when $A_{2}$ and $B_{2}$ are measured simultaneously, an event with $A_{2}=B_{2}=+1$ is detected. It can be seen from Equation~(\ref{eq:2}) that the measurement of $B_{1}$ must yield the output +1, as $A_{2}=1$ can never occur with $B_{1}=-1$, the only option left is $B_{1}=+1$. Following the same logic, from Equation~(\ref{eq:3}), it can be concluded that $A_{1}=+1$ must occur, as the value of $B_{2}=1$ can never be detected with $A_{1}=-1$. Also, because of the locality assumption, the value of $B_{1}$  must be independent of whether Alice measures $A_{1}$ or $A_{2}$. Similarly, the value of $A_{1}$  must be independent of whether Bob measures $B_{1}$ or $B_{2}$. So, it can be concluded from LHV theory that the values of $A_{1}$ and $B_{1}$ must be $+1$. But this is not possible as shown in Equation~(\ref{eq:1}). So, given Equations~(\ref{eq:1})-(\ref{eq:3}) are satisfied, a single occurrence of the event $A_{2}=B_{2}=+1$ can rule out all possibilities that experiment can be described by an LHV theory.

The maximum value of Hardy's probability $ q $ is found to be $ q_{max}=\frac{5\sqrt{5}-11}{2} \approx 0.09017  $ for two qubits~\cite{hardy93,rabelo12}. For the two-qubit system, no MES as well as no PS obey Hardy's non-locality, but all NMES exhibit Hardy's non-locality~\cite{gold94}. This is the specialty of Hardy's test that only a single event can discard all LHV theories. The motivation of this work is to validate this statement for the quantum computer using superconducting qubits. As every MES of three or higher qubits exhibits Hardy's non-locality, we restrict our discussion for two qubits only. 

\subsection{Equivalence between CHSH inequality and Hardy's equations}
 Using simple set-theoretic arguments, one can show that Hardy's equations are a special case of the famous CHSH inequality~\cite{CHSH69}. The CHSH version of Hardy's Equations~\cite{barun08} is described as
\begin{equation}
 \begin{aligned}
 \label{CHSH}
 P(+1,+1|A_{2},B_{2}) -P(+1,+1|A_{1},B_{1})
- P(+1,-1|A_{2},B_{1})  - P(-1,+1|A_{1},B_{2}) \leq 0.
 \end{aligned}
 \end{equation} 
A violation of Equation~(\ref{CHSH}) means a violation of local realism, which supports non-locality. 
Putting the ideal values of the probabilities from Equations~(\ref{eq:1})-(\ref{eq:4}) into Equation~(\ref{CHSH}),  
we get $q \leq 0$. So, $q=0$ supports LHV theory and $q>0$ supports non-locality.

\subsection{Revisiting known experimental results on Hardy's test and their statistics}
In this section, first we revisit important experimental works on Hardy's test in optical set-up. Next, we look into significant experiments performed in IBM quantum computer. 

\subsubsection{Prior works on Hardy's test in optical set-up}

In this section, we will survey the major experiments performed for Hardy's test using optical circuits so far. Their set-up and the results are summarized in Table~\ref{table:2}.

\begin{table} [h!]
\centering
\begin{tabular}{  p{3.8cm}| p{1cm}| p{2.5cm}| p{2.5cm}| p{1.5cm}| p{1.3cm}}
\hline
 Authors & Year  & Distribution of data  & number of samples mentioned? & Form: $\text{mean} \pm \text{SD}$   & CI Representation \\
\hline
 Irvine \textit{et al.} \cite{irvine2005} & 2005 & Not specified & Not specified & Yes & No\\ 
 \hline
  Lundeen \textit{et al.} \cite{Lundeen2009} & 2009 & Poissonian & Not specified & Yes & No\\ 
 \hline
  Yokota \textit{et al.} \cite{Yokota2009} & 2009 & Poissonian & Not specified & Yes & No\\ 
 \hline
  Fedrizzi \textit{et al.}  \cite{fedrizzi2011} & 2011& Not specified & Not specified &Yes & No\\ 
 \hline
  Vallone \textit{et al.}  \cite{vallone2011} & 2011& Not specified & Not specified & Yes& No\\ 
 \hline
  Chen \textit{et al.} \cite{chen2012} & 2012 & Not specified  & Not specified& Yes&No\\ 
 \hline
 karimi \textit{et al.} \cite{karimi2014} & 2014 & Poissonian &Not specified  & Yes&No\\ 
 \hline
 Zhang \textit{et al.}  \cite{zhang2016} &2016 & Not specified  & Not specified& Yes&No\\ 
 \hline
Fan  \textit{et al.} \cite{fan2017}  & 2017 & Poissonian & Not specified& Yes&No\\ 
 \hline
Chen \textit{et al.} \cite{ladder2017} & 2017 & Not specified  & Not specified& Yes&No\\ 
 \hline
 Luo \textit{et al.} \cite{Luo2018} & 2018 & Poissonian  & Not specified& Yes&No\\ 
 \hline
 Yang \textit{et al.} \cite{Yang19} & 2019 & Poissonian  & Not specified& Yes&No\\ 
 \hline
\end{tabular}
\caption{Summary of prior works on Hardy's test in optical set-up where SD=standard deviation and CI=confidence interval.} \label{table:2}
\end{table}

In \cite{irvine2005,fedrizzi2011,vallone2011,chen2012,zhang2016,ladder2017}, the authors have represented their coincidence count in mean $\pm$ standard deviation form but they did not specify the distribution of the sample data and the number of samples taken. They also did not represent the confidence on their data.

In \cite{Lundeen2009,Yokota2009,karimi2014,fan2017,Luo2018,Yang19}, the authors have mentioned that they have assumed that all the error bars follow the Poissonian statistics for the coincidence count and they have represented their data in mean $\pm$ standard deviation form. But again, they did not mention the distribution of samples and number of samples they have used and confidence on their data.

\subsubsection{Significant experiments performed in superconducting qubits}
\begin{table}[h!]
\centering
\begin{tabular}{p{5cm}| p{0.7cm}| p{0.5cm}|p{1cm}| p{1.8cm}| p{1.8cm}}
\hline
 Authors & Year  & $n$ & $s$  & Form:  $\text{mean} \pm \text{SD}$  & CI Representation \\
\hline 
S. J. Devitt \cite{devitt16} & 2016 & 1 &8192 & No  & No\\ 
 \hline
 Alsina \textit{et al.}  \cite{mermin2016} & 2016  & 1 & 8192 & Yes & No\\ 
 \hline
 Berta \textit{et al.} \cite{berta2016} & 2016  & 1 & 8192 & No & No\\  
 \hline
   Huffman \textit{et al.} \cite{huffman2017} & 2017  &  10& 8192 &Yes &No\\ 
 \hline
  Hebenstreit  \textit{et al.} \cite{alsina2017} & 2017  & 1 & 8192 & No & No\\ 
 \hline
Behara \textit{et al.}  \cite{Qcheque} & 2017  & 1& 1024, 4096, 8192 & No&No\\ 
 \hline
   Yal\ifmmode \mbox{\c{c}}\else \c{c}\fi{}\ifmmode \imath \else \i \fi{}nkaya \textit{et al.} \cite{Yalçınkaya17} & 2017 & 5& 8192& Yes & No\\ 
 \hline
     W. Hu   \cite{hu2018} & 2018 &  1 & 8192 & No & No\\ 
 \hline
  Gangopadhyay \textit{et al.}  \cite{DJ_ALGO_Pani_18} & 2018 & 10 & 8192 & Yes & No\\
  \hline
  Lee \textit{et al.}  \cite{Lee19} & 2019 & 1 & 1024 & No & No\\
  \hline
\end{tabular}
\caption{Significant experiments performed in IBM five-qubit quantum computer and their error statistics where $n$= number of times the experiment has been performed, $s$=number of shots in each time, SD=standard deviation, and CI=confidence interval.} \label{table:3}
\end{table} 
 We present a survey of the other experiments done in the IBM quantum computer and their statistical representation of the results which are summarized in Table~\ref{table:3}.
In \cite{devitt16,berta2016,mermin2016}, the authors have represented the error in their experiments by the standard formula  $\sqrt{p(1-p)/8192}$, where $p$ is the estimate of the probability of a given measurement outcome in a given experiment. However, some authors~\cite{mermin2016,huffman2017,Yalçınkaya17,DJ_ALGO_Pani_18} have represented their data in the form of mean $\pm$ standard deviation but none of them have represented their data in confidence interval form.

\section{Quantum entanglement}
Here, we discuss the definition of entanglement of distinguishable and indistinguishable particles as proposed in~\cite{Benatti10,Benatti12,Benatti14} and followed in the rest of the thesis.
\subsection{Definition of entanglement}
Let us consider a many-body system which is represented by the Hilbert space $\mathcal{H}$. The algebra of all bounded operators, which includes all the observables, is represented by $\mathcal{Z(H)}$. In this algebraic framework, the standard notions of states and the tensor product partitioning of $\mathcal{H}$ are changed into the observables and local structures of $\mathcal{Z(H)}$. Now before defining entanglement, we define \textit{algebraic bipartition} and \textit{local operators}. 
\subsubsection{Algebraic bipartition}
An algebraic bipartition of operator algebra $\mathcal{Z(H)}$ is any pair ($\mathcal{A}$, $\mathcal{B}$) of commuting subalgebras of $\mathcal{Z(H)}$ such that $\mathcal{A}$, $\mathcal{B}\in \mathcal{Z(H)}$. If any element of $\mathcal{A}$ commutes with any element of $\mathcal{B}$, then $\left[  \mathcal{A} ,\mathcal{B} \right]  = 0$. 
\subsubsection{Local operators}
 For any algebraic bipartition $(\mathcal{A}, \mathcal{B})$, an operator is called a local operator if it can be represented as the product $A B$, where $A \in \mathcal{A}$ and $B \in \mathcal{B}$. 
\subsubsection{Entangled states}
For any algebraic bipartition $(\mathcal{A}, \mathcal{B})$, a state $\rho$ on the algebra $\mathcal{Z(H)}$ is called separable if the
expectation of any local operator $A B$ can be decomposed
into a linear convex combination of products of local expectations, as follows
\begin{equation}
\begin{aligned}
 Tr(\rho A B) &= \sum_{k} \lambda_{k}Tr(\rho_{k}^{(1)}A) Tr(\rho_{k}^{(2)} B), \\
  \lambda_{k} & \geq  0,  \hspace{1cm} \sum_{k} \lambda_{k}=1,
\end{aligned}
\end{equation}
where $\rho_{k}^{(1)}$ and $\rho_{k}^{(2)}$ are given states on $\mathcal{Z(H)}$; otherwise the
state $\rho$ is said to be entangled with respect to the algebraic bipartition $(\mathcal{A}, \mathcal{B})$.
Note that, this algebraic bipartition can also be spatial modes like distinct laboratories each controlled by Alice and Bob. If any state cannot be written in the above form, then it would certainly violate the CHSH inequality. Therefore, we use the violation of the CHSH inequality as an indicator of entanglement.

\subsection{Types of entanglement}
Quantum entanglement is encoded in the particle's degrees of freedom (DOFs)  like spin, polarization, path, angular momentum, etc. Based on the arrangement entanglement in the DOFs of particles, we have the following types of entanglement structures
\begin{figure}[h!]
\centering
\includegraphics[width=\columnwidth]{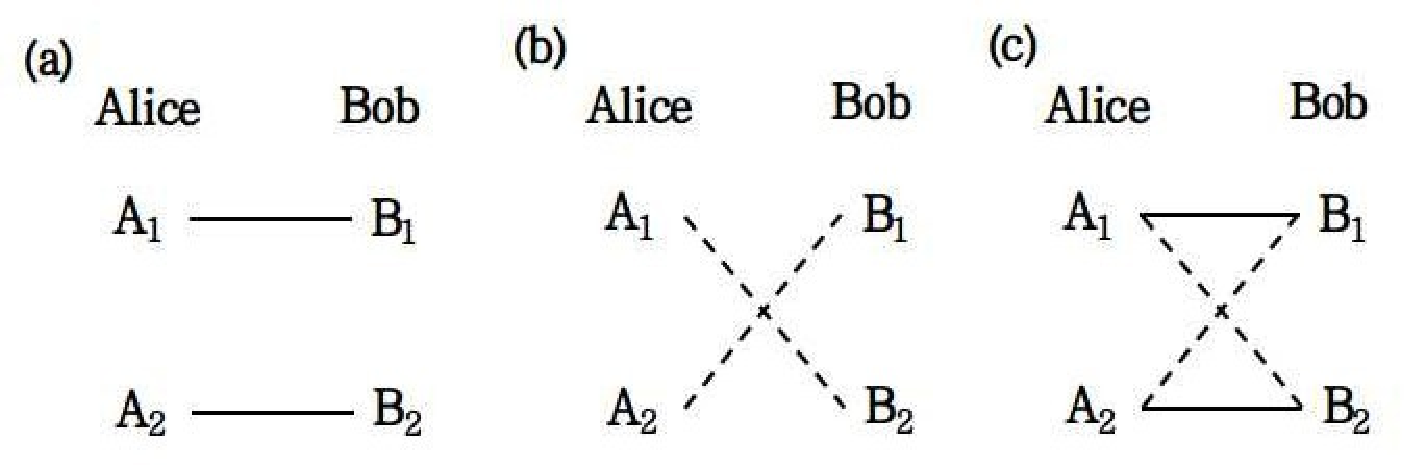} 
\caption{From left to right: (a) hyper-entanglement (solid lines), (b) hybrid-entanglement (dotted lines), and (c) hyper-hybrid entangled state of two qubits with two degrees of freedom.}
\label{fig:HHES}
\end{figure}
\subsubsection{Hyper entanglement} 
Suppose two particles $A$ and $B$, each having two DoFs, are with the possession of Alice and Bob. Hyper-entanglement~\cite{Kwait97} means the simultaneous presence of entanglement in similar kind of multiple DOFs as shown in Fig.~\ref{fig:HHES} (a) where $A_1$ is entangled with $B_1$, and $A_2$ is entangled with $B_2$. It is useful for some tasks like complete Bell state analysis~\cite{Sheng10}, entanglement concentration ~\cite{Ren13}, purification~\cite{Ren13b}, etc.~\cite{Deng17}.  

\subsubsection{Hybrid entanglement} 
If the entanglement is present between different DoFs, then it is known as hybrid entanglement~\cite{Zukowski91,Ma09}. In the Fig.~\ref{fig:HHES} (b), $A_1$ is entangled with $B_2$ and similarly $A_2$ is entangled with $B_1$. Hybrid entanglement is useful in quantum repeaters~\cite{Loock06}, quantum erasers~\cite{Ma13}, quantum cryptography~\cite{Sun11}, etc.~\cite{chen02,Neves09,Andersen15}.

\subsubsection{Hyper-hybrid entanglement}
Hyper-Hybrid entanglement means the the simultaneous presence of these two above types of entanglements. In the Fig.~\ref{fig:HHES} (c), every DoF of one particle is entangled with all other DoFs of the other particle. Hyper-hybrid entanglement is useful for Complete Bell-state analysis~\cite{HHNL}. Although hyper-entanglement and hybrid entanglement were separately known for more than two decades, interestingly, the simultaneous presence of these twohas been proposed very recently by Li \textit{et al.}~\cite{HHNL} in 2018 using using spin and momentum DoF. The above three types of entanglements are shown schematically in Fig.~\ref{fig:HHES}.

\subsection{Representation and degrees of freedom trace-out rule for two distinguishable particles each having $n$ DoFs} \label{Tr_out_rule_dis}
Trace-out operation is very useful tool to represent the reduced density matrix of any density matrix.
In this section, we will represent the general state, density matrix and the degree of freedom (DoF) trace-out rule for two distinguishable particles.

Let two distinguishable particles $A$ and $B$ each having $n$ DoFs. The $i$-th and the $j$-th DoF of $A$ and $B$ are represented by $a_{i}$ and $b_{j}$  respectively where $i, j \in \mathbb{N}_n :=\lbrace 1, 2, \ldots , n \rbrace$.  Suppose each DoF is $d$-dimensional whose eigenvalues are denoted by
$\mathbb{D}_{k}:=\lbrace D_{k_{1}}, D_{k_{2}}, \ldots, D_{k_{d}} \rbrace$ where $k \in \mathbb{N}_n$. The general state of $A$ and $B$ is denoted by
\begin{equation} \label{Statedis}
\begin{aligned}
\ket{\Psi^{(2,n)}}_{AB}= \sum_{a_{1}, a_{2}, \ldots , a_{n},  b_{1}, b_{2}, \ldots, b_{n}}\kappa^{ a_{1} a_{2} \ldots a_{n}}_{  b_{1} b_{2} \ldots b_{n}} \ket{ a_{1} a_{2} \ldots a_{n}} \otimes \ket{  b_{1} b_{2} \ldots b_{n} },
\end{aligned}
\end{equation}
where $a_i \in \mathbb{D}_{i}$, $b_j \in \mathbb{D}_{j}$, and $i,j \in \mathbb{N}_n$.

The general density matrix can be represented as 
\begin{equation} \label{DMdis}
\begin{aligned}
\rho^{(2, n)}_{AB}= \sum_{\substack{ a_{1}, a_{2}, \ldots, a_{n}, \\ u_{1}, u_{2}, \ldots, u_{n}  \\   b_{1}, b_{2}, \ldots, b_{n}, \\ v_{1}, v_{2}, \ldots, v_{n}}} \kappa^{ a_{1} a_{2} \ldots a_{n} u_{1} u_{2} \ldots u_{n} }_{  b_{1} b_{2} \ldots b_{n}  v_{1} v_{2} \ldots v_{n}}  \ket{ a_{1} a_{2} \ldots a_{n}} \ket{  b_{1} b_{2} \ldots b_{n}} \otimes \bra{ u_{1} u_{2} \ldots u_{n}}\bra{  v_{1} v_{2} \ldots v_{n}},
\end{aligned}
\end{equation}
where $a_i, u_i \in \mathbb{D}_{i}$, $b_j, v_j \in \mathbb{D}_{j}$, and $i,j \in \mathbb{N}_n$. 

If $\rho^{(2,n)}_{AB} = \ket{\Psi^{(2,n)}}_{AB}\bra{\Psi^{(2,n)}}_{AB}$ then Eq.~\eqref{DMdis} is represented as
\begin{equation} \label{DMdis_fac}
\begin{aligned}
\rho^{(2, n)}_{AB}= \sum_{\substack{ a_{1}, a_{2}, \ldots, a_{n}, \\ u_{1}, u_{2}, \ldots, u_{n}  \\   b_{1}, b_{2}, \ldots, b_{n}, \\ v_{1}, v_{2}, \ldots, v_{n}}} \kappa^{ a_{1} a_{2} \ldots a_{n}}_{  b_{1} b_{2} \ldots b_{n}} \kappa^{ u_{1} u_{2} \ldots u_{n}*}_{  v_{1} v_{2} \ldots v_{n}} \ket{ a_{1} a_{2} \ldots a_{n}} \ket{  b_{1} b_{2} \ldots b_{n}} \otimes \bra{ u_{1} u_{2} \ldots u_{n}}\bra{  v_{1} v_{2} \ldots v_{n}},
\end{aligned}
\end{equation}
where $*$ denotes complex conjugate.

 If we want to trace-out the $i$-th DoF of particle $A$, then from Eq.~\eqref{DMdis}, the reduced density matrix can be written as
\begin{equation} \label{DisDoFTraceout}
\begin{aligned}
\rho_{a_{\bar{i}}} \equiv & \text{Tr}_{a_{i}} \left( \rho^{(2, n)}_{AB} \right) := \sum_{\substack{ a_{i}, a_{\bar{i}}, u_{i} u_{\bar{i}}, \\ b_{1}, b_{2}, \ldots, b_{n}, \\ v_{1}, v_{2}, \ldots, v_{n}}} \kappa^{ a_{\bar{i}} u_{\bar{i}}}_{ b_{1} b_{2} \ldots b_{n} v_{1} v_{2} \ldots v_{n}} \ket{ a_{\bar{i}}} \ket{   b_{1} b_{2} \ldots b_{n} }  \bra{u_{\bar{i}}} \bra{ v_{1} v_{2} \ldots v_{n}}
 \left\lbrace \braket{a_{i}|c_{i}} \right\rbrace ,
\end{aligned}
\end{equation}
where $a_{\bar{i}}=a_{1}a_{2} \ldots a_{i-1}a_{i+1} \ldots a_{n}$ and similar meaning for $u_{\bar{i}}$. One can show that when the DoF trace-out rule in Eq.~\eqref{DisDoFTraceout} is applied to the same particle for $n$ times, it becomes equivalent to our familiar particle trace-out rule~\cite[Eq. 2.178]{Nielsenbook}.

\section{Measures of entanglement}
 There are various methods to measure quantum entanglement~\cite{Plenio07}.  Let us denote an entanglement measure by $\mathbb{E}$ and a density matrix by $\rho$, then the following properties which any entanglement measure should follow.
\begin{enumerate}
 \item The value of $\mathbb{\rho}$ is zero if $\rho$ is a separable state.
 \item The entanglement measure should be invariant under local local unitary transformations, i.e., for any unitary operator $U_1$ and $U_2$
 \begin{equation}
 \mathbb{E}(\rho) = \mathbb{E}(U_1 \otimes U_2 ~\rho~ U{\dagger}_{1} \otimes U^{\dagger}_{2}).
 \end{equation}
 \item Entanglement cannot be increased by local operations and classical communications.
\item The convexity  property should be obeyed by the entanglement measure, i.e., for any two density matrix $\rho_1$ and $\rho_2$
\begin{equation}
\mathbb{E}(\epsilon \rho_1 + (1-\epsilon) \rho_2) \leq \epsilon \mathbb{E}(\rho_1) + (1-\epsilon) \mathbb{E}(\rho_2)
\end{equation}
 for all $\epsilon \in [0.1]$.
\item They should follow the additivity property, i.e., for $n$ copies of $\rho$ then
\begin{equation}
 \mathbb{E}(\rho^{\otimes n}) = n \mathbb{E}.
 \end{equation}
 
 \end{enumerate}
Now, we will briefly review some of the entanglement measures used in this thesis.
 
 \subsection{Concurrence}
For any density matrix $\rho$, the concurrence $\mathcal{C}$~\cite{Hill97} of $\rho$ can be calculated as
\begin{equation}
\mathcal{C}(\rho) = \max{0, \sqrt{\lambda_4} -\sqrt{\lambda_3} - \sqrt{\lambda_2} - \sqrt{\lambda_1}}
\end{equation}
where $\lambda_{i}$ are the eigenvalues of the matrix
\begin{equation}
\mathbb{R}= \rho~(\sigma_y \otimes \sigma_y \rho^{*} \sigma_y \otimes \sigma_y )
\end{equation}
 in the decreasing order, $*$ denotes complex conjugation and $\sigma_y$ is Pauli matrix. This concurrence measure is very useful for pure and mixed stated in two dimensions. Moreover, concurrence is not additive.
  
\subsection{Negativity}
 The Negativity $\mathcal{N}$~\cite{Zyczkowski98,Vidal02} of any density matrix $\rho$ is defined as
\begin{equation}
\mathcal{N}(\rho)=\dfrac{\mid \mid \rho^{\text{T}_{\text{B}}} \mid \mid_{1} -1}{2}
\end{equation} 
 where $\mid \mid \ldots \mid \mid_{1}$ denotes the trace-norm, i.e., the sum of all the singular values of the partially transposed reduced density matrix. It can be proved that the negatively is convex in nature. Also it is easy to compute. However, like concurrence, it is also not additive. To achieve additivity, let us define logarithmic negativity as
\begin{equation}
 \mathbb{E}_{\mathcal{N}}(\rho)= \text{log}_{2} \mid \mid \rho^{\text{T}_{\text{B}}} \mid \mid_{1}.
\end{equation}  
 However, log-negativity is not convex~\cite{Plenio05}.

\subsection{Other measures of entanglement}
Some other commonly used entanglement measures are  entanglement of formation~\cite{Bennett96}, Tsallis-q entropy~\cite{Kim10,Luo16}, R\'{e}nyi-$\alpha$ entanglement~\cite{Kim_Sanders10,Song16}, Unified-(q, s) entropy~\cite{Kim11,Khan19}, etc.~\cite{Plenio07,Guine09}, one-way distillable entanglement~\cite{Devetak05}, squashed entanglement~\cite{Christandl04,Brandao11}, etc. Most of the measures works in two-qubit systems but not applicable in higher dimensions, specially on the mixed states. A universal measure of entanglement which obeys all the features is the ``Holy Grail" of quantum entanglement theory.

\section{Monogamy of entanglement}
Monogamy of entanglement (MoE) is an unique feature of quantum correlations which is absent in classical correlations.
A bipartite entanglement measure $\mathbb{E}$ that obeys the relation 
\begin{equation} \label{Gen_Monogamy}
\mathbb{E}_{A|B}(\rho_{AB})+\mathbb{E}_{A|C}(\rho_{AC}) \leq \mathbb{E}_{A|BC}(\rho_{ABC}),
\end{equation} 
for all $\rho_{ABC}$ where $\rho_{AB}=\text{Tr}_{C} \left( \rho_{ABC}\right) $, $\rho_{AC}=\text{Tr}_{B} \left( \rho_{ABC}\right) $, $\mathbb{E}_{X|Y}$ measures the entanglement between the systems $X$ and $Y$ of the composite system $XY$, and the vertical bar represents bipartite splitting, is called \textit{monogamous}. Such inequality was first shown for squared concurrence ($\mathcal{C}$)~\cite{Hill97,Wootters98} by Coffman, Kundu and Wootters (CKW) for three parties~\cite{CKW00} and later generalized for $n$ parties~\cite{Osborne06}.


Equation~\eqref{Gen_Monogamy} states that, the sum of the pairwise entanglement between $A$ and the other particles, i.e., $B$ and $C$ cannot exceed the entanglement between $A$ and the remaining particles are taken together as a whole system. From Eq.~\eqref{Gen_Monogamy}, it follows that if $\mathbb{E}_{A|B}(\rho_{AB}) =\mathbb{E}^{max}$, then necessarily $\mathbb{E}_{A|B}(\rho_{AB}) = \mathbb{E}_{A|BC}(\rho_{ABC})$ for any $\rho_{ABC}$ and  $\mathbb{E}_{A|C}(\rho_{AC})=0$ which leads to the qualitative description of monogamy as stated earlier. 

Not all entanglement measures follow Eq.~\eqref{Gen_Monogamy}. Those who follow for all $\rho_{ABC}$ is known to be monogamous in nature. 
Some commonly used monogamous entanglement measures for qubit systems are the entanglement of formation~\cite{Bennett96}, log-negativity~\cite{Zyczkowski98,Vidal02}, Tsallis-q entropy~\cite{Kim10,Luo16}, R\'{e}nyi-$\alpha$ entanglement~\cite{Kim_Sanders10,Song16}, Unified-(q, s) entropy~\cite{Kim11,Khan19}, etc.~\cite{Plenio07,Guine09}. For higher dimensional systems, squared concurrence is known to violate~\cite{Ow07} Eq.~\eqref{Gen_Monogamy}, and only a few entanglement measures are monogamous like one-way distillable entanglement~\cite{Devetak05} and squashed entanglement~\cite{Christandl04,Brandao11}.

\subsection{Coffman, Kundu, and Wootters (CKW) monogamy inequality}

\begin{figure}[h!] 
\centering
\includegraphics[width=\columnwidth]{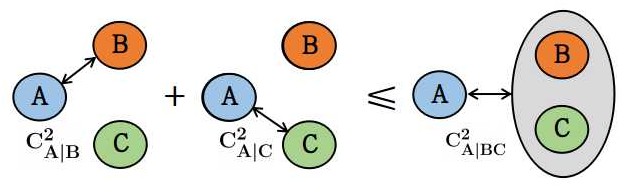} 
\caption{CKW inequality for distinguishable particles using square of the concurrence as entanglement measure.}
\label{Mono}
\end{figure}
Monogamy of entanglement was first introduced by Coffman, Kundu, and Wootters (CKW)~\cite{CKW00} in terms of an inequality which is famously known as CKW inequality for arbitrary states of three qubits using squared concurrence~\cite{Hill97,Wootters98} as entanglement measure~\cite{Plenio07}. Let $A$, $B$, and $C$ are three particles whose joint density matrix is $\rho_{ABC}$. The concurrence between any two particles, let $A$ and $B$, whose reduced density matrix is $\rho_{AB}=\text{Tr}_{C}\left( \rho_{ABC}\right) $, can be calculated as 
\begin{equation}
\mathcal{C}_{A \mid B} \left( \rho_{AB} \right) = \max \lbrace \lambda_{1} - \lambda_{2}- \lambda_{3} - \lambda_{4},0\rbrace,
\end{equation}
where $\lambda_{i}$'s ($i \in \lbrace 1, 2, 3, 4 \rbrace$) are the square root of the eigenvalues of the non-hermitian matrix $\mathbb{R}=\rho_{AB}\tilde{\rho}_{AB}$ in decreasing order, $\tilde{\rho}_{AB}=\left( \sigma_{y} \otimes \sigma_{y} \rho^{*}_{AB} \sigma_{y} \otimes \sigma_{y}  \right) $, $\sigma_{y}$ is Pauli matrix,  the asterisk denotes complex conjugation and the vertical represents bipartite splitting. Similarly, $\mathcal{C}_{A \mid C}\left( \rho_{AC} \right)$ can be calculated. Now the CKW inequality can be written as 
\begin{equation} \label{CKWIq}
\mathcal{C}^{2}_{AB} \left( \rho_{AB} \right) + \mathcal{C}^{2}_{AC}\left( \rho_{AC} \right) \leq \mathcal{C}^{2}_{A \mid BC}\left( \rho_{ABC} \right).
\end{equation}
This inequality states that sum of the square of the pairwise concurrence between $A$ and the other particles, i.e., $B$ and $C$ cannot exceed the square of the concurrence between $A$ and the remaining particles taken together as a whole system. Equation~\eqref{CKWIq} was originally proved for for arbitrary states of three qubits. Later it was generalized for multi-qubit systems in~\cite{Osborne06}. 
All bipartite qubit systems obey Eq.~\eqref{CKWIq}.

\subsection{Equivalence of the monogamy of entanglement and the no-cloning theorem for distinguishable particles} \label{MOEandNC}

\begin{figure}[h!]
\centering
\includegraphics[width=\columnwidth]{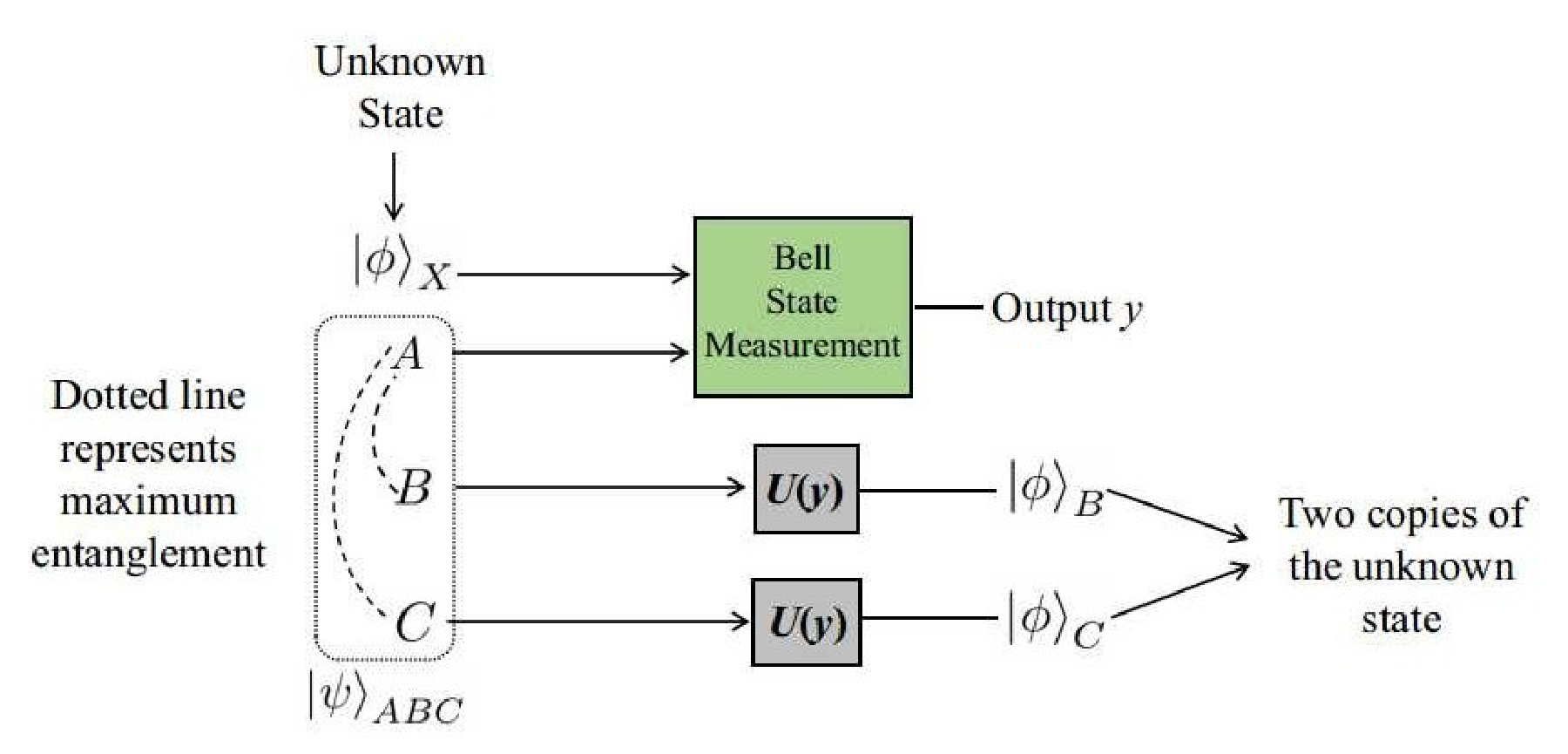} 
\caption{Circuit to get violation of the no-cloning theorem from the maximum violation of MoE.}
\label{fig:BSM}
\end{figure} 
To show that no-cloning implies monogamy of entanglement (MoE), let us prove its contrapositive. When MoE is violated maximally, one can achieve quantum cloning~\cite{QC05,QC14} of any unknown quantum state using standard teleportation protocol~\cite{QT93,QTnat15} as follows. Assume a particle $A$ is maximally entangled with the particles $B$ and $C$ and their joint state is denoted by $\ket{\psi}_{ABC}$ and the particle $X$ is in unknown quantum state $\ket{\phi}_{C}$. To achieve cloning of the state $\ket{\phi}$, one has to perform Bell state measurements (BSM)~\cite{BSM99} jointly on the particles $A$ and $X$. Based on the measurement result denoted by $y$, suitable unitary operations $U_{y}$ have to be performed on the particles $B$ and $C$ so that the state $\ket{\phi}$ appears on each of them, where $U_{y} \in \left\lbrace \mathcal{I}, \sigma_{x}, \sigma_{y}, \sigma_{z}\right\rbrace $, $\mathcal{I}$ being the identity operation and $\sigma_{i}$'s $\left( i = x, y, z \right)$ the Pauli matrices. Thus we can have two copies of the unknown state $\ket{\phi}$ as $\ket{\phi}_{B}$ and $\ket{\phi}_{C}$. 

Next, to show that MoE implies no-cloning, again we prove its contrapositive. Let two particles $A$ and $B$ share a maximally entangled state $\ket{\psi}_{AB}$. If possible, suppose one of them, say, $B$ is cloned and we get a copy $B_{1}$ of $B$, then in the tripartite state $\ket{\psi}_{ABB_{1}}$, $A$ is maximally entangled with both $B$ and $B_{1}$ simultaneously, thus violating the MoE maximally.

\section{Indistinguishability}
 In the last century, physicists were puzzled about whether ``the characteristic trait of Quantum Mechanics"~\cite{Schrodinger}, i.e., entanglement~\cite{epr}, is real and, if so, whether it can show some nontrivial advantages over classical information processing tasks. The answers to both are positive, thanks to several experimentally verified quantum protocols like teleportation~\cite{QT93}, dense coding~\cite{DC92}, quantum cryptography,~\cite{BB84} etc.~\cite{HHHH08}.
 
 In the current century, entanglement of indistinguishable particles and its similarity with as well as difference from that of distinguishable ones have been extensively studied~\cite{Li01,You01,John01,Zanardi02,Ghirardhi02,Wiseman03,Ghirardi04,Vedral03,Barnum04,Barnum05,Zanardi04,Omar05,Eckert02,Grabowski11,Sasaki11,Tichy13,Kiloran14,Benatti17,LFC16,Braun18,LFC18}. Here, indistinguishable particles means independently prepared identical particles like bosons or fermions~\cite{Feynman94,Sakurai94}, where each particle cannot be addressed individually, i.e., a label cannot be assigned to each.
Experiments on quantum dots~\cite{Petta05,Tan15}, Bose-Einstein condensates~\cite{Morsch06,Esteve08}, ultracold atomic gases~\cite{Leibfried03}, etc., support the existence of entanglement of indistinguishable particles. 

\subsection{How to distinguish two identical particles?} \label{IDP_dis_back}
The interesting question comes that if two particles are identical, then can we distinguish them? Let us start with two classical identical particles. Although it is not possible to differentiate them by any of their properties, it is always possible to assign them two different labels, like $A$ and $B$, and using that it is possible, at least in principle, to always keep track their trajectories. In this way, we can in principle, always distinguish two classical identical particles. 

 \begin{figure}[h!] 
\centering
\includegraphics[width=\columnwidth]{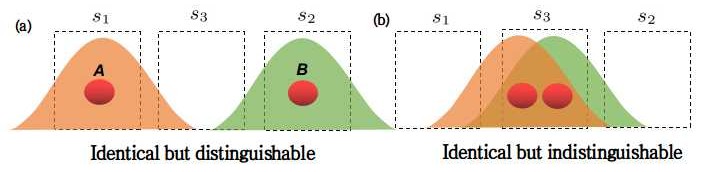} 
\caption{Creation of indistinguishable particles from initially separated identical particles. Here $s_1$, $s_2$, and $s_3$ are three distinct spatial locations. (a) Initially, two identical particles with are present in $s_1$ and $s_2$ in such a way that their wave-functions do not overlap. We label the particles at $s_1$ and $s_2$ as $A$ and $B$ respectively. Thus the particles are identical but are distinguishable via their spatial locations. (b) The particles are brought close to each other so that their wave-functions overlap and they become indistinguishable. Now, they cannot be identified by their spatial locations. If the measurement is done in the overlapped region, i.e., $s_3$, then it is not possible to detect which particle is measured. Even if the particles are again moved apart, they can no longer be labeled. The information about which of $A$ and $B$ appears at $s_1$ or $s_2$ is lost and they remain indistinguishable until they are measured again.}%
\label{Wave_overlap}
\end{figure}

But, unfortunately this method is not possible due to Heisenberg's uncertainty principle~\cite{Heisenberg27} which states that it is not possible to uniquely determine the position and momentum exactly at the same time. Thus to distinguish two identical quantum particle, we need to use their spatial locations. Suppose two identical quantum particles $A$ and $B$  are present in distinct spatial locations $s_1$ and $s_2$ in such a way that their wave-functions do not overlap. Thus it is possible to keep tract those two particles by their spatial locations as long as their wave functions do not overlap as shown in Fig.~\ref{Wave_overlap} (a). Now if they are brought close enough that their wave functions overlap, in spatial region $s_3$ in Fig.~\ref{Wave_overlap} (b). If the measurement is done in the overlapped region, i.e., $s_3$, then it is not possible to detect which particle is measured. Even if the particles are again moved apart, they can no longer be labeled. The information about which of $A$ and $B$ appears at $s_1$ or $s_2$ is lost and they remain indistinguishable until they are measured again. However, this method does not applicable for classical particles as their wave functions are so small that they cannot overlap.

Particle exchange phase is an intrinsic property for states of indistinguishable particles when swapping single-particle states~\cite{Y&S92PRA,Y&S92PRL}. This is a basic scenario based on no-which way information is preserved. The direct measurement of the particle exchange phase (for photons, also including simulations for the case of fermions
and anyons) is demonstrated for the first time in~\cite{Tschernig2021,LoFranco2021,Wang2022}.
This method is extended and demonstrated more recently where one can use independent identical particles and make them entangled by making them spatially overlapped in separated sites~\cite{LFC16,LFC18}. This has been demonstrated experimentally in~\cite{Sun2020,Barros2020,Wang2021}. In~\cite{Sun2020}, the authors have designed an experiment  where they tuned the remote spatial indistinguishability of two independent photons. This is done by individually controlling their spatial distribution in two distant regions. This results polarization entanglement from uncorrelated photons.
In~\cite{Barros2020}, the authors investigate the entanglement between two indistinguishable bosons that is created by spatial overlap.
 In~\cite{Wang2021}, the authors show that it is possible to activate  and distribute of entanglement between two photons in their polarization DoFs that does not require a pair of entangled photons and the Bell-state measurements.

\section{Entanglement of indistinguishable particles}
 Experiments on quantum dots~\cite{Petta05,Tan15}, Bose-Einstein condensates~\cite{Morsch06,Esteve08}, ultracold atomic gases~\cite{Leibfried03}, etc., support the existence of entanglement of indistinguishable particles. 
 The notion of entanglement for distinguishable particles is well studied in the literature~\cite{HHHH08}, where the standard bipartite entanglement is  measured by Schmidt coefficients~\cite{Nielsenbook}, von Neumann entropy~\cite{Bennett96}, concurrence~\cite{CKW00}, log negativity~\cite{Vidal02}, etc.~\cite{Plenio07}. 
Indistinguishability, on the other hand, is represented and analyzed via particle-based first quantization approach~\cite{Li01,You01,John01,Zanardi02,Ghirardhi02,Wiseman03,Ghirardi04} or mode-based second-quantization approach~\cite{Vedral03,Barnum04,Barnum05,Zanardi04}. 
Entanglement in such a scenario requires measures~\cite{Eckert02,Grabowski11,Sasaki11,Tichy13,Kiloran14,Benatti17,LFC16} different from those of distinguishable particles, but there is no consensus on this in the scientific community~\cite[Sec.~III]{Braun18}, particularly on the issues of physicality~\cite{Zanardi02,Barnum04}, accessibility~\cite{Esteve08,Eckert02}, and usefulness~\cite{Tichy13,Kiloran14} of such entanglement. Very recently, the resource theory of indistinguishable particles~\cite{LFC16,LFC18} has been proposed aiming to settle this debate.
 
 
\subsection{Lo Franco \textit{et al}.'s approach to represent indistinguishable particles} \label{LFC_app}
 If the state vector of two indistinguishable particles are labeled by $\phi$ and $\psi$, then the two-particle state is represented by a single entity $\ket{\phi,\psi}$. The  two-particle probability amplitudes is represented by
\begin{equation}\label{inner_P}
\braket{\varphi,\zeta|\phi,\psi} := \braket{\varphi|\phi}\braket{\zeta|\psi} + \eta \braket{\varphi|\psi}\braket{\zeta|\phi},
\end{equation}
where $\varphi,\zeta$ are one-particle states of another global two-particle state vector and $\eta = 1$ for bosons and $\eta=-1$ for fermions. The right hand side of Eq.~\eqref{inner_P} is symmetric if one-particle state position
is swapped with another, i.e., $\ket{\phi,\psi}=\eta \ket{\psi,\phi}$.
From Eq.~\eqref{inner_P}, the probability of finding two particles in the same state $\ket{\varphi}$ is $\braket{\varphi,\varphi|\phi,\psi} = (1+\eta) \braket{\varphi|\phi}\braket{\varphi|\psi}$ which is zero for fermions due to Pauli exclusion principle~\cite{Pauli25} and maximum for bosons. 
As Eq.~\eqref{inner_P} follows symmetry and linearity property, the symmetric inner product of states
with spaces of different dimensionality is defined as
\begin{equation}\label{SinplePunN}
\bra{\psi_{k}}\cdot \ket{\varphi_{1},\varphi_{2}} \equiv \braket{\psi_{k} \mid \varphi_{1},\varphi_{2}} = \braket{\psi_{k}|\varphi_{1}}\ket{\varphi_{2}} + \eta \braket{\psi_{k}|\varphi_{2}}\ket{\varphi_{1}},
\end{equation} 
where $\ket{\tilde{\Phi}}=\ket{\varphi_{1},\varphi_{2}}$ is the un-normalized state of two indistinguishable particles and $\ket{\psi_{k}}$ is a single-particle state. Equation~\eqref{SinplePunN} can be interpreted as a projective measurement where the two-particle un-normalized state $\ket{\tilde{\Phi}}$ is projected into a single particle state $\ket{\psi_{k}}$. Thus, the resulting normalized pure-state of a single particle after the projective measurement can be written as
\begin{equation}
\ket{\phi_{k}}=\frac{\braket{\psi_{k}|\Phi}}{\sqrt{\braket{\Pi^{(1)}_{k}}}_{\Phi}}, 
\end{equation}
where $\ket{\Phi}:=\frac{1}{\sqrt{\mathbb{N}}}\ket{\tilde{\Phi}}$ with $\mathbb{N}=1+\eta \mid \braket{\varphi_{1}|\varphi_{2}} \mid^2$ and $\Pi^{(1)}_{k}=\ket{\psi_{k}} \bra{\psi_{k}}$ is the one-particle projection operator. The one-particle identity operator can be defined as $\mathbb{I}^{(1)}:=\sum_{k}\Pi^{(1)}_{k}$. So, using the linearity property of projection operators, one can write similar to Eq.~\eqref{SinplePunN}: 
\begin{equation}
\ket{\psi_{k}} \bra{\psi_{k}} \cdot  \ket{\varphi_{1},\varphi_{2}} =  \braket{\psi_{k}|\varphi_{1}}\ket{\psi_{k},\varphi_{2}} + \eta \braket{\psi_{k}|\varphi_{2}}\ket{\varphi_{1},\psi_{k}}.
\end{equation} 
Note that
\begin{equation} \label{Identity}
\mathbb{I}^{(1)} \ket{\Phi} = 2 \ket{\Phi},
\end{equation}
where the probability of resulting the state $\ket{\psi_{k}}$ is $p_{k}=\braket{\Pi^{(1)}_{k}}_{\Phi}/2$. The partial trace in this method is can be written as 
\begin{equation}\label{tracedef}
\begin{aligned}
\rho^{(1)} =& \frac{1}{2} \text{Tr}^{(1)} \ket{\Phi} \bra{\Phi} \\ =&\frac{1}{2} \sum_{k} \braket{\psi_{k}|\Phi} \braket{\Phi|\psi_{k}}\\ =& \sum_{k} p_{k} \ket{\phi_{k}} \bra{\phi_{k}},
\end{aligned}
\end{equation}
where the factor $1/2$ comes from Eq.~\eqref{Identity}. 

Another useful concept that of \textit{localized partial trace}~\cite{LFC16}, which means that local measurements are being performed on a region of space $M$ where  the particle has a non-zero probability of being found. So, performing the localized partial trace on a region $M$, we get
\begin{equation}
\rho^{(1)}_{M}=\frac{1}{\mathbb{N}_{M}}\text{Tr}^{(1)}_{M} \ket{\Phi} \bra{\Phi},
\end{equation}
where $\mathbb{N}_{M}$  is a normalization constant such that $\text{Tr}^{(1)}\rho^{(1)}_{M}=1$. The entanglement entropy can be calculated as
\begin{equation} \label{IndVNE}
E_{M}(\ket{\Phi}) := S(\rho^{(1)}_{M}) = -\sum_{i} \lambda_{i}\text{ln}\lambda_{i},
\end{equation} 
where $S(\rho)=-\text{Tr}(\rho \text{ln} \rho)$ is the von Neumann entropy and $\lambda_{i}$ are the eigenvalues of $\rho^{(1)}_{M}$. We will call the state as entangled state if we get a non-zero value of Eq.~\eqref{IndVNE}.

\subsection{Hyper-hybrid entanglement using indistinguishable particles}\label{HHNL}
The circuit of Yurke \textit{et al.}~\cite{Y&S92PRA,Y&S92PRL} to generate  quantum entanglement between the same DoFs of two indistinguishable particles (bosons and fermions) is extended by Li \textit{et al.}~\cite{HHNL} to generate inter-DoF entanglement between two indistinguishable bosons. Details of their generation scheme are as follows.

For bosons, the second quantization formulation deals with bosonic operators $b_{i,\textbf{p}}$ with $\ket{i,\textbf{p}}=b^{\dagger}_{i,\textbf{p}}\ket{0}$, where $\ket{0}$ is the vacuum and $\ket{i,\textbf{p}}$ describes a particle with spin $\ket{i}$ and momentum $\textbf{p}$. These operators satisfy the canonical commutation relations:
\begin{equation}
\left[  b_{i,\textbf{p}_{i}}, b_{j,\textbf{p}_{j}} \right]  = 0, \hspace{0.2cm} \left[  b_{i,\textbf{p}_{i}}, b^{\dagger}_{j,\textbf{p}_{j}}\right]  = \delta(\textbf{p}_{i}-\textbf{p}_{j})\delta_{ij}.
\end{equation} 

\begin{figure}[h!]
\centering
\includegraphics[width=\textwidth]{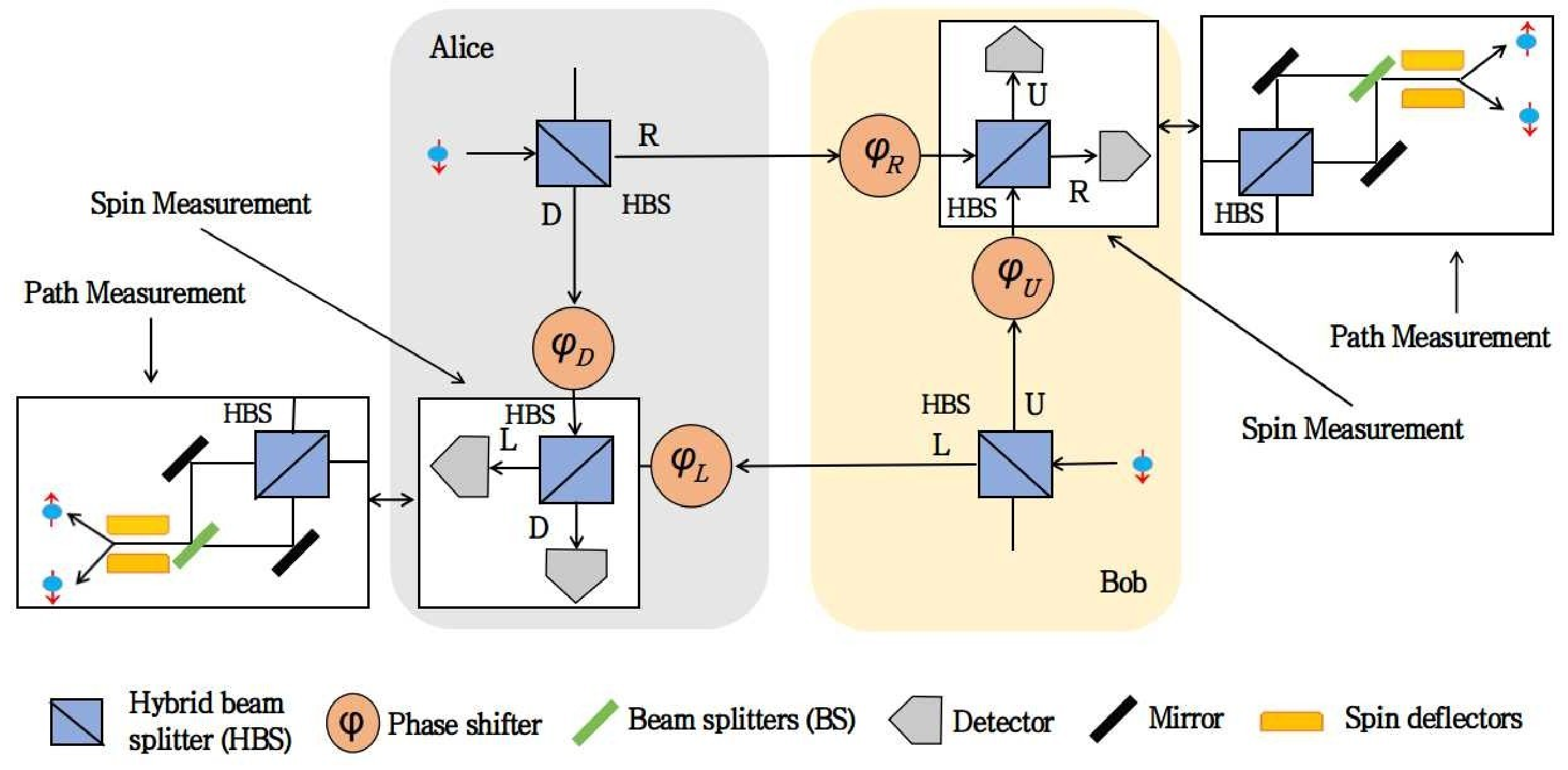} 
\caption{Circuit to generate hyper-hybrid entangled state as proposed by Li \textit{et al.}~\cite{HHNL}. Here the bi-directional arrow represents the measurement is done either in spin DoF or in Path DoF.}
\label{fig:Lietal}
\end{figure} 

Analysis of the circuit of Li \textit{et al.}~\cite{HHNL} for bosons involves an array of hybrid beam splitters (HBS)~\cite[Fig. 3]{HHNL}, phase shifts, four orthogonal external modes $L$, $D$, $R$ and $U$ and two orthogonal internal modes $\uparrow$ and $\downarrow$ as shown in Fig.~\ref{fig:Lietal}. Here, particles exiting through the modes $L$ and $D$ are received by Alice (A) who can control the phases $\varphi_{L}$ and $\varphi_{D}$, whereas particles exiting through the modes $R$ and $U$  are received by Bob (B) who can control the phases $\varphi_{R}$ and $\varphi_{U}$. 

In this circuit, two particles, each with spin $\ket{\downarrow}$, enter the set up in the mode $R$ and $L$ for Alice and Bob respectively. The initial state of the two particles is $\ket{\Psi_{0}}=b^{\dagger}_{\downarrow,R} b^{\dagger}_{\downarrow,L}\ket{0}$. Now, the particles are sent to HBS such that one output port of HBS is sent to other party ($R$ or $L$) and the other port remains locally accessible ($D$ or $U$). Next, each party applies state-dependent (or spin-dependent) phase shifts. Lastly, the output of local mode and that received from the other party is mixed with HBS and then the measurement is performed in either  external or internal modes. The final state can be written as
\begin{equation} \label{Final_state_boson}
\begin{aligned}
\ket{\Psi}=& \frac{1}{4} \left[ e^{i\varphi_{R}}\left( b^{\dagger}_{\downarrow,R}+ib^{\dagger}_{\uparrow,U}\right) +ie^{i\varphi_{D}} \left( b^{\dagger}_{\uparrow,D} + i b^{\dagger}_{\downarrow,L}\right)  \right] 
\\ & \otimes  \left[ e^{i\varphi_{L}} \left( b^{\dagger}_{\downarrow,L}+ib^{\dagger}_{\uparrow,D}\right)  + i e^{i\varphi_{U}}\left( b^{\dagger}_{\uparrow,U}+ib^{\dagger}_{\downarrow,R}\right) \right] \ket{0}. 
\end{aligned}
\end{equation}

\section{Quantum teleportation} \label{Teleport}
The verbal description of quantum teleportation is presented in Section~\ref{Quant_Ent}. In this section, we present the details mathematics of the teleportation process.
 \subsection{Teleportation using distinguishable particles} 
\begin{figure}[h!] 
\centering
\includegraphics[width=8.6cm]{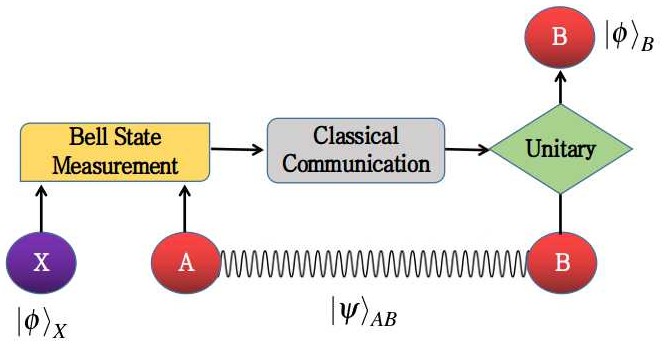} 
\caption{Standard teleportation protocol }
\label{standard_teleport}
\end{figure}
One of the major breakthrough application of quantum theory is teleportation of unknown quantum state of a particle~\cite{QT93,QTnat15}. The brief description of quantum teleportation protocol is as follows:
\begin{enumerate}
\item Alice possess a particle $X$ with an unknown quantum state $\ket{\phi}_{X}$ which as to be teleported to Bob who is at distant from Alice. Also Alice and Bob shares a maximally entangled state $\ket{\psi}_{AB}$ where particle $A$ is with Alice and particle $B$ is with Bob. 
\item Alice performs Bell state measurements (BSM)~\cite{QT93} jointly in her particles $A$ and $X$ and sends the measurement result to Bob via classical communication channels.
\item Based on the measurement result of Alice, Bob performs some unitary operation on his particle $B$. Now the unknown state is teleported from the particle $X$ to $B$.
\end{enumerate} 
One may argue that we get a cloned version of the state $\ket{\phi}$ but after BSM, the original particle is destroyed so that we can have only one copy of the state $\ket{\phi}_{B}$ at Bob's side. This process is shown schematically in Fig.~\ref{standard_teleport}.

 \subsubsection{Fidelity calculation} 

The fidelity of quantum teleportation ($f$)~\cite{Popescu94} is the overlap between the initial quantum state before teleportation denoted by the density matrix $\rho_{in}$ and the density matrix obtain after teleportation denoted by $\rho_{out}$ which is defined  as 
\begin{equation}
f:= \text{Tr} \sqrt{\sqrt{\rho_{\text{in}}} \rho_{\text{out}} \sqrt{\rho_{\text{in}}} } .
\end{equation}
 If the shared entanglement between Alice and Bob is maximally entangled then we get unit fidelity quantum teleportation (UFQT) and if the entanglement is non-maximal then we get the fidelity in between 0 and $\frac{2}{3}$. If there is no entanglement between Alice and Bob, i.e., for separable states, the fidelity is $\frac{2}{3}$. 
 
 \subsection{Relation between teleportation fidelity and singlet fraction}
 Let us assume Alice will teleport an unknown state to Bob. They share a singlet state of two spin-$s$ particles denoted by
\begin{equation}
P_{+}=\ket{\psi_{+}}\bra{\psi_{+}}, \hspace{0.2cm} \ket{\psi_{+}}=\dfrac{1}{\sqrt{d}}\sum^{d}_{i=0} \ket{i}\ket{i}, \hspace{0.2cm} d=s2+1.  
\end{equation} 

Teleportation fidelity~\cite{Jozsa94} measures the closeness between the initial state $\rho^{in}$ that we want to teleport and the final state $\rho^{out}$ obtained after the teleportation protocol. It is given by
\begin{equation} \label{Tf}
 f:= \text{Tr}\sqrt{\sqrt{\rho^{in}}\rho^{out}\sqrt{\rho^{in}}}.
\end{equation} 
On the other hand, singlet fraction~\cite{Horodecki99} of a state $\rho$ measures the maximum overlap of $\rho$ with maximally entangled states. It is given by
\begin{equation} \label{SFold}
F:=\max_{\psi} \braket{\psi \mid \rho \mid \psi},
\end{equation}
where $\ket{\psi}$ varies over all maximally entangled states.  

Now we will derive the relation between teleportation fidelity and singlet fraction for the one-parameter family of states as given in~\cite{Horodecki99} which is given by
\begin{equation} \label{rho_p}
\rho_p =pP_{+} + (1-p)\frac{I \otimes I}{d^2}, \hspace{1cm} 0 \leq p \leq 1
\end{equation}
This is called noisy singlets which is the most natural generalizations of the $2 X 2$ Werner states. Now we will calculate teleportation fidelity and singlet fraction for the state given in Eq.~\eqref{rho_p}. 

For standard teleportation protocol~\cite{QT93}, the fidelity is 1 for singlet state. For a completely random noise represented by the second term of the right hand side of Eq.~\eqref{rho_p}, i.e., $\dfrac{I \otimes I}{d^2}$ as the average final state after teleportation will be $\dfrac{I}{d}$ at the receiver the final state after teleportation will not depend on the initial state before teleportation. Thus the fidelity will be  $\dfrac{I}{d}$. Thus for the state in Eq.~\eqref{rho_p}, the teleportation fidelity is given by
\begin{equation} \label{f_rho_p}
f= p + (1-p)\dfrac{1}{d}, \hspace{1cm} \dfrac{1}{d} \leq p \leq 1.
\end{equation}
The singlet fraction for the state in Eq.~\eqref{rho_p} can be calculated using Eq.~\eqref{SFold}. Clearly, for singlet state, the value of $F$ is 1 and for the completely random noise, the value of $F$ is $\dfrac{1}{d^2}$. Thus the value of $F$ for the state in Eq.~\eqref{rho_p} is given by
\begin{equation} \label{F_rho_p}
F = p + (1)-p)\dfrac{1}{d^2}, \hspace{1cm} \dfrac{1}{d^2} \leq p \leq 1.
\end{equation}
Thus for Eq.~\eqref{f_rho_p} and~\eqref{F_rho_p}, we get
\begin{equation} \label{f_F_par}
f=\frac{Fd+1}{d+1}.
\end{equation}
We can now formate that the state $\rho_p$ in Eq~\eqref{rho_p} is separable \textit{if and only if} 
\begin{equation}
0 \leq p \leq \frac{1}{\left(d+1 \right) }, \hspace{0.5cm} \text{or}\hspace{0.5cm} \dfrac{1}{d^2} \leq p \leq \dfrac{1}{d }, \hspace{0.5cm} \text{or} \hspace{0.5cm} \dfrac{1}{d} \leq f \leq \dfrac{2}{\left( d+1\right) }.
\end{equation}
 

\section{Entanglement swapping} \label{ES}
A brief overview of entanglement swapping process is described in Section~\ref{Quant_Ent}. In this section, we present the details mathematical background of the entanglement swapping process. 
\subsection{Entanglement swapping using distinguishable particles} \label{ES_dis}
Entanglement swapping is the process of creating entanglement between two particles who have never interacted before. The entanglement swapping protocol is as follows:
\begin{figure}[h!] 
\centering
\includegraphics[width=8.6cm]{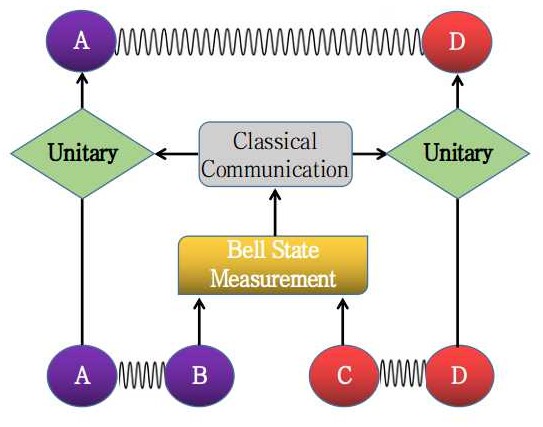} 
\caption{Standard Entanglement Swapping protocol }
\label{standard_ES}
\end{figure}
\begin{enumerate}
\item Suppose two pair of particles $A$ and $B$ is maximally entangled and similarly another pair of particles $C$ and $D$ is also maximally entangled. Now the particles $A$ and $D$ is never interacted before. Our aim is to create entanglement between the particles $A$ and $D$.
\item Particles $B$ and $C$ are brought to a lab and BSM is performed jointly in the particles $B$ and $C$ and the measurement result is send to the labs having particles $A$ and $D$ via classical communication channels.
\item Now based on the measurement result, suitable unitary operation is performed to the particles $A$ and $D$. After that, the particles $A$ and $D$ becomes entangled. 
\end{enumerate} 
 This process is shown schematically in Fig.~\ref{standard_ES}. This is very useful for quantum repeaters.
 

\section{Some quantum cryptographic protocols}
In this section, we discuss three cryptographic protocols, namely, Quantum private query~\cite{QPQ07,DIQPQ},  Local Clauser-Horne-Shimony-Holt (CHSH) protocol~\cite{Lim13}, and Quantum pseudo-telepathy protocol~\cite{Brassard05,Jyoti18}. 
\subsection{Quantum private query protocol} \label{QPQ}
Quantum private query (QPQ) is a cryptographic protocol which deals with the communication between a database owner and its clients. Registered client can query about required entry in the database and the server returns appropriate answer to that query. The security in this protocol depends upon two things:
\begin{enumerate}
\item The server should not reveal any extra information to the clients other than their queries.
\item The server should not gain any extra information about the queries of the clients. 
\end{enumerate}
Conventionally, it is assumed Bob as the database owner or the server and Alice as the client. Like other cryptographic protocols, the communication will start with the key establishment part. But there is a major difference with normal quantum key distribution protocols where the whole key is shared between Alice and Bob. Here, the key is distributed between Alice and Bob such that
\begin{enumerate}
\item Bob, the database owner knows the whole key.
\item Alice, the client knows only a part of the key or the value of some  specific positions of the key.
\item The database owner has no information about the positions of the key known to the clients.	
\end{enumerate}
As a result, unlike normal quantum key distribution protocols, there is no need for any external advisories. Here, Alice and Bob both may work like adversary to each other. The motivation of Alice will be to gain more information about the database whereas Bob will try to know the position of the bits of the key known to Alice and hence gain information about her queries. 

Here, we will discuss the QPQ protocol as proposed in~\cite{QPQ07,DIQPQ} which is based on B92 quantum key distribution~\cite{B92} scheme. There are two phases 
\begin{enumerate}
\item \textit{Key generation:} in this phase, a secure key is established between Bob and Alice. This can be done in two ways:
\begin{enumerate}
\item The source shares entangled pair of particles to Bob and Alice in the specific form
\begin{equation} \label{QPQstate}
\ket{\psi}_{BA}=\frac{1}{\sqrt{2}}\left( \ket{0}_{B}\ket{\phi_{0}}_{A} + \ket{1}_{B}\ket{\phi_{1}}_{A}\right)
\end{equation}
where
\begin{eqnarray*}
\ket{\phi_{0}}_{A} =& \text{cos} \left( \frac{\theta}{2}\right) \ket{0} + \text{sin} \left( \frac{\theta}{2}\right) \ket{1} \\
\ket{\phi_{1}}_{A} =& \text{cos} \left( \frac{\theta}{2}\right) \ket{0} - \text{sin}\left( \frac{\theta}{2}\right) \ket{1}
\end{eqnarray*}
and $0 < \theta < \frac{\pi}{2}$.
\item Now, after sharing Bob measures his qubits in $\left\lbrace \ket{0}_{B}, \ket{1}_{B} \right\rbrace $ basis and Alice measures her qubits in either $\left\lbrace \ket{\phi_{0}}_{A}, \ket{\phi_{0}^{\perp}}_{A} \right\rbrace $ basis or $\left\lbrace \ket{\phi_{1}}_{A}, \ket{\phi_{1}^{\perp}}_{A} \right\rbrace $ basis.
\end{enumerate}
By simple calculations, it can be concluded if Alice's measurement output is $\ket{\phi_{0}^{\perp}}$ or $\ket{\phi_{1}^{\perp}}$, then Bob's measurement output must be $1$ or $0$ respectively. After classical post-processing, by this way, a key can be established between Alice and Bob such a way that Alice can get only one or more bits of information of the whole key whereas Bob has the knowledge of the whole key but Bob has no information about the bit or bits of the key is known to Alice.
 
\item \textit{Private query:} After the key is established let $K$, then let us assume Alice knows the $j$ th bit of key and she make a query about the $i$th element of the database. Then she calculated an inter $s = (j - i)$. Alice sends $s$ to Bob. He then shifts the key $K$ by $s$ amount and generates a new key, say $\bar{K}$. Using this new key $\bar{K}$, Bob encrypts his database using a one-time pad. Bob then transmits the whole database to Alice. Alice can easily get the $j$th bit by decrypting the database.
\end{enumerate}
\subsubsection{Device  independent tests}
The security of the above protocol lies in the fact that the source share the specific state to the Bob as given in Eq.~\eqref{QPQstate} for a specific value of $\theta$. It can be shown that if the state is not exactly the same, i.e., for some value $\theta + \epsilon$ where $\epsilon \neq 0$, the Alice can always generate more information~\cite{DIQPQ}. To mitigate this problem, Bob has to has to remove the trust on the source and her should test the correctness of the state given to him at his end. There are several tests to check the correctness of the given state. The test is chooses which gives the maximum success probability. 

\subsection{Local Clauser-Horne-Shimony-Holt (CHSH) test} \label{L_CHSH}

Bob has to perform this local CHSH test on some of the randomly chosen $n$  pairs. The main steps are
\begin{enumerate}
\item Bob choose two random bit strings $x_i, y_i \in \lbrace0, 1\rbrace$ where $i \in \lbrace1, 2, \ldots, n\rbrace$.
\item If $x_i = 0$, the Bob measures the first particle in $\lbrace \ket{0}, \ket{1}\rbrace$ basis, else in $\lbrace \ket{+}, \ket{-}\rbrace$ basis.
\item Similarly, if $y_i=0$, the the Bob measures the first particle in $\lbrace \ket{\psi_1}, \ket{\psi^{\perp}_1}\rbrace$ basis, else in $\lbrace \ket{\psi_2}, \ket{\psi^{\perp}_2}\rbrace$ basis where
\begin{equation}
\begin{aligned}
\ket{\psi_1} =& \text{cos} \left( \frac{\psi_1}{2}\right) \ket{0} + \text{sin} \left( \frac{\psi_1}{2}\right) \ket{1}, \\ \ket{\psi_2} =& \text{cos} \left( \frac{\psi_2}{2}\right) \ket{0} + \text{sin} \left( \frac{\psi_2}{2}\right) \ket{1}.
\end{aligned}
\end{equation}
\item The measurement result is stored in another bit stings $a_i, b_i \in \lbrace0, 1\rbrace$ such that if the measurement result of fist particle is $\ket{0}$ or $\ket{+}$, then $a_i=0$, else $a_i=1$. 
\item Similarly, if  the measurement result of second particle is $\ket{\psi_1}$ or $\ket{\psi_2}$, then $a_i=0$, else $a_i=1$.
 \item The test is called successful if $a_i \otimes b_i = x_i \wedge y_i$.
\end{enumerate} 
The success probability of this test is 
\begin{equation}
P_{s} =\frac{1}{8} \left(\text{sin}\theta \left(\text{sin} \psi_1 + \text{sin} \psi_2 \right) + \text{cos}\psi_1 - \text{cos}\psi_2\right)+\frac{1}{2} 
\end{equation}
 which is dependent on the values of $\theta$, $\psi_1$, and $\psi_2$. The maximum value of $P_{s}$ is 0.85.
\subsection{Quantum pseudo-telepathy test} \label{Pseudo_tele}

 Using Quantum pseudo-telepathy test~\cite{Brassard05,Jyoti18} the success probability can reach upto unity. The steps of this tests are given below
 \begin{enumerate}
 \item With the help of an ancilla qubit $X$, the state in Eq.~\eqref{QPQstate} can be transformed into the following state
 \begin{small}
 \begin{equation} \label{pseudo_state}
 \ket{\psi}_{BAX} = \frac{1}{\sqrt{2}} \left( \text{cos}\frac{\theta}{2} \ket{000}_{BAX}  + \text{sin}\frac{\theta}{2} \ket{010}_{BAX} + \text{cos}\frac{\theta}{2} \ket{111}_{BAX} - \text{sin}\frac{\theta}{2} \ket{100}_{BAX} \right). 
 \end{equation}
  \end{small}
 \item Now Bob will randomly select a three bit numbers such that there is an even number of 1's in the string. Now based of the input, he performs some specific measurements in the state in Eq.~\ref{pseudo_state}. Based on the outputs, he has to produce output that contains an even number of 1`s if and only if the number
of 1`s in the input is divisible by 4. 
 \end{enumerate}
Now using some specific measurements as given in~\cite{Jyoti18} it can be shown that the success probability is
\begin{equation}
P_{s}=\frac{1}{4} \left(3 + \text{cos} \theta \right).
\end{equation}
 This value goes to 1 asymptomatically.

\chapter{Hardy's non-locality in superconducting qubits} \label{Chap3}
In this chapter, we will introduce a new error-modeling for superconducting qubits and its experimental verification in IBM quantum experience. Here, we argue that for practical verification of Hardy's test, the error-modeling used for optical circuits cannot be used for superconducting qubits. So, we propose a new error-modeling and a new method to estimate the lower bound on Hardy's probability for superconducting qubits. 
We also point out that the earlier tests performed in optical circuits and in the IBM quantum computer have not analyzed the test results in a statistically correct and coherent way. We analyze our data using Student's t-distribution~\cite{feller} which is the statistically correct way to represent the test results. 
 We experimentally verify Hardy's paradox for two qubits on a quantum computer based on superconducting circuits. Our statistical analysis leads to the conclusion that any two-qubit non-maximally entangled state (NMES) gives a nonzero value of Hardy's probability, whereas any two-qubit maximally entangled state (MES) as well as any product state (PS) yields a zero value of Hardy's probability. We identify the difficulties associated with the practical implementation  of quantum protocols based on Hardy's paradox  and discuss how to overcome them. We propose two performance measures for any two qubits of any quantum computer based on superconducting qubits.
 
 This chapter is based on the work in~\cite{Hardy_soumya}.
\section{Practical verification of Hardy's test} \label{Practical Hardy's test} 
In the section~\ref{Hardy_gen} of Chapter~~\ref{Chap2}, we have discussed the standard Hardy's paradox. In this section, we will discuss how Hardy's paradox can be verified practically.

For any experimental set-up, it is quite obvious that the joint probabilities described in Equations~(\ref{eq:1})-(\ref{eq:4}) in the section~\ref{Hardy_gen} of Chapter~~\ref{Chap2} of may not be zero due to errors caused by any external environment or internal device or both. So, Equations~(\ref{eq:1})-(\ref{eq:4}) can be written with some error parameter $\epsilon$~\cite{ghirardi06}.
Here we present the error-model in a slightly different manner so that the result of the practical experiment on an unknown state can be interpreted in a statistically correct and coherent way.
\begin{align}
P(+1,+1|A_{1},B_{1})= & \epsilon_{1},\label{eq:5}\\ 
P(+1,-1|A_{2},B_{1})=& \epsilon_{2},\label{eq:6}\\
P(-1,+1|A_{1},B_{2})= & \epsilon_{3},\label{eq:7}\\
P(+1,+1|A_{2},B_{2})=& \epsilon_{5}=\epsilon_{4}+ q,  \hspace{0.2cm}\text{where} \hspace{0.2cm}
\begin{cases}
       q = 0 & \text{for LHV theory,} \label{eq:8}\\
       q > 0 & \text{for non-locality,}
  \end{cases}
\end{align}
 and $0 \leq \epsilon_{i} \leq 1$, $ \forall$ $ i \in \left\lbrace 1,2,3,5\right\rbrace$.  The bounds of $\epsilon_{4}$ become  $0 \leq \left( \epsilon_{4}+q \right) \leq 1$ or $-q \leq \epsilon_{4} \leq \left( 1-q\right) $. For every MES and every PS of two qubits, the right-hand side of Equation~(\ref{eq:8}) is $\epsilon_{5}=\epsilon_{4}$, i.e., $q=0$, which supports LHV theory. But for every NMES, it is $\epsilon_{5}=\epsilon_{4}+q$ where $q>0$, which supports non-locality.
Thus, by inspecting the values of $q$ in an experiment, it may be possible to infer whether the underlying state is MES/PS or NMES.

\subsection{Connection to the CHSH inequality}
Using simple set-theoretic arguments, one can show that Hardy's equations are a special case of the famous CHSH inequality~\cite{CHSH69}. The CHSH version of Hardy's Equations~\cite{barun08} is described as
\begin{equation}
 \begin{aligned}
 \label{CHSH_to Hardy}
 P(+1,+1|A_{2},B_{2}) -P(+1,+1|A_{1},B_{1})
- P(+1,-1|A_{2},B_{1})  - P(-1,+1|A_{1},B_{2}) \leq 0.
 \end{aligned}
 \end{equation} 
A violation of Equation~(\ref{CHSH_to Hardy}) means a violation of local realism, which supports non-locality. 
Putting the ideal values of the probabilities from Equations~(\ref{eq:1})-(\ref{eq:4}) into Equation~(\ref{CHSH_to Hardy}),  
we get $q \leq 0$. So, $q=0$ supports LHV theory and $q>0$ supports non-locality.
 But when the practical values of the probabilities from Equations~(\ref{eq:5})-(\ref{eq:8}) are put into Equation~(\ref{CHSH_to Hardy}), we get
\begin{equation} \label{CHSH_noise_2}
\epsilon_{5} - \epsilon_{1}- \epsilon_{2} -\epsilon_{3} \leq 0, \hspace{0.3cm} \text{or}  \hspace{0.3cm} \epsilon_{5} \leq \epsilon_{1} + \epsilon_{2} + \epsilon_{3}.
\end{equation}
i.e.,
\begin{equation}
\label{CHSH_noise_2}
\epsilon_{5} \leq \epsilon_{1} + \epsilon_{2} + \epsilon_{3}.
\end{equation}

\section{Our proposed error modeling in superconducting qubits} \label{Pro_error}
In this section, we have first elaborated the difference between the error distribution in optical and superconducting qubits. Then we have proposed a new error modeling for superconducting qubits. Finally, we have discussed bounds in errors in optical circuits vs superconducting qubits.
\subsection{Error distributions in optical circuits vs superconducting qubits} \label{Error distributions}
If we perform Hardy's experiment in optical set-up \cite{Giuseppe1997,irvine2005,Lundeen2009,Yokota2009,vallone2011,chen2012,karimi2014,zhang2016,fan2017,ladder2017,Luo2018,Yang19}, errors can occur in many different ways, such as: (i) the preparation of the ideal quantum state, (ii)  in the construction of the measurement operators $A_{i}$ and $B_{j}$ where $i , j \in \left\lbrace 1,2\right\rbrace $ (as defined in Section \ref{Hardy_gen} in Chapter~\ref{Chap2}), (iii) due to the detection problems of the particles (this includes particle loss), etc. Introducing an error in Equation~(\ref{eq:1}) due to the above-mentioned reasons has a direct impact on Equation~(\ref{eq:4}), i.e., the logic of Hardy's argument ceases to work. Similarly, when Equation~(\ref{eq:2}) and Equation~(\ref{eq:3}) are non-zero, they make a contribution to the right-hand side of Equation~(\ref{eq:4}).  So, in an optical set-up, if Equation~(\ref{CHSH_noise_2}) is violated, it leads to the violation of local realism. No estimation of $q$ is required. The work \cite{irvine2005} does exactly this check of Equation~(\ref{CHSH_noise_2}) in its optical circuits.

But in the case of superconducting qubits, there are three types of errors, namely gate error, readout error, and multi-qubit gate error \cite{IBM}. Unlike in optical circuits, these errors are common to all superconducting qubits circuit and not specific to the circuits to test Hardy's paradox.
So, having an error in Equation~(\ref{eq:1}) does not have any impact on the right-hand side of Equation~(\ref{eq:4}) and Hardy's argument still works with this error. Similar reason applies when Equation~(\ref{eq:2}) and Equation~(\ref{eq:3}) are not zero. So, in this case, to test the violation of local realism, we will estimate the value of $q$ in Equation~(\ref{eq:8}) by a new method which will be described in the next section.

\subsection{Our proposed model for verifying whether Hardy's probability greater than zero in superconducting qubits} \label{Estimate q}
We recall from Equation~(\ref{eq:8}) that $\epsilon_{5}=\epsilon_{4}+ q$. While doing the experiment, we can observe only the values of  $\epsilon_{5}$. 
 To estimate the value of $q$, we need to get the values of $\epsilon_{4}$. Now, the values of $\epsilon_{4}$ can be observed from the experiment directly for every MES as well as PS of two qubits as $q=0$, but it cannot be observed directly for NMES as $q>0$.
 
However, because of the nature of errors in superconducting qubits as explained in Section \ref{Error distributions}, the values of $\epsilon_{4}$ in both the cases (for $q=0$ \& $q > 0$) will follow the same distribution. So, its maximum value, say $\Sigma_{4}$, can be estimated from a large number of known MES and PS (with $q=0$), and then this estimate can be used in Equation~(\ref{eq:8}) to infer about $q$ for any unknown state as follows:
\begin{equation} \label{q_bound}
\epsilon_{5} \leq \Sigma_{4} + q, \hspace{0.3cm}  \text{or} \hspace{0.3cm}  q \geq \epsilon_{5} - \Sigma_{4}.
\end{equation} 
or
\begin{equation}
\label{q_bound}
 q \geq \epsilon_{5} - \Sigma_{4}.
\end{equation} 
From Equation~(\ref{q_bound}), we can define the lower bound on $q$ as
\begin{equation} \label{q_lb}
q_{lb}=\epsilon_{5} - \Sigma_{4}. 
\end{equation}
 If  $q_{lb} > 0$, then Equation~(\ref{q_bound}) implies that Hardy's probability  $q > 0$.   

\subsection{Bounds in errors in optical circuits vs superconducting qubits}
For Hardy's test in optical circuits, if $\epsilon_{1}=\epsilon_{2}=\epsilon_{3}=\epsilon$, then from Equation~(\ref{CHSH_noise_2}), we can get the bounds in errors, i.e., $0 \leq \epsilon < \frac{1}{3}$ \cite{rabelo12}. But if $\epsilon_{1} \neq \epsilon_{2}\neq\epsilon_{3}$, then these bounds are not valid because there may be a case where any of $\epsilon_{1},\epsilon_{2},\epsilon_{3}$ can be close to one and the rest close to zero. Then theoretically the bounds in errors are between zero and one.

For Hardy's test in superconducting qubits, as explained in Section \ref{Error distributions}, all the error parameters $\epsilon_{1},\epsilon_{2},\epsilon_{3},\epsilon_{4},\epsilon_{5}$ follow the same distribution. So, the bounds in $\epsilon_{1},\epsilon_{2},\epsilon_{3},\epsilon_{4},\epsilon_{5}$ will also be the same, i.e., in between zero and one. But to verify Hardy's circuit, the maximum value of $\epsilon_{4}$, i.e., $\Sigma_{4}$ needs to be bounded. From Equation~(\ref{q_bound}), the theoretical bound of $\Sigma_{4}$ is $\left(1-q_{max}\right)=\left(1 - 0.09017\right) \approx 0.90983$. So, this bound is also valid for $\epsilon_{1},\epsilon_{2},\epsilon_{3}$. In the case of optical circuits, when $\epsilon_{1} \neq \epsilon_{2}\neq\epsilon_{3}$, the bounds are trivial, but for the case of superconducting qubits, the error bounds are defined independently, i.e., whether all of them takes the same value or not. It may be noted that if we get a value of any of $\epsilon_{1},\epsilon_{2},\epsilon_{3}$ greater than $0.90983$, then that circuit cannot be used for Hardy's test using superconducting qubits. But in optical circuits, in theory, if we get one of the values of $\epsilon_{1},\epsilon_{2},\epsilon_{3}$  greater than $0.90983$ and the rest zero, we still can perform Hardy's test.

The basic motivation for performing Hardy's test in superconducting qubits is to show the violation of local realism in superconducting qubits. Though Hardy's paradox is already tested in optical circuits~\cite{Giuseppe1997,irvine2005,Lundeen2009,Yokota2009,fedrizzi2011,vallone2011,chen2012,karimi2014,zhang2016,fan2017,ladder2017,Luo2018,Yang19}, none of them have been able to estimate the lower bound on Hardy's probability, i.e., $q_{lb}$ from their experiments. The advantage of performing Hardy's test in superconducting qubits is that we can estimate the lower bound on of Hardy's probability $q_{lb}$ as discussed in Section~\ref{Estimate q}.

\section{Circuits for Hardy's test using superconducting qubits} \label{Circuits for Hardy's equations}
We perform a series of experiments to check Hardy's non-locality for two qubits in the IBM quantum computer~\cite{IBM}. For simplicity, we use the \textit{ibmqx4}\footnote[1]{IBM has several chips and a subset of those becomes available for experiments from time to time. Currently, the \textit{ibmqx4} chip is under maintenance, while the available chips are \textit{ibmqx2}, \textit{ibmq\_vigo}, \textit{ibmq\_ourense} and \textit{ibm\_16\_melbourne}.} chip which is five-qubit, as we only need two qubits for our experiment. This experiment can also be done using other chips consisting of any other number of qubits (more than or equal to two).  It uses a particular physical type of qubit called a superconducting transmon qubit made from superconducting materials niobium and aluminum, patterned on a silicon substrate. During all the experiments, the fridge temperature is maintained at $0.021$ K. Any experiment in the IBM quantum computer can be performed for 1 shot, 1024 shots, 4096 shots or 8192 shots in every run.

In the current IBM \textit{ibmqx4} chip topology~\cite{IBM}, for using multi-qubit gates like $CNOT$, there is a restriction, i.e., not all pairs of qubits can be used for circuit implementation. The list of possible combinations are given in details in the IBM website~\cite{IBM} and it is also discussed in Section \ref{Other_CNOT_Gate} in details. It should be noted that all the qubits are subject to different types of errors as given in the IBM website~\cite{IBM}. Initially, we  implement our circuit by choosing any possible pair of qubits and then validate the results for the rest of the possible combinations of qubits. 

In~\cite{barun08}, a circuit consisting of two coupled electronic Mach-Zehnder (MZ) interferometers has been proposed for Hardy's test which is similar to the Hardy's original thought experiment~\cite{hardy92}. We implement this circuit in the IBM quantum computer. As described in~\cite{barun08}, there are three important parameters of this experiment: beam splitters $U_{B}(\theta)=\left(\begin{smallmatrix}$cos$\theta&-$sin$\theta\\ $sin$\theta& $cos$\theta\end{smallmatrix}\right)$, phase shifter $ U_{P}(\phi)=\left(\begin{smallmatrix}1&0\\0&e^{i\phi}\end{smallmatrix}\right)$, and the coupling $U_{C}(\phi)=\left(\begin{smallmatrix}1& & & \\ &1& & \\ & &1& \\ & & &e^{i2\phi}\end{smallmatrix}\right)$ which can be expressed as

\begin{gather}
\begin{aligned}
U_{B}(\theta)=&U_{3}(2\theta,0,0), \\
U_{P}(\phi)=&U_{1}(\lambda),\\
U_{C}(\phi)=&M_{3} \cdot CNOT \cdot M_{2} \cdot CNOT \cdot M_{1},
\end{aligned} \label{eq:12}
\end{gather}

where
\begin{gather}
\begin{aligned}
U_{1}(\lambda)=&\left(\begin{matrix}1&0\\0&e^{\lambda i}\end{matrix}\right), \\
U_{3}(\theta,\lambda,\phi)=& \left(\begin{matrix} \text{cos} \frac{\theta}{2} &-e^{-\lambda i} \text{sin} \frac{\theta}{2}\\ e^{-\lambda i} \text{sin} \frac{\phi}{2} & e^{i\left( \phi +\lambda \right) }$ cos$\frac{\theta}{2}\end{matrix}\right),\\
M_{1}=&Id \otimes U_{1}(-\lambda), \\ M_{2}=&U_{1}(\lambda) \otimes U_{1}(-\lambda), \\ M_{3}=&Id \otimes U_{1}(2\lambda).
\end{aligned}
\end{gather}

\noindent $CNOT$ is a controlled-NOT gate and $ Id $ is the identity gate. Here $U_{1}(\lambda) $, $U_{3}(\theta,\lambda,\phi)$, $CNOT$, and $Id$ are available as standard gates provided by the IBM quantum computer~\cite{IBM}. We decompose the coupling $U_{C}(\phi)$ by the standard IBM gates is shown in Equation~(\ref{eq:12}). But it can also be decomposed in different ways such that the total number of gates are reduced further and that is left as future work.

\begin{figure}[h!]
\centering
\includegraphics[width=\textwidth]{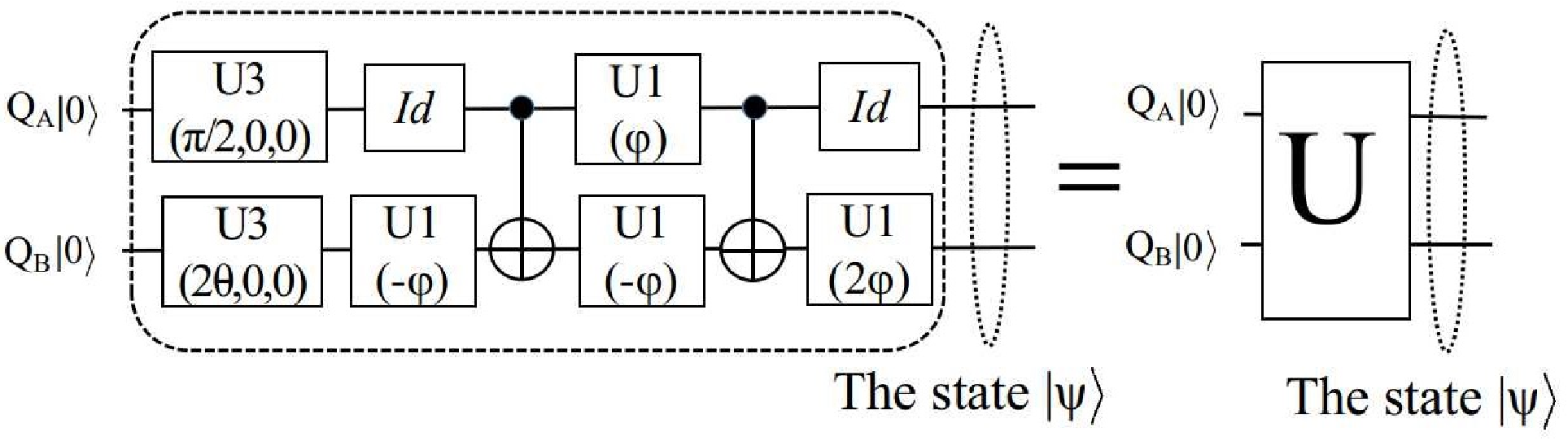}
\caption{The state $|\psi\rangle$ for Equation~(\ref{eq:13}).}
\label{fig:1}
\end{figure}

 For Hardy's test, the state  $\ket{\psi} $ for Alice and Bob is considered in~\cite{barun08} as
\begin{equation}
|\psi\rangle = V_{0}(V_{1} \otimes V_{2} )|00\rangle = \dfrac{\text{cos}\theta}{\sqrt{2}}(|00\rangle) + |10\rangle)+ \dfrac{\text{sin}\theta}{\sqrt{2}}(|01\rangle) + e^{i2\phi}|11\rangle), \label{eq:13}
\end{equation}
where  $V_{0}= U_{C}(\phi)$, $V_{1}=U_{B}\left( \dfrac{\pi}{4}\right)$, and $V_{2}=U_{B}(\theta)$.  The state $\ket{\psi}$ is expressed by the IBM gates is shown in Figure~\ref{fig:1} where $Q_{A}$ and $Q_{B}$ are the qubits for Alice and Bob respectively. The values of $\theta$ and $\phi$ in between $0$ to $90$ degrees for which $|\psi\rangle$ is found to be MES as well as PS are given in Table~\ref{table:1}. The measurements for Alice and Bob are described as follows. 

\begin{figure}[h!]
\centering
\includegraphics[width=0.7\textwidth]{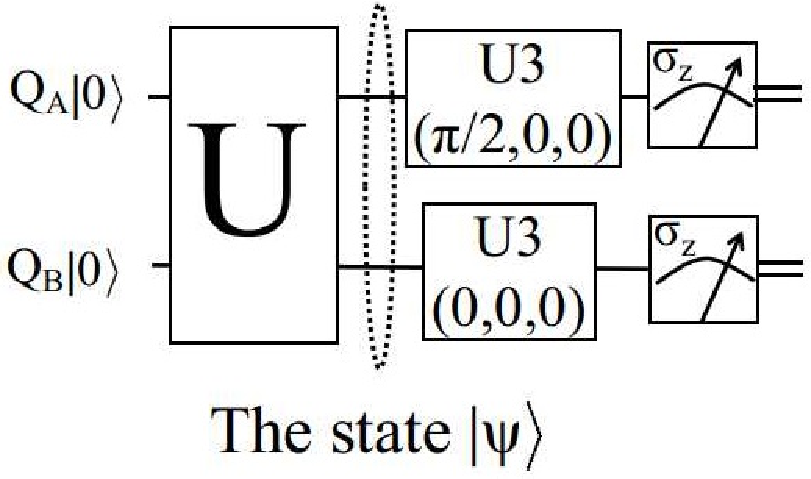}
\caption{Quantum circuit and measurement for $P(+1,+1|A_{1},B_{1})$ for Equation~(\ref{eq:5}).}
\label{fig:2}
\end{figure}

\begin{figure}[h!]
\centering
\includegraphics[width=\textwidth]{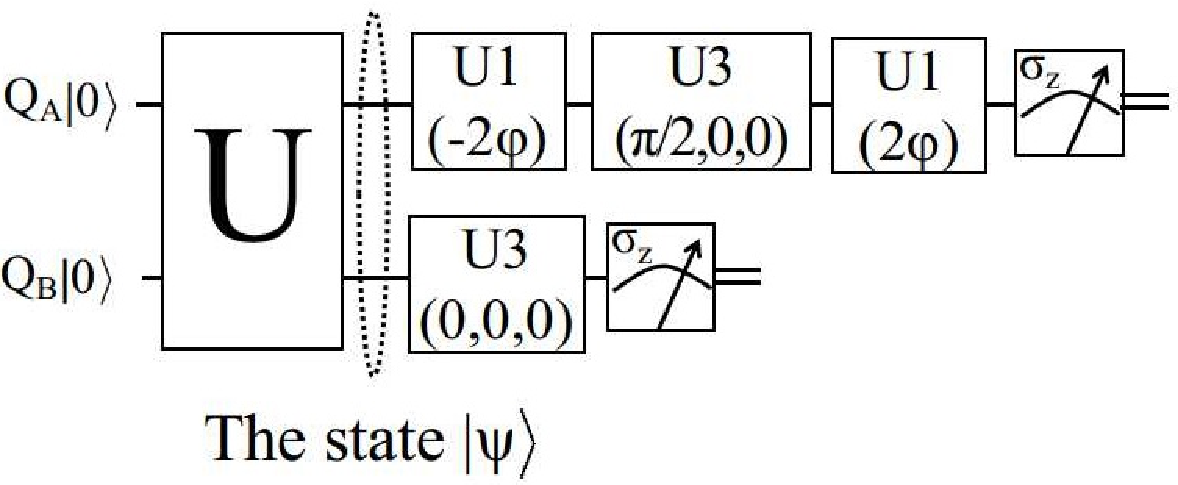} 
\caption{Quantum circuit and measurement for $P(+1,-1|A_{2},B_{1})$ for Equation~(\ref{eq:6}).}
\label{fig:3}
\end{figure}

\begin{figure}[h!]
\centering
\includegraphics[width=\textwidth]{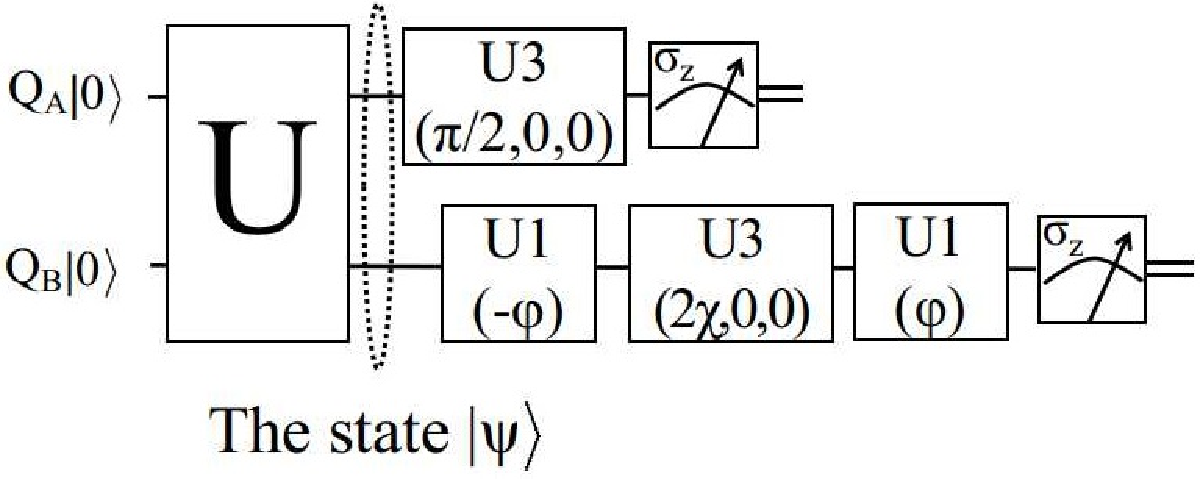} 
\caption{Quantum circuit and measurement for $P(-1,+1|A_{1},B_{2})$ for Equation~(\ref{eq:7}).}
\label{fig:4}
\end{figure}

\begin{figure}[h!]
\centering
\includegraphics[width=\textwidth]{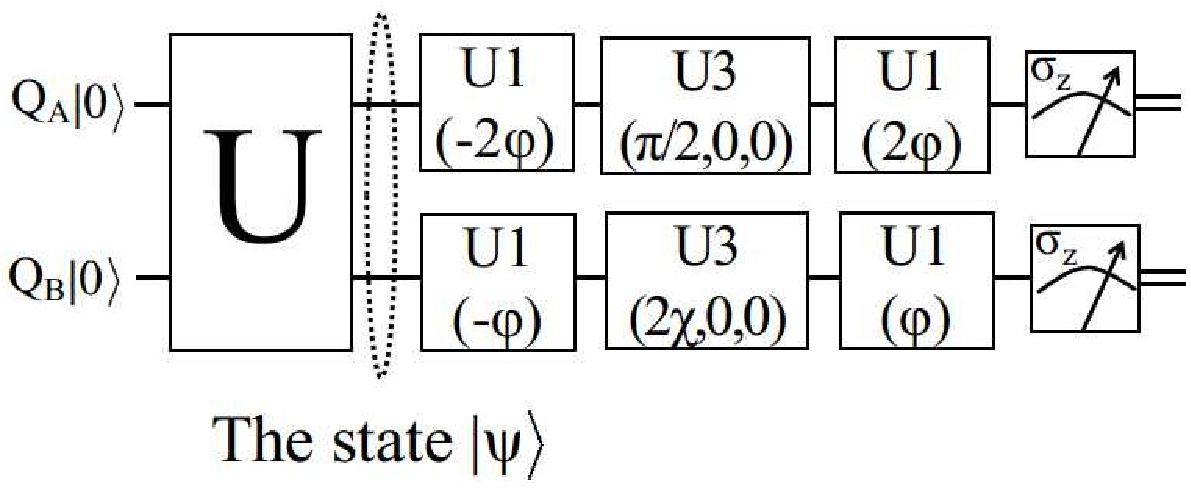} 
\caption{Quantum circuit and measurement for $P(+1,+1|A_{2},B_{2})$ for Equation~(\ref{eq:8}).}
\label{fig:5}
\end{figure}

\begin{table}[t!]
\centering
\begin{tabular}{ p{1.5cm}   p{1.5cm}  p{0.8cm}  p{5cm}    }
 \hline
   $\theta$ & $\phi$ & state & $|\psi\rangle $\\
 \hline 
 0  & any value & PS &  $\dfrac{1}{\sqrt{2}} (|0\rangle + |1\rangle) \otimes |0\rangle $ \\

 any value  & 0 & PS & $\dfrac{1}{\sqrt{2}} (|0\rangle + |1\rangle) \otimes $(cos$\theta |0\rangle + $sin$ \theta |1\rangle ) $\\

 90  & any value & PS & $\dfrac{1}{\sqrt{2}} (|0\rangle + e^{i2 \phi }|1\rangle) \otimes |1\rangle $\\

 45 & 90 & MES& $\dfrac{1}{2}(|00\rangle + |01\rangle + |10\rangle - |11\rangle)$ \\
\hline
\end{tabular}
\caption{Maximally entangled states (MES) and product states (PS) based on different values of $ \theta $ and $ \phi $ in between $0$ to $90$ degrees.} \label{table:1}
\end{table}

\begin{gather} \label{measurements}
\begin{aligned}
A_{1}=& U_{B}\left( \dfrac{\pi}{4}\right) =U_{3}\left( \dfrac{\pi}{2},0,0\right) ,\\
B_{1}=& U_{B}\left( 0\right) =U_{3}\left( 0,0,0\right), \\
A_{2}=& U_{P}\left( 2\phi\right)  U_{B}\left( \dfrac{\pi}{4}\right)  U_{P}\left( -2\phi\right)
    =U_{1}(2\lambda) U_{3}\left( \dfrac{\pi}{2},0,0\right)  U_{1}(-2\lambda),\\
B_{2}=& U_{P}(\phi) U_{B}(\chi) U_{P}(-\phi)
    = U_{1}(\lambda) U_{3}\left( 2\chi,0,0\right) U_{1}(-\lambda), 
\end{aligned}
\end{gather}
where $\text{cot}\chi = \text{tan}\theta\text{cos}\phi$.  The measurements are done in $\sigma_{z}$ basis. The experimental circuits of Equations~(\ref{eq:1}), (\ref{eq:2}), (\ref{eq:3}), and~(\ref{eq:4}) for the state $\ket{\psi}$ using the above measurements in the IBM quantum computer are given in Figures~\ref{fig:2},~\ref{fig:3},~\ref{fig:4}, and~\ref{fig:5} respectively. The theoretical value of $P(+1,+1|A_{2},B_{2})$  is given by $\left|\braket{\psi|A_{2} \otimes B_{2} | \psi} \right| ^{2}$, which using Equation~(\ref{eq:13}) and Equation~(\ref{measurements}) becomes
\begin{equation}
q=\left|\frac{1}{2}\text{cos} \theta \text{cos} \chi \left( 1-e^{-2i\phi}\right) \right| ^{2}.
\label{eq:15}
\end{equation}

\begin{figure}[t!]
\centering
\includegraphics[width=1.0\textwidth]{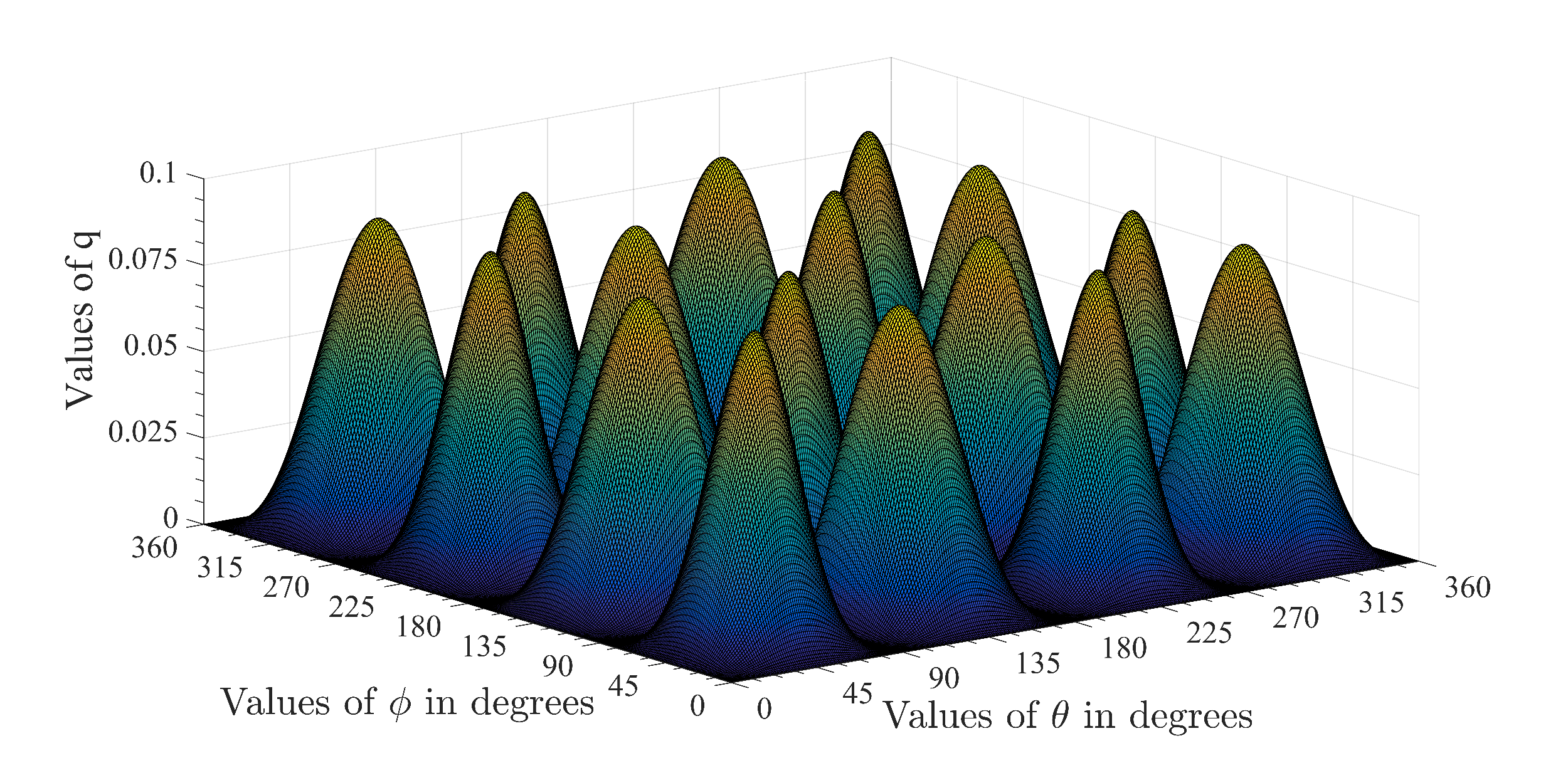}
\caption{The variation of $\theta$ and $\phi$ in degrees vs the values of $q$ from Equation~(\ref{eq:15}). }
\label{fig:6}
\end{figure}

The maximum value of $q$, i.e., $q_{max}$ is found from Equation~(\ref{eq:15}) when
\begin{equation}
\text{cos}(2\theta)=\text{cos}(2\phi)=2-\sqrt{5}. \label{eq:14}
\end{equation}
 If the variation of $\theta$ and $\phi$ are carried out in between $0$ to $90$ degrees, then $q_{max}$ occurs at $\theta=\phi=51.827$ degrees approximately. But in general similar analysis can be done for any values of $\theta$ and $\phi$. In Figure~\ref{fig:6}, we plot\footnote[2]{We use MATLAB\textsuperscript{\textregistered} to get these values.} the values of $q$  from Equation~(\ref{eq:15}) by varying $\theta$ and $\phi$ from $0$ to $360$ degrees. In this figure, we can see that $q_{max}$ is achieved for $\theta=\phi=51.827$ degrees and also for other values of $\theta$ and $\phi$ such that Equation~(\ref{eq:14}) is satisfied. 

For $\phi=90$ and $\theta \neq \left\lbrace 0,45,90 \right\rbrace $ degrees, from Equation~(\ref{eq:13}), we get $|\psi\rangle$ as NMES. This means, if we perform Hardy's test, a non-zero value of $q$ in Equation~(\ref{eq:4}) has to be found. But when $\phi=90$ degree, we get $\chi=90$ degree which means the right-hand side of Equation~(\ref{eq:15}) is zero. So, in this experimental set-up, Hardy's test fails for all NMES for the values of $\phi=90$ and $\theta \neq \left\lbrace 0,45,90 \right\rbrace $ degrees.

\subsubsection{Boundary values of $\chi$}
\label{app1}
In Section \ref{Circuits for Hardy's equations}, $\chi$  is defined as $\text{cot} \chi = \text{tan} \theta \text{cos} \phi$. When $\theta = 90$ degree and $\phi= 90$  degree, we get $\text{tan} \theta = \infty $ and $\text{cos} \phi = 0$, which leads to  $\text{cot} \chi = \infty \cdot 0$. This is an indeterminate form. 

When $\theta \to 90+$ and $\phi \to 90+$, the value of $\chi$ is positive.
When $\theta \to 90+$ and $\phi \to 90-$, the value of $\chi$ is negative.
When $\theta \to 90-$ and $\phi \to 90+$, the value of $\chi$ is negative.
When $\theta \to 90-$ and $\phi \to 90-$, the value of $\chi$ is positive.

So, the limit does not exist. More formally,
\begin{equation}
\begin{aligned} 
& \lim_{\left(\theta, \phi \right) \to \left( \frac{\pi}{2}, \frac{\pi}{2} \right)} \text{tan} \theta \text{cos} \phi \\
=& \lim_{\left( x, y \right) \to \left( 0, 0 \right)} \text{tan} \left( x +\frac{\pi}{2}\right)  \text{cos} \left( y + \frac{\pi}{2}\right) \\
=&\lim_{\left( x, y \right) \to \left( 0, 0 \right)} \text{cot} x \text{sin} y,
\end{aligned}
\end{equation}
where $x=\left(\theta-\frac{\pi}{2} \right) $ and  $y=\left(\phi-\frac{\pi}{2} \right) $.
Now changing the co-ordinate system from rectangular to polar co-ordinate system and substituting $x=r$ cos$\varphi$ and $y=r$ sin$\varphi$, we can write the above expression as 

\begin{equation}
\begin{aligned}
 & \lim_{ r  \to 0 }  \text{cot}\left(  r \text{cos} \varphi \right) \text{sin} \left( r \text{sin} \varphi\right)\\
=&\lim_{r  \to 0 } \frac{\text{cos} \left(  r \text{cos} \varphi \right)}{\text{sin} \left(  r \text{cos} \varphi \right)} \text{sin} \left( r \text{sin} \varphi\right)\\
 =&  \lim_{r  \to 0 }  \frac{\left( 1 - \left( r \text{cos}  \varphi\right)^{2} + \cdots \right) }{\left( r \text{cos} \varphi - \frac{\left(r \text{cos} \varphi ^{3} \right)}{3!} + \cdots  \right)}{\left( r \text{sin} \varphi - \frac{\left(r \text{sin} \varphi ^{3} \right)}{3!} + \cdots  \right)}\\
=&\lim_{r  \to 0 }  \frac{\left( 1 - \left( r \text{cos}  \varphi\right)^{2} + \cdots \right) }{r \left(  \text{cos} \varphi - \frac{r ^{2}\left( \text{cos} \varphi ^{3} \right)}{3!} + \cdots  \right)}{r \left(  \text{sin} \varphi - \frac{r^{2}\left( \text{sin} \varphi ^{3} \right)}{3!} + \cdots  \right)} \\
=& \frac{\text{sin} \varphi}{\text{cos} \varphi}=\text{tan} \varphi.
\end{aligned}
\end{equation}
 So, $\chi$ doesn't have any definite value, but it depends on $\varphi$.

\subsection{How to calculate the difference between two Student's t-distributed variables?}
 \label{student's t calculation}
In the section~\ref{Stat} of Chapter~\ref{Chap2}, we have discussed how to perform analysis of any experimental results. Using that we will now calculate $q_{lb}$.
 
To calculate $q_{lb}$ from Equation~(\ref{q_lb}), we need to calculate the statistics for the difference of two random variables $X$ and $Y$ corresponding to the experimental values of $\epsilon_{5}$ and $\Sigma_{4}$ respectively.
Here both $X$ and $Y$ are assumed to follow standard normal distributions. Also, for $n$ number of samples, let the sample means of $X$ and $Y$ be $\bar{X}$ and $\bar{Y}$ respectively and the sample standard deviations be $S_{X}$ and $S_{Y}$ respectively. Then for some value of $\alpha$, let $ \left( \bar{X} \pm t_{X} \right) $ where $t_{X}=t_{\frac{\alpha}{2}}\frac{S_{X}}{\sqrt{n}} $ represents the $(1-\alpha)\%$ confidence interval  around the mean of $X$. Similarly, let $ \left( \bar{Y} \pm t_{Y} \right) $  where $t_{Y}=t_{\frac{\alpha}{2}}\frac{S_{Y}}{\sqrt{n}} $ represents the same for the random variable $Y$. Now if we want to calculate the value of another Student's t-distributed random variable $W$ such that $W=X-Y$, then the sample mean $\bar{W}$ of $W$ is given by $\bar{W}=\left( \bar{X}-\bar{Y}\right)$ and the sample standard deviation is given by $S_{W}=\sqrt{S_{X}^{2}+S_{Y}^{2}}$. Then the $(1-\alpha)\%$ confidence interval around mean of $W$ is $ \left( \bar{W} \pm t_{W} \right) $  where $t_{W}=t_{\frac{\alpha}{2}}\frac{S_{W}}{\sqrt{n}} $. Now the value of $W_{lb}$ will be 
\begin{equation} \label{zmin}
W_{lb}=\left( \bar{W}-t_{W}\right) = \left( \bar{X}-\bar{Y}-t_{\frac{\alpha}{2}}\frac{\sqrt{S_{X}^{2}+S_{Y}^{2}}}{\sqrt{n}}\right).
\end{equation}
We can use Equation~(\ref{zmin}) to estimate $\hat{q}_{lb}$ of $q_{lb}$ as defined in Equation~(\ref{q_lb}) as
\begin{equation} \label{hatqlb}
\hat{q}_{lb} = \left( \bar{\epsilon}_{5}-\bar{\Sigma}_{4}-\Delta\right) \hspace{0.3cm} \text{where} \hspace{0.3cm} \Delta=t_{\frac{\alpha}{2}}\frac{\sqrt{S_{\epsilon_{5}}^{2}+S_{\Sigma_{4}}^{2}}}{\sqrt{n}}.
\end{equation}
Here $\bar{\epsilon}_{5}$ and $S_{\epsilon_{5}}$ are the sample mean and sample standard deviation of $\epsilon_{5}$ respectively over $n$ number of samples. Similarly, $\bar{\Sigma}_{4}$ and $S_{\Sigma_{4}}$ are the corresponding quantities for $\Sigma_{4}$.
When the value of $\alpha$ increases, the confidence in data $(1-\alpha)$ decreases. So, the value of $t_{W}$ increases because the value of $t_{\frac{\alpha}{2}}$ is computed from the minimum side of the distribution and as per Equation~(\ref{zmin}), the value of $W_{lb}$ decreases. Thus, in Equation~(\ref{hatqlb}), as $\alpha$ increases, $\Delta$ increases, and so $\hat{q}_{lb}$ decreases. Note that, all the statistical analysis is done for certain fixed data~\cite{feller}.

Given an unknown state, to find whether that state is NMES or not, create an experimental set-up where Equations~(\ref{eq:1})-(\ref{eq:3}) are satisfied theoretically and Equations~(\ref{eq:5})-(\ref{eq:7}) are validated experimentally. Then we use a two-phase procedure as presented in Algorithm~\ref{alg1}. 
\RestyleAlgo{boxed}
\begin{algorithm}[h!] 
\textbf{OFF-LINE PHASE (Estimation of $\bar{\Sigma}_{4}$ from unknown MES \& PS)}: \\
\nl Do the experiment for Equation~(\ref{eq:8}) for $k$ numbers of known MES as well as PS to get the values of $\bar{\epsilon}_{4}$.\\
\nl Then from that data, calculate the values of $\bar{\Sigma}_{4}= \max \lbrace\bar{\epsilon}_{4}\rbrace$ and $S_{\Sigma_{4}}$.\\ \ \\
\textbf{ON-LINE PHASE (Estimation of $\hat{q}_{lb}$ from the unknown state)}:\\ 
\nl  Do the experiment for Equation~(\ref{eq:8}) to calculate $\bar{\epsilon}_{5}$ and $S_{\epsilon_{5}}$ for the unknown state. \\
\nl Calculate the value of $\Delta$ from Equation~(\ref{hatqlb}). \\
\nl Plug-in the values of $\bar{\epsilon}_{5}$, $\bar{\Sigma}_{4}$ and $\Delta$ in Equation~(\ref{hatqlb}) to get an estimate $\hat{q}_{lb}$ of the lower bound $q_{lb}$ on $q$.\\
\nl \If{$\hat{q}_{lb} > 0$ (i.e., if $\bar{\epsilon}_{5} > \bar{\Sigma}_{4} + \Delta$ )} {
the unknown state is NMES and the value of Hardy's probability $q$ of that unknown state is greater than or equals to the value of $\hat{q}_{lb}$, i.e., $q \geq \hat{q}_{lb}$} 
\Else{
no decision can be made about the unknown state.}
\caption{Estimation for the lower bound $q_{lb}$ on $q$, i.e., $\hat{q}_{lb}$ for superconducting qubits.}
\label{alg1}
\end{algorithm}

In this way, we can identify whether an unknown state is NMES or not with some confidence. If it is NMES, then we can estimate a lower bound on Hardy's probability of that NMES.
The two-phase process is similar to the scenario of classical error-control coding \cite{shanon47} wherein off-line phase channel noise is estimated using known messages and subsequently that estimation is used in the on-line phase to correct the transmission error of unknown messages. 

\section{Our experimental results and discussion}  \label{Experimental Results and Discussion}

The confidence interval around the mean for Student's t-distribution depends upon two quantities. 
\begin{enumerate}
\item The degrees of freedom for Student's t-distribution which is one less than the number of samples taken for the experiment.
\item The percentage of confidence we need on our data.
\end{enumerate} 
\noindent Based on these two quantities, we get the confidence interval around the mean for any data. 

In the IBM five-qubit quantum computer, any experiment can be performed for 1 shot or one of 1024, 4096 and 8192 (which is the maximum available) shots. It means that these many number of times the experiment is performed internally and the average of that is reported as output. But if we repeat any experiment with 8192 shots for a few times, we may see a significant deviation among each of the average values. Further, if we perform any experiment for 8192 shots, then the IBM quantum experience does not provide us with all the samples of those shots, rather it only provides the mean value. So if we perform any experiment only once with 8192 shots and make a conclusion based on the result, then that is not a statistically correct way. Instead, if someone does the experiment $n$ number of times with 8192 shots/time and then estimates the mean and the standard deviation assuming Student's t-distribution with required confidence interval, then that would be statistically more accurate. We will show the variation of the same result by
 varying $n$ and the percentage of confidence interval for Hardy's experiment.
 
 \begin{figure}[t!]
\centering
\includegraphics[width=\columnwidth]{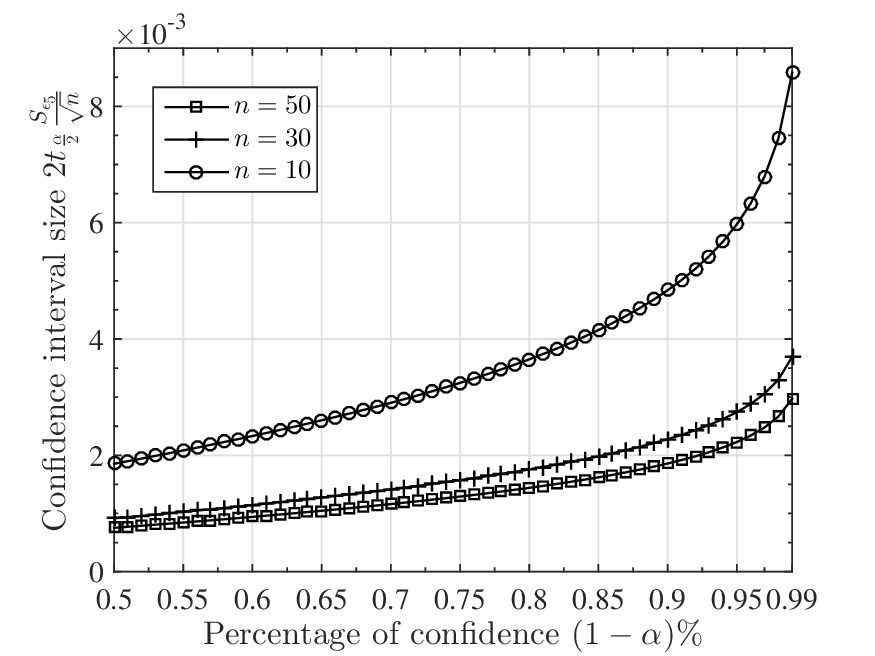}
\caption{The variation of the confidence interval size $\left( 2t_{\frac{\alpha}{2}}\frac{S_{\epsilon_{5}}}{\sqrt{n}} \right) $ with the percentage of confidence $\left( 1 - \alpha\right) \% $ with varying number of samples is $n \in \lbrace10,30,50\rbrace$.}
\label{fig:7}
\end{figure} 

In Figure~\ref{fig:7}, we show the variation of the confidence interval size $ 2t_{\frac{\alpha}{2}}\frac{S_{\epsilon_{5}}}{\sqrt{n}}  $ with the percentage of confidence $\left( 1 - \alpha\right) \% $ with varying number of samples $n \in \lbrace10,30,50\rbrace$. Each of the samples is the mean of Hardy's experiment run for 8192 shots. It can be seen that the confidence interval size increases with a decreasing number of samples. Also, for a fixed number of samples, the interval size increases with an increasing percentage of confidence in data. From this result, it can be concluded that if an experiment in the IBM quantum computer is done only on one sample for 8192 shots, then that would not be a statistically correct way of representation since each of the samples can deviate from the mean significantly. 
As we have limited access to the IBM quantum computer, we take $n=10$ and the experiments are run for MES, PS, and NMES for some values of $\theta$ and $\phi$ in between $0$ to $90$ degrees as shown in Table~\ref{table:4} and Table~\ref{table:5}. 

We perform experiments with different combinations of qubits for Alice and Bob and discuss the results below. 

\subsection{Experiments for Hardy's non-locality using two specific qubits for Alice and Bob}
In this section, we check Hardy's non-locality by taking a specific combination of qubits from all available combinations in the IBM quantum computer. Later we will discuss rest of the possible combinations.

\subsubsection{Experimental validation of the circuit for Hardy's test}
We perform the experiments for Equations~(\ref{eq:5}),~(\ref{eq:6}), and~(\ref{eq:7}) (Figures~\ref{fig:2},~\ref{fig:3} and~\ref{fig:4} respectively) for the values of $\theta$ and $\phi$ in degrees. The values of $\bar{\epsilon}_{1}$, $\bar{\epsilon}_{2}$ and $\bar{\epsilon}_{3}$ in each of the experiments are found to be less than $0.1$, implying that Equations~(\ref{eq:5}),~(\ref{eq:6}) and~(\ref{eq:7}) are satisfied. These results indicate that this experimental set-up is now valid for Hardy's test. As discussed in Section \ref{Practical Hardy's test}, for practical verification of Hardy's test using superconducting qubits, only one experiment for  Equation~(\ref{eq:8}) (Figure~\ref{fig:5}) has to be performed to establish non-locality, that is why the details of the results for $\bar{\epsilon}_{1}$, $\bar{\epsilon}_{2}$ and $\bar{\epsilon}_{3}$ are not presented. The experimental results of Equation~(\ref{eq:8}) are presented in Table~\ref{table:4} and Table~\ref{table:5} for some selected values of $\theta$ and $\phi$. 

\begin{table} [t!]
\centering 
\begin{tabular}{ p{0.7cm}| p{0.9cm}| p{1.1cm}| p{0.9cm}|  p{1.2cm}| p{1.2cm}| p{1.2cm}| p{1.2cm}} 
\hline
 State & $\theta$, $\phi$  & $\bar{\epsilon}_{4}$ & SD & \multicolumn{4}{c}{ $t_{\frac{\alpha}{2}}\frac{S_{\epsilon_{4}}}{\sqrt{n}}$ for different CIs}\\
 \cline{5-8} & & ($q=0$) & ($S_{\epsilon_{4}}$)&  $99\%$ & $95\%$ & $90\%$ & $80\%$ \\
\hline 
 MES  & 45, 90 & 0.0807 & 0.0037 & 0.003802	& 0.002647 &	0.002145 &	0.001618\\
\hline 
PS & 0, 0  &0.0193 & 0.0014 & 0.001439 & 0.001001 & 0.000812 & 0.000612 \\
\hline
PS & 90, 0  &0.0209 & 0.0015 & 0.001542 &0.001073  & 0.00087 & 0.000656\\
\hline
PS & 45, 0  & 0.0217 & 0.0019 & 0.001953  & 0.001359  & 0.001101 & 0.000831\\
\hline
PS & 90, 45  & 0.0282 & 0.0013 & 0.001336 & 0.00093 & 0.000754 & 0.000569 \\
\hline
\end{tabular}
\caption{Results of $\bar{\epsilon}_{4}$, standard deviations (SD) and the values of $  t_{\frac{\alpha}{2}}\frac{S_{\epsilon_{4}}}{\sqrt{n}}  $ for different confidence intervals (CIs) with $n=10$ for some values of $\theta$ and $\phi$ for which  MES as well as PS are created for the pair of qubits $\left( Q_{3}, Q_{4}\right) $.}
\label{table:4}
\end{table}

\subsubsection{Test of non-locality when $q=q_{max}$} \label{TestofNL_q=qmax}
From Table~\ref{table:4}, for the MES, i.e., $\theta=45$ and $\phi=90$ degrees, we get $\bar{\epsilon}_{4}=0.0807$, $S_{\epsilon_{4}}=0.0037$, and the values of $t_{\frac{\alpha}{2}}\frac{S_{\epsilon_{4}}}{\sqrt{n}}$ for different confidence intervals. We take four possible confidence intervals: $99\%$, $95\%$, $90\%$, and $80\%$. Similarly, for the PS, we get $\bar{\epsilon}_{4}$ to be less than $0.03$ (there are only one MES possible but ideally an infinite number of PS possible as shown in Table~\ref{table:1}. But due to limited access to the IBM quantum computer, we take $k=50$ number of PS as indicated in Algorithm 1. As for all the PS, we get $\bar{\epsilon}_{4}$ to be less than $0.03$, we present some of the values of the PS in Table~\ref{table:4}). As stated earlier, when $\theta=\phi=51.827$ degrees, we get $q=q_{max}$ for NMES. So, to test the non-locality when $q=q_{max}$, we have to check whether $\hat{q}_{lb}>0$ as given in Equation~\eqref{hatqlb}. From Table~\ref{table:4}, we get $\bar{\Sigma}_{4}=0.0807$ and $S_{\Sigma_{4}}=0.0037$ which is value we get from the off-line phase as stated in Algorithm~\ref{alg1}. This value is constant for the pair of qubits $\left(Q_{3}, Q_{4}  \right) $ and will be different for other pairs. Now we will calculate the value of $\hat{q}_{lb}$.

From the experiment, we get $\bar{\epsilon}_{5}=0.1281$ with $S_{\epsilon_{5}}=0.0039$ and the values of confidence intervals are given for $\left( 1-\alpha\right) \in \left\lbrace 0.99,0.95,0.90,0.80\right\rbrace $. To calculate  $\hat{q}_{lb}$ with different confidence intervals, we will use Equation~(\ref{hatqlb}).  
From Table~\ref{table:6}, when  $\theta=\phi=51.827$ degrees, we get  $\hat{q}_{lb} > 0$ for different confidence intervals. 
The error in estimating the value of $\hat{q}_{lb}$ is approximately $52\%$. If we increase the number of samples in the experiment, i.e., the value of $n$, then $\hat{q}_{lb}$ will also increase as seen in Figure~\ref{fig:7}. But despite the errors in the experiment, when $q=q_{max}$, we get the lower bound  on Hardy's probability greater than zero with $99\%$ confidence on the data. 

\begin{table}[t!]
\centering
\begin{tabular}{ p{1.5cm}| p{1.5cm}| p{1.7cm}| p{1.5cm}| p{1.5cm}| p{1.5cm}| p{1.5cm}| p{1.5cm}}
\hline
 State & $\theta$, $\phi$  & $\bar{\epsilon}_{5}$ & SD & \multicolumn{4}{c}{ $t_{\frac{\alpha}{2}}\frac{S_{\epsilon_{5}}}{\sqrt{n}}$ for different CIs}\\
 \cline{5-8} & & $(q > 0)$ & ($S_{\epsilon_{5}}$)&  $99\%$ & $95\%$ & $90\%$ & $80\%$ \\
\hline 
NMES & 51.827, 51.827  &0.1281 & 0.0039 & 0.004008 & 0.00279  &  0.002261 & 0.001706 \\
\hline
 NMES & 55, 55  &0.1273 &  0.0045 & 0.004625 &  0.003219 & 0.002609 & 0.001968  \\
  \hline
NMES & 45, 45  &0.1041 &  0.0044 & 0.004522 & 0.003148 &  0.002551 &  0.001924 \\
 \hline 
 NMES & 30, 60   &0.0832 & 0.0052 &  0.005344 &  0.00372 & 0.003014 & 0.002274 \\
 \hline
   NMES & 60, 30  &0.0553 & 0.0028 & 0.002878 & 0.00200 & 0.001623 & 0.001225\\
\hline
NMES & 10, 80 & 0.067 & 0.0038 & 0.00391 & 0.00272 & 0.0022 & 0.00166 \\ 
 \hline
  NMES & 80, 10 &0.0241 & 0.0016 & 0.001644 & 0.001145 & 0.000927 &  0.0007\\
 \hline
\end{tabular}
 \caption{Results of $\bar{\epsilon}_{5}$, standard deviations (SD) and the values of $ t_{\frac{\alpha}{2}}\frac{S_{\epsilon_{5}}}{\sqrt{n}} $ for different confidence intervals (CIs) with $n=10$ for some values of $\theta$ and $\phi$ for which NMES are created for the pair of qubits $\left( Q_{3}, Q_{4}\right) $.} \label{table:5}
\end{table}

\subsubsection{Test of non-locality when $q < q_{max}$} \label{TestofNL_q<qmax}
From Table~\ref{table:6}, when $\theta=\phi=55$ degrees ($q=0.0886$), by a similar analysis, we get $\hat{q}_{lb}$ to be around $0.042$, for different confidence intervals, i.e., $\hat{q}_{lb}>0$. The error in estimating the value of $\hat{q}_{lb}$, in this case, is around $52.5\%$.
Similarly, when $\theta=\phi=45$ ($q=0.0833$), the value of $\hat{q}_{lb}$ is around $0.02$. But the error in this case in estimating $\hat{q}_{lb}$ is around $76\%$. Clearly these results support a non-zero value of Hardy's probability.

But when $\theta=30$ and $\phi=60$ ($q=0.0433$), we estimate $\hat{q}_{lb}<0$. Similar result is obtained  when $\theta=60$ and $\phi=30$ ($q=0.0433$). So, form these results we cannot conclude that the state is really an NMES or not when $q=0.0433$.

When the value of $q$ is decreased further, when $q=0.00088$, the values of $\hat{q}_{lb}$ again become less than zero. So, form these results, we cannot conclude about the state as discussed above.
\subsubsection{Summary of the above two experiments} \label{summary_of_two_exp}
We know that in Hardy's test, we should get Hardy's probability $q > 0 $ for all NMES. But, from the above experimental data, we get for some NMES, the estimated lower bound on Hardy's probability $\hat{q}_{lb} \leq 0$. The mismatch between theoretical value and the estimated value from the experiments can be explained with respect to Equation~(\ref{q_bound}) as follows:
\begin{table}[t!]
\centering
\begin{tabular}{ p{1.5cm}| p{1.3cm}| p{1.3cm}|  p{2cm}| p{2cm}| p{2cm}| p{2cm}}
\hline
 State & $\theta$, $\phi$  & $q$  & \multicolumn{4}{c}{ $\hat{q}_{lb}$ for different CIs }\\
 \cline{4-7} & & &  $99\%$ & $95\%$ & $90\%$ & $80\%$  \\
\hline
NMES & 51.827, 51.827  & 0.09017  & 0.041876 & 0.043554 & 0.044283 & 0.045049 \\
\hline
 NMES & 55, 55  & 0.0886  & 0.040613 & 0.042432 & 0.043222 & 0.044052 \\
 \hline 
NMES & 45, 45  & 0.0833  & 0.017492 & 0.019287 &  0.020067 & 0.020886 \\
 \hline 
 NMES & 30, 60   &0.0433  & -0.004058 & -0.002066 & -0.001199 & -0.000290\\
 \hline
   NMES & 60, 30  &0.0433  & -0.030168 & -0.028718 & -0.02809 & -0.027429 \\
\hline
NMES & 10, 80 &  0.00088 & -0.019154 & -0.017495 & -0.016773 &  -0.016018 \\ 
 \hline
  NMES & 80, 10 & 0.00088 & -0.060742 & -0.059484 & -0.058937 &  -0.058363\\
 \hline
\end{tabular}
\caption{Comparison of the results of $q$ and $\hat{q}_{lb}$ for some values of $\theta$ and $\phi$ for which NMES is created for different confidence interval (CIs) for the pair of qubits $\left( Q_{3}, Q_{4}\right) $}
\label{table:6}
\end{table}

\begin{itemize}
\item for those NMES when $q$ is larger than $ \bar{\Sigma}_{4}$, the distinction between $\bar{\epsilon}_{5}$  and $\bar{\Sigma}_{4}$ is clear and $\hat{q}_{lb} > 0$.
\item But for those NMES when $q$ is less than $\bar{\Sigma}_{4}$, it is hard to distinguish between $\bar{\epsilon}_{5}$ and $ \bar{\Sigma}_{4}$, and we have $\hat{q}_{lb} \leq 0$.
\end{itemize}
we can observe that for those NMES when $q$ is larger than $ \bar{\Sigma}_{4}$, the distinction between $\bar{\epsilon}_{5}$ for those NMES and $\bar{\Sigma}_{4}$ for all MES as well as PS is clear, i.e., $\hat{q}_{lb} > 0$. But for those NMES when the value of $q$ is less than $\bar{\Sigma}_{4}$ for all MES as well as PS, it is hard to distinguish between $\bar{\epsilon}_{5}$ for those NMES and $ \bar{\Sigma}_{4}$ for all MES as well as PS, i.e., $\hat{q}_{lb} \leq 0$.

\begin{figure}[t!]
\centering
\includegraphics[width=\columnwidth]{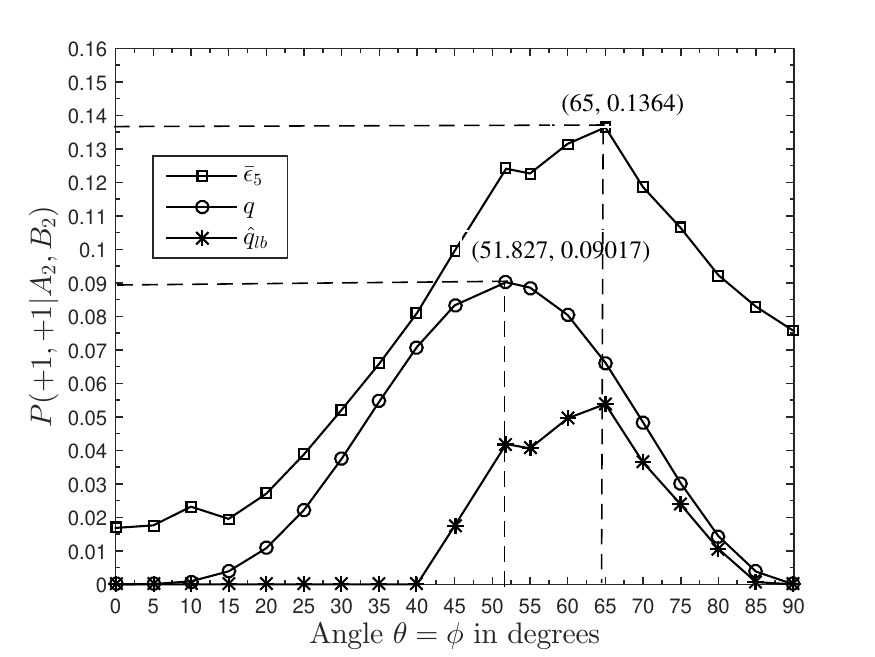}
\caption{The variation of $q$, $\epsilon_{5}$ and $\hat{q}_{lb}$ against
$\theta=\phi$ for CI=$99\%$ and $n=10$ with $Q_{3}$ as control qubit and $Q_{4}$ as target qubit.} 
\label{fig:8}
\end{figure}

\subsubsection{Consistency check} \label{ShiftQ3Q4}
The above summary indicates that as $q$ decreases, so does $\bar{\epsilon}_{5}$. Thus it is a natural question to ask, whether at $q=q_{max}$, we get $\bar{\epsilon}_{5}=\bar{\epsilon}_{5}^{max}$ or not, where
\begin{equation}\label{Epsilon_5max}
\bar{\epsilon}^{max}_{5}=\max_{0 \leq \theta=\phi \leq 90} \lbrace \bar{\epsilon}_{5_{\theta=\phi}} \rbrace.
\end{equation} 
 We conduct another set of experiments to investigate this. If we take  $\theta=\phi$ and vary it from $0$ to $90$ degrees, a bell-shaped curve is obtained with a peak at $\theta=\phi=51.827$ degrees for $q$ as shown in Figure~\ref{fig:8} (a). For limited control of the IBM quantum computer, we plot $\bar{\epsilon}_{5} $,  $q$  and  $\hat{q}_{lb}$ with $\left( 1-\alpha\right)=0.99$ for  $\theta=\phi$ varying from $0$ to $90$ degrees with an increment of $5$ degrees, i.e.,  $\theta=\phi=5i$, where $i \in \lbrace 0,1,\ldots,18\rbrace$ (when $i=18$ we get $\theta=\phi=90$, then the value of $\chi$ is undefined as shown in the Section~\ref{app1}. So we take $\theta=\phi=89.99$ for this case).


From Figure~\ref{fig:8} (a), we observe $ \bar{\epsilon}^{max}_{5} $, as defined in Equation~(\ref{Epsilon_5max}), is shifted right where $\theta=\phi=65$ degrees, instead of $\theta=\phi=51.827$ degrees. To get a more accurate result of $\bar{\epsilon}_{5}^{max}$, we repeat this experiment for $75 \geq \theta=\phi \geq 55$ degrees with an increment of one degree. The result shows that $\bar{\epsilon}_{5}^{max}$ occurs where $\theta=\phi=62$ degrees (not shown in Figure~\ref{fig:8} (a)). This procedure can be repeated again  if the more precise occurrence of $\bar{\epsilon}_{5}^{max}$ is needed. We also observe that $\hat{q}_{lb}>0$ when $85 \geq \theta=\phi \geq 40$ degrees and zero elsewhere, whereas $q > 0$ for $90 > \theta=\phi > 0$ degrees. This mismatch happens because of the reasons stated in Section~\ref{summary_of_two_exp}. From these results it can be concluded that although in the IBM quantum computer, we get a non-zero value of the lower bound on Hardy's probability for NMES, but the errors occurred in the computer need to be more stable.

\subsubsection{Summary}
In summary, for the parameters $\theta$ and $\phi$ that lead to $q = 0$, our experimental outcome gives us $ \bar{\epsilon}_{4} $ for MES and PS. For the parameters ($\theta$ and $\phi$) that lead to $q > 0$, our experimental outcome gives us $\bar{\epsilon}_{5} $ for NMES. From Equation~(\ref{hatqlb}), the value of $\bar{\epsilon}_{5}$ for NMES is greater than $\left( \bar{\Sigma}_{4} + \Delta \right) $ implying  $\hat{q}_{lb}>0$. But when the value of $\bar{\epsilon}_{5}$ is in the same range or less than $\bar{\epsilon}_{4}^{max}$, then it implies $\hat{q}_{lb} \leq 0$.

\subsection{Other possible combinations of multi-qubit gate} \label{Other_CNOT_Gate}
In this section, we check Hardy's non-locality for the rest of the available combinations of qubits in the IBM quantum computer.

 \begin{table} [h!]
\centering
\begin{tabular}{ p{1.5cm}   p{1.5cm}  p{2cm}  p{2cm}   p{2cm}}
 \hline   Control  & Target & \multicolumn{3}{c}{values of $\bar{\epsilon}_{5}$ for $n=10$ for $\theta=\phi$} \\
\cline{3-5} qubit & qubit & \hspace{4mm} 45 &  \hspace{1mm} 51.827 &  \hspace{4mm} 55\\
 \hline 
  $Q_{2}$ & $Q_{0}$   &0.101022 &  0.111367 & 0.116217 \\

  $Q_{3}$ & $Q_{2}$   & 0.094067  & 0.112275 &   0.114400 \\
 $Q_{1}$ & $Q_{0}$   & 0.099931  & 0.119311 & 0.119497 \\

 $Q_{2}$ & $Q_{1}$   & 0.138400 & 0.136175 & 0.134289
 \\

 $Q_{2}$ & $Q_{4}$   & 0.138505 & 0.150186 & 0.157253\\
\hline
\end{tabular}
 \caption{The values of $\bar{\epsilon}_{5}$ for $n=10$ for all possible pairs of qubits other than $\left( Q_{3},Q_{4}\right)$ when $\theta=\phi \in \left\lbrace 45,51.827,55\right\rbrace $ in degrees.}  \label{table:7}
\end{table}
\subsubsection{Check for non-locality}
There are currently six combinations of multi-qubit gate implementation available  in \textit{ibmqx4}~\cite{IBM}. For the multi-qubit $CNOT$ gate that we use, the possible control qubit and target qubit pairs other than $\left( Q_{3},Q_{4}\right)$, are summarized in Table~\ref{table:7}. We measure $\bar{\epsilon}_{5}$ for $\theta=\phi \in \left\lbrace 45,51.827,55\right\rbrace $ degrees with all combinations of pairs of qubits. It can be seen, when $\theta=\phi=51.827$ degrees, the value of $\bar{\epsilon}_{5}$ is minimum for the pair $\left( Q_{2},Q_{0}\right)$ and maximum for the pair $\left( Q_{2},Q_{4}\right)$. All the experiments described earlier have been performed using all combinations of these pair of qubits as described in Table~\ref{table:7} and similar conclusions of the non-locality are obtained as we get for the pair $\left( Q_{3},Q_{4}\right)$.

\subsubsection{Consistency check and shift of the peaks} \label{Shiftother}
We want to see, for other possible two-qubit pairs, whether $ \bar{\epsilon}_{5}^{max} $ (as defined in Equation~(\ref{Epsilon_5max})) occurs when $q=q_{max}$  or not. We also want to see, in case there is any shift, whether it is in the same direction as shown in Figure~\ref{fig:8} for the pair $\left( Q_{3},Q_{4}\right)$ or not.

\begin{figure}[t!]
\centering
\includegraphics[width=\columnwidth]{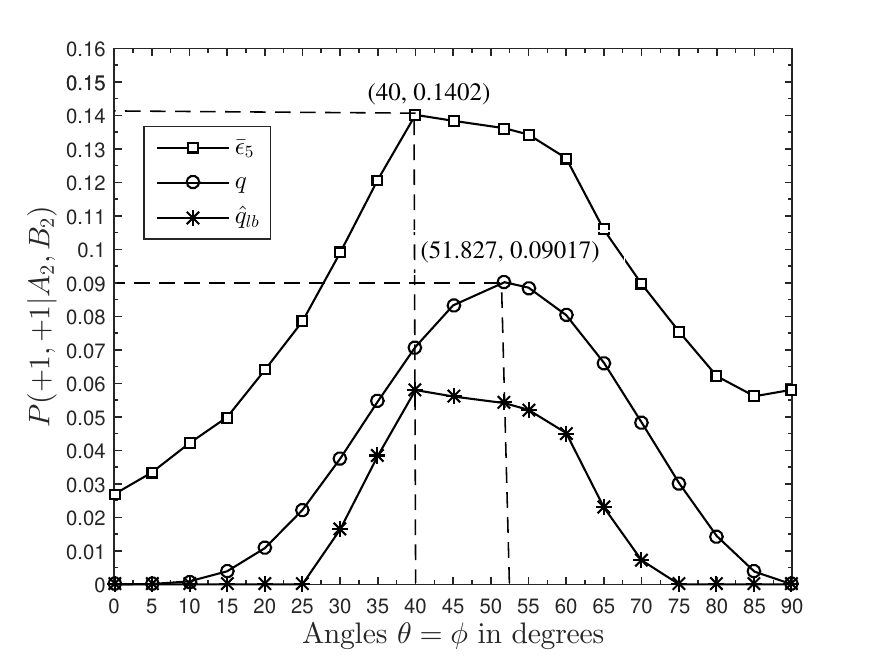}
\caption{The variation of  $q$, $\epsilon_{5}$ and $\hat{q}_{lb}$  
against $\theta=\phi$ for CI=$99\%$ and $n=10$ with
$Q_{2}$ as control qubit and $Q_{1}$ as target qubit.}
\label{fig:9}
\end{figure}

From Table~\ref{table:7}, it can be seen for the pair $\left( Q_{2},Q_{1}\right)$, when $\theta=\phi=51.827$ degrees, the value of $\bar{\epsilon}_{5}$  is less than the  value when $\theta=\phi=45$ degrees and greater than the value when $\theta=\phi=55$ degrees, i.e., $ \bar{\epsilon}_{5_{\theta=\phi=45}}>\bar{\epsilon}_{5_{\theta=\phi=51.827}} > \bar{\epsilon}_{5_{\theta=\phi=55}} $. To verify this result, we perform a similar experiment for the pair $\left( Q_{2},Q_{1}\right)$ as we did  for the pair $\left( Q_{3},Q_{4}\right)$ (as shown in Figure~\ref{fig:8}). Results are shown in Figure~\ref{fig:9} which indicates that there is a shift of the value of $\bar{\epsilon}_{5}^{max}$ to the left for the pair $\left( Q_{2},Q_{1}\right)$ and it occurs when $\theta=\phi=40$ degrees. 
For the rest of the pairs, we find that  $\bar{\epsilon}_{5}^{max}$ is shifted to the right, i.e., $ \bar{\epsilon}_{5_{\theta=\phi=55}}>\bar{\epsilon}_{5_{\theta=\phi=51.827}} > \bar{\epsilon}_{5_{\theta=\phi=45}} $ as shown in Table~\ref{table:7}. So, we can conclude that $\bar{\epsilon}^{max}_{5}$ did not occur at $q=q_{max}$, rather it is shifted to the right or to the left due to the errors.

\subsection{Application of the shift of the peaks in quantum  protocols using Hardy's test} \label{App_shift}
For some of the protocols like quantum  Byzantine agreement~\cite{QBA}, it is necessary to check whether Hardy's state (the state for which Equations~\eqref{eq:1}-\eqref{eq:4} are satisfied) is actually prepared or not. Suppose, theoretically  $q=q_{max}$ is achieved for a specific value $\rho$ of the relevant parameter (of which $q$ is a function, like $\theta$ and $\phi$ in our experiment). Because of the peak shifts as described in Section~\ref{ShiftQ3Q4} and~\ref{Shiftother}, in practical experiments, $q_{max}$ should not be estimated corresponding to the exact value of $\rho$, but in an interval around $\rho$. 

During the experiment, let the value of $q_{max}$ with the addition of errors be $Q_{max}$ (like $\bar{\epsilon}^{max}_{5}$ in our experiment). Now if the errors in the experiment are not stable, it is expected that $Q_{max}$ will lie in an interval around $\rho$, i.e., $\left\lbrace \rho-\delta,\rho+\delta \right\rbrace$ for some $\delta > 0$. For example, in our experimental set-up, $\delta$ is found to be $12$ degrees when $\rho=\left( 51.827, 51.827 \right) $ in degrees of the parameter  $ \left( \theta , \phi \right) $.  

\subsection{Verifying whether reducing the number of gates reduces the error in the circuit}
\begin{figure}[h!]
\centering
\subfloat[Quantum circuit for the PS $\ket{\psi} = \frac{1}{\sqrt{2}}(\ket{0} + \ket{1}) \otimes \ket{0}$ when  $\theta=\phi=0$ degrees.]{\includegraphics[width = 0.7\linewidth]{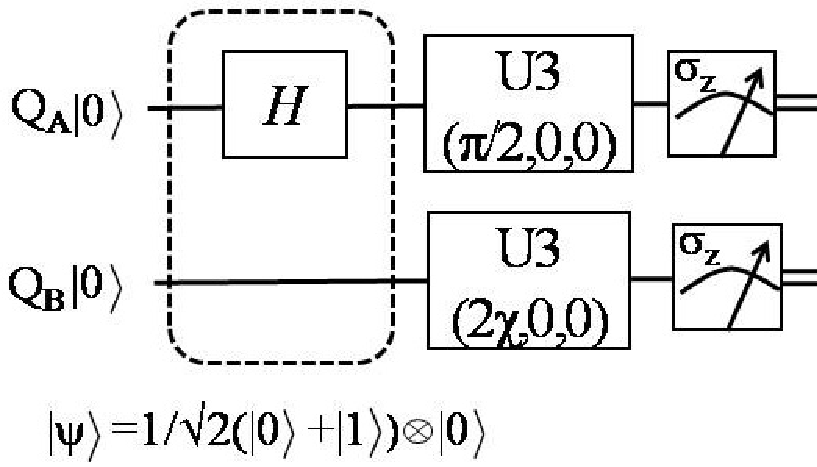}} \\
\subfloat[Quantum circuit for the PS $|\psi\rangle = \frac{1}{\sqrt{2}}(|0\rangle + |1\rangle) \otimes |1\rangle $ when  $\theta=90$ and $\phi=0$ degrees.]{\includegraphics[width = 0.7\linewidth]{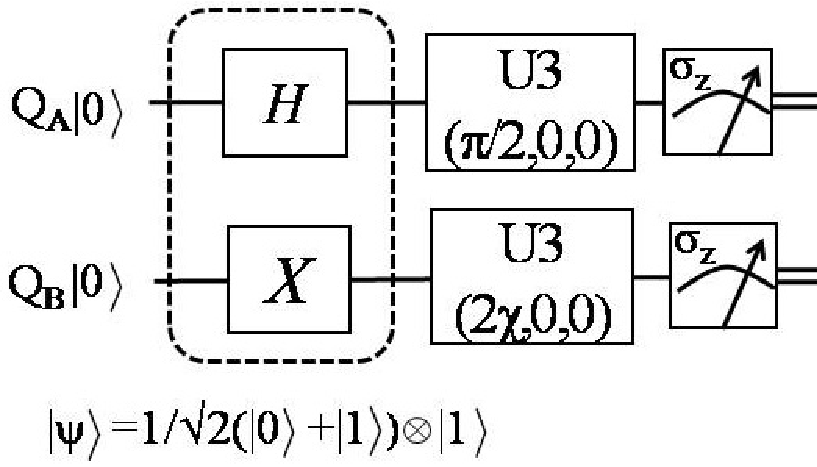}}
        \caption{Quantum circuit and measurement for $P(+1,+1|A_{1},B_{2})$ when number of gates are reduced significantly. }\label{fig:10}
\end{figure}

To verify whether reducing the number of gates in the circuit reduces the error or not, we perform another series of experiments for $\left( Q_{3},Q_{4}\right)$ pair of qubits. For  $\theta=\phi=0$, we get a PS, i.e., $|\psi\rangle = \frac{1}{\sqrt{2}}(|0\rangle + |1\rangle) \otimes |0\rangle $. This state can be created easily by using a Hadamard gate $H$ in Alice's qubit and $U_{1}$ and $U_{3}$ used  in Figure~\ref{fig:5}, becomes the identity ($Id$) gate. So, the number of gates is reduced significantly. For the modified circuit as shown in Figure~\ref{fig:10} (a), the value of $\bar{\epsilon}_{4}=0.0084$ for $n=10$ is less than the previous value (as shown in Table~\ref{table:4}), i.e., $ 0.0193 $ 

Also, for $\theta=90,\phi=0$,  we get a PS $|\psi\rangle = \frac{1}{\sqrt{2}}(|0\rangle + |1\rangle) \otimes |1\rangle $. For this state, a Hadamard gate on Alice's qubit and a bit flip gate $X$ on Bob's qubit is required as shown in Figure~\ref{fig:10} (b). Experimental results show that $\bar{\epsilon}_{4}=0.0079$ for $n=10$ is again less than the previous value (as shown in Table~\ref{table:4}), i.e., $0.0209$.

These experiments are repeated for the rest of the pairs of qubits and similar results are obtained. We can conclude that reducing the number of gates reduces the error in the circuit of the IBM quantum computer. 
 
\subsection{Study of change of errors with respect to time in superconducting qubits}
 \begin{table} [h!]
\centering
\begin{tabular}{p{1.5cm} p{1.5cm}   p{1.5cm}  p{1.5cm}  p{1.5cm}   p{1.5cm}  p{1.5cm}}
 \hline  Time  & $\bar{\epsilon}_{5}$ & SD & \multicolumn{4}{c}{ $t_{\frac{\alpha}{2}}\frac{S_{\epsilon_{5}}}{\sqrt{n}}$ for different CIs} \\
 \cline{4-7}line & &($ S_{\epsilon_{5}}$) & \hspace{3mm} $99\%$ & \hspace{3mm} $95\%$  &  \hspace{3mm} $90\%$  &  \hspace{3mm} $80\%$    \\
 \hline 
 Before & 0.16254 & 0.0078 & 0.154524 &	0.15696 &	0.158018 &	0.159129 \\
 After & 0.1281 &	0.0039 & 0.124092 & 0.12531	& 0.125839	& 0.126394 \\
\hline
\end{tabular}
\caption{The values of $\bar{\epsilon}_{5}$ for $n=10$ for  the pair $\left( Q_{3},Q_{4}\right)$ before and after one month when $\theta=\phi=51.827$ in degrees, SD=standard deviation, and CI=confidence interval.} \label{table:8}
\end{table} 
During the experiments, the IBM quantum computer has undergone maintenance for nearly one month when some of the experiments were done. To present all the results in the same time-line, we  repeat all the previous experiments. When we compare the data of one month earlier experiments, we find a significant change of values, similar to what was reported in~\cite{mermin2016}. For  the pair $\left( Q_{3},Q_{4}\right)$ when $\theta=\phi=51.827$ degrees, one month earlier, we got the value of $\bar{\epsilon}_{5}=0.16254$, $S_{\epsilon_{5}}= 0.0078$ and the values of $t_{\frac{\alpha}{2}}\frac{S_{\epsilon_{5}}}{\sqrt{n}}$ for different values of confidence intervals are given in Table~\ref{table:8}. From this, we can see that the value of $\bar{\epsilon}_{5}$ one month earlier is greater than the values of what we get one month later. A similar trend is noticed for the rest of the data.

\subsection{Benchmarking of superconducting quantum devices using Hardy's paradox} \label{Ben_hardy}
Benchmarking is the process by which the performance of any computing device is evaluated. In the current era of Noisy Intermediate-Scale Quantum (NISQ)~\cite{Preskill2018}, choosing a universal metric for benchmarking is extremely difficult, because computing can be performed using various quantum technologies such as 
 superconducting qubits~\cite{Gambetta2017,Krantz2019}, ion trap~\cite{HAFFNER2008}, optical lattices~\cite{Treutlein2006}, quantum dots~\cite{Loss1998}, nuclear magnetic resonance (NMR)~\cite{lu2016nmr}, etc. Also, how the noise affects the devices and their characterization are not well-understood.

Researchers have proposed different metrics for benchmarking, for instance, fidelity~\cite{Beale2018}, unitarity~\cite{Wallman2015}, quantum volume~\cite{Cross2019}, quantum chemistry~\cite{McCaskey2019}, etc., each having their own drawbacks~\cite{1912.00546}. In~\cite{mermin2016}, the authors have also demonstrated non-locality in the case of Mermin polynomials for three, four and five qubits. They have concluded that the fidelity of the quantum computer decreases when the number of qubits is increased from three. However, they do not mention anything for two qubits. From the experimental test of Hardy's paradox, we propose two metrics for benchmarking of two qubits of any superconducting quantum device as discussed below.

First, from Section~\ref{TestofNL_q=qmax} and~\ref{TestofNL_q<qmax}, it can be seen that, for those states where $q > \bar{\Sigma}_{4}$,  we get $\hat{q}_{lb} > 0$ which supports a non-zero value of Hardy's probability. When $\bar{\epsilon}_{4}^{max}< q <q_{max}$, the previous conclusion is still valid.  But for those states where  $q < \bar{\Sigma}_{4}$, no conclusion can be drawn about the value of  Hardy's probability as $\hat{q}_{lb} \leq 0$. So, the minimum value of $q$ from which we get $\hat{q}_{lb} > 0$ (as illustrated in Table~\ref{table:6}), can be the performance measure of a quantum computer. Lesser the value of $q$ which supports $\hat{q}_{lb} > 0$, the better the performance of the quantum computer. This parameter measures how well the device realizes NMES states.

Second, although we get $q_{max}$ at $\theta=\phi=51.827$ degrees, during the experiment, due to unstable errors, the value of $\bar{\epsilon}_{5}^{max}$ may be shifted to any nearby value, as shown in Section~\ref{ShiftQ3Q4} and~\ref{Shiftother}. In our case, it shifted to the left for the pair $\left( Q_{2},Q_{1}\right)$ and to the right for the rest of the pairs of qubits. For our experiment when $\theta=\phi$, the amount of shift is $12$ degrees. So, the amount of shift can be considered as a performance measure of a quantum computer. The smaller the shift, the better is the performance. This parameter measures the precision of the quantum device.
\chapter{Quantum teleportation and hyper-hybrid entangled state}\label{Chap4}
In this Section, we show that some properties and applications of distinguishable and indistinguishable entanglement. First, we show that hyper-hybrid entanglement can exists for two indistinguishable fermions. Then I have generalized it for both bosons and fermions. Next, we prove that \textit{quantum particles (either distinguishable or indistinguishable) can simultaneously produce and perform hyper hybrid entangled state and unit fidelity quantum teleportation respectively then using that cloning of any arbitrary quantum state is possible}. As the no-cloning theorem cannot be violated using in quantum theory, the logical conclusion from this statement can be written in the form of two no-go theorems; (i) \textit{no-hyper hybrid entangled state for distinguishable particles} and (ii) \textit{no-unit fidelity quantum teleportation for indistinguishable particles}. Finally we show the overall picture of the properties and applications of distinguishable and indistinguishable particles.  

 This chapter is based on the work in~\cite{Das20}.

\section{Hyper-hybrid entangled state for two indistinguishable fermions} \label{HHESfermions} 
Yurke and Stolar~\cite{Y&S92PRA,Y&S92PRL} had proposed an optical circuit to generate quantum entanglement between the same DoFs of two identical particles (bosons and fermions) from initially separated independent sources. Some recent
experiments realizing entanglement between the same DoFs of identical particles whose degree of spatial indistinguishability can be arbitrarily harnessed~\cite{Sun2020,Sun2022,Wang2021,Wang2022}. The experiments are realized not only with photons but also simulating fermions.
Recently, the above method has been extended by Li \textit{et al.}~\cite{HHNL} to generate hyper-hybrid entangled state between two independent bosons among their internal (e.g., spin) DoFs, external (e.g., momentum) DoFs, and across. We show that their circuit can also be used for independent fermions obeying the Pauli exclusion principle~\cite{Pauli25}, albeit with different detection probabilities.

For fermions, the second quantization formulation deals with fermionic creation operators $f_{i,\textbf{p}}$ with $\ket{i,\textbf{p}}=f^{\dagger}_{i,\textbf{p}}\ket{0}$, where $\ket{0}$ is the vacuum and $\ket{i,\textbf{p}}$ describes a particle with spin $\ket{i}$ and momentum $\textbf{p}$. These operators satisfy the canonical anticommutation relations:
\begin{equation} 
\left\lbrace f_{i,\textbf{p}_{i}}, f_{j,\textbf{p}_{j}}\right\rbrace = 0, \hspace{0.2cm} \left\lbrace f_{i,\textbf{p}_{i}}, f^{\dagger}_{j,\textbf{p}_{j}}\right\rbrace = \delta\left( \textbf{p}_{i}-\textbf{p}_{j}\right)\delta_{ij}.
\end{equation}

Analysis of the circuit of Li \textit{et al.}~\cite[Fig. 2]{HHNL} as shown in Fig.~\ref{fig:Lietal} for fermions involves an array of hybrid beam splitters (HBSs)~\cite[Fig. 3]{HHNL}; phase shifter; four orthogonal external modes $L$, $D$, $R$, and $U$; and two orthogonal internal modes $\uparrow$ and $\downarrow$. Here, particles exiting through the modes $L$ and $D$ are received by Alice (A), who can control the phases $\phi_{L}$ and $\phi_{D}$, whereas particles exiting through the modes $R$ and $U$  are received by Bob (B), who can control the phases $\phi_{R}$ and $\phi_{U}$. 

In this circuit~\cite[Fig. 2]{HHNL} as shown in Fig.~\ref{fig:Lietal}, two particles, each with spin $\ket{\downarrow}$, enter the setup in the mode $R$ and $L$ for Alice and Bob, respectively. The initial state of the two particles is $\ket{\Psi_{0}}=f^{\dagger}_{\downarrow,R} f^{\dagger}_{\downarrow,L}\ket{0}$. Now, the particles are sent to the HBS such that one output port of the HBS is sent to the other party ($R$ or $L$) and the other port remains locally accessible ($D$ or $U$). Next, each party applies path-dependent phase shifts. Lastly, the output of the local mode and that received from the other party are mixed with the HBS and then the measurement is performed in either  external or internal modes. The final state can be written as
\begin{equation} \label{Final_state_fermion}
\begin{aligned}
&\ket{\Psi}=\frac{1}{4} \left[ e^{i\phi_{R}}\left( f^{\dagger}_{\downarrow,R}+if^{\dagger}_{\uparrow,U}\right) +ie^{i\phi_{D}} \left( f^{\dagger}_{\uparrow,D} + i f^{\dagger}_{\downarrow,L}\right)  \right] \\
& \otimes \left[ e^{i\phi_{L}} \left( f^{\dagger}_{\downarrow,L}+if^{\dagger}_{\uparrow,D}\right)  + i e^{i\phi_{U}}\left( f^{\dagger}_{\uparrow,U}+if^{\dagger}_{\downarrow,R}\right) \right] \ket{0}.
\end{aligned}
\end{equation} 
Alice and Bob can perform coincidence measurements both in external DoFs or both in internal DoFs or with one party in the internal DoF and the other in the external DoF.  Now from Eq.~\eqref{Final_state_fermion}, the detection probabilities when each party gets exactly one particle where both Alice and Bob measure in external DoFs are given by 
\begin{equation}\label{path_path}
  \begin{tabular}{c | c c} 
  & $B$ : $R$ & $B$ : $U$  \\  
 \hline
 $A$ : $D$ \hspace{0.0005cm} & \hspace{0.0005cm} $\frac{1}{4}\text{cos}^2\phi$  & $\frac{1}{4}\text{sin}^2\phi$ \\ 
 $A$ : $L$ \hspace{0.0005cm} & \hspace{0.0005cm} $\frac{1}{4}\text{sin}^2\phi$  & $\frac{1}{4}\text{cos}^2\phi$ \\
\end{tabular},
\end{equation}
where $\phi=\left( \phi_{D} -\phi_{L} - \phi_{R}+ \phi_{U}\right) / 2$.

Now we assign dichotomic variables $+1$ and $-1$ for the detection events $\lbrace L,U \rbrace$ and $\lbrace D,R \rbrace$, respectively. Let $Pr_{mn}$ denote the probabilities of the coincidence events for Alice and Bob obtaining $m=\pm 1$ and $n=\pm 1$, respectively. The normalized expectation value is then given by
\begin{align} \label{Normal}
    E\left( \phi_{A},\phi_{B} \right)= &\dfrac{Pr_{++}-Pr_{-+}-Pr_{+-}+Pr_{--}}{Pr_{++}+Pr_{-+}+Pr_{+-}+Pr_{--}} \nonumber \\
    =& \text{cos} \left( \phi_{A} -\phi_{B} \right),
\end{align}
\noindent where $\phi_{A}=\left(\phi_{D}-\phi_{L}\right) $ and $\phi_{B}=\left( \phi_{U}-\phi_{R} \right)$. Now the CHSH~\cite{CHSH} inequality can be written as 
\begin{equation} \label{CHSH}
\vert E\left( \phi^{0}_{A},\phi^{0}_{B} \right) + E\left( \phi^{1}_{A},\phi^{0}_{B} \right)+ E\left( \phi^{0}_{A},\phi^{1}_{B} \right) - E\left( \phi^{1}_{A},\phi^{1}_{B} \right) \vert \leq 2,
\end{equation}
where the superscripts $0$ and $1$ stand for two detector settings for each particles. Now for $\phi^{0}_{A}=0$, $\phi^{1}_{A}=\pi$, $\phi^{0}_{B}=\frac{\pi}{4}$, and $\phi^{1}_{B}=-\frac{\pi}{4}$, Eq.~\eqref{CHSH} can be violated maximally by obtaining Tsirelson's bound $2\sqrt{2}$~\cite{Tsirelson}.

Now if Alice and Bob both measure	 in internal DoFs, then  the detection probabilities can be written as
\begin{equation}\label{spin_spin}
   \begin{tabular}{c | c c} 
  & $B$ : $\downarrow$ & $B$ : $\uparrow$  \\  
 \hline\
 $A$ : $\downarrow$  & \hspace{0.0005cm}  $\frac{1}{4}\text{sin}^2\phi$ \hspace{0.1cm} & $\frac{1}{4}\text{cos}^2\phi$\\ 
 \hspace{0.001cm} $A$ : $\uparrow$  & \hspace{0.0005cm}  $\frac{1}{4}\text{cos}^2\phi$ \hspace{0.1cm} & $\frac{1}{4}\text{sin}^2\phi$ \\
\end{tabular}.
\end{equation}
If Alice measures in the internal DoF and Bob measures in  the external DoF, then  the detection probabilities can be written as 
\begin{equation} \label{spin_path}
\begin{tabular}{c | c c} 
  & $B$ : $R$ & $B$ : $U$  \\  
 \hline
 $A$ : $\downarrow$  & \hspace{0.0005cm} $\frac{1}{4}\text{sin}^2\phi$ \hspace{0.1cm} & $\frac{1}{4}\text{cos}^2\phi$\\ 
 $A$ : $\uparrow$ & \hspace{0.0005cm} $\frac{1}{4}\text{cos}^2\phi$ \hspace{0.1cm} & $\frac{1}{4}\text{sin}^2\phi$ \\
\end{tabular}.
\end{equation}
If Alice measures in the external DoF and Bob measures in the internal DoF, then the detection probabilities can be written as 
\begin{equation} \label{path_spin}
\begin{tabular}{c | c c} 
  & $B$ : $\downarrow$ & $B$ : $\uparrow$  \\  
 \hline\
 $A$ : $D$  & \hspace{0.0005cm} $\frac{1}{4}\text{cos}^2\phi$ \hspace{0.1cm} & $\frac{1}{4}\text{sin}^2\phi$\\ 
 $A$ : $L$  & \hspace{0.0005cm} $\frac{1}{4}\text{sin}^2\phi$  \hspace{0.1cm} & $\frac{1}{4}\text{cos}^2\phi$ \\
\end{tabular}.
\end{equation}
 Now by applying similar analysis for Eqs.~\eqref{spin_spin},~\eqref{spin_path}, and~\eqref{path_spin}  as performed for Eqs.~\eqref{Normal} and~\eqref{CHSH}, one can show maximal violation of Bell's inequality. 

\subsection{Generalized Hyper-Hybrid entangled state} \label{Gen_boson_fermion}
Interestingly, following the approach by Yurke and Stolar~\cite{Y&S92PRA},
we can generalize the detection probabilities of hyper-hybrid entangled state for indistinguishable bosons and fermions into a single formulation as shown below. Let
\begin{equation}
\begin{aligned}
\phi_{1}=& \phi_{D}-\phi_{L}, \\ 
\phi_{2} =&
\begin{cases} 
      -\left( \phi_{R}-\phi_{U}\right)  & \text{for bosons} \\
      -\left( \phi_{R}-\phi_{U}\right)  + \frac{\pi}{2} & \text{for fermions} .
   \end{cases}
\end{aligned}
\end{equation}
The generalized detection probabilities of Eqs.~\eqref{path_path},~\eqref{spin_spin},~\eqref{spin_path}, and~\eqref{path_spin} are, respectively, given by 
\begin{eqnarray}
 \begin{tabular}{c | c c} 
  & $B$ : $R$ & $B$ : $U$  \\  
 \hline
 $A$ : $D$ & \hspace{0.1cm} $\frac{1}{4}\text{cos}^2(\phi_{1}-\phi_{2})$ \hspace{0.1cm} & $\frac{1}{4}\text{sin}^2(\phi_{1}-\phi_{2})$\\ 
 $A$ : $L$ & \hspace{0.1cm}  $\frac{1}{4}\text{sin}^2(\phi_{1}-\phi_{2})$ \hspace{0.1cm} & $\frac{1}{4}\text{cos}^2(\phi_{1}-\phi_{2})$ \\
\end{tabular},\\
 \begin{tabular}{c | c c} 
  & $B$ : $\downarrow$ & $B$ : $\uparrow$  \\  
 \hline
 $A$ : $\downarrow$ & \hspace{0.1cm} $\frac{1}{4}\text{sin}^2(\phi_{1}-\phi_{2})$ \hspace{0.1cm} & $\frac{1}{4}\text{cos}^2(\phi_{1}-\phi_{2})$\\ 
 $A$ : $\uparrow$ & \hspace{0.1cm} $\frac{1}{4}\text{cos}^2(\phi_{1}-\phi_{2})$ \hspace{0.1cm} & $\frac{1}{4}\text{sin}^2(\phi_{1}-\phi_{2})$ \\
\end{tabular},\\
 \begin{tabular}{c | c c} 
  & $B$ : $R$ & $B$ : $U$  \\  
 \hline
 $A$ : $\downarrow$ &\hspace{0.1cm} $\frac{1}{4}\text{sin}^2(\phi_{1}-\phi_{2})$ \hspace{0.1cm} & $\frac{1}{4}\text{cos}^2(\phi_{1}-\phi_{2})$\\ 
 $A$ : $\uparrow$ & \hspace{0.1cm} $\frac{1}{4}\text{cos}^2(\phi_{1}-\phi_{2})$ \hspace{0.1cm} & $\frac{1}{4}\text{sin}^2(\phi_{1}-\phi_{2})$ \\
\end{tabular},\\
 \begin{tabular}{c | c c} 
  & $B$ : $\downarrow$ & $B$ : $\uparrow$  \\  
 \hline
 $A$ : $D$ & \hspace{0.1cm} $\frac{1}{4}\text{cos}^2(\phi_{1}-\phi_{2})$ \hspace{0.1cm} & $\frac{1}{4}\text{sin}^2(\phi_{1}-\phi_{2})$\\ 
 $A$ : $L$ & \hspace{0.1cm} $\frac{1}{4}\text{sin}^2(\phi_{1}-\phi_{2})$ \hspace{0.1cm} & $\frac{1}{4}\text{cos}^2(\phi_{1}-\phi_{2})$ \\
\end{tabular}.
\end{eqnarray}

Computations, following Eqs.~\eqref{Normal} and~\eqref{CHSH}, lead to the maximum violation of Bell's inequality.

\section{Does the scheme of Li \textit{et al.}~\cite{HHNL} work for distinguishable particles?} \label{DisHHES}
We are interested to see whether the circuit of Li \textit{et al.}~\cite[Fig. 2]{HHNL} as shown in Fig.~\ref{fig:Lietal} gives the same results for two distinguishable particles. Let us calculate the term in the first row and first column of Eq.~\eqref{path_path} for fermions. It says that the probability of Alice detecting a particle in detector $D$ and Bob detecting a particle in detector $R$ is given by
\begin{equation}
\left| \frac{1}{4} \left[  e^{i\left( \phi_{R}+ \phi_{L} \right) } + e^{i\left( \phi_{D}+ \phi_{U} \right) } \right] \right|^{2}=\frac{1}{4} \text{cos}^{2}\phi.
\end{equation}
If the particles are made distinguishable, this probability is calculated as
\begin{equation}
 \left| \frac{1}{4} e^{i\left( \phi_{R}+ \phi_{L} \right) } \right|^{2} + \left| \frac{1}{4} e^{i\left( \phi_{D}+ \phi_{U} \right) } \right|^{2} =\frac{1}{8}.
\end{equation}
As for other terms of Eq.~\eqref{path_path}, each term of Eqs~\eqref{spin_spin},~\eqref{spin_path}, and~\eqref{path_spin} reduces to $\frac{1}{8}$. From that, one can easily show that the right hand side of Eq.~\eqref{Normal} becomes zero. Thus the Bell violation is not possible by the CHSH test. Similar calculations for the bosons lead to the same conclusion. So, the circuit of~\cite{HHNL} would not work for distinguishable particles.

\section{Signaling using unit fidelity quantum teleportation and hyper-hybrid entangled state} \label{Sig_UFQT_HHES}
In this section we will prove out main theorem and also derive two corollaries from that theorem.
\begin{theorem} \label{Th1}
If quantum particles (either distinguishable or indistinguishable) can simultaneously produce and perform hyper hybrid entangled state and unit fidelity quantum teleportation respectively then using that cloning of any arbitrary quantum state is possible.
\end{theorem}

 It is well-known that unit fidelity quantum teleportation~\cite{Popescu94} for distinguishable particles is possible using BSM and LOCC. Here, we show that if hyper-hybrid entangled state for distinguishable particles could exist, then one could construct a universal quantum cloning machine (UQCM)~\cite{QC05,QC14} using unit fidelity quantum teleportation and hyper-hybrid entangled state, and further, use that UQCM to achieve signaling. Throughout this thesis, by signaling we mean faster-than-light or superluminal communication across spacelike separated regions.
\begin{proof} 
Now we give a details proof of the Theorem~\ref{Th1}.
\subsection{Description of the protocol} Our signaling protocol works in three phases, as follows.

 \begin{figure}[h!] 
\centering
\includegraphics[width=0.8\textwidth]{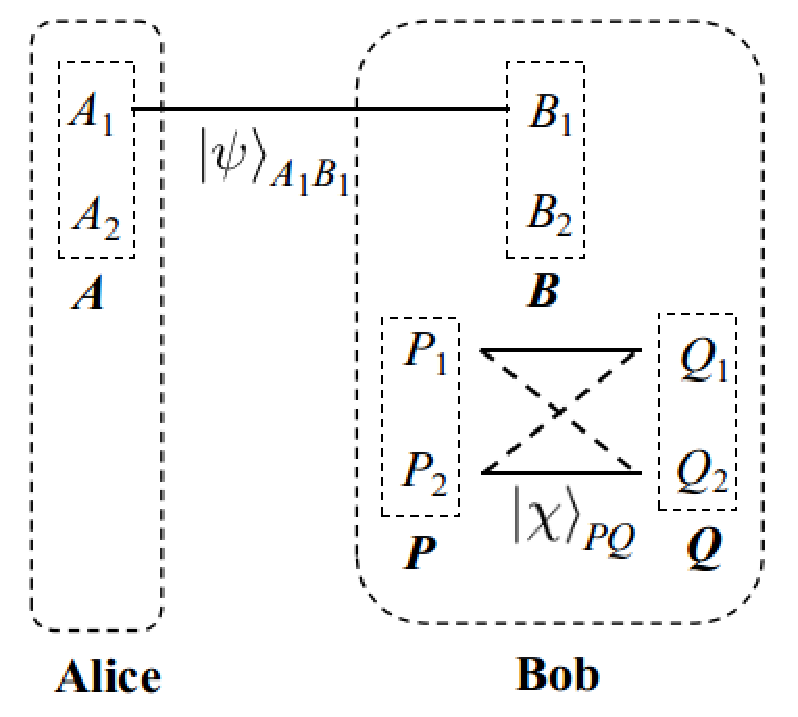} 
\caption{The singlet state $\ket{\psi}_{A_{1}B_{1}}$ is shared between DoF 1 of Alice and that of Bob, whereas hyper-hybrid entangled state $\ket{\chi}_{PQ}$ is kept by Bob.}
\label{fig:Alice_Bob}
\end{figure}

\subsubsection{First phase: initial set-up.} 
Suppose there are four particles $A$, $B$, $P$, and $Q$ each having two DoFs $1$ and $2$. The particle $A$ is with Alice and the remaining three are with Bob, who is spacelike separated from Alice. The pair $\left\lbrace A,B \right\rbrace$ is in the singlet state in DoF $1$  denoted by $\ket{\psi}_{A_{1}B_{1}}$ and the pair $\left\lbrace P,Q \right\rbrace$ is in hyper-hybrid entangled state using both the DoFs $1$ and $2$ denoted by $\ket{\chi}_{PQ}$. To access the DoF $i$ of particle $X$, we use the notation $X_i$, where $X \in \{A, B, P, Q\}$ and $i \in \{1, 2\}$. The situation is depicted in Fig.~\ref{fig:Alice_Bob}. 
Note that $\ket{\psi}_{A_{1}B_{1}}$ can be expressed in any orthogonal basis. We have taken only $Z$ basis or computational basis $\left\lbrace \ket{0}, \ket{1}\right\rbrace$ and $X$ basis or Hadamard basis $\left\lbrace \ket{+}, \ket{-}\right\rbrace$ such that 
\medmuskip=2mu
\thinmuskip=2mu
\thickmuskip=2mu 
\begin{equation}
\begin{aligned}
   \ket{\psi}_{A_{1}B_{1}}&= \frac{1}{\sqrt{2}} \left( \ket{01}_{A_{1}B_{1}}-\ket{10}_{A_{1}B_{1}}\right)\\ & =  \frac{1}{\sqrt{2}} \left(\ket{+-}_{A_{1}B_{1}}-\ket{-+}_{A_{1}B_{1}}\right). 
\end{aligned}
\end{equation}
\medmuskip=2mu
\thinmuskip=2mu
\thickmuskip=2mu 
Alice wants to transfer binary information instantaneously to Bob. Before going apart, Alice and Bob agree on the following convention.
\begin{enumerate}
\item If Alice wants to send zero to Bob, then she would measure in $Z$ basis on the DoF 1 of her particle so that the state of the DoF 1 of the particle at Bob's side would be either $\ket{0}$ or $\ket{1}$. 
\item If Alice wants to send $1$ to Bob, then she would measure in $X$ basis on the DoF 1 of her particle so that the state of the DoF 1 of the particle at Bob's side would be either $\ket{+}$ or $\ket{-}$.
\end{enumerate}

\subsubsection{Second phase: cloning of any unknown state.} 
There are two steps of our proposed UQCM as follows.
\begin{enumerate}
\item Alice does measurement on DoF 1 of her particle $A$, i.e., $A_1$ in either $Z$ basis or $X$ basis. After this measurement, the state on DoF 1 of particle $B$, i.e., $B_1$ on Bob's side, is in an unknown state $\ket{\phi} \in \left\lbrace \ket{0}, \ket{1}, \ket{+}, \ket{-} \right\rbrace$ and it is  denoted  by $\ket{\phi}_{B_1}$. 
\item After Alice's measurement, Bob performs BSM on DoF 1 of particles $B$ and $P$,  i.e., on $B_1$ and $P_1$. 
This results in an output $k$ as one of the four possible Bell states (as seen in standard teleportation protocol~\cite{QT93}). Based on this output $k$, suitable unitary operations $U_{k}$ are applied on both the DoFs of $Q$, i.e., $Q_1$ and $Q_2$, where $U_{k} \in \left\lbrace \mathcal{I}, \sigma_{x}, \sigma_{y}, \sigma_{z}\right\rbrace $, $\mathcal{I}$ being the identity operation and $\sigma_{i}$'s $\left( i = x, y, z \right)$ the Pauli matrices.
As the first DoF of particle $P$, $P_1$ is maximally entangled with both the DoFs of $Q$, i.e., $Q_1$ and $Q_2$; thus, using BSM on DoFs $B_1$ and $P_1$ and suitable unitary operations on DoFs $Q_1$ and $Q_2$, the unknown state $\ket{\phi}$ on DoF ${B_1}$ is copied to both the DoFs $Q_1$ and $Q_2$. This part of the circuit, shown in Fig.~\ref{fig:NO_CLONING}, acts as a UQCM. 
\end{enumerate}

\begin{figure}[t!] 
\centering
\includegraphics[width=0.8\textwidth]{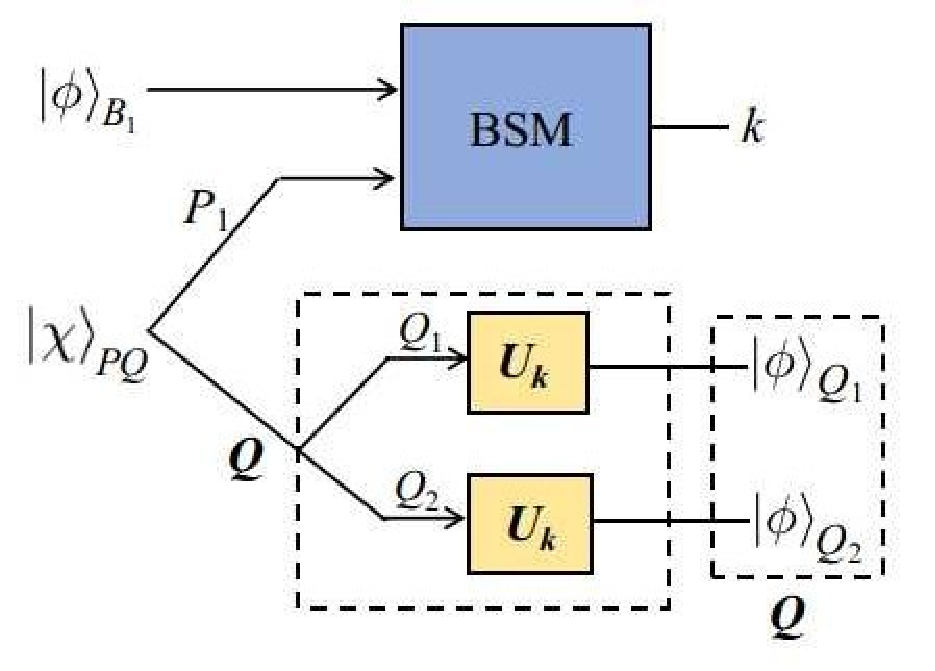} 
\caption{Our proposed universal quantum cloning machine to demonstrate the impossibility of hyper-hybrid entangled state using distinguishable particles. Inputs to this cloning machine are the unknown quantum state $\ket{\phi}$ of DoF 1 of the particle $B$ (denoted by $\ket{\phi}_{B_{1}}$) and the hyper-hybrid entangled state $\ket{\chi}_{PQ}$, as shown in Fig.~\ref{fig:Alice_Bob}. More specifically, $\ket{\phi}_{B_{1}}$ is the state unknown to Bob generated on Bob's side after Alice does measurement on $A_1$ in either $Z$ basis or $X$ basis. After that, Bob performs BSM on $B_1$ and $P_1$, resulting in one of the four possible Bell states as output, denoted by $k$. Based on this output $k$, suitable unitary operations $U_{k}$ are applied on both the DoFs of $Q$, i.e., $Q_1$ and $Q_2$, where $U_{k} \in \left\lbrace \mathcal{I}, \sigma_{x}, \sigma_{y}, \sigma_{z}\right\rbrace $, $\mathcal{I}$ being the identity operation and $\sigma_{i}$'s $\left( i = x, y, z \right)$ the Pauli matrices. As a result, the unknown state $\ket{\phi}$ of $B_1$ is copied to both $Q_1$ and $Q_2$.}
\label{fig:NO_CLONING}
\end{figure}

\subsubsection{Third phase: decoding Alice's measurement basis.}
Now from the two copies of the unknown state $\ket{\phi}$ on the two DoFs of $Q$,  i.e., $\ket{\phi}_{Q_1}$ and $\ket{\phi}_{Q_2}$, Bob tries to discriminate the measurement bases of Alice, so that he can decode the information sent to him. For that, Bob measures both the DoFs of $Q$ in $Z$ basis, resulting in either $0$ or  $1$ in each of the DoFs. Now there are two possibilities.
\begin{enumerate}
\item If Alice has measured in $Z$ basis, then Bob's possible measurement results on the two DoFs of $Q$ are $\{00, 11\}$. 
\item On the other hand, if Alice has measured in $X$ basis, then Bob's possible measurement results on the two DoFs of $Q$ are $\{00, 01, 10, 11\}$.
\end{enumerate}
Suppose, Bob adopts the following strategy. Whenever his measurement results are all zero or all one (i.e., 00 or 11), then he concludes that Alice has sent a zero, and whenever he measures otherwise (i.e., 01 or 10) then he concludes that Alice has sent a 1.

\subsection{Computation of the signaling probability}
Let the random variables $X_A$ and $X_B$ denote the bit sent by Alice and the bit decoded by Bob, respectively. Hence, under the above strategy, Bob's success probability of decoding, which is also the probability of signaling, is given by
\begin{equation}
\begin{aligned}
P_{sig}  = & \Pr(X_A = 0 \wedge X_B = 0) + \Pr(X_A = 1 \wedge X_B = 1)\\
 = & \Pr(X_A = 0)\cdot\Pr(X_B = 0 \mid X_A = 0) \\
& + \Pr(X_A = 1)\cdot\Pr(X_B = 1 \mid X_A = 1)\\
 = & \frac{1}{2}\cdot 1 + \frac{1}{2}\cdot\frac{2}{4} = 0.75.
\end{aligned}
\end{equation}

To increase $P_{sig}$ further, Bob can use hyper-hybrid entangled state involving $N$ DoFs of $P$ and $Q$, with $N \geq 3$. Then he can make $N$ copies of the unknown state $\ket{\phi}$ into the DoFs of $Q$. Analogous, to the strategy above for the case $N=2$, here also if all the measurement results of Bob in the $N$ DoFs of $Q$ in $Z$ basis are the same, i.e., all-zero case or the all-one case, then Bob concludes that Alice has sent a zero; otherwise, he concludes that Alice has sent a 1. Thus, the above expression of $P_{sig}$ changes to
$$\frac{1}{2}\cdot 1 + \frac{1}{2}\cdot\frac{2^N-2}{2^N}.$$
In other words, 
\begin{equation}
\label{sigprob}
P_{sig}=1 - \frac{1}{2^N}.
\end{equation}   
By making $N$ larger and larger,  $P_{sig}$ can be made arbitrarily close to 1.
\end{proof}

\subsection{Experimental realization using optical circuits}
 For the experimental realization of the above protocol, we propose a circuit with DoF sorters, such as a	 spin sorter (SS), path sorter (PS), etc. (A spin sorter can be realized in an optical system using a polarizing beam splitter for sorting between the horizontal $\ket{H}$ and the vertical $\ket{V}$ polarizations of a photon. For an
alternative implementation in atomic systems using Raman process, one can see~\cite[Fig.~3]{HHNL}.)

\begin{figure}[t!] 
\centering
\includegraphics[width=0.8\textwidth]{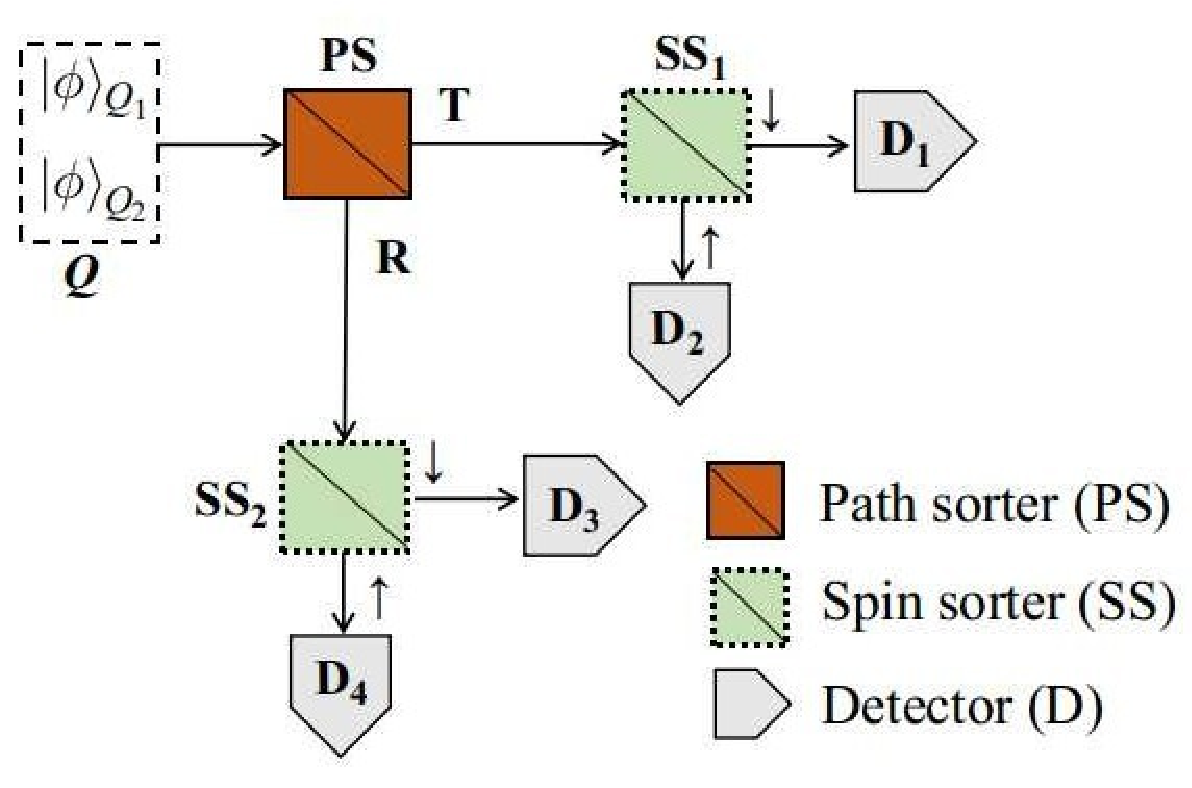} 
\caption{Bob's circuit for distinguishing between $Z$ and $X$ bases using two types of DoF sorters, i.e., a spin sorter (SS) and a path sorter (PS), and four detectors. Here the output of Fig.~\ref{fig:NO_CLONING} is used as an input in this circuit. Here, $\ket{\downarrow}$ and $\ket{\uparrow}$ denote the down- and the up-spin states of the particles, respectively. Further, $\ket{T}$ denotes the transverse mode and $\ket{R}$ denotes the reflected modes of a PS.}
\label{SLS_N_2}
\end{figure}

Suppose DoFs $1$ and $2$  are spin and path, respectively, with the two output states $\left\lbrace \ket{\downarrow}, \ket{\uparrow} \right\rbrace $ and $\left\lbrace \ket{T},\ket{R}\right\rbrace $. Here, $\ket{\downarrow}$ and $\ket{\uparrow}$ denote the down and the up spin states of the particles and $\ket{T}$ and $\ket{R}$ denote the transverse and the reflected modes of a PS, respectively. Without loss of generality, we take 
\begin{equation}
\begin{aligned}
\ket{0}=&\ket{\downarrow}=\ket{T}, \hspace{0.2cm} \ket{1}=\ket{\uparrow}=\ket{R}, \\  \ket{+}=&\frac{1}{\sqrt{2}}\left( \ket{\downarrow}+\ket{\uparrow}\right) =\frac{1}{\sqrt{2}}\left(\ket{T}+\ket{R}\right), \\
\ket{-}=&\frac{1}{\sqrt{2}}\left(\ket{\downarrow}-\ket{\uparrow}\right) =\frac{1}{\sqrt{2}}\left(\ket{T}-\ket{R}\right).
\end{aligned}
\end{equation}
The circuit, shown in Fig.~\ref{SLS_N_2}, takes as input particle $Q$ with DoFs $1$ and $2$, each having the cloned state $\ket{\phi}$ from the output of the circuit in Fig.~\ref{fig:NO_CLONING}. Bob places a path sorter $PS$ followed by two spin sorters $SS_{1}$ and $SS_{2}$ on two output modes of $PS$. Let $ D_{1}$ ($ D_{3}$) and $D_{2}$ ($ D_{4}$) be the detectors at the two output ports of $SS_{1}$ ($SS_{2}$).

If Alice measures in $Z$ basis, Bob detects the particles in $\left\lbrace D_{1},D_{4}\right\rbrace$ with unit probability. On the other hand, if she measures in $X$ basis, the particles would be detected in each of the detector sets $\left\lbrace D_{1},D_{4}\right\rbrace$ and $\left\lbrace D_{2},D_{3}\right\rbrace $ with a probability of 0.5. When Bob detects the particles in either $D_2$ or $D_3$, he instantaneously knows that the measurement basis of Alice is $X$. In this case, the signaling probability is 0.75, which can be obtained by putting $N=2$ in Eq.~\eqref{sigprob}.

\begin{figure}[t!] 
\centering
\includegraphics[width=0.8\textwidth]{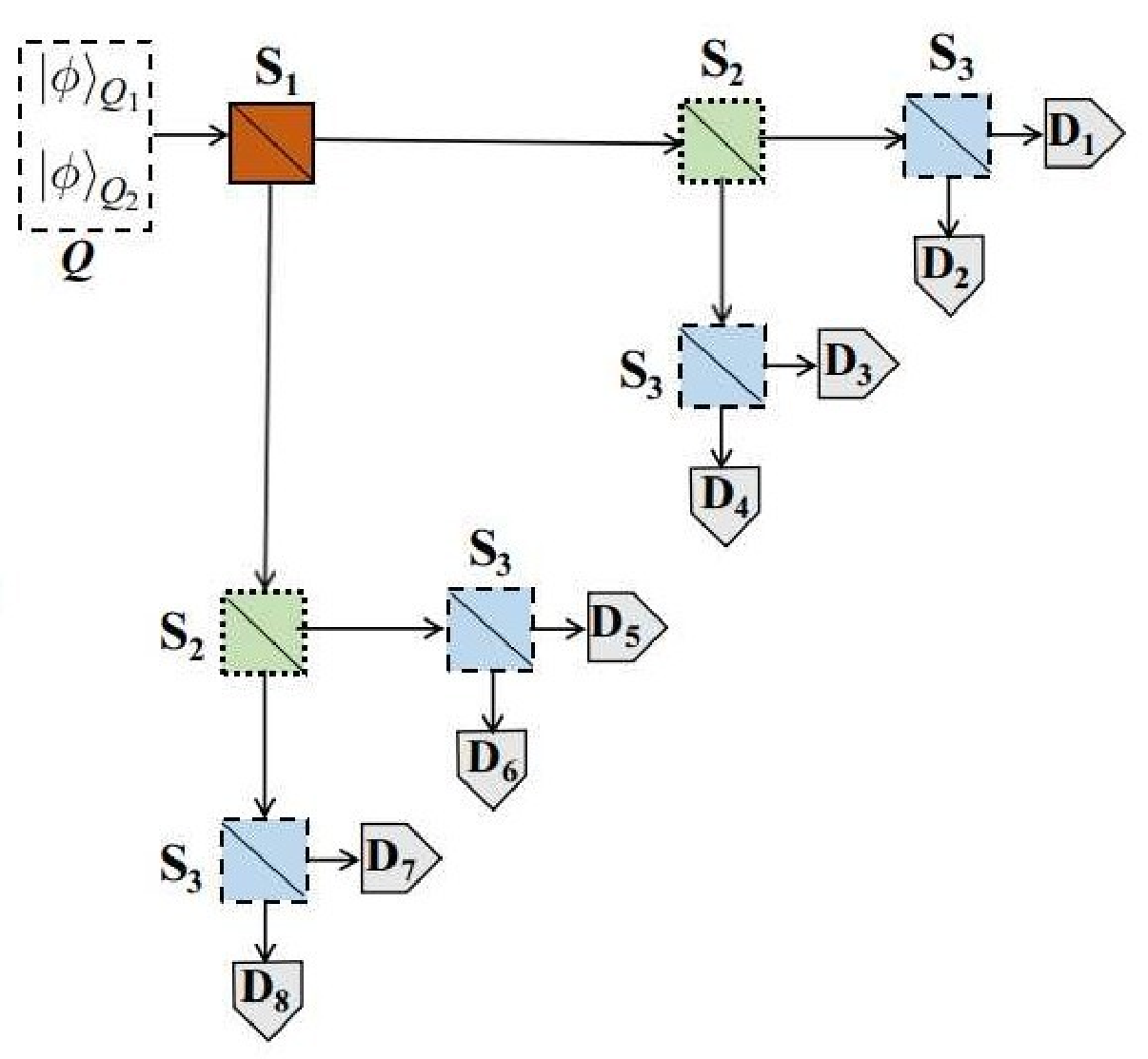} 
\caption{Bob's circuit for distinguishing between $Z$ and $X$ bases using three types of DoF sorters, i.e., $S_1$, $S_2$, and $S_3$, and eight detectors. Here the output of Fig.~\ref{fig:NO_CLONING} is used as an input in this circuit.}
\label{SLS_N_3}
\end{figure}

For better signaling probability, one can use three DoFs $1$, $2$, and $3$, instead of two, in the joint state $\ket{\chi}_{PQ}$. The scheme for three DoFs is shown in Fig.~\ref{SLS_N_3}, where $S_{i}$ represents the sorter for DoF $i$, for $i \in \{1, 2, 3\}$. Now, if Alice  measures in $Z$ basis, Bob detects the particles in $\left\lbrace D_{1},D_{8}\right\rbrace $ with probability 1. But if she measures in $X$ basis, the particles would be detected in $\left\lbrace D_{1},D_{8}\right\rbrace $ with probability $\frac{2}{2^3}$ and in  $\left\lbrace D_{2}, \cdots , D_{7}\right\rbrace $ with probability $\frac{2^3 -2}{2^3}=0.75$. In this case, the signaling probability is 0.875, which can be obtained by putting $N=3$ in Eq.~\eqref{sigprob}.

We can generalize the above schematic as follows. Suppose, each of $P$ and $Q$ has $N$ DoFs (each degree having two eigenstates), numbered 1 to $N$, in hyper-hybrid entangled state $\ket{\chi}_{PQ}$. We also need to use $N$ corresponding types of DoF sorters. Now, if Alice measures in $Z$ basis, Bob detects the particles in $\left\lbrace D_{1},D_{2^N}\right\rbrace $ with probability 1. But if she measures in $X$ basis, the particles are detected in $\left\lbrace D_{1},D_{2^N}\right\rbrace $ with probability $\frac{2}{2^N}$ and in detectors  $\left\lbrace D_{2}, \cdots , D_{2^N-1}\right\rbrace $ with probability $\frac{2^N-2}{2^N}$. For $N$ DoFs, the signaling probability is given in Eq.~\eqref{sigprob}.

\subsubsection{Increasing signaling probability without increasing the number of DoFs}
From Eq.~\eqref{sigprob}, it is clear that the signaling is possible only when $N$ is infinitely large. The existence of such a huge number of accessible DoFs may be questionable. Interestingly, we devise an alternative schematic that can drive the asymptotic success probability to 1 with only two DoFs but using many copies of the singlet state shared between Alice and Bob and the same number of copies of hyper-hybrid entangled state at Bob's disposal.

Suppose, Alice and Bob share $M$ copies of the singlet state \\
 $\left\lbrace  \ket{\psi^{(1)}}_{A_{1}B_{1}}, \ket{\psi^{(2)}}_{A_{1}B_{1}}, \cdots, \ket{\psi^{(M)}}_{A_{1}B_{1}}  \right\rbrace $ and Bob also has an equal number of copies of hyper-hybrid entangled states $ \left\lbrace \ket{\chi^{(1)}}_{PQ}, \ket{\chi^{(2)}}_{PQ}, \cdots, \ket{\chi^{(M)}}_{PQ}  \right\rbrace $ (a single copy of the singlet state and hyper-hybrid entangled state is shown in Fig.~\ref{fig:Alice_Bob}). Now the cloning can be performed in the following two steps.
\begin{enumerate}
\item  Alice performs measurement in her preferred basis on each DoF 1 of her $M$ particles so that on the DoF 1 of each of the $M$ particles on Bob's side, a copy of the unknown state $\ket{\phi}$ is obtained.
\item After that, Bob performs BSM on each of the $M$ pairs of $\left\lbrace B_1, P_1 \right\rbrace $ and suitable unitary operations so that the unknown state $\ket{\phi}$ is copied to each of the $M$ pairs of $\left\lbrace Q_1, Q_2 \right\rbrace $.
\end{enumerate}

Now Bob passes each of the $M$ copies of $Q$ as shown in Fig.~\ref{SLS_N_2} and adopts the following strategy. If Bob receives each of the $M$ particles in $D_{1}$ or $D_{4}$, then he concludes that Alice has measured in $Z$ basis. On the other hand, if Bob observes any one of the $M$ particles in $D_{2}$ or $D_{3}$, then he concludes that Alice has measured in $X$ basis. Under this strategy, Bob encounters a decoding error whenever Alice has measured in $X$ basis, but he receives all the $M$ particles in $D_{1}$ or $D_{4}$. 
In this case, the probability that a single particle is detected in the detector set $\left\lbrace D_{1},D_{4}\right\rbrace$ is $\frac{1}{2}$ and hence the probability that all the $M$ particles are detected in the above set is $\frac{1}{2^M}$. Hence, the corresponding success probability of signaling is given by 
\begin{equation}
    P_{sig}=1-\frac{1}{2^M},
\end{equation}
which also asymptotically goes to 1.

Thus we have the following logical version of Theorem~\ref{Th1}
\begin{equation} \label{Cloning}
\text{HHES} \wedge \text{UFQT} \Rightarrow \text{Signaling}.
\end{equation}

\section{First no-go theorem: no hyper-hybrid entangled state for distinguishable particles} \label{NoHHES}
\begin{proof}

Unit fidelity quantum teleportation can be performed by distinguishable particles  ~\cite{QT93}. Thus if hyper-hybrid entangled state is also possible for distinguishable particles, it would lead to signaling as shown in Section~\ref{Sig_UFQT_HHES}. Thus a logical conclusion is that in a world where special relativity holds barring signaling, hyper-hybrid entangled state is not possible using distinguishable particles. Thus we can write this no-go theorem logically  as
\begin{equation} \label{no-HHES}
\text{No-signaling} \Rightarrow \text{no-HHES} \wedge \text{UFQT}.
\end{equation}
\end{proof}

\section{Second no-go theorem: no unit fidelity quantum teleportation for indistinguishable particles} \label{NoUFQT}

\begin{proof}

 Earlier, we have shown that signaling for distinguishable particles can be achieved using unit fidelity quantum teleportation and hyper-hybrid entangled state as black-boxes. unit fidelity quantum teleportation for distinguishable particles is already known~\cite{QT93}, and so we have concluded that hyper-hybrid entangled state for distinguishable particles must be an impossibility.

Using massive identical particles, Marzolino and Buchleitner~\cite{Ugo15} have shown that unit fidelity quantum teleportation is not possible using a finite and fixed number of indistinguishable particles, due to the particle number conservation superselection rule (SSR)~\cite{Wick52,SSR07}. Interestingly, several independent works~\cite{SSR07,Suskind67,Paterek11} have already established that this superselection rule can be bypassed. So, an obvious question is: whether it is possible to perform unit fidelity quantum teleportation for indistinguishable particles bypassing the superselection rule. This question is also answered in the negative in~\cite{Ugo15}. 

Very recently, for indistinguishable particles, Lo Franco and Compagno~\cite{LFC18} have achieved a quantum teleportation fidelity of $5/6$, overcoming the classical teleportation fidelity bound $2/3$~\cite{HHH99}. But they have not proved whether this value is optimal or whether unit fidelity can be achieved or not.

Systematic calculations show that our earlier scheme of quantum cloning and signaling (Figs.~\ref{fig:Alice_Bob} and Fig.~\ref{fig:NO_CLONING}) would still work, even if one replaces the unit fidelity quantum teleportation and hyper-hybrid entangled state tools for distinguishable particles with those of the indistinguishable ones (assuming that such tools exist). As signaling is not possible in quantum theory, we have concluded that hyper-hybrid entangled state for distinguishable particles is not possible. But, for indistinguishable particles, the creation of hyper-hybrid entangled state is possible~\cite{HHNL}. Thus, a logical conclusion is that, to prevent signaling, unit fidelity quantum teleportation must not be possible for indistinguishable particles. So, this no-go theorem can be represented logically as
\begin{equation}\label{no-UFQT}
\text{No-signaling} \Rightarrow \text{HHES} \wedge \text{no-UFQT}.
\end{equation}
\end{proof}

\section{Physical significance of the two no-go theorems} \label{Sepresult}
\begin{figure}[h!]
\centering
\includegraphics[width=0.8\textwidth]{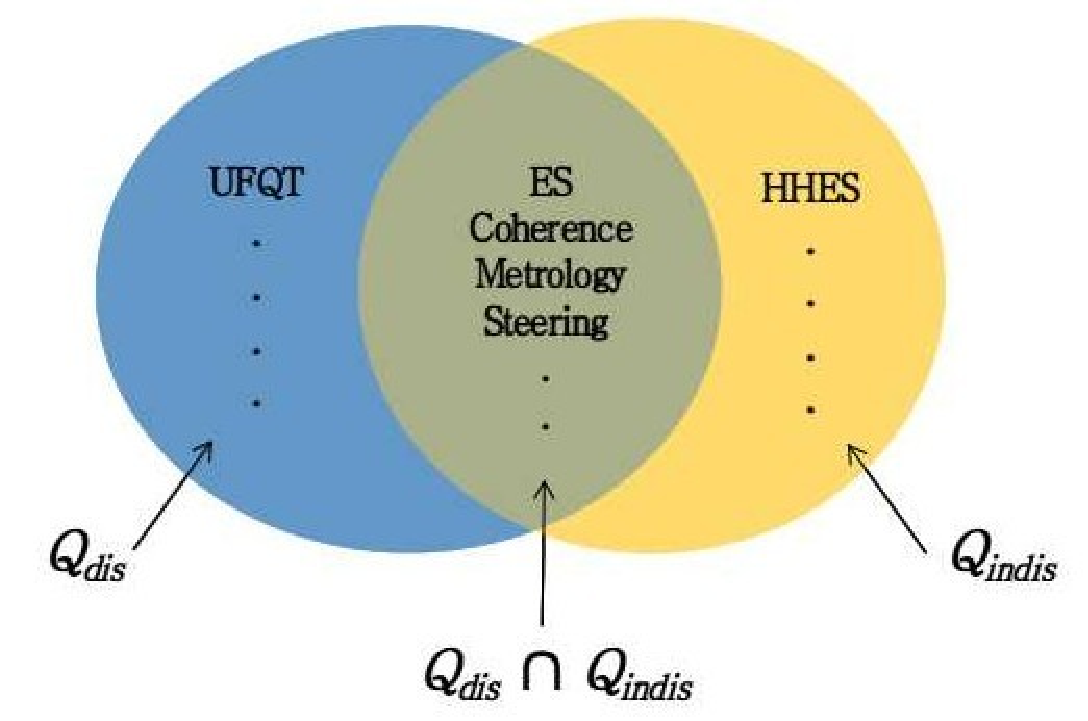} 
\caption{The two sets $Q_{dis}$ (consisting of quantum properties and applications of distinguishable particles), $Q_{indis}$ (consisting of quantum properties and applications of indistinguishable particles), and their intersection (UFQT, HHES, and ES stand for unit fidelity quantum teleportation, hyper-hybrid entangled state, and entanglement swapping, respectively).} 
\label{fig:Overall}
\end{figure}

From the above two no-go results, we can establish a separation result between distinguishable and indistinguishable particles.  
Let $Q_{dis}$ and $Q_{indis}$ be the two sets consisting of quantum properties and applications of distinguishable and indistinguishable particles, respectively, as shown in Fig.~\ref{fig:Overall}.

Several earlier works have attempted extending many results on one of these two sets to the other. For example, quantum teleportation was originally proposed for distinguishable particles~\cite{QT93,QTnat15}. But recent works~\cite{LFC18,Ugo15} have extended it for indistinguishable particles. Similarly, duality of entanglement as proposed in~\cite{Bose13} was thought to be a unique property of indistinguishable particles. But later its existence for distinguishable particle was shown in~\cite{M.Karczewski16}. Another unique property of quantum correlation is quantum coherence, which was proposed for distinguishable particles in~\cite{Baumgratz14} and later for indistinguishable particles in~\cite{Sperling17}. Einstein-Podolsky-Rosen steering~\cite{Schrodinger} was extended from distinguishable particles to a special class of indistinguishable particles called Bose-Einstein condensates~\cite{Fadel18}. Entanglement swapping, originally proposed for distinguishable particles in~\cite{ES93,Pan10,Pan19}, was also shown for indistinguishable particles in~\cite{LFCES19}.

To the best of our knowledge, there is no known quantum correlation or application that is unique for distinguishable particles only and does not hold for indistinguishable particles, and vice versa. In Section~\ref{NoHHES} of this paper, we have shown that hyper-hybrid entangled state is unique to the set $Q_{indis}$, and in Section~\ref{NoUFQT} we have established that unit fidelity quantum teleportation is unique to the set $Q_{dis}$. Thus, we demonstrate a clear separation between these two sets.

\chapter{Degree of freedom trace-out rule for indistinguishable particles} \label{Chap5}
\indent The representation and the trace-out rule of the state of two indistinguishable particles each having single DoF is discussed in Chapter~\ref{LFC_app}. In this Chapter, we first discuss the representation of the general state of indistinguishable particles each having multiple DoFs using the notions of~\cite{LFC16,LFC18}.  Next, we discuss our proposed DoF  trace-out operation of indistinguishable particles when each particle have multiple DoFs. Finally we discuss the physical significance of the proposed DoF trace-out rule.

 This chapter is based on the work in~\cite{Paul21,MoE}.

\section{Representation of the general state of indistinguishable particles}
Lo Franco \textit{et al.}~\cite{LFC16,LFC18} have defined the partial trace-out rule for indistinguishable particles where each particle has a spatial label and a single DoF. But as the notions of~\cite{LFC16,LFC18} for indistinguishable particles is not readily applicable for particles having multiple DoFs, we have extended their notions to overcome this problem. 

\subsection{Two indistinguishable particles each having two DoFs}  \label{2P2DoF}
Assume two indistinguishable particles each having two DoFs are associated with two spatial labels  $\alpha^{1}$ and $\alpha^{2}$. Here  $a^{i}_{j}$ ranges over $\mathbb{D}_{j}:=\lbrace D_{j_{1}}, D_{j_{2}}, \cdots, D_{j_{k_{j}}} \rbrace$, represents the eigenvalue of the $j$-th DoF of the particle in the $\alpha^{i}$-th localized region where $i, j \in \mathbb{N}_{2} =\lbrace 1, 2 \rbrace$ and $k_{j} \geq 2$, since each DoF must have at least two distinct eigenvalues.  The general state of such a system is written as 
\begin{equation} \label{IDstate}
\begin{aligned}
\ket{\Psi^{(2,2)}} := \sum_{\alpha^{1},\alpha^{2}, a^{1}_{1}, a^{1}_{2},  a^{2}_{1}, a^{2}_{2}} \eta^{u}\kappa^{\alpha^{1},  \alpha^{2} }_{a^{1}_{1}, a^{1}_{2}, a^{2}_{1}, a^{2}_{2}} \ket{\alpha^{1} a^{1}_{1} a^{1}_{2}, \alpha^{2}  a^{2}_{1} a^{2}_{2}},
\end{aligned}
\end{equation}
where $\alpha^{1}, \alpha^{2}$ ranges over $\mathbb{S}^P:= \lbrace s^1, s^2, \cdots, s^P \rbrace$ which refers to distinct spatial modes with $P \geq 2$. Here $u$ represents the summation of parity of the cyclic permutations of all the $n$ DoFs. Thus $u$ can be represented as $u=u_1+u_2+ \ldots + u_n=\sum_i u_j$ where $u_j$ is the parity of $j$-th DoF. The value of $\eta$ is $+1$ for bosons and $-1$ for fermions. If we have the following condition that 
\begin{equation} \label{PauliPure}
\left( \alpha^{i}=\alpha^{i^\prime} \right)  \wedge \left( a^{i}_{j}=a^{i^{\prime}}_{j} \right)
\end{equation} 
 for any $i \neq i^{\prime}$ where $\alpha^i , \alpha^{i^{\prime}} \in \mathbb{S}^{P} $ and $j \in \mathbb{N}_{2}$, then we get $\eta=0$ for fermions due to Pauli exclusion principle~\cite{Pauli25}.

The value of $k_{j}$ may vary with $j$. For example, consider two DoFs: polarization and optical orbital angular momentum (OAM), associated with a system of indistinguishable photons. Generally, the polarization belongs to a two-dimensional Hilbert space, whereas the orbital angular momentum  lies in an infinite-dimensional Hilbert space governed by the azimuthal index $l$. In practical implementations, this mismatch in Hilbert space dimensions between the two DoFs is taken care of by mapping the larger dimensional space to the lower dimensional one~\cite{Karimi10,Bhatti15}. For orbital angular momentum , the infinite-dimensional Hilbert space is generally mapped into a two-dimensional one with the eigenvalues $ \lbrace 2l, 2l+1 \rbrace$ or $\lbrace +l, -l \rbrace$.  Also, the Hilbert space is sometimes restricted to smaller dimensions by proper state engineering in which case only certain chosen values of $l$ are allowed.

Following the above notations, the general density matrix of two indistinguishable particles each having two DoFs is expressed as 
\begin{equation} \label{IDDM_gen}
\begin{aligned}
\rho^{(2, 2)}:=  \sum_{\substack{\alpha^{i}, \beta^{i}, a^{i}_{j}, b^{i}_{j} }} \eta^{(u+\bar{u})} 
 \kappa^{\alpha^{1}, \alpha^{2},\beta^{1}, \beta^{2}}_{a^{1}_{1}a^{1}_{2}, a^{2}_{1}a^{2}_{2},b^{1}_{1}b^{1}_{2}, b^{2}_{1}b^{2}_{2}} 
\ket{\alpha^{1} a^{1}_{1}a^{1}_{2}, \alpha^{2} {a^{2}_{1}a^{2}_{2}  }} \bra{\beta^{1} b^{1}_{1}b^{1}_{2}, \beta^{2} {b^{2}_{1}b^{2}_{2} }},
\end{aligned}
\end{equation}
\noindent where $\alpha^{i}, \beta^{i}$ ranges over $\mathbb{S}^P$, $a^{i}_{j}, b^{i}_{j}$ ranges over $\mathbb{D}_{j}$, and $i, j \in \mathbb{N}_2$. 
Here $u$ is as defined in Eq.~\eqref{IDstate} and $\bar{u}$ comes due to density matrix. If we have the following condition that 
\begin{equation} \label{PauliDen}
\left\lbrace  \left( \alpha^{i}=\alpha^{i^\prime} \right) 
\wedge \left( a^{i}_{j}=a^{i^{\prime}}_{j} \right) \right\rbrace \vee \left\lbrace  \left( \beta^{i}=\beta^{i^\prime} \right)   \wedge  \left( b^{i}_{j}=b^{i^{\prime}}_{j} \right) \right\rbrace 
\end{equation} 
for any $i \neq i^{\prime}$ where $i , i^{\prime} \in \mathbb{N}_2$   and $j \in \mathbb{N}_{2}$, then we get $\eta=0$ for fermions due to Pauli exclusion principle~\cite{Pauli25}. 

The general density matrix of the state in Eq.~\eqref{IDstate} is represented as 
\begin{equation} \label{IDDM}
\begin{aligned}
\rho^{(2, 2)}:=  \sum_{\substack{\alpha^{i}, \beta^{i}, a^{i}_{j}, b^{i}_{j} }} \eta^{(u+\bar{u})} 
 \kappa^{\alpha^{1}, \alpha^{2}}_{a^{1}_{1}a^{1}_{2}, a^{2}_{1}a^{2}_{2}} \kappa^{\beta^{1}, \beta^{2}*}_{b^{1}_{1}b^{1}_{2}, b^{2}_{1}b^{2}_{2}} 
\ket{\alpha^{1} a^{1}_{1}a^{1}_{2}, \alpha^{2} {a^{2}_{1}a^{2}_{2}  }} \bra{\beta^{1} b^{1}_{1}b^{1}_{2}, \beta^{2} {b^{2}_{1}b^{2}_{2} }},
\end{aligned}
\end{equation}
where the notations are same as Eq.~\eqref{IDDM_gen}.

In the next section, we extend this notions to two indistinguishable particles each having $n$ DoFs.

\subsection{Two indistinguishable particles each having $n$ DoFs} \label{2PnDoF}

Lets us consider two indistinguishable particles each having $n$ DoFs. Following the same notations of Eq.~\eqref{IDstate}, the general state of Eq.~\eqref{IDstate} of two particles from two DoFs to $n$ DoFs can be represented as
\begin{equation} \label{State}
\begin{aligned}
\ket{\Psi^{\left( 2,n\right) }}:= \sum_{\alpha^{i},  a^{i}_{j} } \eta^u \kappa^{\alpha^{1}, \alpha^{2}}_{a^{1}_{1}a^{1}_{2} \cdots a^{1}_{n}, a^{2}_{1}a^{2}_{2} \cdots a^{2}_{n}} 
\ket{\alpha^{1} a^{1}_{1}a^{1}_{2} \cdots a^{1}_{n}, \alpha^{2} {a^{2}_{1}a^{2}_{2} \cdots a^{2}_{n} }},
\end{aligned}
\end{equation}
where $a^{i}_{j}$ ranges over $\mathbb{D}_{j}$, $i \in \mathbb{N}_{2} =\lbrace 1, 2 \rbrace$, and $ j \in \mathbb{N}_{n} =\lbrace 1, 2, \ldots, n \rbrace$.

 The general density matrix of two indistinguishable particles each having $n$ DoFs following Eq.~\eqref{IDDM_gen} can be described as
 \medmuskip=0mu
\thinmuskip=0mu
\thickmuskip=0mu
\begin{scriptsize}
\begin{equation} \label{DM}
\begin{aligned}
\rho^{(2, n)}:=  \sum_{\substack{\alpha^{i}, \beta^{i}, a^{i}_{j}, b^{i}_{j} }} \eta^{(u+\bar{u})} 
 \kappa^{\alpha^{1}, \alpha^{2},\beta^{1}, \beta^{2}}_{a^{1}_{1}a^{1}_{2} \cdots a^{1}_{n}, a^{2}_{1}a^{2}_{2} \cdots a^{2}_{n},b^{1}_{1}b^{1}_{2} \cdots b^{1}_{n}, b^{2}_{1}b^{2}_{2} \cdots b^{2}_{n}} 
\ket{\alpha^{1} a^{1}_{1}a^{1}_{2} \cdots a^{1}_{n}, \alpha^{2} {a^{2}_{1}a^{2}_{2} \cdots a^{2}_{n},  }} \bra{\beta^{1} b^{1}_{1}b^{1}_{2} \cdots b^{1}_{n}, \beta^{2} {b^{2}_{1}b^{2}_{2} \cdots b^{2}_{n} }},
\end{aligned}
\end{equation}
\end{scriptsize}
 \medmuskip=2mu
\thinmuskip=2mu
\thickmuskip=2mu
where where $a^{i}_{j}, b^{i}_{j}, $ ranges over $\mathbb{D}_{j}$, $i \in \mathbb{N}_{2} =\lbrace 1, 2 \rbrace$, and $ j \in \mathbb{N}_{n} =\lbrace 1, 2, \ldots, n \rbrace$.

In the next section, we describe the most general state of $p$ indistinguishable particles each having $n$ DoFs.

\subsection{$p$ indistinguishable particles each having $n$ DoFs} \label{pPnDoF}
In this section, we describe the most general state of $p$ indistinguishable particles each having $n$ DoF. For better understanding, we are describing the notations once again. 

 The $P$ spatial labels are represented by $\alpha^{i}$ where  $\alpha^{i}$ ranges over $\mathbb{S}^P:= \lbrace s^1, s^2, \cdots, s^P \rbrace$. We write the set  $ \lbrace1, 2, \ldots , n \rbrace$  as $\mathbb{N}_n $. Here  $a^{i}_{j}$ ranges over $\mathbb{D}_{j}:=\lbrace D_{j_{1}}, D_{j_{2}}, \cdots, D_{j_{k_{j}}} \rbrace$, represents the eigenvalue of the $j$-th DoF of the particle in the $\alpha^{i}$-th localized region where $j \in \mathbb{N}_n$.  Thus the general state of $p$ indistinguishable particles each having $n$ DoFs is defined as
\begin{equation} \label{PureGen}
\ket{\Psi^{\left( p,n\right) }}:= \sum_{\alpha^{i},  a^{i}_{j} } \eta^u \kappa^{\alpha^{1}, \alpha^{2}, \cdots, \alpha^{p}}_{a^{1}_{1}a^{1}_{2} \cdots a^{1}_{n}, a^{2}_{1}a^{2}_{2} \cdots a^{2}_{n},\cdots, a^{p}_{1}a^{p}_{2} \cdots a^{p}_{n}} 
\ket{\alpha^{1} a^{1}_{1}a^{1}_{2} \cdots a^{1}_{n}, \alpha^{2} {a^{2}_{1}a^{2}_{2} \cdots a^{2}_{n}, \cdots, \alpha^{p} a^{p}_{1} a^{p}_{2} \cdots a^{p}_{n} }}.
\end{equation}
Here $u$ represents the summation of parity of the cyclic permutations of all the $n$ DoFs. Thus $u$ can be represented as $u=u_1+u_2+ \ldots + u_n=\sum_i u_j$ where $u_j$ is the parity of $j$-th DoF. The value of $\eta$ is $+1$ for bosons and $-1$ for fermions. If we have the following condition that if
\begin{equation} \label{PauliPure}
\left( \alpha^{i}=\alpha^{i^\prime} \right)  \wedge \left( a^{i}_{j}=a^{i^{\prime}}_{j} \right)
\end{equation} 
 for any $i \neq i^{\prime}$ where $\alpha^i , \alpha^{i^{\prime}} \in \mathbb{S}^{P} $ and $j \in \mathbb{N}_{n}$, then we get $\eta=0$ for fermions due to Pauli exclusion principle~\cite{Pauli25}.

Following the above notations, the general density matrix of $p$ indistinguishable particles each having $n$ DoFs is defined as
 \medmuskip=-0mu
\thinmuskip=-0mu
\thickmuskip=-0mu 
\begin{scriptsize}
\begin{equation}	\label{DenGen}
\rho^{(p, n)}:=  \sum_{\substack{\alpha^{i}, \beta^{i}, a^{i}_{j}, b^{i}_{j} }} \eta^{(u+\bar{u})} 
 \kappa^{ \alpha^{(p)},\beta^{(p)}}_{a_{(n)},b_{(n)} }  
\ket{\alpha^{1} a^{1}_{1}a^{1}_{2} \cdots a^{1}_{n}, \alpha^{2} {a^{2}_{1}a^{2}_{2} \cdots a^{2}_{n}, \cdots, \alpha^{p} a^{p}_{1} a^{p}_{2} \cdots a^{p}_{n} }} \bra{\beta^{1} b^{1}_{1}b^{1}_{2} \cdots b^{1}_{n}, \beta^{2} {b^{2}_{1}b^{2}_{2} \cdots b^{2}_{n}, \cdots, \beta^{p} b^{p}_{1} b^{p}_{2} \cdots b^{p}_{n} }},
\end{equation}
\end{scriptsize}
 \medmuskip=2mu
\thinmuskip=2mu
\thickmuskip=2mu 
where 
\begin{equation}
\begin{aligned}
\kappa^{\alpha^{(p)},\beta^{(p)}}_{a_{(n)},b_{(n)} }=&\kappa^{\alpha^{1}, \alpha^{2}, \cdots, \alpha^{p}, \beta^{1}, \beta^{2}, \ldots, \beta^{p}}_{a^{1}_{1}a^{1}_{2} \cdots a^{1}_{n}, a^{2}_{1}a^{2}_{2} \cdots a^{2}_{n}, \cdots, a^{p}_{1}a^{p}_{2} \cdots a^{p}_{n}, b^{1}_{1}b^{1}_{2} \ldots b^{1}_{n}, b^{2}_{1}b^{2}_{2} \ldots b^{2}_{n}, \ldots, b^{p}_{1}b^{p}_{2} \ldots b^{p}_{n}}
\end{aligned}
\end{equation}
and $\alpha^{i}, \beta^{i}$ ranges over $\mathbb{S}^p$, $a^{i}_{j}, b^{i}_{j}$ ranges over $\mathbb{D}_{j}$, $i \in \mathbb{N}_P$ and $j \in \mathbb{N}_n$. 
Here $u$ is as defined in Eq.~\eqref{PureGen} and $\bar{u}$ comes due to density matrix. If we have the following condition that 
\begin{equation} \label{PauliDen}
\left\lbrace  \left( \alpha^{i}=\alpha^{i^\prime} \right) \vee \left( \beta^{i}=\beta^{i^\prime} \right) \right\rbrace  \wedge \left\lbrace  \left( a^{i}_{j}=a^{i^{\prime}}_{j} \right) \vee \left( b^{i}_{j}=b^{i^{\prime}}_{j} \right) \right\rbrace 
\end{equation} 
for any $i \neq i^{\prime}$ where $i , i^{\prime} \in \mathbb{N}_P$   and $j \in \mathbb{N}_{n}$, then we get $\eta=0$ for fermions due to Pauli exclusion principle~\cite{Pauli25}.

The density matrix of Eq.~\eqref{PureGen} is described as
 \medmuskip=-0mu
\thinmuskip=-0mu
\thickmuskip=-0mu 
\begin{scriptsize}
\begin{equation}	\label{DenGenn}
\rho^{(p, n)}:=  \sum_{\substack{\alpha^{i}, \beta^{i}, a^{i}_{j}, b^{i}_{j} }} \eta^{(u+\bar{u})} 
 \kappa^{ \alpha^{(p)}}_{a_{(n)} }  \kappa^{ \beta^{(p)}* }_{ b_{(n)} }
\ket{\alpha^{1} a^{1}_{1}a^{1}_{2} \cdots a^{1}_{n}, \alpha^{2} {a^{2}_{1}a^{2}_{2} \cdots a^{2}_{n}, \cdots, \alpha^{p} a^{p}_{1} a^{p}_{2} \cdots a^{p}_{n} }} \bra{\beta^{1} b^{1}_{1}b^{1}_{2} \cdots b^{1}_{n}, \beta^{2} {b^{2}_{1}b^{2}_{2} \cdots b^{2}_{n}, \cdots, \beta^{p} b^{p}_{1} b^{p}_{2} \cdots b^{p}_{n} }},
\end{equation}
\end{scriptsize}
 \medmuskip=2mu
\thinmuskip=2mu
\thickmuskip=2mu 
where 
\begin{equation}
\begin{aligned}
\kappa^{\alpha^{(p)}}_{a_{(n)} }=&\kappa^{\alpha^{1}, \alpha^{2}, \cdots, \alpha^{p}}_{a^{1}_{1}a^{1}_{2} \cdots a^{1}_{n}, a^{2}_{1}a^{2}_{2} \cdots a^{2}_{n}, \cdots, a^{p}_{1}a^{p}_{2} \cdots a^{p}_{n}},  \\ \kappa^{\beta^{(p)}}_{b_{(n)} }=&\kappa^{\beta^{1}, \beta^{2}, \ldots, \beta^{p}}_{b^{1}_{1}b^{1}_{2} \ldots b^{1}_{n}, b^{2}_{1}b^{2}_{2} \ldots b^{2}_{n}, \ldots, b^{p}_{1}b^{p}_{2} \ldots b^{p}_{n}}, 
\end{aligned}
\end{equation}
and rest of the notations are the same as Eq.~\eqref{DenGen}.

\section{DoF trace-out for indistinguishable particles} \label{DoFTrOut}
In this section, we propose a DoF trace-out rule for indistinguishable particles where each particle has more than one DoF. In~\cite{LFC16,LFC18}, they propose the trace-out rule for indistinguishable particles having single DoF. One may be
tempted to think that the same rule can trace-out a single DoF for indistinguishable particles having single DoF, will work for indistinguishable particles having multiple DoFs. However, this is not so straightforward. When particles
become indistinguishable with multiple DoFs, performing the partial trace-out
of a particular DoF is challenging, because a DoF cannot
be associated with a specific particle. To trace out DoFs of such particles, particularly when the particles are entangled in multiple DoFs, we have proposed a solution. Also, our DoF trace-out rule can treat the cases of both distinguishable and indistinguishable particles under a uniform mathematical framework.

\subsection{Two indistinguishable particles each having two DoFs} \label{Tr2p2DoF}
If we want to perform partial trace in only one region, say $s^{x} \in \mathbb{S^P}$, then the non-normalized density matrix can be written as
 
\begin{equation} \label{Trace2p_Simple}
\begin{aligned}
\tilde{\rho}^{(1)}_{M} = \text{Tr}_{M} \left( \rho^{(2,2)} \right) =\sum_{m_{1},m_{2}, \cdots, m_{n}} \braket{s^{x} m_{1} m_{2} \cdots m_{n} | \rho^{(2,2)} | s^{x} m_{1} m_{2} \cdots m_{n}},
\end{aligned}
\end{equation}
where $m_{j}$ span $ \mathbb{D}_{j}$, where $j \in  \mathbb{N}_2$. 

To perform  DoF trace-out of the $j$-th DoF, $j \in \mathbb{N}_2$, of spatial region $s^x \in \mathbb{S}^{P}$, we define the reduced density matrix of Eq.~\eqref{IDDM_gen} as

\frenchspacing
 \medmuskip=0mu
\thinmuskip=0mu
\thickmuskip=0mu 
\begin{equation} \label{DoFTrRule}
\begin{aligned}
\rho_{s^{x}_{\bar{j}}} \equiv&  \text{Tr}_{s^{x}_{j}} \left(  \rho^{(2,2)} \right) \equiv \sum_{m_{j} \in \mathbb{D}_{j}} \braket{ s^x m_j \mid \rho^{(2,2)} \mid s^x m_j }\\ :=\eta^{(u+\bar{u})}& \sum_{m_{j}} \bigg \lbrace   \sum_{\substack{ \alpha^1, \alpha^2, a^{1}_{j}, a^{1}_{\bar{j}},  a^{2}_{1}, a^{2}_{2}, \\ \beta^1, \beta^2, b^{1}_{j}, b^{1}_{\bar{j}}, b^{2}_{1}, b^{2}_{2}} } \kappa^{\alpha^1 \alpha^2, \beta^1, \beta^2}_{  a^{1}_{j}, a^{1}_{\bar{j}},  a^{2}_{1}, a^{2}_{2},b^{1}_{j}, b^{1}_{\bar{j}}, b^{2}_{1}, b^{2}_{2}}  \braket{s^x m_j \mid \alpha^1 a^{1}_j} \braket{\beta^1 b^{1}_{j}  \mid s^x m_j}  \ket{\alpha^1 a^{1}_{\bar{j}}, \alpha^2  a^{2}_{1} a^{2}_{2}}\bra{\beta^1 b^{1}_{\bar{j}}, \beta^2  b^{2}_{1} b^{2}_{2}}  \\
& +  \eta \sum_{\substack{ \alpha^1, \alpha^2 ,  a^{1}_{1}, a^{1}_{2}, a^{2}_{j}, a^{2}_{\bar{j}}, \\ \beta^1 , \beta^2 , b^{1}_{j}, b^{1}_{\bar{j}}, b^{2}_{1}, b^{2}_{2}} } \kappa^{\alpha^1 \alpha^2, \beta^1, \beta^2}_{a^{1}_{1}, a^{1}_{2}, a^{2}_{j}, a^{2}_{\bar{j}}, b^{1}_{j}, b^{1}_{\bar{j}}, b^{2}_{1}, b^{2}_{2}} \braket{s^x m_j \mid \alpha^2 a^{2}_j }  \braket{\beta^1 b^{1}_j  \mid s^x m_j}  \ket{\alpha^1 a^{1}_{1} a^{1}_{2}, \alpha^2 a^{2}_{\bar{j}}} \bra{\beta^1 b^{1}_{\bar{j}}, \beta^2  b^{2}_{1} b^{2}_{2}} \\
& + \eta  \sum_{\substack{ \alpha^1, \alpha^2 , a^{1}_{j}, a^{1}_{\bar{j}},  a^{2}_{1}, a^{2}_{2}, \\ \beta^1 , \beta^2 , b^{1}_{1}, b^{1}_{2} , b^{2}_{j}, b^{2}_{\bar{j}}} } \kappa^{\alpha^1 \alpha^2, \beta^1, \beta^2}_{a^{1}_{j}, a^{1}_{\bar{j}},  a^{2}_{1}, a^{2}_{2}, b^{1}_{1}, b^{1}_{2} , b^{2}_{j}, b^{2}_{\bar{j}}}  \braket{s^x m_j \mid \alpha^1 a^{1}_j}  \braket{\beta^2 b^{2}_j  \mid s^x m_j}  \ket{\alpha^1 a^{1}_{\bar{j}}, \alpha^2 a^{2}_{1} a^{2}_{2}}\bra{\beta^1 b^{1}_{1} b^{1}_{2}, \beta^2  b^{2}_{\bar{j}}}   \\
 & + \sum_{\substack{ \alpha^1, \alpha^2 ,  a^{1}_{1}, a^{1}_{2}, a^{2}_{j}, a^{2}_{\bar{j}}, \\ \beta^1 , \beta^2 , b^{1}_{1}, b^{1}_{2}, b^{2}_{j}, b^{2}_{\bar{j}}} } \kappa^{\alpha^1 \alpha^2, \beta^1, \beta^2}_{a^{1}_{1}, a^{1}_{2}, a^{2}_{j}, a^{2}_{\bar{j}}, b^{1}_{1}, b^{1}_{2}, b^{2}_{j}, b^{2}_{\bar{j}}}   \braket{s^x m_j \mid \alpha^2 a^{2}_j } \braket{\beta^2 b^{2}_j  \mid s^x m_j}   \ket{\alpha^1 a^{1}_{1} a^{1}_{2}, \alpha^2 a^{2}_{\bar{j}}} \bra{\beta^1 b^{1}_{1} b^{1}_{2}, \beta^2  b^{2}_{\bar{j}}} \bigg \rbrace,
\end{aligned}
\end{equation}
\frenchspacing
 \medmuskip=2mu
\thinmuskip=2mu
\thickmuskip=2mu 
\noindent where $\bar{j}:=(3-j)$. The parameter  $\eta$ is $+1$ ($-1$) for bosons (fermions). Similarly, the trace-out operation of the state in Eq.~\eqref{IDDM} using the above rule.

Equation~\eqref{DoFTrRule} can be generalized for $n$ DoFs and it includes particle trace-out as a special case for $n=1$ as shown in the next section. 
 Similarly, to get the description of the reduced system $\rho_{s^{x}_{i}}$ consisting of a single DoF, we  have to apply our trace-out rule $\left(n-1 \right)$ times.

\subsection{Two indistinguishable particles each having $n$ DoFs} \label{Tr2PnDoF}

Next we  define DoF trace-out rule for indistinguishable particles from the general density matrix of two particles as defined in Eq.~\eqref{DM}. Suppose we want to trace-out the $j$-th DoF of location $s^x \in \mathbb{S}^{P}$. Then the DoF reduced density matrix is
\frenchspacing
 \medmuskip=0mu
\thinmuskip=0mu
\thickmuskip=0mu 
\begin{equation} \label{DoFTrRule_n_dof}
\begin{aligned}
\rho_{s^{x}_{\bar{j}}} \equiv&  \text{Tr}_{s^{x}_{j}} \left(  \rho^{(2,n)} \right) \equiv \sum_{m_{j} \in \mathbb{D}_{j}} \braket{ s^x m_j \mid \rho^{(2,n)} \mid s^x m_j }\\ :=& \sum_{m_{j}} \bigg \lbrace   \sum_{\substack{ \alpha^1, \alpha^2, a^{1}_{j}, a^{1}_{\bar{j}},  a^{2}_{1}, a^{2}_{2}, \ldots, a^{2}_{n}  \\ \beta^1, \beta^2, b^{1}_{j}, b^{1}_{\bar{j}}, b^{2}_{1}, b^{2}_{2}, \ldots, b^{2}_{n} } } \kappa_p  \braket{s^x m_j \mid \alpha^1 a^{1}_j} \braket{\beta^1 b^{1}_{j}  \mid s^x m_j}  \ket{\alpha^1 a^{1}_{\bar{j}}, \alpha^2  a^{2}_{1} a^{2}_{2}  \ldots a^{2}_{n} }\bra{\beta^1 b^{1}_{\bar{j}}, \beta^2  b^{2}_{1} b^{2}_{2} \ldots b^{2}_{n}}  \\
& +  \eta \sum_{\substack{ \alpha^1, \alpha^2,   a^{1}_{1}, a^{1}_{2}, \ldots, b^{1}_{n} a^{2}_{j}, a^{2}_{\bar{j}}, \\ \beta^1 , \beta^2 , b^{1}_{j}, b^{1}_{\bar{j}}, b^{2}_{1}, b^{2}_{2}}, \ldots, b^{2}_{n} }  \braket{s^x m_j \mid \alpha^2 a^{2}_j }  \braket{\beta^1 b^{1}_j  \mid s^x m_j}  \ket{\alpha^1 a^{1}_{1} a^{1}_{2} \ldots a^{1}_{n} \alpha^2 a^{2}_{\bar{j}}} \bra{\beta^1 b^{1}_{\bar{j}}, \beta^2  b^{2}_{1} b^{2}_{2} \ldots b^{2}_{n}} \\
& + \eta  \sum_{\substack{ \alpha^1, \alpha^2, a^{1}_{j}, a^{1}_{\bar{j}},  a^{2}_{1}, a^{2}_{2}, \ldots, a^{2}_{n} \\ \beta^1 , \beta^2 , b^{1}_{1}, b^{1}_{2}, \ldots, b^{1}_{n} b^{2}_{j}, b^{2}_{\bar{j}}} }  \braket{s^x m_j \mid \alpha^1 a^{1}_j}  \braket{\beta^2 b^{2}_j  \mid s^x m_j}  \ket{\alpha^1 a^{1}_{\bar{j}}, \alpha^2 a^{2}_{1} a^{2}_{2} \ldots a^{2}_{n}}\bra{\beta^1 b^{1}_{1} b^{1}_{2} \ldots b^{1}_{n}, \beta^2  b^{2}_{\bar{j}}}   \\
 & + \sum_{\substack{ \alpha^1, \alpha^2 ,  a^{1}_{1}, a^{1}_{2}, \ldots, a^{1}_{n}, a^{2}_{j}, a^{2}_{\bar{j}}, \\ \beta^1 , \beta^2 , b^{1}_{1}, b^{1}_{2} , \ldots, b^{1}_{n} b^{2}_{j}, b^{2}_{\bar{j}}} }   \braket{s^x m_j \mid \alpha^2 a^{2}_j } \braket{\beta^2 b^{2}_j  \mid s^x m_j}   \ket{\alpha^1 a^{1}_{1} a^{1}_{2} \ldots a^{1}_{n},  \alpha^2 a^{2}_{\bar{j}}} \bra{\beta^1 b^{1}_{1} b^{1}_{2} \ldots b^{1}_{n} \beta^2  b^{2}_{\bar{j}}} \bigg \rbrace,
\end{aligned}
\end{equation}
\frenchspacing
 \medmuskip=2mu
\thinmuskip=2mu
\thickmuskip=2mu 
where
\begin{equation}
\begin{aligned}
\kappa_p=&\kappa^{\alpha^1 \alpha^2, \beta^1, \beta^2}_{  a^{1}_{j}, a^{1}_{\bar{j}},  a^{2}_{1}, a^{2}_{2} , \ldots, a^{2}_{n},b^{1}_{j}, b^{1}_{\bar{j}}, b^{2}_{1}, b^{2}_{2}, \ldots, b^{2}_{n}},\\
\kappa_q=& \kappa^{\alpha^1 \alpha^2, \beta^1, \beta^2}_{a^{1}_{1}, a^{1}_{2}, \ldots, b^{1}_{n} a^{2}_{j}, a^{2}_{\bar{j}}, b^{1}_{j}, b^{1}_{\bar{j}}, b^{2}_{1}, b^{2}_{2}, \ldots, b^{2}_{n}} ,\\
\kappa_r=& \kappa^{\alpha^1 \alpha^2, \beta^1, \beta^2}_{a^{1}_{j}, a^{1}_{\bar{j}},  a^{2}_{1}, a^{2}_{2}, \ldots, a^{2}_{n}, b^{1}_{1}, b^{1}_{2} , \ldots, b^{1}_{n}, b^{2}_{j}, b^{2}_{\bar{j}}} ,\\
\kappa_s=& \kappa^{\alpha^1 \alpha^2, \beta^1, \beta^2}_{a^{1}_{1}, a^{1}_{2}, \ldots, a^{1}_{n}, a^{2}_{j}, a^{2}_{\bar{j}}, b^{1}_{1}, b^{1}_{2}, \ldots, b^{1}_{n}, b^{2}_{j}, b^{2}_{\bar{j}}} ,\\
\end{aligned}
\end{equation}
 \medmuskip=0mu
\thinmuskip=0mu
\thickmuskip=0mu

It may be noted that for $n=2$, the DoF trace-out rule defined in Eq.~\eqref{DoFTrRule_n_dof} reduces to Eq.~\eqref{DoFTrRule}. For $n=1$,  this becomes equivalent to the  particle trace-out rule as defined Eq.~\eqref{tracedef}. On the other hand, for $n > 1$, if we apply DoF trace-out rule of Eq.~\eqref{DoFTrRule} $n$ times, the effect will not be the same as the particle trace-out in Eq.~\eqref{tracedef}. The reason behind this is as follows. For indistinguishable particles, the particle trace-out operation vanishes all the DoFs together for one particle; whereas each DoF trace-out operation leaves an expression with many terms each of which vanishes the corresponding DoF from one particle at a time and retains the same DoF in the remaining particles.

\subsection{$p$ indistinguishable particles each having $n$ DoFs} \label{TrpPnDoF}
Finally, we  define DoF trace-out rule for $p$ indistinguishable particles  each having $n$ DoFs from the general density matrix Eq.~\eqref{DenGen}. Suppose we want to trace-out the $j$-th DoF of location $s^x \in \mathbb{S}^{P}$, then the DoF reduced density matrix is
\begin{equation} \label{DoFTrRule_n_dof_p_particle}
\begin{aligned}
\rho_{s^{x}_{\bar{j}}} \equiv&  \text{Tr}_{s^{x}_{j}} \left(  \rho^{(p,n)} \right) \equiv \sum_{m_{j} \in \mathbb{D}_{j}} \braket{ s^x m_j \mid \rho^{(p,n)} \mid s^x m_j }.
\end{aligned}
\end{equation}
This can be expanded similarly as shown in Eq.~\eqref{DoFTrRule_n_dof}.

\section{Physical significance of the proposed DoF trace-out rule} \label{DoF_Tr_Phy_Sig}
Our DoF trace-out rule plays a very critical role with respect to the recently introduced complex systems with inter-DoF entanglements~\cite{HHNL,camalet17,camalet18}. When such entanglement exists, measuring or non-measuring one of the participating DoFs would influence the measurement results of the other participating DoFs. In an experiment involving such systems, the DoF trace-out can operationally be implemented simply by choosing to measure a particular DoF in a particular spatial location while ignoring the others. However, the statistics so obtained cannot be predicted using the existing trace-out rules in either the first or the second quantization notations.

 The physical meaning of the term $\rho_{s^{x_{\bar{j}}}}$ in Eq.~\eqref{DoFTrRule} is that it represents the description of the reduced system after measuring the whole system in the $j$-th DoF in the spatial region $s^x$. On the other hand, the physical interpretation of the term $\rho_{s^{x_{j}}}$ is that if someone measures only the $i$-th DoF in the spatial region $s_x$, then the measure statistics would be equivalent to the system $\rho_{s_{x_{i}}}$. It can be noted that our DoF trace-out rule in Eq.~\eqref{DoFTrRule} is order independent. Our framework expressed in Eq.~\eqref{DoFTrRule} can deal with all such systems with inter-DoF correlations in indistinguishable particles, leading to the prediction of perfect measurement statistics. 

 Further, it generalizes the standard existing trace-out rule and is therefore suitable for such entanglement structures of distinguishable particles as well. More specifically, for distinguishable particles, tracing out a single DoF of a particle is analogous to tracing out a whole particle; for indistinguishable particles, on the other hand, tracing out a single DoF is performed for a specific spatial location where wave-functions of multiple particles might be overlapping. These overlaps are taken care of in the inner-product terms in the expression of Eq.~\eqref{DoFTrRule}.

 In short, the physical significance of our result is that it extends the standard density matrix approach of quantum information to systems of indistinguishable particles entangled in multiple degrees of freedom.

\chapter{Generalized relation between teleportation fidelity and singlet fraction} \label{Chap6}
In this chapter, we generalize the existing relation between teleportation fidelity and singlet fraction for both distinguishable and indistinguishable particles where each particle have multiple DoFs. We also describe the physical significance of the proposed generalized relation. Then we  propose an optical scheme to generate inter-DoF entangled state for distinguishable particles where the non-triviality of the above relation is explained. 

This chapter is based on the work in~\cite{Das_SF_21}.

\section{Generalized teleportation fidelity} \label{Gen_TF}
For particles having a single DoF, there is only one teleportation channel. But, when multiple DoFs, say $n$, are available for each particle, then up to  $n^2$ teleportation channels are possible. 
So the previous definition of teleportation fidelity will not work for distinguishable particles and indistinguishable particles each having multiple DoFs.
For any teleportation protocol, the motivation is to maximize the information transfer. So we define the generalized teleportation fidelity in such a way that it captures the maximum  fidelity over all possible channels. 
\begin{figure}[t!] 
\centering
\includegraphics[width=0.7\columnwidth]{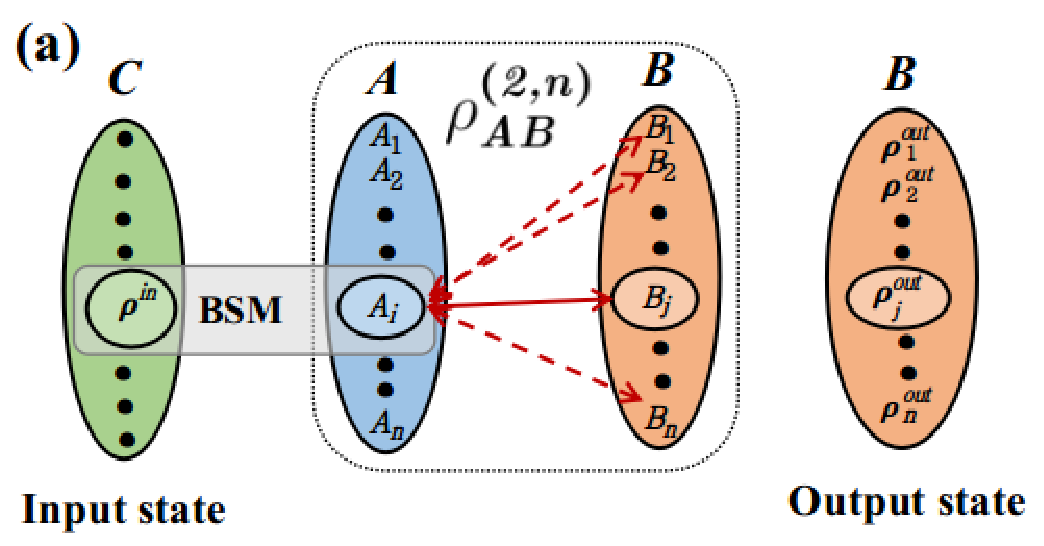}  \includegraphics[width=\columnwidth]{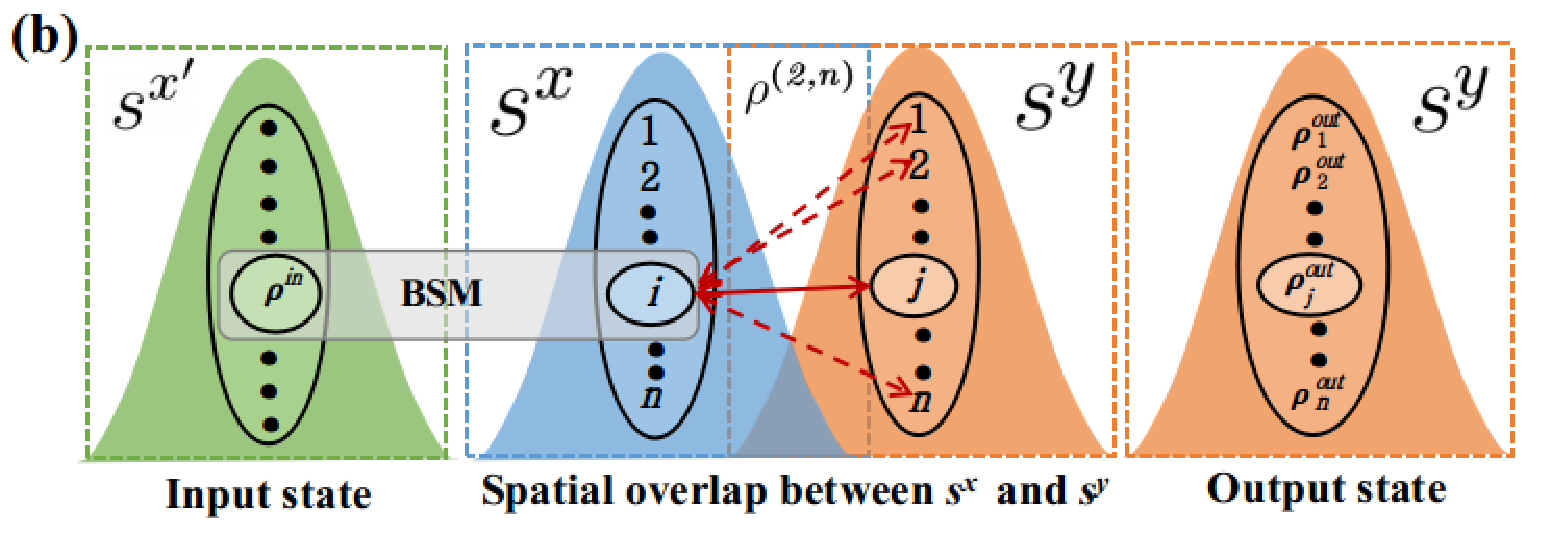} 
\caption{Teleportation operation using (a) two distinguishable particles and (b) two indistinguishable particles, each having $n$ DoFs.}
\label{teleport}
\end{figure}

Suppose $\rho^{(2,n)}_{AB}$ denotes the joint state of two distinguishable particles $A$ and $B$, each having $n$ DoFs following Eq.~\eqref{DMdis}.  Another particle $C$ having $n$ DoFs nearby to $A$, has an unknown state in any DoF denoted by $\rho^{in}$ which has to be teleported to $B$. For teleportation operation, Bell state measurement (BSM) is performed with the DoF of $C$ having the unknown state $\rho^{in}$ and  the $i$-th DoF of particle $A$. After BSM, the unknown state is teleported to all the DoFs of $B$ as shown in Fig.~\ref{teleport} (a). We denote the teleported state on the $j$-th DoF of the particle $B$ by $\rho^{out}_{j}$ where $j \in \mathbb{N}_n = 1, 2, \ldots, n \rbrace$. The teleportation fidelity  between the $i$-th DoF of $A$ and the $j$-th DoF of $B$ is given by 
\begin{equation} \label{f_i_j}
f^{i}_{j}:=\text{Tr}\sqrt{\sqrt{\rho^{in}}\rho^{out}_{j}\sqrt{\rho^{in}}}.
\end{equation} 
As the goal of any teleportation protocol is to maximize the fidelity, so we define the generalized  teleportation fidelity for the state $\rho^{(2,n)}_{AB}$ as
\begin{equation} \label{fg}
f_{g}:=\max_{ i,j  \in \mathbb{N}_n  } \left\lbrace  f^{i}_{j} \right\rbrace.
\end{equation}

The teleportation operation using indistinguishable particles~\cite{Ugo15,LFC18} is almost the same as that of using distinguishable particles.  Consider two indistinguishable particles with density matrix $\rho^{(2,n)}$ as defined in~Eq.~\eqref{DM}, located in the spatial regions $s^x$ and $s^y$ where $s^x, s^y \in \mathbb{S}^P$. The particle $C$, distinguishable from them, is located in the spatial region $s^{x^{\prime}} \in \mathbb{S}^P$ nearby to $s^x$, having an unknown state $\rho^{in}$ in one of its DoF. 
This unknown state has to be teleported in the spatial region $s^y$. Now
Bell state measurement is performed on the $i$-th DoF of the spatial region $s^x$ and the DoF of $C$ having the unknown state $\rho^{in}$. After this operation,  $\rho^{in}$ is teleported (probabilistically) to all the DoFs of the spatial region $s^y$ as shown in Fig.~\ref{teleport} (b). We denote the teleported state of the $j$-th DoF of the spatial region $s^y$ by $\rho^{out}_{j}$ and the  teleportation fidelity between the $i$-th DoF of spatial location $s^x$ and the $j$-th DoF of the spatial region $s^y$ is denoted by  $f^{i}_{j}:=\text{Tr}\sqrt{\sqrt{\rho^{in}}\rho^{out}_{j}\sqrt{\rho^{in}}}.$ With this notation in the case of indistinguishable particles, the generalized  teleportation fidelity for any arbitrary $\rho^{(2,n)}$ is calculated in the same way as Eq.~\eqref{fg}.

\section{Generalized singlet fraction} \label{Gen_SF}
  For two distinguishable particles, say $A$ and $B$, each having $n$ DoFs as in Eq.~\eqref{DMdis}, if  $a_i$ is maximally entangled with $b_j$, then all the other pairs of the form $\lbrace a_i, b_k \rbrace$, $k\neq j$, as well as $\lbrace a_l, b_j \rbrace$, $l\neq i$, become separable due to monogamy of entanglement between DoFs~\cite{camalet17,camalet18}. For two indistinguishable particles each having $n$ DoFs as in Eq.~\eqref{DM}, it is possible to have all $\lbrace a_i, b_j \rbrace$ pairs are maximally entangled as shown in~\cite{Paul21}. The current definition of singlet fraction does not capture all such scenarios for distinguishable particles and indistinguishable particles. So  the motivation behind our  generalized singlet fraction is to capture all the entangled or separable structures between two distinguishable particles or indistinguishable particles. 
 So we take the summation of all the singlet fraction possible for any particular DoF. Then we maximized that value for all DoFs of the both particles.

For two distinguishable particles $A$ and $B$ with the joint state $\rho^{(2,n)}_{AB}$ as shown in~\eqref{DMdis}, we define the generalized singlet fraction as
\begin{equation} \label{SFGen}
 F^{(n)}_{g}:=\max \left\{  \max_{i} \lbrace F(i) \rbrace,   \max_{j} \lbrace F(j) \rbrace \right\},
\end{equation}
\medmuskip=2mu
\thinmuskip=2mu
\thickmuskip=2mu
where
\begin{equation}
\begin{aligned}
 F(i):=& \max_{\psi_{a_{i}b_{{j}}}} { \sum^{n}_{j=1} P_{a_{i}b_{{j}}} }, \\
 F(j):=& \max_{\psi_{a_{i}b_{{j}}}} {  \sum^{n}_{i=1} P_{a_{i}b_{{j}}}  }, \\
 P_{a_{i}b_{j}}:=& \braket{\psi_{a_{i}b_{j}} \mid \rho_{a_{i}b_{j}} |\psi_{a_{i}b_{j}}}, \\
 \rho_{a_{i}b_{j}}:=& \text{Tr}_{a_{\bar{i}}{b_{\bar{j}}}} ( \rho^{(2, n)}_{AB} ). 
\end{aligned}
\end{equation}
Here $a_{\bar{i}}=a_{1}a_{2} \cdots a_{i-1}a_{i+1} \cdots a_{n}$ and similar meaning for $b_{\bar{i}}$. 
In the terms \\ $\max_{\psi_{a_{i}b_{{j}}}} \big\lbrace \sum^{n}_{j=1} P_{a_{i}b_{{j}}} \big\rbrace$, the $i$-th DoF of $A$ is kept fixed and $\ket{\psi_{a_{i}b_{{j}}}}$ spans all possible maximally entangled states between the $i$-th DoF of $A$ and the $j$-th DoF of $B$. Similarly for the other term, the $j$-th DoF of $B$ is kept fixed. 
 
\medmuskip=2mu
\thinmuskip=2mu
\thickmuskip=2mu 
Next consider two indistinguishable particles with joint state $\rho^{(2, n)}$ as in~\eqref{DM}, each having $n$ DoFs. The generalized singlet fraction for two spatial regions $s^x, s^y \in \mathbb{S}^P$ as shown in Fig.~\ref{teleport} (b) can be calculated similarly  using Eq.~\eqref{SFGen} where 
\begin{equation}
\begin{aligned}
F(i)=&\max_{\psi_{s^{x}_{i}s^{y}_{j}}} { \sum^{n}_{j=1} P_{s^{x}_{i}s^{y}_{{j}}}}, \\
 F(j)=&\max_{\psi_{s^{x}_{i}s^{y}_{j}}} {  \sum^{n}_{i=1} P_{s^{x}_{i}s^{y}_{{j}}}}, \\
 P_{s^{x}_{i}s^{y}_{{j}}}:=& \braket{\psi_{s^{x}_{i}s^{y}_{{j}}}|\rho_{s^{x}_{i}s^{y}_{{j}}} |\psi_{s^{x}_{i}s^{y}_{{j}}}}, \\
 \rho_{s^{x}_{i}s^{y}_{{j}}}:=& \text{Tr}_{s^{x}_{\bar{i}}s^{y}_{\bar{j}}} ( \rho^{(2, n)} )
\end{aligned}
\end{equation}
\frenchspacing
 \medmuskip=2mu
\thinmuskip=2mu
\thickmuskip=2mu 
\frenchspacing
 \medmuskip=2mu
\thinmuskip=2mu
\thickmuskip=2mu 
Here $\ket{\psi_{s^{x}_{i}s^{y}_{{j}}}}$ spans all possible maximally entangled states between the $i$-th DoF of $s^x$ and the $j$-th DoF of $s^y$ and $s^x, s^y \in  \mathbb{S}^P $, following the DoF trace-out rule in Chapter~\ref{Chap5}.

\section{Relation between the generalized teleportation fidelity and the generalized singlet fraction} \label{TF_ST_Rel}
Consider a two-parameter family of states for two particles each having $n$ DoFs as
\begin{equation} \label{twopara}
\rho^{(2,n)}_{p}:= p \mathbb{P}^{(2,n)} + (1-p) \frac{\mathbb{I}^{n} \otimes \mathbb{I}^{n} }{d^{2n}}, \hspace{1cm} 0 \leq p \leq 1.
\end{equation} 
This equation is applicable for both distinguishable and indistinguishable particles.

 For two distinguishable particles $A$ and $B$, the state $\rho^{(2, n)}_{p}$ is an arbitrary $\rho^{(2, n)}_{AB}$ as in Eq.~\eqref{DMdis} and  $ \mathbb{P}^{(2, n)}$ is a special form of $ \rho^{(2, n)}_{AB}$ such that for every $a_i$, there exists atleast one $b_j$ with $\mathbb{P}_{a_{i}b_{j}}=\text{Tr}_{a_{\bar{i}}b_{\bar{j}}} \left( \mathbb{P}^{(2,n)} \right)$ is $d$-dimensional maximally entangled where $i, j \in \mathbb{N}_n $. 
 
  For two indistinguishable particles, $\rho^{(2,n)}_{p}$ is an arbitrary state $\rho^{(2,n)}$ of Eq.~\eqref{DM}, and $\mathbb{P}^{(2, n)} $ is a special form of $\rho^{(2,n)}$  such that $\mathbb{P}_{s^{x}_{i}s^{y}_{{j}}} = \text{Tr}_{s^{x}_{\bar{i}}s^{y}_{\bar{j}}} \left( \mathbb{P}^{(2,n)} \right)$ is maximally entangled for all $i, j \in \mathbb{N}_n$. 
This type of state is possible using indistinguishable particle~\cite{Paul21}. But for distinguishable particles, such type of state is not possible as they obey monogamy of entanglement~\cite{camalet17}. 

First, we calculate the generalized  teleportation fidelity for the state $\rho^{(2,n)}_{p}$ of Eq.~\eqref{twopara} and for that we have to calculate the same separately for both $\mathbb{P}^{(2,n)}$ and  completely random noise $\frac{\mathbb{I}^{n} \otimes \mathbb{I}^{n} }{d^{2n}}$. 
We denote $f_{max}$ to be the value of $f_{g}$ for the state $\mathbb{P}^{(2,n)} $.
 Note that, the difference between $f_{g}$ and $f_{max}$ is that the first one is the maximum value of $f^{i}_{j}$ for an arbitrary state $\rho^{(2,n)}_{AB}$  or $\rho^{(2, n)}$  but the second one is the maximum value of $f^{i}_{j}$ for the special state $\mathbb{P}^{(2, n)} $.
 
 For distinguishable particles, $\max_{(i,j)} \lbrace f^{i}_{j} \rbrace=1$  occurs for the pair $(i, j)$ such that $\mathbb{P}_{a_{i}b_{{j}}}$ is maximally entangled. For indistinguishable particles, $f_{max} < 1$ due to the \textit{no-go} theorem of~\cite{Das20}.  The value of $f_g$ for  $\frac{\mathbb{I}^{n} \otimes \mathbb{I}^{n} }{d^{2n}}$  is the same for both distinguishable particles and indistinguishable particles which is  $\frac{1}{d}$   as  $\rho^{out}=\frac{\mathbb{I}}{d}$ independent of the initial state $\rho^{in}$ before teleportation. Thus the value of $f_g$ for Eq.~\eqref{twopara} is
Thus we calculate the generalized  teleportation fidelity applicable  for both distinguishable and indistinguishable particles for the state in Eq.~\eqref{twopara} as
\begin{equation} \label{fgparam}
f_{g} = p f_{max} + (1-p)\frac{1}{d},
\end{equation}
where $f_{max}=1$ for distinguishable particles and $f_{max} < 1$ for indistinguishable particles.

Next, we calculate the generalized singlet fraction for the state $\rho^{(2,n)}_{p}$ of Eq.~\eqref{twopara}. We denote  $F^{(n)}_{max}$ be the value of $F^{(n)}_{g}$ for the state $\mathbb{P}^{(2,n)}$. We know that distinguishable particles obey monogamy of  entanglement. So, let us fix a particular $i$, then  
$P_{a_{i}b_{j}}=\braket{\psi_{a_{i}b_{j}} \mid \mathbb{P}^{(2,n)}_{a_{i}b_{j}} \mid \psi_{a_{i}b_{j}}}$ would be 1 for a particular $j$, say $j^{\prime}$, as $\mathbb{P}^{(2,n)}_{a_{i}b_{j^{\prime}}}$ is maximally entangled. When $j \neq j^{\prime}$, we get $P_{a_{i}b_{j}}=\frac{1}{d}$ as $\mathbb{P}^{(2, n)}_{a_{i}b_{j}}$ is separable, where $i, j , j^{\prime} \in \mathbb{N}_n$. So, for the rest of the $(n-1)$ DoFs, we get $P_{a_{i}b_{j}}=\frac{1}{d}$. Thus the value of $F^{(n)}_{max}$ for distinguishable particles is $ \left( 1+\frac{n-1}{d} \right) $.

It is proved in~\cite{Paul21} that indistinguishable particles does not obey monogamy. So, if we fix any particular $i$, then the value of $P_{s^{x}_{i}s^{y}_{{j}}}$ can be 1 for all the values of $j$, as $\mathbb{P}_{s^{x}_{i}s^{y}_{{j}}}$ can be maximally entangled for all the values of $j$. So, for indistinguishable particles the value of $\displaystyle \max_{i} \lbrace F(i) \rbrace$  is $n$, and similarly the value of $ \displaystyle \max_{j} \lbrace F(j) \rbrace=n$, leading to $F^{(n)}_{max}=n$.
For  $\frac{\mathbb{I}^{n} \otimes \mathbb{I}^{n} }{d^{2n}}$, we  get $P_{a_{i}b_{j}}=P_{s^{x}_{i}s^{y}_{j}}=\frac{1}{d^{2}}$ for all $i$ and $j$. So, the value of $F^{(n)}_{g}$ is $\frac{n}{d^2}$ which is the same for distinguishable particles and indistinguishable particles.
Thus, the value of $F^{(n)}_{g}$ in Eq.~\eqref{twopara} is 
 \begin{equation} \label{Fn}
  F^{(n)}_{g} = p F^{(n)}_{max} + (1-p)\frac{n}{d^{2}}.
 \end{equation}

Using Eq.~\eqref{fgparam} and Eq.~\eqref{Fn}, we get the relation between the generalized teleportation fidelity and the generalized singlet fraction as
\begin{equation} \label{fandF}
f_{g}=\frac{\left( F^{(n)}_{g} - \frac{n}{d^{2}} \right) \left( f_{max} -\frac{1}{d}\right)  }{\left( F^{(n)}_{max}-\frac{n}{d^{2}} \right)}  + \frac{1}{d},
\end{equation}
where $f_{g} \in \left[\frac{1}{d}, f_{max} \right] $ and $F^{(n)}_{g} \in \left[ \frac{n}{d^{2}}, F^{(n)}_{max} \right] $. Whether the particles are distinguishable or indistinguishable, the same Eq.~\eqref{fandF} holds, only the values of $f_{max}$ and $F^{(n)}_{max}$ varies.
For particles having single DoF, i.e., $n=1$, Eq.~\eqref{fandF}, reduces to the standard particle version of Eq.~\eqref{f_F_par}.

\section{Illustration of the proposed generalized relation to some special states for distinguishable and indistinguishable particles} \label{Optical_ckt}
If two particles each having a single DoF are maximally entangled, then the value of generalized singlet fraction is one. However, the converse is not necessarily true which we illustrate by proposing an optical circuit  using two distinguishable particles each  having two DoFs, polarization and orbital angular momentum (OAM).

\begin{figure}[h!] 
\centering
\includegraphics[width=\columnwidth]{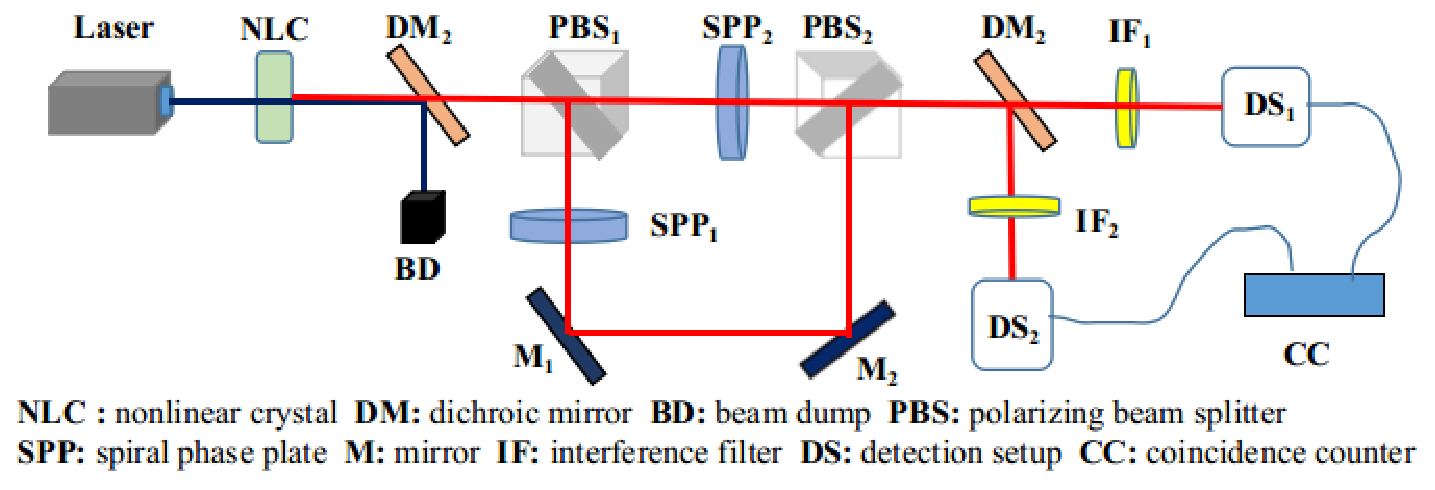} 
\caption{Optical set up to the generation of entangled state between polarization and OAM DoF for distinguishable particles.
NLC: nonlinear crystal, DM: dichroic mirror, BD: beam dump, PBS: polarizing beam spliters, SPP: spiral phase plates, M: mirrors, IF: interference filters, DS: detection setup and C.C.: coincidence counter. The detection setups could be made up of spatial light modulators and single-mode fibers coupled to avalanche photo detectors if measuring in OAM or half-wave plates and PBS if measuring in polarization.
}
\label{Dis_HHES_ckt}
\end{figure}

In Fig.~\ref{Dis_HHES_ckt}, a nonlinear  crystal ($\beta$ barium borate, periodically poled potassium triphosphate etc.) is pumped by a pulsed laser from which a pump photon is absorbed and two photons are generated governed by the phase-matching conditions:  
\begin{equation}
\omega_p=\omega_s + \omega_i, \hspace{0.2cm} \textbf{k}_p = \textbf{k}_s+\textbf{k}_i
\end{equation}
 where $\hbar\omega_j$ is the energy and $\hbar\textbf{k}_j$ is the momentum of the \emph{j}-th photon. The indices \emph{p}, \emph{s} and \emph{i} stand for pump, signal and idler respectively. The signal and idler photon pair produced in this method under non-degenerate type II spontaneous parametric down conversion are correlated in polarization. It is represented in the most general form as 
\begin{equation}
 \vert \Psi_{i} \rangle = \cos\theta \vert H\rangle_{\omega_s}\vert V \rangle_{\omega_i} + e^{i\phi}\sin\theta \vert V\rangle_{\omega_s} \vert H \rangle_{\omega_i},
 \end{equation} where $\ket{H}$ and $\ket{V}$
  denotes horizontal and vertical polarization, $\theta$ controls the normalization factor and $\phi$ is a phase term that arises from birefringence in the nonlinear crystal. Since $\omega_s\neq\omega_i$ (under non-degenerate phase matching) this leads to distinguishability between the signal and idler photons. Let this state be now incident on a broadband polarizing beam splitter (PBS) which allows $\vert H \rangle$ photons to pas through to the transmitted path, while $\vert V \rangle$ photons are directed to the reflected path. Each of the paths contains a spiral phase plate (SPP) of equal topological charge \emph{l} which can take any value from $- \infty$ to $+ \infty$ through $0$.  SPP are optical devices with continuously varying thickness. photons pick up OAM $l$ on passing through them. This results in the output wavefront possessing a helical nature corresponding to the topological charge. The two paths are again combined at another broadband PBS resulting in the state

\begin{equation}
\label{DIsHHES}
\vert \Psi_{f} \rangle = \cos\theta \vert H, +l\rangle_{\omega_s}\vert V, -l \rangle_{\omega_i} + e^{i\phi}\sin\theta \vert V, -l\rangle_{\omega_s} \vert H, +l \rangle_{\omega_i}.
\end{equation}

The handedness of \emph{l} is sensitive to reflections and changes by $\exp(i\pi)$ under each reflection. The difference in the handedness of \emph{l} arises in the above equation due to an unequal number of reflections between the transmitted and reflected arms. 

The generalized singlet fraction of the above state is 1 as shown in Section~\ref{Gen_Fn_Exp}. Similar conclusion can be drawn for two indistinguishable particles as proposed in~\cite{Bhatti15}. 
Note that, our proposed state in Eq.~\eqref{DIsHHES}
 is different from~\cite{Bhatti15} as we use non-degenerate phase-matching to make use of distinguishable photons.
Thus, the singlet fraction alone might not be the best quantifier for the presence of maximal entanglement.

 Another special state is hyper-hybrid entangled state~\cite{HHNL} with two particles each having two DoFs. The maximum value of $F^{\left(2 \right) }_g=2 $ for this state is achieved as each DoF of one particles is maximally entangled with all the other DoFs of other particle as shown in~\cite{Paul21}. 

\begin{figure}[h!] 
\centering
\includegraphics[width=\columnwidth]{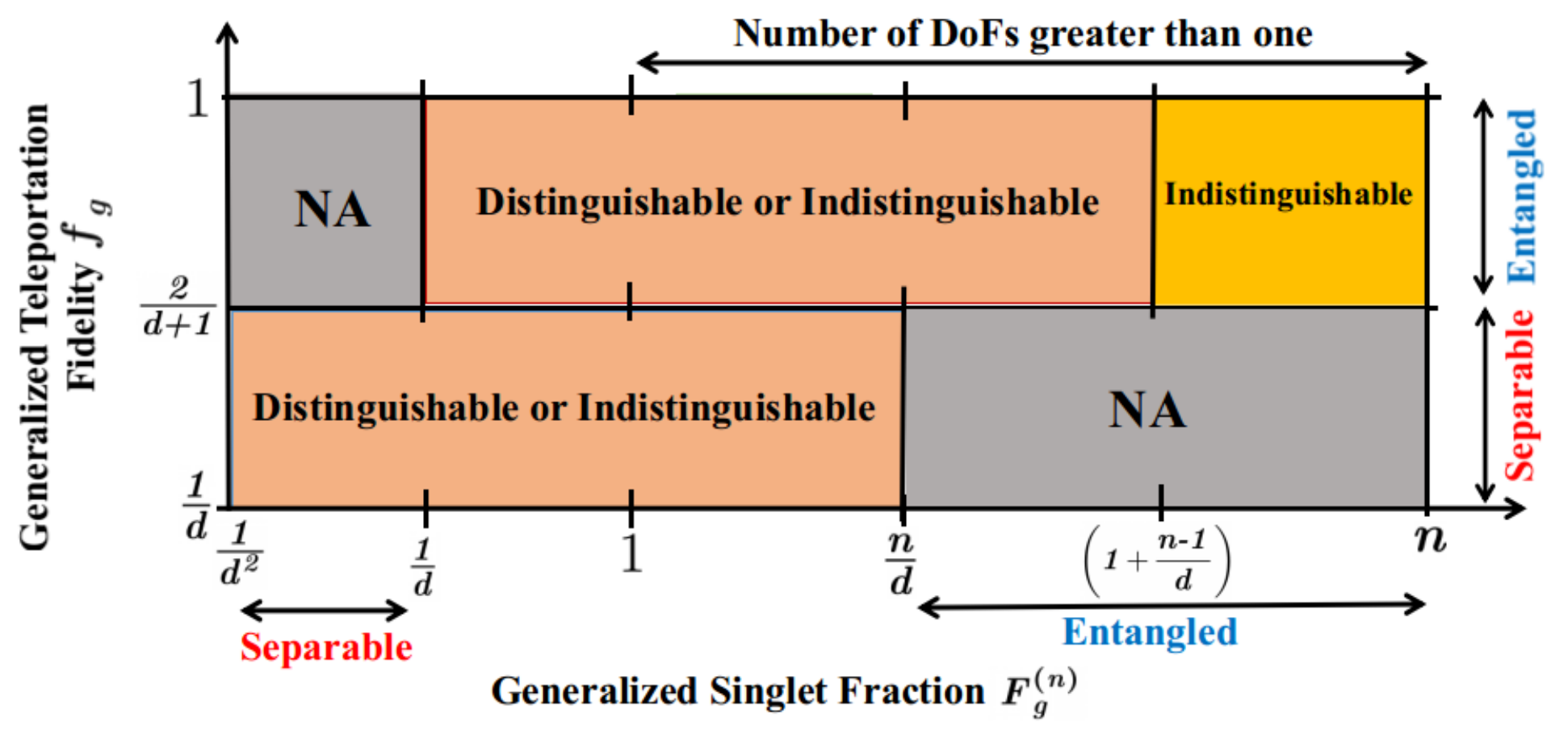} 
\caption{Characterization of the different kinds of states based on their separability, indistinguishability and number of DoFs present using the number of DoFs $n$,  generalized Singlet fraction $F^{(n)}_{g}$ and generalized telepotation fidelity $f_{g}$.}
\label{TF_SF_rel}
\end{figure}   

\section{Generalized singlet fraction for our proposed state} \label{Gen_Fn_Exp}
Here, we will calculate the value of generalized singlet fraction for the state proposed in Eq. (10) of the main text.
The state is written as 
 \medmuskip=2mu
\thinmuskip=2mu
\thickmuskip=2mu 
\begin{equation}
\label{DIsHHES}
\vert \Psi_{f} \rangle= \cos\theta \vert H, +l\rangle_{\omega_s}\vert V, -l \rangle_{\omega_i} + e^{i\phi}\sin\theta \vert V, -l\rangle_{\omega_s} \vert H, +l \rangle_{\omega_i}.
\end{equation}
\medmuskip=2mu
\thinmuskip=2mu
\thickmuskip=2mu 
The density matrix of this state is 
\begin{equation}
\begin{aligned}
\rho_{f} =\vert \Psi_{f} \rangle \bra{\Psi_{f}}= & \cos^{2}\theta  \vert H, +l\rangle_{\omega_s}\vert V, -l \rangle_{\omega_i} \bra{ H, +l}_{\omega_s}\bra{V, -l }_{\omega_i} \\
&+ e^{-i\phi} \cos \theta \sin\theta \vert H, +l\rangle_{\omega_s}\vert V, -l \rangle_{\omega_i} \bra{ V, -l }_{\omega_s} \bra{ H, +l }_{\omega_i} \\
&+e^{i\phi} \cos \theta \sin\theta \ket{ V, -l }_{\omega_s} \ket{ H, +l }_{\omega_i}  \bra{ H, +l}_{\omega_s} \bra{ V, -l }_{\omega_i} \\
&+ \sin^{2}\theta \ket{ V, -l }_{\omega_s} \ket{ H, +l }_{\omega_i} \bra
{ V, -l }_{\omega_s} \bra{ H, +l }_{\omega_i}.
\end{aligned} 
\end{equation}
Now tracing out both the OAM DoFs from the density matrix $\rho_f $ using Eq.~\eqref{DisDoFTraceout}, we get
\begin{equation}\label{rho21wswi}
\begin{aligned}
\rho = & \cos^{2}\theta  \vert H\rangle_{\omega_s}\vert V\rangle_{\omega_i} \bra{ H}_{\omega_s}\bra{V }_{\omega_i} + \sin^{2}\theta \ket{ V}_{\omega_s} \ket{ H }_{\omega_i} \bra
{ V }_{\omega_s} \bra{ H }_{\omega_i}.
\end{aligned}
\end{equation}
Now the singlet fraction for Eq.~\eqref{rho21wswi} is 
\begin{equation}
\max_{\psi} \braket{\psi|\rho|\psi} = \frac{1}{2} (\cos^{2} \theta + \sin^{2} \theta ) = \frac{1}{2}.
\end{equation}
where $\ket{\psi}$ varies over all maximally entangled states.
So, the singlet fraction of the above state is $\frac{1}{2}$. If we calculate singlet fraction of the other entanglement connections by the similar way, it will also be $\frac{1}{2}$. So, the generalized singlet fraction of the state $\vert \Psi_{f} \rangle$ using the Eq.(8) of main text is $F^{(2)}_g =1$. 

\section{Derivation of the upper bound of generalized singlet fraction} \label{up_bound_Fn}
It is already shown in Section~\ref{TF_ST_Rel} that the maximum value of the generalized singlet fraction for indistinguishable particles having $n$ DoFs is $n$. Now we will derive for the maximum value of the generalized singlet fraction for distinguishable particles.

Suppose a particle $A$ is entangled with $n$ other distinguishable particles labeled as $B_1, B_2, \ldots , B_n$, each having a single DoF with dimention $d$. Its joint state is represented as $\rho^{(1)}_{AB_1B_1 \ldots B_n}$. Now we can calculate the singlet fraction between $A$ and $B_j$ for $i \in \mathbb{N}_n=\lbrace 1, 2, \ldots, n \rbrace$ as
\begin{equation}
F_{AB_j} = \max_{\psi} \braket{\psi\mid \rho^{(1)}_{AB_{j}}\mid \psi},
\end{equation}
where  $\rho^{(1)}_{AB_{j}}=\text{Tr}_{B_{\bar{j}}} (\rho^{(1)}_{AB_1B_1 \ldots B_n}) $ and $B_{\bar{j}}=B_1 B_2 \ldots B_{j-1} B_{j+1} \ldots B_n$. Here, $\ket{\psi}$ varies over all maximally entangled states.

 The monogamy relation with respect to $A$ is given in~\cite{Kay09} as
 \begin{equation} \label{Kay}
 \sum^{n}_{j=1} F_{AB_j} \leq \dfrac{d-1}{d} + \dfrac{1}{n+d-1} \left( \sum^{n}_{j=1} \sqrt{F_{AB_j}} \right)^2. 
\end{equation}  

This relation is valid if $B_1, B_2, \ldots , B_n$ are the $n$ DoFs of the particle $B$. Then we want to the bound of $F^{(n)}_g$ for the state $\rho^{(n)}_{AB}$ as defined in Eq.~\eqref{DMdis}.

We take any $n$ numbers of random variable $x_1, x_2 \ldots x_n$ such that any $ 0 \leq x_j \leq 1$ for $j \in \mathbb{N}_n$. Then we have
\begin{equation} \label{Iq}
\begin{aligned}
\left( \sum^{n}_{j=1} \sqrt{x_i} \right)^2 =& \left( \sqrt{x_1}+ \sqrt{x_2}+ \ldots + \sqrt{x_n}\right)^2   \\
=& \sum^{n}_{j=1} x_i + 2 \sum^{n}_{\substack{i, j =1 \\
i > j}} \sqrt{x_ix_j} \\
\leq & n \left( \sum^{n}_{j=1} x_i \right)  \hspace{1cm} [\text{using A.M. $\geq$ G.M. inequality, i.e., $(x_i+x_j) \geq 2 \sqrt{x_ix_j}$}]. \\
\end{aligned}
\end{equation}
Using Eq.~\eqref{Iq}, we have
\begin{equation} \label{Fiq}
\left( \sum^{n}_{j=1} \sqrt{F_{A_iB_j}} \right)^2 \leq n \left( \sum^{n}_{j=1} F_{A_iB_j}\right) 
\end{equation}
for any $i \in \mathbb{N}_n$. Now, substituting Eq.~\eqref{Fiq} in Eq.~\eqref{Kay}, we can write
\begin{equation} \label{SumFij}
\sum^{n}_{j=1} F_{A_iB_j}  \leq \left( 1 + \dfrac{n-1}{d}\right). 
\end{equation}
This bound is valid over all the DoFs of $A$, i.e.,
\begin{equation} \label{ij}
\max_{i}  \left\lbrace \sum^{n}_{j=1} F_{A_iB_j} \right\rbrace   \leq \left( 1 + \dfrac{n-1}{d}\right). 
\end{equation}
Similarly, we can write
\begin{equation} \label{ji}
\max_{j}  \left\lbrace \sum^{n}_{i=1} F_{A_iB_j} \right\rbrace   \leq \left( 1 + \dfrac{n-1}{d}\right). 
\end{equation}
From Eq.~\eqref{ij} and Eq.~\eqref{ji}, we can write
\begin{equation} 
 \max \left\lbrace \max_{i}  \left\lbrace \sum^{n}_{j=1} F_{A_iB_j}\right\rbrace  , \max_{j} \left\lbrace \sum^{n}_{i=1} F_{A_iB_j}\right\rbrace  \right\rbrace \leq \left( 1 + \dfrac{n-1}{d}\right). 
 \end{equation}
From the main text of Eq. (5), we have the bound as
\begin{equation}
F^{(n)}_g = \max \left\lbrace \max_{i}  \left\lbrace \sum^{n}_{j=1} F_{A_iB_j}\right\rbrace  , \max_{j} \left\lbrace \sum^{n}_{i=1} F_{A_iB_j}\right\rbrace  \right\rbrace \leq \left( 1 + \dfrac{n-1}{d}\right).
\end{equation}
for $i, j \in \mathbb{N}_n$. 

\section{Physical significance of the proposed generalized relation} \label{Phy_Sig_TF_SF}
 We investigate the answers to the following questions about any arbitrary two-particle state $\rho$. 
 
 \textit{(i)} \textit{The number of DoFs $n$ in each particle?}
 
\textit{Using $F^{\left(n \right) }_g$:} If  $F^{\left(n \right) }_g > 1$, then $n>1$, because $F^{(1)}_g \leq 1$ from Eq.~\eqref{f_F_par}. 

 \textit{(ii)} \textit{The particles are distinguishable or indistinguishable? }
 
 For distinguishable particles the bound for generalized singlet fraction is $(1+(n-1)/d)$ that can be  proved using the monogamy of singlet fraction~\cite{Kay09} as shown in Section~\ref{up_bound_Fn}. Thus if $F^{\left(n \right) }_g > (1+(n-1)/d)$, then the particles are indistinguishable, else no conclusion can be drawn.

  If $f_g=1$, then the particles are distinguishable because unit fidelity teleportation is not possible for indistinguishable particles~\cite{Das20}, else no conclusion can be drawn.

 \textit{(iii) } \textit{Is any entanglement is present in $\rho$?} 
 
 If $f_g > 2/(d+1)$ or $F^{\left(n \right) }_g > n/d$, then atleast one entanglement structure is present between any pair of DoFs.

  \textit{(iv)}  \textit{How many maximally entangled state is present?}

If $F^{\left(n \right) }_g = n$, then $n$ number of maximally entangled structure is present for any DoF. 

If $f_g=1$,  then the particles are distinguishable, so only one maximally entangled structure is present.

\begin{figure}[h!] 
\centering
\includegraphics[width=\columnwidth]{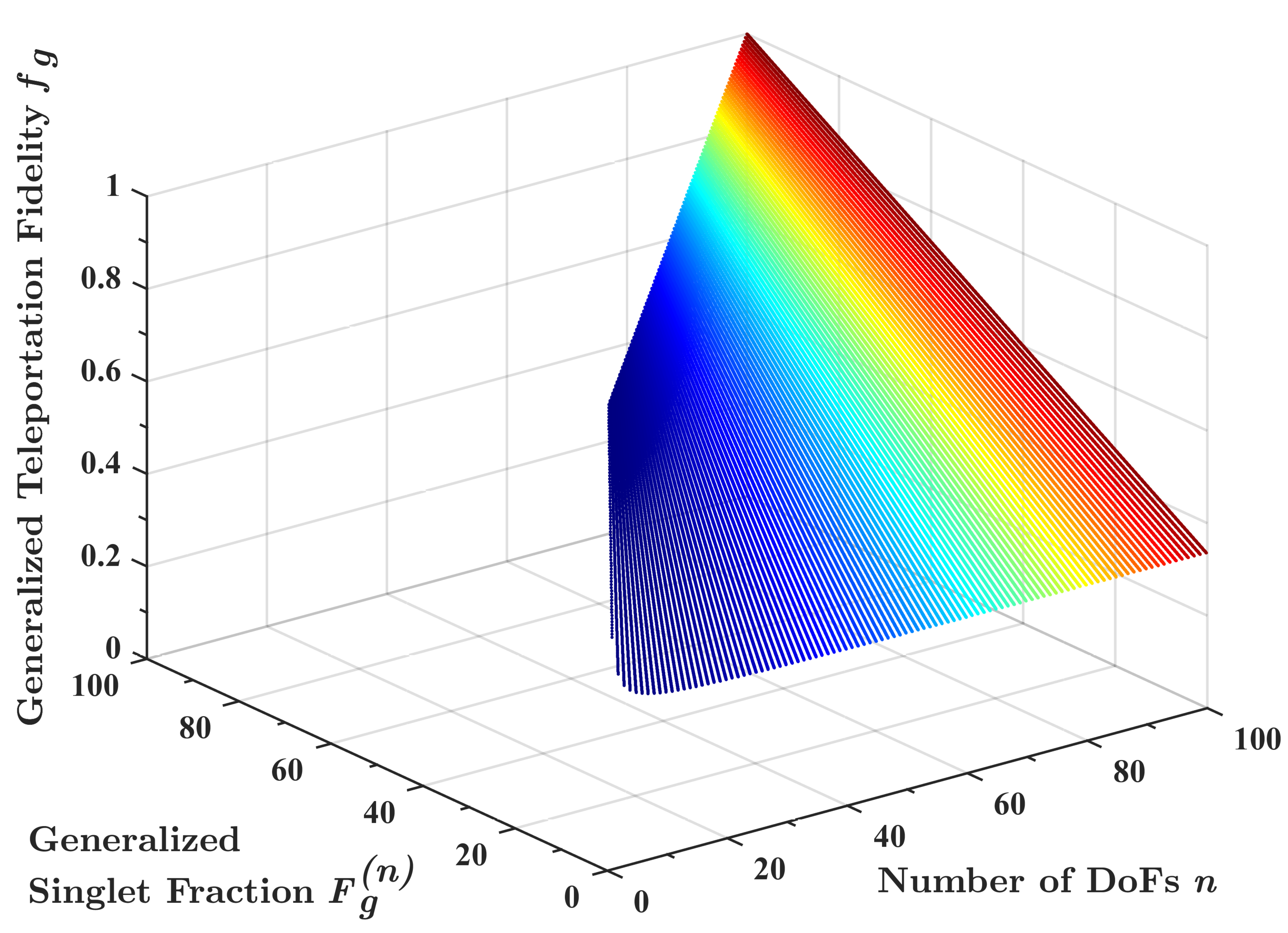} 
\caption{The variation of generalized teleportation fidelity $f_{g}$ and generalized singlet fraction $F^{(n)}_{g}$ with varying the number of DoFs $n$ where the dimension of each DoF is $d=2$.}
\label{TF_SF_graph}
\end{figure}

All these answers are pictorially represented in Fig.~\ref{TF_SF_rel}. For $d=2$, the relation between $n$, $F^{\left(n \right) }_g$, and $f_g$ are plotted in Fig.~\ref{TF_SF_graph} where $1 \leq n \leq 100$ using Eq.~\eqref{fandF}.

\chapter{Violation of monogamy of entanglement for two indistinguishable particles} \label{Chap7}
In this Section, we show that Maximum Violation of monogamy of Entanglement for two indistinguishable particles, each particles having two degrees of freedom is possible using measures which are monogamous for distinguishable particles. To show that, we first derives the condition for maximum violation of monogamy of entanglement. Then we re-write the standard inequality of monogamy of entanglement from particle view to DoF view. Finally we have shown the maximum violation of monogamy of entanglement  using an optical circuit. 

This chapter is based on the work in~\cite{Paul21}.
\section{Violation of no-cloning theorem using the maximum violation of monogamy of entanglement} \label{Con_noclong}
In this section, we briefly overview the monogamy inequality for any general entanglement measure. Then we discuss the condition for the violation of no-cloning theorem using the maximum violation of monogamy of entanglement.

 \begin{figure}[t!] 
\centering
\includegraphics[width=0.8\columnwidth]{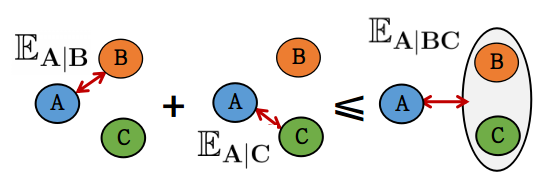} 
\caption{Consider three particles $A$, $B$, $C$, and a bipartite entanglement measure $\mathbb{E}$ where $\mathbb{E}_{X|Y}$ measures the entanglement between the subsystems $X$ and $Y$ of the composite system $XY$ and $\mathbb{E}_{max}$ denotes its maximum value. Using these notations, we show the particle-based monogamy of entanglement obeying Eq.~\eqref{Gen_Monogamy_new}.}
\label{Mono_a}
\end{figure}
 A bipartite entanglement measure $\mathbb{E}$ that obeys the relation 
\begin{equation} \label{Gen_Monogamy_new}
\mathbb{E}_{A|B}(\rho_{AB})+\mathbb{E}_{A|C}(\rho_{AC}) \leq \mathbb{E}_{A|BC}(\rho_{ABC}),
\end{equation} 
for all $\rho_{ABC}$ where $\rho_{AB}=\text{Tr}_{C} \left( \rho_{ABC}\right) $, $\rho_{AC}=\text{Tr}_{B} \left( \rho_{ABC}\right) $, $\mathbb{E}_{X|Y}$ measures the entanglement between the systems $X$ and $Y$ of the composite system $XY$, and the vertical bar represents bipartite splitting, is called \textit{monogamous} as shown in Fig.~\ref{Mono_a}. Such inequality was first shown for squared concurrence ($\mathcal{C}$)~\cite{Hill97,Wootters98} by Coffman, Kundu and Wootters (CKW) for three parties~\cite{CKW00} and later generalized for $n$ parties~\cite{Osborne06}.

Suppose a bipartite entanglement measure $\mathbb{E}$ attains the maximum value $\mathbb{E}_{max}$ for maximally entangled states. Consider a situation when 
\begin{equation}
\begin{aligned}
\mathbb{E}_{A|B}(\rho_{AB}) <& \mathbb{E}_{max}, \\
\mathbb{E}_{A|C}(\rho_{AC}) <& \mathbb{E}_{max}, \\
\mathbb{E}_{A|B}(\rho_{AB}) + \mathbb{E}_{A|C}(\rho_{AC}) > &\mathbb{E}_{max}.\\
\end{aligned}
\end{equation}
 Obviously, this causes a violation of monogamy of entanglement which we call a \textit{non-maximal violation}. Consider another situation, when 
\begin{equation} \label{MaxVio}
\begin{aligned}
\mathbb{E}_{A|B}(\rho_{AB}) =& \mathbb{E}_{max},\\ \mathbb{E}_{A|C}(\rho_{AC}) =& \mathbb{E}_{max},
\end{aligned} 
\end{equation}
i.e., when $A$ is maximally entangled with both $B$ and $C$, we call the corresponding violation as the \textit{maximal violation} of monogamy of entanglement. 
For qubit systems with distinguishable particles, the first situation above would not lead to a violation of the no-cloning theorem~\cite{QC05,QC14}, but the second situation would do as shown in~\ref{MOEandNC}.

\section{Apparent violation of particle-based monogamy of entanglement} \label{App_vio}
\begin{figure}[h!] 
\centering
\includegraphics[width=0.8\columnwidth]{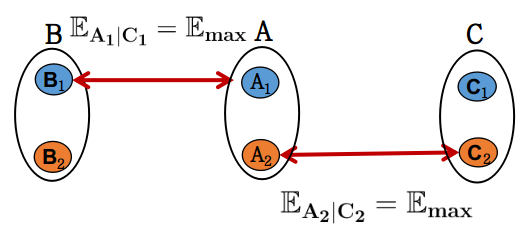} 
\caption{Consider three particles $A$, $B$, $C$, and a bipartite entanglement measure $\mathbb{E}$ where $\mathbb{E}_{X|Y}$ measures the entanglement between the subsystems $X$. Using these notations, we show that $A$ is maximally entangled with $B$ in  DoF 1 (i.e., $\mathbb{E}_{A_{1} \mid B_{1}}=\mathbb{E}_{max} $) and with $C$ in DoF 2 (i.e., $\mathbb{E}_{A_{2} \mid C_{2}}=\mathbb{E}_{max} $). In particle view, apparently monogamy of entanglement is violated; but in DoF view, it is not.}
\label{Mono_b}
\end{figure}

When two particles are entangled, they share correlations in well-defined DoFs such as spin, path,  orbital angular momentum (OAM), etc. Particle-based monogamy of entanglement however sometimes is misleading and incomplete. For example, suppose $A$ is maximally entangled in polarization DoF with $B$ and in OAM DoF with $C$~\cite{chittra15} as shown in Fig.~\ref{Mono_b}. 

This situation apparently violates Eq.~\eqref{Gen_Monogamy_new} and satisfies Eq.~\eqref{MaxVio}. But using this state along with the standard teleportation protocol does not lead to cloning. This contradiction motivates us to re-consider the monogamy with respect to DoFs of each particle.

\section{Inter-DoF monogamy of Entanglement} \label{Inter_DoF_MoE}
 \begin{figure}[h!] 
\centering
\includegraphics[width=0.8\columnwidth]{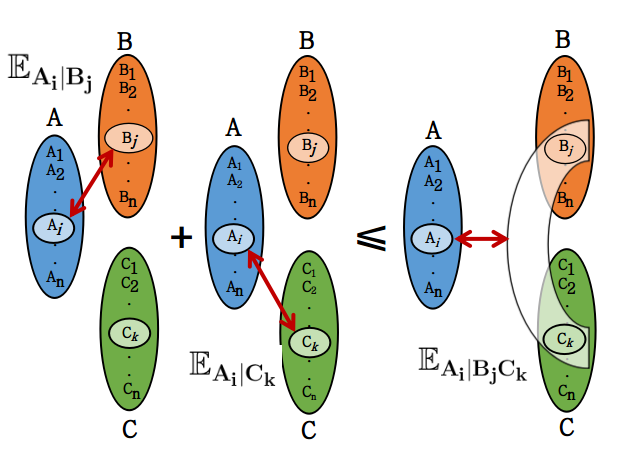} 
\caption{Consider three particles $A$, $B$, $C$, and a bipartite entanglement measure $\mathbb{E}$ where $\mathbb{E}_{X|Y}$ measures the entanglement between the subsystems $X$ and $Y$ of the composite system $XY$ and $\mathbb{E}_{max}$ denotes its maximum value. Using this notations, we generalize the Eq.~\eqref{Gen_Monogamy_new} from particle view to inter-DoF monogamy of entanglement as proposed in Eq.~\eqref{Degree_Monogamy} which resolves the previous apparent violation in Fig.~\ref{Mono_b}.}
\label{Mono_c}
\end{figure}
The above result holds irrespective of whether $A$, $B$ and $C$ are single-DoF particles or DoFs of the same/different particles as shown in Fig.~\ref{Mono_a} and~\ref{Mono_b}. The entanglement measures which are monogamous for distinguishable particles are also so for systems of indistinguishable particles, where $A$, $B$, and $C$ are distinct spatial locations~\cite{Vedral03,Wiseman03,Bose13} with one particle each. However, interesting scenarios might arise when the involved particles are entangled in multiple DoFs which we investigate here.

Here we reformulate Eq.~\eqref{Gen_Monogamy_new} in a more general framework to include multiple DoFs of the same/different particles/entities. Although this is not a contribution, we include it here to establish the background for subsequent analysis.

Consider three entities $A$, $B$, and $C$, each with $n$ DoFs, numbered 1 to $n$. If the joint state of the $i$-th, $j$-th and $k$-th DoFs of $A$, $B$, and $C$ respectively is represented by $\rho_{A_{i}B_{j}C_{k}}$, then the inter-DoF monogamy of entanglement can be formulated as
\begin{equation} \label{Degree_Monogamy}
\mathbb{E}_{A_{i}|B_{j}}(\rho_{A_{i}B_{j}})+\mathbb{E}_{A_{i}|C_{k}}(\rho_{A_{i}C_{k}}) \leq \mathbb{E}_{A_{i}|B_{j}C_{k}}(\rho_{A_{i}B_{j}C_{k}}),
\end{equation} 
where $\rho_{A_{i}B_{j}}=\text{Tr}_{C_{k}} ( \rho_{A_{i}B_{j}C_{k}}) $, $\rho_{A_{i}C_{k}}=\text{Tr}_{B_{j}} ( \rho_{A_{i}B_{j}C_{k}}) $, and $\mathbb{E}_{X_{i}|Y_{j}}$ measures the entanglement between subsystems $X_{i}$ and $Y_{j}$ of the composite system $X_{i}Y_{j}$ as is shown in Fig.~\ref{Mono}~(c). It means that if the $i$-th DoF of $A$ is maximally entangled with the $j$-th DoF of $B$, then it cannot share any correlation with the $k$-th DoF of $C$. 

\begin{figure}[h!] 
\centering
\includegraphics[width=0.8\columnwidth]{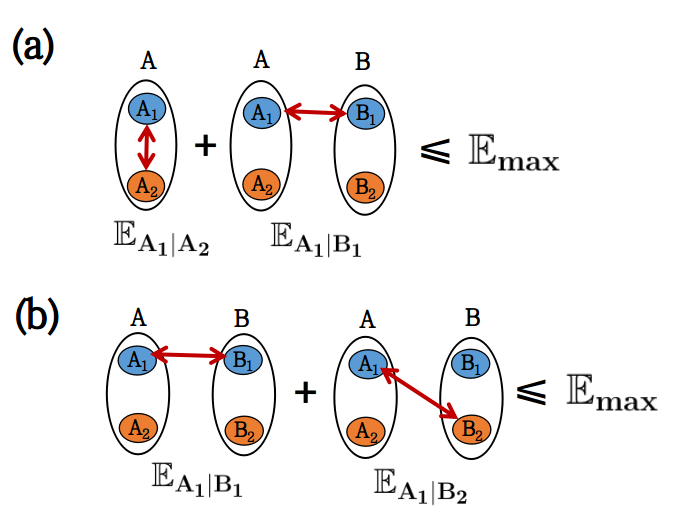} 
\caption{Consider three particles $A$, $B$, $C$, and a bipartite entanglement measure $\mathbb{E}$ where $\mathbb{E}_{X|Y}$ measures the entanglement between the subsystems $X$ and $Y$ of the composite system $XY$ and $\mathbb{E}_{max}$ denotes its maximum value. Now consider the following two scenarios: (a) Two-particle inter-DoF monogamy of entanglement, where $\mathbb{E}_{A_{1} \mid A_{2}}$ measures entanglement between the two DoFs $A_{1}$, $A_{2}$ of $A$ and $\mathbb{E}_{A_{1} \mid B_{1}}$  between $A_{1}$, $B_{1}$ ; and (b) $\mathbb{E}_{A_{1} \mid B_{j}}$ between $A_{1}$ of $A$ and $B_{j}$ of $B$, $j \in \lbrace 1, 2\rbrace$.}
\label{Mono_d_e}
\end{figure}

The inter-DoF monogamy of entanglement of Eq.~\eqref{Degree_Monogamy} is more general than the particle-based monogamy of entanglement of Eq.~\eqref{Gen_Monogamy_new}. The former includes the latter when the three DoFs $i$, $j$, and $k$ belong to three different particles $A$, $B$, and $C$ respectively. However, the inter-DoF monogamy of entanglement can capture many other scenarios that are illustrated in Fig.~\ref{Mono_d_e}~(a) and (b). 
Two interesting types of monogamy of entanglement involving only two particles can also be explained using the inter-DoF formulation. 

(i) \textit{Type I}: Here, monogamy of entanglement is calculated using $\mathbb{E}_{A_{i} \mid A_{j}}$ and $\mathbb{E}_{A_{i} \mid B_{k}}$, as shown in Fig.~\ref{Mono_d_e}~(a). Equation~\eqref{Degree_Monogamy} can capture this scenario by setting $A=B$. The recent analysis for distinguishable particles in~\cite{camalet17,camalet18} is a specific example of this type.

(ii) \textit{Type II}: Here, monogamy of entanglement is calculated using $\mathbb{E}_{A_{i} \mid B_{j}}$ and $\mathbb{E}_{A_{i} \mid B_{k}}$, as shown in Fig.~\ref{Mono_d_e}~(b). Equation~\eqref{Degree_Monogamy} can capture this scenario by setting $B=C$.

This formulation also includes the case of single-particle entanglement~\cite{Zanardi02,Karimi10,Hasegawa03}, when all the three DoFs come from a single particle. Equation~\eqref{Degree_Monogamy} can capture this scenario by setting $A=B=C$. Further, inter-DoF monogamy of entanglement is also valid for indistinguishable particles where the labels $A$, $B$, and $C$ denote spatial locations with each mode containing exactly one particle and $i$, $j$, and $k$ represents the DoFs at each spatial mode.

\section{Violation of monogamy of entanglement by indistinguishable particles} \label{MoE_Vio}
The inter-DoF monogamy of entanglement is not absolute and can be violated maximally by indistinguishable particles. For illustration, consider two-particle inter-DoF entanglement~\cite[Eq. (4)]{HHNL} between spin and path as shown in Eq.~\eqref{Final_state_boson}. It can be represented as $\ket{\Psi^{(2,2)}}$ of Eq.~\eqref{IDstate} with the parameters $\alpha^1, \alpha^2 \in \lbrace s^1, s^2 \rbrace$, $a^{1}_{1}, a^{2}_{1} \in \lbrace L, D, R, U \rbrace$, $a^{1}_{2}, a^{2}_{2} \in \lbrace \uparrow, \downarrow \rbrace$. The coefficients
\begin{equation} \label{Li2Lo}
\begin{aligned}
\kappa^{s^1 L \downarrow}_{s^2 R \downarrow}=&-\kappa^{s^1 D \uparrow}_{s^2 U \uparrow}=\frac{1}{4} \left( \kappa_{1} + \kappa_{2} \right), \\
\kappa^{s^1 D \uparrow}_{s^2 R \downarrow}=&\kappa^{s^1 L \downarrow}_{s^2 U \uparrow}=\frac{i}{4} \left( \kappa_{1} - \kappa_{2} \right), \\
\kappa^{s^2 R \downarrow}_{s^2 R \downarrow}=&\kappa^{s^2 U \uparrow}_{s^2 U \uparrow}=\frac{i \kappa_{1}}{4}, \\
\kappa^{s^1 D \uparrow}_{s^1 D \uparrow}=&\kappa^{s^1 L \downarrow}_{s^1 L \downarrow}=\frac{i \kappa_{2}}{4},
\end{aligned}
\end{equation}
 and the rest are 0, where 
\begin{equation}
\begin{aligned}
\kappa_{1}=&e^{ i \left( \phi_{R} + \phi_{L}\right)}, \\
\kappa_{2}=& e^{ i \left( \phi_{D} + \phi_{U}\right)}.
\end{aligned}
\end{equation}
 
Next we show maximal violation of monogamy of entanglement through squared concurrence measure as follows.

First, we calculate concurrence of the state $\ket{\Psi^{(2,2)}}$ as described in Eq.~\eqref{Li2Lo}.

Projecting $\rho^{(2,2)}=\ket{\Psi^{(2,2)}}\bra{\Psi^{(2,2)}}$ onto the (operational) subspace spanned by the computational basis 
\begin{equation}
\begin{aligned}
\Omega_{s^1s^2}= \lbrace & \ket{ s^1 L \downarrow, s^2 R \downarrow},  \ket{ s^1 L \downarrow, s^2 U \downarrow},  \ket{s^1 L \downarrow, s^2 R \uparrow},  \ket{s^1 L \downarrow, s^2 U \uparrow} 
,  \ket{s^1 L \uparrow, s^2 R \downarrow}, \\ &  \ket{s^1 L \uparrow, s^2 U \downarrow},  \ket{s^1 L \uparrow, s^2 R \uparrow},  \ket{s^1 L \uparrow, s^2 U \uparrow},\ket{s^1 D \downarrow, s^2 R \downarrow}, \ket{s^1 D \downarrow, s^2 U \downarrow}, \\ &  \ket{s^1 D \downarrow, s^2 R \uparrow},  \ket{s^1 D \downarrow, s^2 U \uparrow} 
, \ket{s^1 D \uparrow, s^2 R \downarrow},  \ket{s^1 D \uparrow, s^2 U \downarrow}, \ket{s^1 D \uparrow, s^2 R \uparrow}, \\
& \ket{s^1 D \uparrow, s^2 U \uparrow}
 \rbrace, 
\end{aligned}
\end{equation}
by the projector
\begin{equation}
\begin{aligned}
\Pi_{s^1s^2}= & \sum_{\sigma,\tau = \left\lbrace  \uparrow, \downarrow \right\rbrace , \varsigma = \left\lbrace L, D\right\rbrace,  \upsilon = \left\lbrace  R, U \right\rbrace  } \ket{s^1 \varsigma \sigma, s^2 \upsilon \tau}\bra{s^1 \varsigma \sigma, s^2 \upsilon \tau}, 
\end{aligned}
\end{equation}
one gets the distributed resource state where each localized region $s^1$ and $s^2$ have exactly one particle as
\begin{equation}
\begin{aligned}
\ket{\Psi^{(2,2)}}_{s^1s^2}=& \frac{\Pi_{s^1s^2} \ket{\Psi^{(2,2)}} }{\sqrt{\braket{\Psi^{(2,2)}|\Pi_{s^1s^2}|\Psi^{(2,2)}}}}\\ =& \sum_{ a^{1}_{1} \in \lbrace L, D \rbrace ,  a^{2}_{1} \in \lbrace R, U \rbrace , a^{1}_{2}, a^{2}_{2} \in \lbrace \uparrow, \downarrow \rbrace} \kappa^{s^1 a^{1}_{1} a^{1}_{2}}_{s^2  a^{2}_{1} a^{2}_{2}} \ket{s^1 a^{1}_{1} a^{1}_{2}, s^2  a^{2}_{1} a^{2}_{2}},
\end{aligned}
\end{equation}
where the non-zero coefficients are
\begin{equation}	
\begin{aligned}
\kappa^{s^1 L \downarrow}_{s^2 R \downarrow}=&- \kappa^{s^1 D \uparrow}_{s^2 U \uparrow}=\frac{1}{2 \sqrt{2}} \left(\kappa_{1} + \kappa_{2} \right), 
\\ 
\kappa^{s^1 D \uparrow}_{s^2 R \downarrow}=&\kappa^{s^1 L \downarrow}_{s^2 U \uparrow}=\frac{i}{2 \sqrt{2}} \left(\kappa_{1} - \kappa_{2} \right).
\end{aligned}
\end{equation}
 The density matrix $\rho^{(2,2)}_{s^1s^2}=\ket{\Psi^{(2,2)}}_{s^1s^2}\bra{\Psi^{(2,2)}}$ can also be calculated as
\begin{equation} \label{rhoAB}
\rho^{(2,2)}_{s^1s^2}= \dfrac{\Pi_{s^1s^2} \rho^{(2,2)} \Pi_{s^1s^2}}{\text{Tr}\left( \Pi_{s^1s^2} \rho^{(2,2)}\right)},
\end{equation}
where $\rho^{(2,2)}=\ket{\Psi^{(2,2)}}\bra{\Psi^{(2,2)}}$.

Now from Eq.~\eqref{rhoAB}, if we trace-out the path DoFs of location $s^1$ and $s^2$ using Eq.~\eqref{DoFTrRule} (the order does not matter), we get the reduced density matrix as
\begin{equation} \label{Rho_s1a2_s2b2}
\begin{aligned}
\rho_{s^{1}_{a^{1}_{2}} s^{2}_{a^{2}_{2}}}=&\text{Tr}_{s^{1}_{a^{1}_{1}} s^{2}_{a^{2}_{1}}} \left( \rho^{(2,2)}_{s^1s^2} \right)\\ =&\sum_{ a^{1}_{2},  a^{2}_{2}, b^{1}_{2}, b^{2}_{2}  \in \lbrace \uparrow, \downarrow \rbrace }\kappa^{s^1 a^{1}_{2} }_{s^2 a^{2}_{2}} \kappa^{s^1 b^{1}_{2} *}_{s^2 b^{2}_{2}} \ket{s^1 a^{1}_{2} ,s^2  a^{2}_{2}}\bra{s^1 b^{1}_{2} ,s^2  b^{2}_{2}}, 
\end{aligned}
\end{equation}
where 
\begin{equation}
\begin{aligned}
\kappa^{s^1 \downarrow}_{s^2 \downarrow} =&- \kappa^{s^1 \uparrow}_{s^2 \uparrow}=\frac{1}{2 \sqrt{2}} \left(\kappa_{1} + \kappa_{2} \right), \\  \kappa^{s^1 \uparrow}_{ s^2 \downarrow} =&\kappa^{s^1 \downarrow}_{s^2 \uparrow}  = \frac{i}{2 \sqrt{2}} \left(\kappa_{1} - \kappa_{2} \right),
\end{aligned}
\end{equation}
and rest are zero where complex conjugates are calculated accordingly.

 The physical significance of the reduced density matrix $\rho_{s^{1}_{a^{1}_{2}} s^{2}_{a^{2}_{2}}}$ can be explained as follows: if we measure only the spin DoFs in the spatial regions $s^1$ and $s^2$, then the measurement statistics would be equivalent to the reduced density matrix $\rho_{s^{1}_{a^{1}_{2}} s^{2}_{a^{2}_{2}}}$. This can be obtained by applying our DoF trace-out rule two times by taking the spatial regions first as $s^x=s^1$ and then $s^x=s^2$ (or taking first as $s^x=s^2$ and then $s^x=s^1$, as the order does not matter) in Eq.~\eqref{DoFTrRule} where $m_i \in \lbrace L, D, R, U \rbrace$. On the other hand, when we measure the path DoF in the spatial region $s^1$ and the spin DoF in the spatial region $s^2$, then the measurement statistics would be equivalent to the reduced density matrix $\rho_{s^{1}_{a^{1}_2}s^{2}_{a^{2}_1}}$. This can be obtained by performing our DoF trace-out rule by taking first $s^x=s^1$ where $m_i \in \lbrace L, D, R, U \rbrace$ and then $s^x=s^2$ where $m_i \in \lbrace \uparrow, \downarrow \rbrace$ or vice-versa in Eq.~\eqref{DoFTrRule}.

To calculate the maximum violation using squared concurrence, first we calculate the following:
\begin{equation}
 \widetilde{\rho}_{s^{1}_{a^{1}_{2}} s^{2}_{a^{2}_{2}}}=\sigma^{s^1}_{y} \otimes \sigma^{s^2}_{y} \rho^{*}_{s^{1}_{a^{1}_{2}} s^{2}_{a^{2}_{2}}} \sigma^{s^1}_{y} \otimes \sigma^{s^2}_{y},
\end{equation}
where $\sigma^{X}_{y}=\ket{X}\bra{X} \otimes \sigma_{y}$, $X  \in \lbrace s^1, s^2 \rbrace $,  $\sigma_{y}$  is Pauli matrix and the asterisk denotes complex conjugation. 
So, the expression becomes
\begin{equation}
\begin{aligned}
\widetilde{\rho}_{s^{1}_{a^{1}_{2}} s^{2}_{a^{2}_{2}}}=\sum_{ a^{1}_{2},  a^{2}_{2}, b^{1}_{2}, b^{2}_{2}  \in \lbrace \uparrow, \downarrow \rbrace }\tilde{\kappa}^{s^1 a^{1}_{2} }_{s^2 a^{2}_{2}} \tilde{\kappa}^{s^1 b^{1}_{2} *}_{s^2 b^{2}_{2}} \ket{s^1 a^{1}_{2} ,s^2  a^{2}_{2}}\bra{s^1 b^{1}_{2} ,s^2  b^{2}_{2}}, 
\end{aligned}
\end{equation}
where 
\begin{equation}
\begin{aligned}
 \tilde{\kappa}^{s^1 \downarrow}_{s^2 \downarrow} \tilde{\kappa}^{s^1 \downarrow * }_{s^2 \downarrow} = - \tilde{\kappa}^{s^1 \downarrow}_{s^2 \downarrow} \tilde{\kappa}^{s^1 \uparrow * }_{s^2 \uparrow}=- \tilde{\kappa}^{s^1 \uparrow}_{s^2 \uparrow} \tilde{\kappa}^{s^1 \downarrow * }_{s^2 \downarrow}=\tilde{\kappa}^{s^1 \uparrow}_{s^2 \uparrow} \tilde{\kappa}^{s^1 \uparrow * }_{s^2 \uparrow}  =& \frac{1}{2} \cos^{2} \phi, \\
  \tilde{\kappa}^{s^1 \downarrow}_{s^2 \downarrow} \tilde{\kappa}^{s^1 \uparrow *}_{s^2 \downarrow} = \tilde{\kappa}^{s^1 \downarrow}_{s^2 \downarrow} \tilde{\kappa}^{s^1 \downarrow *}_{s^2 \uparrow} =  - \tilde{\kappa}^{s^1 \uparrow}_{s^2 \uparrow} \tilde{\kappa}^{s^1 \uparrow *}_{s^2 \downarrow} =- \tilde{\kappa}^{s^1 \uparrow}_{s^2 \uparrow} \tilde{\kappa}^{s^1 \downarrow *}_{s^2 \uparrow} =& \frac{1}{2} \cos \phi \sin \phi, \\
   \tilde{\kappa}^{s^1 \uparrow}_{s^2 \downarrow} \tilde{\kappa}^{s^1 \downarrow * }_{s^2 \downarrow} = \tilde{\kappa}^{s^1 \downarrow}_{s^2 \uparrow} \tilde{\kappa}^{s^1 \downarrow * }_{s^2 \downarrow} =\tilde{\kappa}^{s^1 \uparrow}_{s^2 \downarrow} \tilde{\kappa}^{s^1 \uparrow * }_{s^2 \uparrow}  = \tilde{\kappa}^{s^1 \downarrow}_{s^2 \uparrow} \tilde{\kappa}^{s^1 \uparrow * }_{s^2 \uparrow} =& \frac{1}{2} \cos \phi \sin \phi, \\
  \tilde{\kappa}^{ s^1 \uparrow}_{s^2 \downarrow} \tilde{\kappa}^{s^1 \uparrow * }_{s^2 \downarrow}=\tilde{\kappa}^{s^1 \uparrow}_{s^2 \downarrow} \tilde{\kappa}^{s^1 \downarrow * }_{ s^2 \uparrow} = \tilde{\kappa}^{s^1 \downarrow}_{s^2 \uparrow} \tilde{\kappa}^{s^1 \uparrow * }_{s^2 \downarrow} = \tilde{\kappa}^{s^1 \downarrow}_{s^2 \uparrow} \tilde{\kappa}^{s^1 \downarrow * }_{s^2 \uparrow} =& \frac{1}{2} \sin^{2} \phi ,
\end{aligned}
\end{equation}
with $\phi=\frac{1}{2} \lbrace \phi_{D} + \phi_{U} - \phi_{R} -\phi_{L}\rbrace$.
Now we  calculate concurrence as

\begin{equation}
\begin{aligned}
\mathcal{C}_{s^{1}_{a^{1}_{2}} s^{2}_{a^{2}_{2}}}\left( \rho_{s^{1}_{a^{1}_{2}} s^{2}_{a^{2}_{2}}}\right)= \text{max}\left\lbrace 0, \sqrt{\lambda_{4}} - \sqrt{\lambda_{3}} - \sqrt{\lambda_{2}} - \sqrt{\lambda_{1}} \right\rbrace , 
\end{aligned}
\end{equation}
where $\lambda_{i}$ are the eigenvalues, in decreasing order, of the non-Hermitian matrix 
\begin{equation}
\begin{aligned}
\mathbb{R}=\rho_{s^{1}_{a^{1}_{2}} s^{2}_{a^{2}_{2}}}\widetilde{\rho}_{s^{1}_{a^{1}_{2}} s^{2}_{a^{2}_{2}}}=\sum_{ a^{1}_{2},  a^{2}_{2}, b^{1}_{2}, b^{2}_{2}  \in \lbrace \uparrow, \downarrow \rbrace }\bar{\kappa}^{s^1 a^{1}_{2} }_{s^2 a^{2}_{2}} \bar{\kappa}^{s^1 b^{1}_{2} *}_{s^2 b^{2}_{2}} \ket{s^1 a^{1}_{2} ,s^2 a^{2}_{2}}\bra{s^1 b^{1}_{2} , s^2 b^{2}_{2}}, 
\end{aligned}
\end{equation}
where 
\begin{equation}
\begin{aligned}
 \bar{\kappa}^{s^1 \downarrow}_{s^2 \downarrow} \bar{\kappa}^{s^1 \downarrow * }_{ s^2 \downarrow} = - \bar{\kappa}^{s^1 \downarrow}_{s^2 \downarrow} \bar{\kappa}^{s^1 \uparrow * }_{s^2 \uparrow}=- \bar{\kappa}^{s^1 \uparrow}_{s^2 \uparrow} \bar{\kappa}^{s^1 \downarrow * }_{s^2 \downarrow}=\bar{\kappa}^{s^1 \uparrow}_{s^2 \uparrow} \bar{\kappa}^{s^1 \uparrow * }_{s^2 \uparrow}  =& \frac{1}{4} \cos^{2} \phi, \\
  \bar{\kappa}^{s^1 \downarrow}_{s^2 \downarrow} \bar{\kappa}^{s^1 \uparrow *}_{s^2 \downarrow} = \bar{\kappa}^{s^1 \downarrow}_{s^2 \downarrow} \bar{\kappa}^{s^1 \downarrow *}_{s^2 \uparrow} =
   - \bar{\kappa}^{s^1 \uparrow}_{s^2 \uparrow} \bar{\kappa}^{s^1 \uparrow *}_{s^2 \downarrow} =- \bar{\kappa}^{s^1 \uparrow}_{s^2 \uparrow} \bar{\kappa}^{s^1 \downarrow *}_{s^2 \uparrow} 
  = &\frac{1}{4} \cos \phi \sin \phi
  \\  \bar{\kappa}^{s^1 \uparrow}_{s^2 \downarrow} \bar{\kappa
}^{s^1 \downarrow * }_{s^2 \downarrow} = \bar{\kappa}^{s^1 \downarrow}_{ \uparrow} \bar{\kappa}^{ \downarrow * }_{ \downarrow} =\bar{\kappa}^{ \uparrow}_{s^2 \downarrow} \bar{\kappa}^{s^1 \uparrow * }_{s^2 \uparrow}  = \bar{\kappa}^{s^1 \downarrow}_{s^2 \uparrow} \bar{\kappa}^{s^1 \uparrow * }_{s^2 \uparrow} =& \frac{1}{4} \cos \phi \sin \phi, \\
  \bar{\kappa}^{s^1 \uparrow}_{s^2 \downarrow} \bar{ \kappa}^{s^1 \uparrow * }_{ s^2 \downarrow}=\bar{\kappa}^{ s^1 \uparrow}_{s^2 \downarrow} \bar{\kappa}^{s^1 \downarrow * }_{s^2 \uparrow} = \bar{\kappa}^{ s^1 \downarrow}_{s^2 \uparrow} \bar{\kappa}^{s^1 \uparrow * }_{s^2 \downarrow} = \bar{\kappa}^{s^1 \downarrow}_{s^2 \uparrow} \bar{\kappa}^{s^1 \downarrow * }_{s^2 \uparrow} =& \frac{1}{4} \sin^{2} \phi.
\end{aligned}
\end{equation}

So, the eigenvalues of $\mathbb{R}$ are $\lbrace 1, 0, 0, 0 \rbrace$.  Thus 
\begin{equation}
\mathcal{C}_{s^{1}_{a^{1}_{2}} \mid s^{2}_{b^{2}_{2}}}\left( \rho_{s^{1}_{a^{1}_{2}} \mid s^{2}_{b^{2}_{2}}}\right) =1.
\end{equation}
Similar calculations follows that, $\mathcal{C}_{s^{1}_{a^{1}_{2}} \mid s^{2}_{b^{2}_{1}}}\left( \rho_{s^{1}_{a^{1}_{2}} \mid s^{2}_{b^{2}_{1}}}\right)=1$.

Likewise, we can also calculate the log-negativity~\cite{Zyczkowski98,Vidal02} for the density matrix $\rho_{s^{1}_{a^{1}_{2}} s^{2}_{a^{2}_{2}}}$ in Eq.~\eqref{Rho_s1a2_s2b2}. For that, we need the eigenvalues of the density matrix $\rho_{s^{1}_{a^{1}_{2}}} $  after taking the partial transpose with respect to $s^{2}_{a^{2}_{2}}$. The eigenvalues are found to be $\lbrace -\frac{1}{2},\frac{1}{2}, \frac{1}{2}, \frac{1}{2} \rbrace$. Thus the value of negativity is $\frac{1}{2}$ and so the log-negativity is given by 
\begin{equation}
E_{\mathcal{N}}\left( \rho_{s^{1}_{a^{1}_{2}}  s^{2}_{a^{2}_{2}}}\right) =1.
\end{equation}
Similar calculations give $E_{\mathcal{N}}\left(\rho_{s^{1}_{a^{1}_{2}}  s^{2}_{a^{2}_{1}}}\right) =1.$

All other monogamous measures of entanglement for qubit systems such as entanglement of formation~\cite{Bennett96}, log-negativity~\cite{Zyczkowski98,Vidal02}, Tsallis-q entropy~\cite{Kim10,Luo16}, R\'{e}nyi-$\alpha$ entanglement~\cite{Kim_Sanders10,Song16}, Unified-(q, s) entropy~\cite{Kim11,Khan19}, one-way distillable entanglement~\cite{Devetak05}, squashed entanglement~\cite{Christandl04,Brandao11} etc.~\cite{Plenio07} are calculable from the reduced density matrix. If one starts with the same reduced density matrix as in Eq.~\eqref{Rho_s1a2_s2b2}, one can easily show that all the above measures attain their respective maximum value for both the subsystems  $ \lbrace s^{1}_{a^{1}_{2}}, s^{2}_{a^{2}_{2}} \rbrace$ and $ \lbrace s^{1}_{a^{1}_{2}}, s^{2}_{a^{2}_{1}} \rbrace$ simultaneously, thereby violating the monogamy of entanglement. Thus, for any bipartite monogamous entanglement measure $\mathbb{E}$, we get
\begin{equation}
\mathbb{E}_{s^{1}_{a^{1}_{2}} \mid s^{2}_{a^{2}_{2}}}\left( \rho_{s^{1}_{a^{1}_{2}}  s^{2}_{a^{2}_{2}}}\right) = \mathbb{E}_{s^{1}_{a^{1}_{2}} \mid s^{2}_{a^{2}_{1}}}\left( \rho_{s^{1}_{a^{1}_{2}}  s^{2}_{a^{2}_{1}}}\right) = 1.
\end{equation}

Interestingly, this violation is irrespective of any particular entanglement measure like squared concurrence. It can be shown that such a violation happens in indistinguishable particles by any monogamous bipartite entanglement measure for qubit systems. This leads to the following result.
\begin{theorem}\label{thm1}
{In qubit systems, indistinguishability is a necessary criterion for maximum violation of monogamy of  entanglement by the same measures that are monogamous for distinguishable particles}.
\end{theorem}

\section{Physical significance of the maximum violation of monogamy} \label{Phy_sig_moe_vio}
Monogamy of entanglement is widely regarded as one of the basic principles of quantum physics~\cite{Terhal04}. Qualitatively, it is always expected to hold, as a maximal violation will have consequences for the no-cloning theorem. So much so, that a quantitative violation is interpreted as the non-monogamistic nature of the entanglement measure and not of the system of particles itself~\cite{Lancien16}. Some of those non-monogamous measures can be elevated to be monogamous through convex roof extension~\cite{Kim09}.

Through Theorem~\ref{thm1}, we establish a qualitative violation of monogamy of entanglement which was hitherto unheard of. We show that using quantum indistinguishability, it is possible to maximally violate the monogamy of entanglement for all such entanglement measures which are known to be monogamous for distinguishable systems. To establish this theorem, we first needed to modify the qualitative definition of monogamy of entanglement itself, transiting from the particle-view to the DoF-view and had to introduce the DoF trace-out rule for indistinguishable particles. Thus, this is a non-trivial extension of the well-known monogamy of entanglement.

Further, our framework also takes into account the recently introduced inter-DoF entanglement~\cite{HHNL}.  Quantum physics dictates the measurement results of a particular DoF when it is correlated with another DoF. Taking a partial trace while keeping the rule of quantum physics intact is extremely non-trivial and requires a rigorous mathematical treatment. Our framework, therefore, captures these nuances of quantum physics better than any other existing framework.

Theorem~\ref{thm1} unveils a non-trivial difference between distinguishable and indistinguishable systems. For distinguishable systems, monogamy of entanglement and no-cloning theorem imply one another as shown in~\ref{MOEandNC}. The significance of our result is that for indistinguishable systems, the no-cloning theorem remains more fundamental than monogamy of entanglement and the former does not necessarily imply the latter. In fact, no-cloning is derived from the linearity of quantum mechanics~\cite{Wootters82} and hence even indistinguishable particles are also bound to follow it. It appears that the only way to reconcile the co-existence of monogamy of entanglement violation and no-cloning for indistinguishable particles is to consider that such particles do not yield unit fidelity in quantum teleportation~\cite{Ugo15,LFC18,Das20}

Moreover, indistinguishability is not a sufficient condition for violation of monogamy of entanglement. There can be scenarios where indistinguishable subsystems may be maximally entangled, respecting monogamy. Only specific entanglement structures (such as the circuit discussed in this work) can lead to a maximum violation of monogamy of entanglement. That is why we call indistinguishability a necessary criterion for maximum violation of monogamy of entanglement.

Theorem~\ref{thm1} raises a few fundamental questions on the properties of entanglement for indistinguishable particles. 

(i) There are several applications of monogamy of entanglement for distinguishable particles such as~\cite{Pawlowski10,Tomamichel13,Johnston16,Dur00,Bae06,Chiribella06,Lloyd14}. In particular, for cryptographic applications~\cite{Pawlowski10,Tomamichel13,Johnston16}, monogamy of entanglement provides security in the distinguishable scenario. What happens to such applications in the indistinguishable case?

(ii) Can there be a new application of sharability of maximal entanglement among indistinguishable particles that are not possible for distinguishable ones? 

(iii) Do indistinguishable particles also exhibit maximum violation of monogamy for general quantum correlations~\cite{Streltsov12} such as discord~\cite{Braga12,Prabhu12,Bai13,Liu14}, coherence~\cite{Radhakrishnan16,Basso20}, steering~\cite{Reid13,Milne15,Cheng16}, etc.

\chapter{Monogamy of entanglement for three or more indistinguishable particles} \label{Chap8}
For distinguishable particles, MoE is known to hold, irrespective of whether the DoFs involved come from two particles~\cite{camalet17,camalet18,Paul21} or more~\cite{CKW00,Osborne06}. For two indistinguishable particles, it has been shown that monogamy does not necessarily hold and can be violated maximally~\cite{Paul21} as shown in Chapter~\ref{Chap7}. So a natural question arises, whether MoE always holds for three or more indistinguishable particles or not? 

There are fundamental difference between the physicality of entanglement of distinguishable particles and that of indistinguishable ones. For example, two distinguishable particles with orthogonal eigenstates in one of the DoFs are separable as they can be written in tensor product. However, two indistinguishable
particles can become entangled even under the conditions of orthogonal eigenstates, differently from two distinguishable particles which remain in a product state~\cite[Methods]{LFC16}. So, if three or more particles becomes indistinguishable in the same/different localized regions in their same/different eigenstates of same/different DoFs in an arbitrary manner, whether MoE holds or not is not immediately obvious and needs non-trivial analysis. This is the motivation behind this chapter. 
Here, we use concurrence as entanglement measure to calculate the monogamy of entanglement. 

This chapter is based on the work in~\cite{MoE}.

\section{Calculation of concurrence between two spatial regions between any DoFs} \label{Cal_con}
In the general state given in Eq.~\eqref{DenGen}, it is possible that each localized region have more than one particle. To calculate the concurrence, we have to ensure that each of the localized regions $s^1, s^2, \ldots, s^p$ has only one particle. For that we have to apply a projector as following. 

Projecting $\rho^{(p, n)}$ onto the operational subspace spanned by the basis 
\begin{equation}
\begin{aligned} \label{ProjGenPure}
&\mathcal{B}^{s^1 s^2 \ldots s^p}\\ =& \lbrace \ket{s^1 D^{1}_{1_{1}} \ldots D^{1}_{n_{1}}, s^2 D^{2}_{1_{1}}  \ldots D^{2}_{n_{1}}, \ldots s^p D^{p}_{1_{1}}  \ldots D^{p}_{n_{1}}}, \\ &
  \ket{s^1 D^{1}_{1_{2}}  \ldots D^{1}_{n_{1}}, s^2 D^{2}_{1_{1}}  \ldots D^{2}_{n_{1}}, \ldots s^p D^{p}_{1_{1}}  \ldots D^{p}_{n_{1}}}, \\
  & \vdots \\
  &\ket{s^1 D^{1}_{1_{k_{1}}}  \ldots D^{1}_{n_{k_{n}}}, s^2 D^{2}_{1_{k_{1}}} , \ldots s^p D^{p}_{1_{k_{1}}} D^{p}_{2_{k_{2}}} \ldots D^{p}_{n_{k_{n}}}}\rbrace
\end{aligned}
\end{equation}
by the projector
 \medmuskip=-1mu
\thinmuskip=-1mu
\thickmuskip=-1mu 
\begin{footnotesize}
\begin{equation} \label{ProjGenDen}
\mathcal{P}_{s^1 s^2 \ldots s^p}=\sum_{x^{i}_{j} \in  \mathbb{D}_{j}, i \in \mathbb{N}_p,  j \in \mathbb{N}_{n}}  \ket{s^{1} x^{1}_{1}x^{1}_{2} \ldots x^{1}_{n}, s^{2} x^{2}_{1}x^{2}_{2} \ldots x^{2}_{n}, \ldots, s^{p} x^{p}_{1}x^{p}_{2} \ldots x^{p}_{n}  } \bra{s^{1} x^{1}_{1}x^{1}_{2} \ldots x^{1}_{n}, s^{2} x^{2}_{1}x^{2}_{2} \ldots x^{2}_{n}, \ldots, s^{p} x^{p}_{1}x^{p}_{2} \ldots x^{p}_{n}  }
\end{equation}
\end{footnotesize}

 \medmuskip=2mu
\thinmuskip=2mu
\thickmuskip=2mu 
results in 
\begin{equation} \label{AfterProjGen}
\rho^{(p, n)}_{s^1 s^2 \ldots s^p}=\dfrac{\mathcal{P}_{s^1 s^2 \ldots s^p} \rho^{(p,n)} \mathcal{P}_{s^1 s^2 \ldots s^p}}{\text{Tr} \left(\mathcal{P}_{s^1 s^2 \ldots s^p}  \rho^{(p,n)} \right) }.
\end{equation}

To calculate the concurrence between two spatial regions, we have to trace out other $(p-2)$ regions using the method described in \cite{Paul21}. The trace out rule for tracing out say $s^{h} \in \mathbb{S}^p$ region can be described as
\begin{equation}
\begin{aligned}
&\rho^{(p-1,n)}_{\left(  \mathbb{S}^p - \lbrace s^{h} \rbrace \right) }=\text{Tr}_{s^{h}} \left(\rho^{(p, n)} \right)=\sum_{m^{h}_{1},m^{h}_{2}, \ldots, m^{h}_{n}} \braket{s^{h} m^{h}_{1} m^{h}_{2} \ldots m^{h}_{n} \mid \rho^{(p, n)} \mid s^{h} m^{h}_{1} m^{h}_{2} \ldots m^{h}_{n} },
\end{aligned}
\end{equation}
where $m^{h}_{j}$ span $\mathbb{D}_{j}$ for $j \in \mathbb{N}_{n}$.

This if we trace out $k$ number of particles from the localized regions $s^{h_{1}},s^{h_{1}}, \ldots, s^{h_{k}}$, then the reduced density matrix is be represented as 
 \medmuskip=-0mu
\thinmuskip=-0mu
\thickmuskip=-0mu 
\begin{footnotesize}
\begin{equation} \label{GenReduced}
\begin{aligned}
&\rho^{(p-h,n)}_{\left(  \mathbb{S}^p - \lbrace s^{h_{1}},s^{h_{1}}, \ldots, s^{h_{k}}\rbrace \right) } =\text{Tr}_{s^{h_{1}},s^{h_{2}}, \ldots, s^{h_{k}}} \left( \rho^{(p, n)} \right) \\
 =& \sum_{ \substack{s^{h_{i}} \in \mathbb{S}^p, m^{h_{i}}_{j} \in \mathbb{D}_{j} }} \braket{s^{h_{1}} m^{h_{1}}_{1} m^{h_{1}}_{2} \ldots m^{h_{1}}_{n}, \ldots, s^{h_{k}} m^{h_{k}}_{1} m^{h_{k}}_{2} \ldots m^{h_{k}}_{n} \mid \rho^{(p, n)} \mid s^{h_{1}} m^{h_{1}}_{1} m^{h_{1}}_{2} \ldots m^{h_{1}}_{n},  \ldots , s^{h_{k}} m^{h_{k}}_{1} m^{h_{k}}_{2} \ldots m^{h_{k}}_{n}}.
\end{aligned}
\end{equation}
\end{footnotesize}
 \medmuskip=2mu
\thinmuskip=2mu
\thickmuskip=2mu 

 Suppose we want to calculate the concurrence between the particle in the location $s^r$ and the particle in the location $s^t$ where $s^r, s^t \in \mathbb{S}^p$, we apply the DoF trace-out rule as defined in~\cite{Paul21}. Thus the reduced density matrix is
\begin{equation}
\rho^{\left( 2,n \right) }_{s^{r},s^{t}}=\text{Tr}_{ \left( \mathbb{S} -\lbrace s^{r}, s^{t}\rbrace \right) } \left(\rho^{(p, n)} \right).
\end{equation}

To calculate the concurrence between the $v$-th DoF of the particle in the location $s^r$ and the $w$-th DoF of the particle in the location $s^t$ where $1\leq v, w \leq n$, we have to trace-out all the other non-contributing DoFs from these two locations using the DoF trace-out rule as defined in~\cite{Paul21}. So, the reduced density matrix of the $v$-th and $w$-th DoF of the locations $s^r$ and $s^t$ respectively is given by
\begin{equation}
\rho^{\left( 2,1 \right) }_{s^{r}_{v},s^{t}_{w}} = \text{Tr}_{\left(  s^{r}_{\bar{v}},s^{t}_{\bar{w}}\right) } \left( \rho^{\left( 2,n \right) }_{s^{r},s^{t}}\right) =\sum_{ m^{r}_{j}, m^{t}_{j} \in \mathbb{D}_{j}} \braket{ \psi^{s^{r}}_{m_{\bar{v}}}, \psi^{s^{t}}_{m_{\bar{w}}}  \mid \rho^{\left( 2,n \right) }_{s^{r},s^{t}} \mid \psi^{s^{r}}_{m_{\bar{v}}}, \psi^{s^{t}}_{m_{\bar{w}}} },
\end{equation}
where $\ket{\psi^{s^{r}}_{m_{\bar{v}}}}=\ket{s^{r} m^{r}_{1}m^{r}_{2} \ldots m^{r}_{(v-1)} m^{r}_{(v+1)} \ldots m^{r}_{n}}$\, and \\ $\ket{\psi^{s^{t}}_{m_{\bar{w}}}}$=$\ket{ s^{t} m^{t}_{1}m^{t}_{2} \ldots m^{t}_{(w-1)} m^{t}_{(w+1)} \ldots m^{t}_{n}}$.

To calculate the concurrence of $\rho^{\left( 2,1 \right) }_{s^{r}_{v}, s^{t}_{w}} $, i.e., $\mathcal{C}_{s^{r}_{v}\mid s^{t}_{w}}$,  we have to calculate the following
\begin{equation}
\widetilde{\rho}_{s^{r}_{v},s^{t}_{w}}=\sigma^{s^r}_{y} \otimes \sigma^{s^t}_{y}  \rho^{*}_{s^{r}_{v},s^{t}_{w}} \sigma^{s^r}_{y} \otimes \sigma^{s^t}_{y}.
\end{equation}
where $\sigma^{s^r}_{y} = \ket{s^r}\bra{s^r} \otimes \sigma_{y}$, and similarly $\sigma^{s^t}_{y} = \ket{s^t}\bra{s^t} \otimes \sigma_{y}$,	 and $\sigma_{y}$ is Pauli matrix and the asterisk denotes complex conjugation. 

Now we have to calculate the eigenvalues of the non-hermitian matrix 
\begin{equation}
\mathcal{R}_{s^{r}_{v},s^{t}_{w}}=\rho_{s^{r}_{v},s^{t}_{w}} \widetilde{\rho}_{s^{r}_{v},s^{t}_{w}}.
\end{equation}
Finally the concurrence is calculated as the 
\begin{equation}
\mathcal{C}_{s^{r}_{v}\mid s^{t}_{w}}=\text{max} \left\lbrace  0, \sqrt{\lambda_{4}}-\sqrt{\lambda_{3}}-\sqrt{\lambda_{2}}-\sqrt{\lambda_{1}} \right\rbrace, 
\end{equation}
where $\lambda_{i}$'s are the eigenvalues of $\mathcal{R}_{s^{r}_{v}, s^{t}_{w}}$ in decreasing order.

\section{Monogamy of $p$ indistinguishable particles each having $n$ DoFs} \label{Pure_MoE}
 As the state-space structure of distinguishable and indistinguishable particles are completely different and so the proof for MoE shown for distinguishable particles in~\cite{CKW00} is not applicable for indistinguishable particles. So, we calculate the MoE for all the possible ways in which indistinguishability can occur. First, we calculate it for three particles each having three DoFs, for example spin, OAM, and path DoF having eigenstates $\lbrace \ket{\uparrow}, \ket{\downarrow} \rbrace$, $\lbrace \ket{+l}, \ket{-l} \rbrace$, and $\lbrace \ket{R}, \ket{L} \rbrace$ respectively in three localized regions  $\mathbb{S}^3$. We describe the first five cases where one of the eigenstates of the DoFs contributed for entanglement, and the other non-contributing DoFs can take arbitrary values. Then we consider the other cases where contributing DoFs for entanglement can be in arbitrary superposition of their eigenstates.  Finally, we generalize it for $p$ indistinguishable particles each having $n$ DoFs.

Suppose there are $p$ number of indistinguishable particles, each having $n$ DoFs. Recall that, the $k$-th eigenvalue of the $j$th DoF of a particle is represented by $\mathcal{D}_{j_{k}} \in \mathbb{D}_{j}$ (the set of eigenvalues of the $j$th DoF). As we are considering squared concurrence measure, so we take only two eigenstates of each DoF. For any eigenvalue $\lambda$, we use the notion $\ket{\lambda}$ for the corresponding eigenstate. In Table~\ref{Tab:1}, we summarize the list of possible combinations to create indistinguishability using three indistinguishable particles, each having three DoFs denoted by $j$, $j^{\prime}$, and $j^{\prime \prime}$, localized in three regions $s^1$, $s^2$, and $s^3$. Calculations for concurrences are done using the method described in Section~\ref{Cal_con}. These cases can be extended for $p$ number of indistinguishable particles as shown below.
 \medmuskip=2mu
\thinmuskip=2mu
\thickmuskip=2mu
\begin{sidewaystable} 
\begin{landscape}
\centering
\begin{scriptsize}
\setlength\tabcolsep{1pt}
\begin{tabular}{|c|c|c|c|c|c|c|c|c|c|c|c|c|} 
\toprule
\ \ &\textbf{DoF} & \textbf{Eigenstate} & \textbf{1st particle} & \textbf{2nd particle} & \textbf{3rd particle} & \textbf{Relations} &  $s^1$ \hspace{0.3cm}& $s^2$ \hspace{0.05cm} &$s^3$&$\mathcal{C}^{2}_{s^{1} \mid s^{2}}$ & $\mathcal{C}^{2}_{s^{1} \mid s^{3}}$& $\mathcal{C}^{2}_{s^{1} \mid s^{2}s^{3}}$ \\
\cline{8-10}
 & & & & & && \multicolumn{3}{c|}{\textbf{measures}} & & & \\ 
 & & & & & &&\multicolumn{3}{c|}{\textbf{in the DoF}} & & & 
\tabularnewline
\bottomrule
 \textbf{1} & Same & Same   & $\ket{\mathcal{D}}_{j_{k}}$ & $\ket{\mathcal{D}}_{j_{k}}$ & $\ket{\mathcal{D}}_{j_{k}}$ & Nil & $j$ & $j$ & $j$ &0 &0 &0 \tabularnewline
\midrule
 \textbf{2} & Same & Different & $\ket{\mathcal{D}}_{j_{k}}$ & $\ket{\mathcal{D}}_{j_{k}}$ & $\ket{\mathcal{D}}_{j_{k^\prime}}$ &  $\mathcal{D}_{j_{k}}, \mathcal{D}_{j_{k^\prime}} \in \mathbb{D}_{j}$, $\ket{\mathcal{D}}_{j_{k^\prime}}=\ket{\mathcal{D}}^{\perp}_{j_{k}} $ & $j$ & $j$ & $j$ & $\geq 0$ & $\geq 0$ &$\geq 0$ \tabularnewline
    \midrule
 \textbf{3 }& Different & Different & $\ket{\mathcal{D}}_{j_{k}}$ & $\ket{\mathcal{D}}_{j_{k}}$  & $\ket{\mathcal{D}}_{j^{\prime}_{l}}$ & $j \neq j^{\prime}$, $\mathcal{D}_{j_{k}} \in \mathbb{D}_{j}$, $\mathcal{D}_{j^{\prime}_{l}} \in \mathbb{D}_{j^{\prime}}$ & $j$ & $j$ & $j^{\prime}$ &0 &0 &0 \tabularnewline
    \midrule
\textbf{ 4}& Different & Different& $\ket{\mathcal{D}}_{j_{k}}$ & $\ket{\mathcal{D}}_{j_{k^{\prime}}}$ & $\ket{\mathcal{D}}_{j^{\prime}_{l}}$ & $\ket{\mathcal{D}}_{j_{k^\prime}}=\ket{\mathcal{D}}^{\perp}_{j_{k}} $,  $\mathcal{D}_{j^{\prime}_{l}} \in \mathbb{D}_{j^{\prime}} $  & $j$ & $j$ & $j^{\prime}$ & $\geq 0$ &0 & $\geq 0$  \tabularnewline
    \midrule
 \textbf{5} & Different& Different & $\ket{\mathcal{D}}_{j_{k}}$ & $\ket{\mathcal{D}}_{j^{\prime \prime}_{h}}$ & $\ket{\mathcal{D}}_{j^{\prime}_{l}}$ & $j \neq j^{\prime} \neq j^{\prime \prime} $, $\mathcal{D}_{j^{\prime \prime}_{h}} \in \mathbb{D}_{j^{\prime \prime}}$ & $j$ & $j^{\prime \prime}$ & $j^{\prime}$ &0 &0 &0  \tabularnewline
      \midrule
      \midrule
 \textbf{6} & Same & Different & $\ket{\mathcal{D}}_{j_{k}} $& $\ket{\mathcal{D}}_{j_{k}} $ & $\kappa_{j_{k}} \ket{\mathcal{D}}_{j_{k}} + \kappa_{j_{k^{\prime}}} e^{i \phi} \ket{\mathcal{D}}_{j_{k^{\prime}}} $ & $\kappa^{2}_{j_{k}} + \kappa^{2}_{j_{k^{\prime}}}=1$ & $j$ & $j$ & $j$ & $\geq 0$ & $\geq 0$ &$\geq 0$  \tabularnewline
      \midrule
      \textbf{7} & Same & Different & $\ket{\mathcal{D}}_{j_{k}} $& $\ket{\mathcal{D}}_{j_{k^{\prime}}} $ & $\kappa_{j_{k}} \ket{\mathcal{D}}_{j_{k}} + \kappa_{j_{k^{\prime}}} e^{i \phi} \ket{\mathcal{D}}_{j_{k^{\prime}}} $ & $\ket{\mathcal{D}}_{j_{k^\prime}}=\ket{\mathcal{D}}^{\perp}_{j_{k}} $, $\kappa^{2}_{j_{k}} + \kappa^{2}_{j_{k^{\prime}}}=1$ & $j$ & $j$ & $j$ & $\geq 0$ & $\geq 0$ &$\geq 0$  \tabularnewline
      \midrule
 \textbf{8} & Same & Same & $\kappa_{j_{k}} \ket{\mathcal{D}}_{j_{k}} + \kappa_{j_{k^{\prime}}} e^{i \phi} \ket{\mathcal{D}}_{j_{k^{\prime}}} $& $\kappa_{j_{k}} \ket{\mathcal{D}}_{j_{k}} + \kappa_{j_{k^{\prime}}} e^{i \phi} \ket{\mathcal{D}}_{j_{k^{\prime}}} $ & $\kappa_{j_{k}} \ket{\mathcal{D}}_{j_{k}} + \kappa_{j_{k^{\prime}}} e^{i \phi} \ket{\mathcal{D}}_{j_{k^{\prime}}} $ & $\kappa^{2}_{j_{k}} + \kappa^{2}_{j_{k^{\prime}}}=1$ & $j$ & $j$ & $j$ & $ 0$ & $ 0$ &$0$ \\
  & & superposition  &  & & & &&&&&& \tabularnewline
     \midrule
    \textbf{ 9} & Same & Different & $\kappa_{j_{k}} \ket{\mathcal{D}}_{j_{k}} + \kappa_{j_{k^{\prime}}} e^{i \phi_{1}} \ket{\mathcal{D}}_{j_{k^{\prime}}} $& $\kappa^{\prime}_{j_{k}} \ket{\mathcal{D}}_{j_{k}} + \kappa^{\prime}_{j_{k^{\prime}}} e^{i \phi_{2}} \ket{\mathcal{D}}_{j_{k^{\prime}}} $ & $\kappa^{\prime\prime}_{j_{k}} \ket{\mathcal{D}}_{j_{k}} + \kappa^{\prime\prime}_{j_{k^{\prime}}} e^{i \phi_{3}} \ket{\mathcal{D}}_{j_{k^{\prime}}} $ & $\phi_1 \neq \phi_2 \neq \phi_3$& $j$ & $j$ & $j$ & $\geq 0$ & $\geq 0$ &$\geq 0$  \\
   & & superposition  & where & where & where & $\kappa_{j_{k}} \neq \kappa^{\prime}_{j_{k}} \neq \kappa^{\prime \prime}_{j_{k}}$ &&&&&&\\
     & & &$\kappa^{2}_{j_{k}} + \kappa^{2}_{j_{k^{\prime}}}=1$  &$\kappa^{\prime^{ 2}}_{j_{k}} + \kappa^{\prime^{2}}_{j_{k^{\prime}}} =1$ & $\kappa^{\prime \prime^{2}}_{j_{k}} +\kappa^{\prime \prime^{2}}_{j_{k^{\prime}}}=1$ & $\kappa_{j_{k^{\prime}}} \neq \kappa^{\prime}_{j_{k^{\prime}}} \neq \kappa^{\prime \prime}_{j_{k^{\prime}}}$ &&&&&& \tabularnewline
\midrule
\midrule
 \textbf{10} & Different  & Different & $\ket{\mathcal{D}}_{j_{k}} $& $\ket{\mathcal{D}}_{j_{k}} $ & $\kappa_{j^{\prime}_{l}} \ket{\mathcal{D}}_{j^{\prime}_{l}} + \kappa_{j^{\prime}_{l^{\prime}}} e^{i \phi} \ket{\mathcal{D}}_{j^{\prime}_{l^{\prime}}} $ & $\kappa^{2}_{j^{\prime}_{l}} + \kappa^{2}_{j^{\prime}_{l^{\prime}}}=1$ & $j$ & $j$ & $j^{\prime}$ & $ 0$ & $ 0$ &$ 0$  \tabularnewline
      \midrule
      \textbf{11} & Different & Different & $\ket{\mathcal{D}}_{j_{k}} $& $\ket{\mathcal{D}}_{j_{k^{\prime}}} $ & $\kappa_{j^{\prime}_{l}} \ket{\mathcal{D}}_{j^{\prime}_{l}} + \kappa_{j^{\prime}_{l^{\prime}}} e^{i \phi} \ket{\mathcal{D}}_{j^{\prime}_{l^{\prime}}} $ & $\kappa^{2}_{j^{\prime}_{l}} + \kappa^{2}_{j^{\prime}_{l^{\prime}}}=1$ & $j$ & $j$ & $j^{\prime}$ & $\geq 0$ & $ 0$ &$\geq 0$   \tabularnewline
      \midrule
      \textbf{12} & Different & Different & $\ket{\mathcal{D}}_{j_{k}} $& $\kappa_{j_{k}} \ket{\mathcal{D}}_{j_{k}} + \kappa_{j_{k^{\prime}}} e^{i \phi} \ket{\mathcal{D}}_{j_{k^{\prime}}} $ & $\kappa_{j^{\prime}_{l}} \ket{\mathcal{D}}_{j^{\prime}_{l}} + \kappa_{j^{\prime}_{l^{\prime}}} e^{i \phi} \ket{\mathcal{D}}_{j^{\prime}_{l^{\prime}}} $ & $j \neq j^{\prime}$ & $j$ & $j$ & $j^{\prime}$ & $\geq 0$ & $\geq 0$ &$\geq 0$  \\
       & & superposition  &  & where & where & &&&&&& \\
     & & & &$\kappa^{2}_{j_{k}} + \kappa^{2}_{j_{k^{\prime}}} =1$ & $\kappa^{2}_{j^{\prime}_{l}} + \kappa^{2}_{j^{\prime}_{l^{\prime}}}=1$ &&&&&&&  \tabularnewline
     \midrule
     \midrule
\textbf{13} & Different & Different& $\kappa_{j_{k}} \ket{\mathcal{D}}_{j_{k}} + \kappa_{j_{k^{\prime}}} e^{i \phi} \ket{\mathcal{D}}_{j_{k^{\prime}}} $ & $\kappa_{j^{\prime \prime}_{h}} \ket{\mathcal{D}}_{j^{\prime \prime}_{h}} + \kappa_{j^{\prime \prime}_{h^{\prime}}} e^{i \phi^{\prime \prime}} \ket{\mathcal{D}}_{j^{\prime \prime}_{h^{\prime}}} $ & $\kappa_{j^{\prime}_{l}} \ket{\mathcal{D}}_{j^{\prime}_{l}} + \kappa_{j^{\prime}_{l^{\prime}}} e^{i \phi^{\prime}} \ket{\mathcal{D}}_{j^{\prime}_{l^{\prime}}} $ & $j \neq j^{\prime} \neq j^{\prime \prime} $& $j$ & $j^{\prime \prime}$ & $j^{\prime}$ & $ 0$ & $ 0$ &$ 0$\\
  & & superposition  & where & where & where &$\mathcal{D}_{j_{k^{\prime}}},   \in \mathbb{D}_{j}$, $\mathcal{D}_{j^{\prime \prime}_{h^{\prime}}} \in \mathbb{D}_{j^{\prime \prime}}$&&&&&& \\
     & & &$\kappa^{2}_{j_{k}} + \kappa^{2}_{j_{k^{\prime}}}=1$  &$\kappa^{2}_{j^{\prime \prime}_{h}} + \kappa^{2}_{j^{\prime \prime}_{h^{\prime}}} =1$ & $\kappa^{2}_{j^{\prime}_{l}} +\kappa^{2}_{j^{\prime}_{l^{\prime}}}=1$ & $\mathcal{D}_{j^{\prime}_{l^{\prime}}} \in \mathbb{D}_{j^{\prime}}$ &&&&&&  \tabularnewline
\bottomrule
\end{tabular}
\end{scriptsize}
\label{Tab:1}
 \medmuskip=2mu
\thinmuskip=2mu
\thickmuskip=2mu
\caption{List of possible combinations to create indistinguishability using three indistinguishable particles localized in three regions $s^1$, $s^2$, and $s^3$, each having three DoFs denoted by $j$, $j^{\prime}$, $j^{\prime \prime}$. Here the second column denotes whether entanglement is calculated in same DoFs or different DoFs of all particles; the third column denotes whether the eigenstate of the contributing DoFs in entanglement is the same or not or in superposition; the fourth, fifth and sixth columns describe the eigenstates of the three particles in the corresponding DoFs; the seventh column describes the relations between the eigenstates of the for entanglement. The eighth, ninth, and tenth columns describe the DoFs numbers (e.g., $j$ means the $j$th DoF) in which the measurements are done in the localized regions $s^1$, $s^2$, and $s^3$ respectively; the rest of the columns represent of the squared concurrences are zero or $\geq 0$.}
\end{landscape}
\end{sidewaystable}

\textbf{Case 1:} Entanglement is calculated in the same DoF of all  particles. Each particle is in the eigenstate $\ket{\mathcal{D}}_{j_{k}}$ of the $j$th DoF. Then after calculation, we get $\mathcal{C}^{2}_{s^{1} \mid s^{2}} =0$, $\mathcal{C}^{2}_{s^{1} \mid s^{3}}= 0$, and $\mathcal{C}^{2}_{s^{1} \mid s^{2}s^{3}} = 0$. Similar result holds for $p$ indistinguishable particles having the eigenstate $\ket{\mathcal{D}}_{j_{k}}$ of the $j$th DoF.

\textbf{Case 2:} Entanglement is calculated in the same DoF for all particles. For three indistinguishable particles, if two  of them are in the eigenstate $\ket{\mathcal{D}}_{j_{k}}$ and one is in the eigenstate $\ket{\mathcal{D}}_{j_{k^{\prime}}}$ where $\ket{\mathcal{D}}_{j_{k^{\prime}}}=\ket{\mathcal{D}}^{\perp}_{j_{k}}$, then $\mathcal{C}^{2}_{s^{1} \mid s^{2}} \geq 0$, $\mathcal{C}^{2}_{s^{1} \mid s^{3}}\geq 0$, and $\mathcal{C}^{2}_{s^{1} \mid s^{2}s^{3}} \geq 0$ as shown in Section~\ref{AppC}. Similar result holds for $p$ indistinguishable particles in $\mathbb{S}^p$ locations with each particle having $n$ DoFs where $(q+r)$ number of particles are in the eigenstate $\ket{\mathcal{D}}_{j_{k}}$ and rest of $(p-q-r)$ number of particles are in the eigenstate $\ket{\mathcal{D}}_{j_{k^{\prime}}}$. 

\subsection{Monogamy of three indistinguishable particles in spin DoF where two particles are in $\ket{\uparrow}$ eigenstate and one particles in $\ket{\downarrow}$ eigenstate} \label{AppC}
 \begin{figure}[t!] 
\centering
\includegraphics[width=\columnwidth]{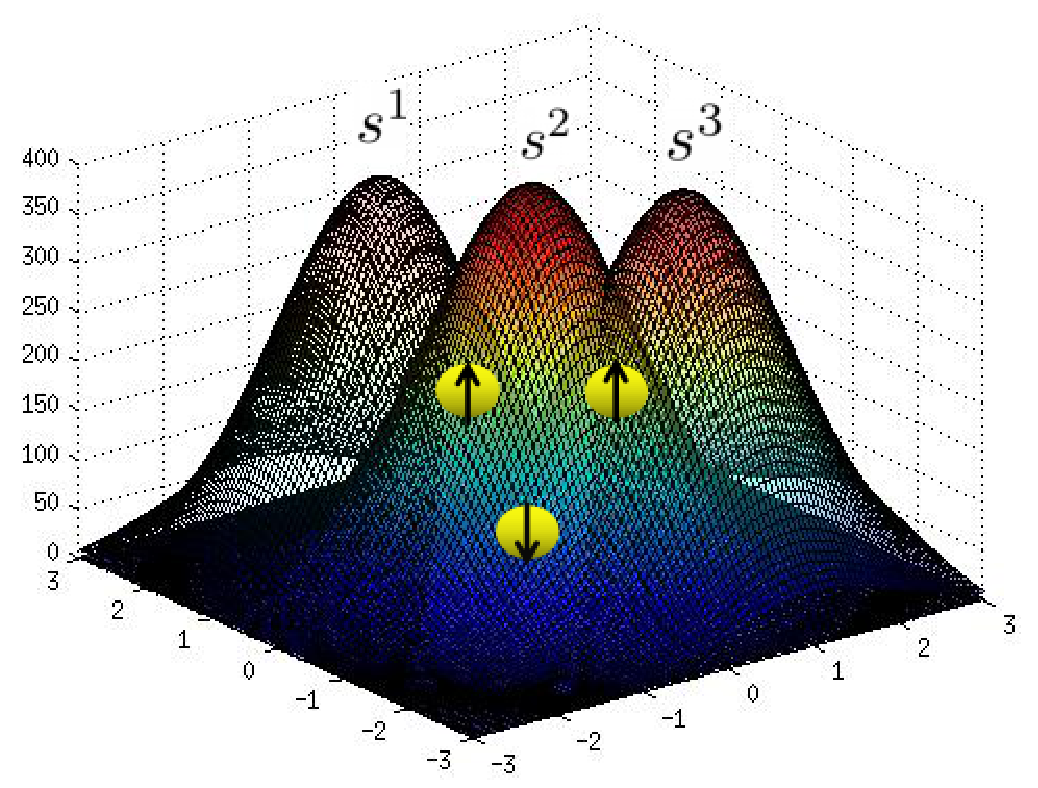} 
\caption{Overlapping of the wave-function of three indistinguishable particles in the localized regions $s^1$, $s^2$, and $s^3$ where two of them have $\ket{\uparrow}$ eigenstate and one having $\ket{\downarrow}$ eigenstate in spin degrees of freedom.}
\label{s1s2s3}
\end{figure} 
In this section, we  calculate the monogamy of entanglement using three indistinguishable particles each having two DoFs, are localized in three spatial regions $s^1$, $s^2$, and $s^3$ which we denote as $\mathbb{S}^3$ as shown in Fig.~\ref{s1s2s3}. We consider two particles with $\ket{\uparrow}$ eigenstate and one particles with $\ket{\downarrow}$ eigenstate in their spin DoF as we  calculate entanglement with only spin DoF. The other DoF of each particle can take any arbitrary eigenvalues. Thus the general state can be written as
\begin{equation}
\begin{aligned} \label{Pure3p2dof}
\ket{\Psi^{\left( 3,2 \right)}}=&\sum_{\alpha^{i} \in \mathbb{S}^3 , i \in \mathbb{N}_3} \eta^u \kappa^{\alpha^{1}, \alpha^{2}, \alpha^{3}}_{a^{1}_{1} a^{1}_{2}, a^{2}_{1} a^{2}_{2}, a^{3}_{1} a^{3}_{2}} \ket{\alpha^{1} a^{1}_{1} a^{1}_{2}, \alpha^{2} a^{2}_{1} a^{2}_{2}, \alpha^{3} a^{3}_{1} a^{3}_{2} } \\
=& \sum_{a^{i}_{2} \in \mathbb{D}_{2}} \eta^{u_{2}} \kappa^{\alpha^{1}, \alpha^{2}, \alpha^{3}}_{\uparrow a^{1}_{2}, \uparrow a^{2}_{2},\downarrow a^{3}_{2}} \ket{\alpha^1 \uparrow a^{1}_{2}, \alpha^2 \uparrow a^{2}_{2}, \alpha^3 \downarrow a^{3}_{2}} \\&
+  \sum_{a^{i}_{2} \in \mathbb{D}_{2}} \eta^{(1+u_{2})}  \kappa^{\alpha^{1}, \alpha^{2}, \alpha^{3}}_{ \uparrow a^{1}_{2},  \downarrow a^{2}_{2}, \uparrow a^{3}_{2}} \ket{\alpha^1 \uparrow a^{1}_{2}, \alpha^2 \downarrow a^{2}_{2}, \alpha^3 \uparrow a^{3}_{2}} \\
&+ \sum_{a^{i}_{2} \in \mathbb{D}_{2}} \eta^{u_{2}}  \kappa^{\alpha^{1}, \alpha^{2}, \alpha^{3}}_{\downarrow a^{1}_{2}, \uparrow a^{2}_{2},\uparrow a^{3}_{2}}  \ket{\alpha^1 \downarrow a^{1}_{2}, \alpha^2 \uparrow a^{2}_{2}, \alpha^3 \uparrow a^{3}_{2}}.
\end{aligned}
\end{equation}
Here $a^{i}_{1} \in \lbrace \ket{\uparrow}, \ket{\downarrow} \rbrace$, $a^{i}_{2} \in \mathbb{D}_{2}$ for $i \in \lbrace 1, 2, 3 \rbrace$ such that $a^{i}_{1} \neq a^{i^{\prime}}_{1}$ for all $i \neq i^{\prime}$ and  if $ \ket{\uparrow}=-\frac{1}{2}, \ket{\downarrow}=+\frac{1}{2}$ then $ \sum a^{i}_{1} = -\frac{1}{2}$. The value of $\eta=0$ if  $\left( \alpha^{i} = \alpha^{\prime}\right) \wedge \left(a^{i}_{j}=a^{i^{\prime}}_{j} \right) $ for all $i=i^{\prime}$ where $j \in \mathbb{N}_2  $.

 The density matrix of Equation~\eqref{Pure3p2dof} can be written as 
\begin{footnotesize}
 \begin{equation} \label{den3p2dof}
 \begin{aligned}
 \rho^{\left( 3, 2\right) } = &\sum_{\substack{\alpha^{i}, \beta^{i} \in \mathbb{S}^3  \& i \in \mathbb{N}_3   }} \eta^{\left(  u + \bar{u} \right) } \kappa^{\alpha^{1},\alpha^{2},\alpha^{3}}_{a^{1}_{1} a^{1}_{2}, a^{2}_{1} a^{2}_{2},a^{3}_{1} a^{3}_{2}}  \kappa^{\beta^{1}, \beta^{2}, \beta^{3}*}_{b^{1}_{1} b^{1}_{2},b^{2}_{1} b^{2}_{2},b^{3}_{1} b^{3}_{2}} \ket{\alpha^{1} a^{1}_{1} a^{1}_{2}, \alpha^{2} a^{2}_{1} a^{2}_{2}, \alpha^{3} a^{3}_{1} a^{3}_{2} } \bra{\beta^{1} b^{1}_{1} b^{1}_{2}, \beta^{2} b^{2}_{1} b^{2}_{2}, \beta^{3} b^{3}_{1} b^{3}_{2} }. \\
 \end{aligned}
 \end{equation}
\end{footnotesize}

Here $a^{i}_{1}, b^{i}_{1} \in \lbrace \ket{\uparrow}, \ket{\downarrow} \rbrace$, $a^{i}_{2}, b^{i}_{2} \in \mathbb{D}_{2}$ for $i \in \lbrace 1, 2, 3\rbrace$ such that $a^{i}_{1} \neq a^{i^{\prime}}_{1}$ and $ b^{i}_{1} \neq b^{i^{\prime}}_{1}$ for all $i \neq i^{\prime}$. Also if we take $ \ket{\uparrow}=-\frac{1}{2}, \ket{\downarrow}=+\frac{1}{2}$ then $ \sum a^{i}_{1} =\sum b^{i}_{1}= -\frac{1}{2}$. The value of $\eta=0$ if  $$\left\lbrace \left( \alpha^{i} = \alpha^{\prime}\right) \vee \left( \beta^{i} = \beta^{\prime}\right)\right\rbrace \rbrace \wedge \left\lbrace  \left(a^{i}_{j}=a^{i^{\prime}}_{j} \right) \vee \left(b^{i}_{j}=b^{i^{\prime}}_{j} \right) \right\rbrace  $$ for all $i=i^{\prime}$ where $j \in \mathbb{N}_2  $.
Here the normalization condition is
\begin{equation}
\sum_{\alpha^{i}, \beta^{i} \in \mathbb{S}^3, a^{i}_{1}, b^{i}_{1} \in \lbrace \uparrow, \downarrow \rbrace, a^{i}_{2}, b^{i}_{2} \in \mathbb{D}_{2}} \kappa^{\alpha^{1},\alpha^{2},\alpha^{3}}_{a^{1}_{1} a^{1}_{2}, a^{2}_{1} a^{2}_{2},a^{3}_{1} a^{3}_{2}}  \kappa^{\beta^{1}, \beta^{2}, \beta^{3}*}_{b^{1}_{1} b^{1}_{2},b^{2}_{1} b^{2}_{2},b^{3}_{1} b^{3}_{2}}  =1,
\end{equation}
where $\alpha^{i}=\beta^{i}$,  $a^{i}_{j}=b^{i}_{j}$ for all $i \in \lbrace 1, 2, 3 \rbrace$ and $j \in \lbrace 1, 2 \rbrace$.

Next step, we have to apply the projector $\mathcal{P}_{s^1s^2s^3}$ so that in each of the location $s^{1}$, $s^{2}$, and $s^{3}$ have exactly one particle which is defined as 
\begin{equation} \label{Projector3p2dof}
\begin{aligned}
\mathcal{P}_{s^1s^2s^3}= &\sum_{ \substack{ x^{i}_{1} \in \lbrace \uparrow, \downarrow \rbrace,  x^{i}_{2} \in \mathbb{D}_{2} }} \ket{s^{1} x^{1}_{1} x^{1}_{2}, s^{2} x^{2}_{1} x^{2}_{2}, s^{3} x^{3}_{1} x^{3}_{2} } \bra{s^{1} x^{1}_{1} x^{1}_{2}, s^{2} x^{2}_{1} x^{2}_{2}, s^{3} x^{3}_{1} x^{3}_{2} }.
\end{aligned}
\end{equation}

Thus after applying the projector, we get the density matrix as
\begin{equation}
\begin{aligned} \label{Denafterprojec}
\rho^{\left(3,2 \right)}_{s^1s^2s^3} =& \frac{\mathcal{P}_{s^1s^2s^3} \rho^{\left( 3, 2\right) } \mathcal{P}_{s^1s^2s^3}}{\text{Tr}\left(\mathcal{P}_{s^1s^2s^3} \rho^{\left( 3, 2\right) } \right) }\\
=&\sum_{\substack{ a^{i}_{2}, b^{i}_{2},x^{i}_{2} \in \mathbb{D}_{2} } } \dfrac{   \sum_{h,k \in \lbrace1,2,3\rbrace} \eta^{\left(k+  u_2 + \bar{u_2}-1 \right) } z_{h}z^{*}_{k}\rho^{\left( 3,2\right)}_{hk} }{   \sum_{h \in \lbrace 1,2,3\rbrace }   z_{h}z^{*}_{h} },
\end{aligned}
\end{equation}
where $a^{i}_{2}=b^{i}_{2}=x^{i}_{2}$. The values of 
\begin{equation}
\begin{aligned}
z_1=&\kappa^{s^1,s^2,s^3}_{\uparrow a^{1}_{2},\uparrow a^{2}_{2},\downarrow a^{3}_{2}}, 
\\ z_2=&\kappa^{s^1,s^2,s^3}_{\uparrow a^{1}_{2}, \downarrow a^{2}_{2}, \uparrow a^{3}_{2}}, 
\\ z_3=&\kappa^{s^1,s^2,s^3}_{\downarrow a^{1}_{2},\uparrow a^{2}_{2},\uparrow a^{3}_{2}},
\end{aligned}
\end{equation}
and the complex conjugates of $z_j$ for $j \in \lbrace1, 2, 3\rbrace$ can be calculated accordingly.

Also $\rho^{\left( 3,2\right)}_{hk}=\ket{\psi}^{(3,2)}_{h}\bra{\psi}^{(3,2)}_{k}$
 where 
\begin{equation}
\begin{aligned}
\ket{\psi}^{(3,2)}_{1}=&\ket{s^{1} \uparrow x^{1}_{2}, s^{2} \uparrow x^{2}_{2}, s^{3} \downarrow x^{3}_{2} }, \\ \ket{\psi}^{(3,2)}_{2}=&\ket{s^{1} \uparrow x^{1}_{2}, s^{2} \downarrow x^{2}_{2}, s^{3} \uparrow x^{3}_{2} }, 
\\ \ket{\psi}^{(3,2)}_{3}=&\ket{s^{1} \downarrow x^{1}_{2}, s^{2} \uparrow x^{2}_{2}, s^{3} \uparrow x^{3}_{2} },
\end{aligned}
\end{equation}
and the complex conjugates of $\ket{\psi}^{(3,2)}_{j}$ for $j \in \lbrace1, 2, 3\rbrace$ can be calculated accordingly. 

Now we have to trace out the particle at the region $s^3$. So, we get the reduced density matrix as
\begin{equation}
\begin{aligned}
\rho^{\left( 2, 2\right) }_{s^1 s^2 }=&\text{Tr}_{s^3} \left( \rho^{\left( 3, 2\right) }_{s^1 s^2 s^3 } \right)\\ =&\sum_{\substack{ m^{3}_{1}, m^{3}_{1} \in \lbrace \uparrow, \downarrow \rbrace, m^{3}_{2} \in \mathbb{D}_{2} }} \braket{s^3 m^{3}_{1} m^{3}_{2} \mid \rho^{\left( 3, 2\right) }_{s^1 s^2 s^3 } \mid s^3 m^{3}_{1} m^{3}_{2}} \\
=& \sum_{\substack{ a^{i}_{2}, b^{i}_{2},x^{i}_{2} \in \mathbb{D}_{2}}} \dfrac{  \sum_{h,k \in \lbrace1,2,3\rbrace} \eta^{\left(k+  u_2 + \bar{u_2}-1 \right) } z_{h}z^{*}_{k}\rho^{\left( 2,2\right)}_{hk}  }{ \sum_{h \in \lbrace 1,2,3\rbrace }   z_{h}z^{*}_{h}},  
\end{aligned}
\end{equation}
where $a^{i}_{2}=b^{i}_{2}=x^{i}_{2}$, and $m^{3}_{2}=x^{3}_{2} $. The values of $\rho^{\left( 2,2\right)}_{hk}=\ket{\psi}^{(2,2)}_{h}\bra{\psi}^{(2,2)}_{k}$
 where 
\begin{equation}
\begin{aligned}
\ket{\psi}^{(2,2)}_{1}=&\ket{s^{1} \uparrow x^{1}_{2}, s^{2} \uparrow x^{2}_{2} }, \\ \ket{\psi}^{(2,2)}_{2}=&\ket{s^{1} \uparrow x^{1}_{2}, s^{2} \downarrow x^{2}_{2},  }, \\ \ket{\psi}^{(2,2)}_{3}=&\ket{s^{1} \downarrow x^{1}_{2}, s^{2} \uparrow x^{2}_{2}}, 
 \\  \rho^{\left(2,2\right) }_{12}=&\rho^{\left(2,2\right) }_{13}=\rho^{\left(2,2\right) }_{21}=\rho^{\left(2,2\right) }_{31}=0,
 \end{aligned}
\end{equation}
and the complex conjugates of $\ket{\psi}^{(2,2)}_{j}$ for $j \in \lbrace1, 2, 3\rbrace$ can be calculated accordingly. 

Finally Tracing out the second DoF of each particle we have 

\begin{equation} \label{rho_2_1}
\begin{aligned}
\rho^{\left( 2, 1\right) }_{s^{1}_1,s^{2}_{1} }=& \sum_{ m^{1}_{2}, m^{2}_{2} \in \mathbb{D}_{2}  } \braket{s^1 m^{1}_{2} , s^2 m^{2}_{2} \mid \rho^{\left( 2, 2\right) }_{s^{1}s^{2}}  \mid s^1 m^{1}_{2} , s^2 m^{2}_{2}} \\
=& \sum_{\substack{ a^{i}_{2}, b^{i}_{2},x^{i}_{2} \in \mathbb{D}_{2}}} \dfrac{  \sum_{h,k \in \lbrace1,2,3\rbrace} \eta^{\left(k+  u_2 + \bar{u_2}-1 \right) } z_{h}z^{*}_{k}\rho^{\left( 2,1\right)}_{hk}  }{ \sum_{h \in \lbrace 1,2,3\rbrace }   z_{h}z^{*}_{h}}, 
\end{aligned}
\end{equation}

where  $a^{i}_{2}=b^{i}_{2}=x^{i}_{2}= m^{i}_{2}$. The values of $\rho^{\left( 2,1\right)}_{hk}=\ket{\psi}^{(2,1)}_{h}\bra{\psi}^{(2,1)}_{k}$
 where 
\begin{equation}
\begin{aligned}
\ket{\psi}^{(2,1)}_{1}=&\ket{s^{1} \uparrow , s^{2} \uparrow  }, \\ \ket{\psi}^{(2,1)}_{2}=&\ket{s^{1} \uparrow , s^{2} \downarrow   }, \\ \ket{\psi}^{(2,1)}_{3}=&\ket{s^{1} \downarrow , s^{2} \uparrow }, 
 \\  \rho^{\left(2,1\right) }_{12}=&\rho^{\left(2,1\right) }_{13}=\rho^{\left(2,1\right) }_{21}=\rho^{\left(2,1\right) }_{31}=0,
 \end{aligned}
\end{equation}
and the complex conjugates of $\ket{\psi}^{(2,1)}_{j}$ for $j \in \lbrace1, 2, 3\rbrace$ can be calculated accordingly. 

To calculate concurrence for $\rho^{\left( 2, 1\right) }_{s^{1}_1,s^{2}_{1} }$, we have to calculate the following
\begin{equation}
\widetilde{\rho}^{\left( 2, 1\right) }_{s^{1}_1,s^{2}_{1} } ~= ~\sigma^{s^1}_{y} \otimes  \sigma^{s^2}_{y}\rho^{\left( 2, 1\right) *}_{s^{1}_1,s^{2}_{1} } \sigma^{s^1}_{y} \otimes  \sigma^{s^2}_{y},
\end{equation}
where $\sigma^{s^1}_{y} = \ket{s^1}\bra{s^1} \otimes \sigma_{y}$, $\sigma^{s^2}_{y} = \ket{s^2}\bra{s^2} \otimes \sigma_{y}$. Here $\sigma_{y}$ is the Pauli matrix and the asterisk denotes complex conjugation. 
Finally, we have to calculate the eigenvalues of  $\mathcal{R}=\rho^{\left( 2, 1\right) }_{s^{1}_1,s^{2}_{1} }\widetilde{\rho}^{\left( 2, 1\right) }_{s^{1}_1,s^{2}_{1} }$. 

So, the value of square of the concurrence $\mathcal{C}^{2}_{s^1 \mid s^2}$ is
\begin{equation}
\begin{aligned}
\mathcal{C}^{2}_{s^1\mid s^2} ~=~ 2|z_2z_3|^2+z^{2}_2z^{*2}_3+z^{*2}_2 z^{2}_3 - 2|z_2z^{*}_{3}z^{*}_2z_3-z^{2}_2z^{2}_3|^2.
\end{aligned}
\end{equation}

Similarly, to calculate the squared concurrence $\mathcal{C}^{2}_{s^1s^3}$, 
first step is to trace out the particle at the region $s^2$ from $\rho^{\left(3,2 \right)}_{s^1s^2s^3}$ as shown in Eq.~\eqref{Denafterprojec}. So, we get the reduced density matrix as
\begin{equation}
\begin{aligned}
\rho^{\left( 2, 2\right) }_{s^1 s^3 }~=&\text{Tr}_{s^2} \left( \rho^{\left( 3, 2\right) }_{s^1 s^2 s^3 } \right) \\ =&\sum_{\substack{ m^{2}_{1} \in \lbrace \uparrow, \downarrow \rbrace, \\ m^{2}_{2} \in \mathbb{D}_{2} }} \braket{s^2 m^{2}_{1} m^{2}_{2} \mid \rho^{\left( 3, 2\right) }_{s^1 s^2 s^3 } \mid s^2 m^{2}_{1} m^{2}_{2}} \\
=&  \sum_{\substack{ a^{i}_{2}, b^{i}_{2},x^{i}_{2} \in \mathbb{D}_{2}}} \dfrac{  \sum_{h,k \in \lbrace1,2,3\rbrace} \eta^{\left(k+  u_2 + \bar{u_2}-1 \right) } z_{h}z^{*}_{k}\rho^{\left( 2,2\right)}_{hk}  }{ \sum_{h \in \lbrace 1,2,3\rbrace }   z_{h}z^{*}_{h}}, 
\end{aligned}
\end{equation}
where $a^{i}_{2}=b^{i}_{2}=x^{i}_{2}$, and $m^{2}_{2}=x^{2}_{2} $. The values of $\rho^{\left( 2,2\right)}_{hk}=\ket{\psi}^{(2,2)}_{h}\bra{\psi}^{(2,2)}_{k}$
 where 
\begin{equation}
\begin{aligned}
\ket{\psi}^{(2,2)}_{1}=&\ket{s^{1} \uparrow x^{1}_{2}, s^{3} \uparrow x^{3}_{2} }, \\ \ket{\psi}^{(2,2)}_{2}=&\ket{s^{1} \uparrow x^{1}_{2}, s^{3} \downarrow x^{3}_{2},  }, \\ \ket{\psi}^{(2,2)}_{3}=&\ket{s^{1} \downarrow x^{1}_{2}, s^{3} \uparrow x^{3}_{2}}, 
 \\  \rho^{\left(2,2\right) }_{12}=&\rho^{\left(2,2\right) }_{21}=\rho^{\left(2,2\right) }_{23}=\rho^{\left(2,2\right) }_{32}=0, 
 \end{aligned}
\end{equation}
and the complex conjugates of $\ket{\psi}^{(2,2)}_{j}$ for $j \in \lbrace1, 2, 3\rbrace$ can be calculated accordingly. 

Now following similar calculations as above we get
square of the concurrence between $s^1$ and $s^3$ is 
\begin{equation}
\begin{aligned}
\mathcal{C}^{2}_{s^1 \mid s^3}~=~ 2|z_1z_3|^2+z^{2}_{1}z^{*2}_{3}+z^{*2}_{1}z^{2}_{3} - 2|z_1z^{*}_{3}z^{*}_{1}z_3-z^{2}_{1}z^{2}_{3}|^2.
\end{aligned}
\end{equation}

Thus the monogamy relation is
\begin{equation}
\mathcal{C}^{2}_{s^1\mid s^2}+\mathcal{C}^{2}_{s^1 \mid s^3} ~=~ 4(1-|z_3|^{2})|z_3|^{2}  \leq 1.
\end{equation}. 

If we further trace-out the particle at $s^2$ from Eq.~\eqref{rho_2_1}, we get

\begin{equation} \label{rho_1_1}
\begin{aligned}
\rho^{\left( 1, 1\right) }_{s^{1}_1}=& \sum_{ m^{2}_{1} \in \mathbb{D}_{2}  } \braket{s^2 m^{2}_{1} \mid \rho^{\left( 2, 1\right) }_{s^{1}_1,s^{2}_{1} } \mid s^2 m^{2}_{1} } \\
=&~ \left( \mid z_1\mid^2+|z_2|^2 \right) \ket{s^1 \uparrow}\bra{s^1 \uparrow}+ |z_1|^3 \ket{s^1 \downarrow}\bra{s^1 \downarrow}.
\end{aligned}
\end{equation}

Thus, as Eq.~\eqref{Pure3p2dof} is a pure state so,  we have  
\begin{equation}
\mathcal{C}^{2}_{s^1\mid s^2 s^3} = 4 \text{det}(\rho^{\left( 1, 1\right) }_{s^{1}_1})= 4(1-|z_3|^{2})|z_3|^{2} \leq 1. 
\end{equation}

So, we get 
\begin{equation}
\mathcal{C}^{2}_{s^1 \mid s^2} + \mathcal{C}^{2}_{s^1 \mid s^3} ~=~ \mathcal{C}^{2}_{s^1 \mid s^2s^3}.
\end{equation}

\textbf{Case 3:} Entanglement is calculated between two different DoFs. 
Here, if two particles are in the eigenstate $\ket{\mathcal{D}}_{j_{k}}$ of the $j$th DoF and one particle is in the eigenstate $\ket{\mathcal{D}}_{j^{\prime}_{l}}$ of the $j^{\prime}$th DoF where $j \neq j^{\prime}$, then $\mathcal{C}^{2}_{s^{1} \mid s^{2}} = 0$, $\mathcal{C}^{2}_{s^{1} \mid s^{3}} = 0$, and $\mathcal{C}^{2}_{s^{1} \mid s^{2}s^{3}} = 0$. Similar result holds for $p$ indistinguishable particles in $\mathbb{S}^p$ locations with each particle having $n$ DoFs where $(q+r)$ number of particles are in the eigenstate $\ket{\mathcal{D}}_{j_{k}}$ of the $j$th DoF and rest of $(p-q-r)$ number of particles are in the eigenstate $\ket{\mathcal{D}}_{j^{\prime}_{l}}$ of the $j^{\prime}$th DoF. 

\textbf{Case 4:} Entanglement is calculated between two different DoFs.
 Here, if two particles are in the eigenstate $\ket{\mathcal{D}}_{j_{k}}$ and $\ket{\mathcal{D}}_{j_{k^{\prime}}}$  of the $j$th DoF respectively and one particle is in the eigenstate $\ket{\mathcal{D}}_{j^{\prime}_{l}}$ of the $j^{\prime}$th DoF where $j \neq j^{\prime}$ and $\ket{\mathcal{D}}_{j_{k^{\prime}}}=\ket{\mathcal{D}}^{\perp}_{j_{k}}$, then $\mathcal{C}^{2}_{s^{1} \mid s^{2}} \geq 0$, $\mathcal{C}^{2}_{s^{1} \mid s^{3}} = 0$, and $\mathcal{C}^{2}_{s^{1} \mid s^{2}s^{3}} \geq  0$ as shown in the Section~\ref{up_down_oam}. 
Similar result holds for $p$ indistinguishable particles in $\mathbb{S}^p$ locations with each particle having $n$ DoFs where $q$ and $r$ number of particles are in the eigenstate $\ket{\mathcal{D}}_{j_{k}}$ and $\ket{\mathcal{D}}_{j_{k^{\prime}}}$ respectively of the $j$th DoF and rest of $(p-q-r)$ number of particles are in the eigenstate $\ket{\mathcal{D}}_{j^{\prime}_{l}}$ of the $j^{\prime}$th DoF.

\subsection{Monogamy of three indistinguishable particles where two particles are in  $\ket{\uparrow}$ and $\ket{\downarrow}$ eigenstate respectively in spin DoF and one particle is in $\ket{+l}$ eigenstate in OAM DoF.} \label{up_down_oam}
Consider two particles with spin DoF having $\ket{\uparrow}$ and $\ket{\downarrow}$ eigenstates respectively and one particle with orbital angular momentum DoF with $\ket{+l}$ eigenstate. The eigenvalues of spin DoF and OAM DoF are represented by $a^{i}_{1} \in \mathbb{D}_{1}= \lbrace \ket{\uparrow}, \ket{\downarrow} \rbrace$ and $a^{i}_{2} \in \mathbb{D}_{2} =\lbrace \ket{+l} , \ket{-l}\rbrace$ respectively where $ i \in  \lbrace 1, 2 ,3 \rbrace$. The other non-contributing DoFs in entanglement of each particle can take any arbitrary eigenvalues. Thus the general state can be written as

\begin{equation}
\begin{aligned} \label{Pure3p2dof_OAM}
\ket{\Psi^{\left( 3,2 \right)}}
=& \sum_{a^{3}_{1} \in \mathbb{D}_{1},a^{1}_{2}, a^{2}_{2}  \in \mathbb{D}_{2}} \eta^{0} \kappa^{\alpha^{1}, \alpha^{2}, \alpha^{3}}_{\uparrow a^{1}_{2}, \downarrow a^{2}_{2}, a^{3}_{1} +l} \ket{\alpha^1 \uparrow a^{1}_{2}, \alpha^2 \downarrow a^{2}_{2}, \alpha^3 a^{3}_{1} +l} \\
&+  \sum_{a^{2}_{1} \in \mathbb{D}_{1},a^{1}_{2}, a^{3}_{2}  \in \mathbb{D}_{2}} \eta^{1}  \kappa^{\alpha^{1}, \alpha^{2}, \alpha^{3}}_{ \uparrow a^{1}_{2},   a^{2}_{1} +l, \downarrow a^{3}_{2}} \ket{\alpha^1 \uparrow a^{1}_{2}, \alpha^2 a^{2}_{1} +l, \alpha^3 \downarrow a^{3}_{2}} \\
&+ \sum_{a^{3}_{1} \in \mathbb{D}_{1},a^{1}_{2}, a^{2}_{2}  \in \mathbb{D}_{2}} \eta^{2} \kappa^{\alpha^{1}, \alpha^{2}, \alpha^{3}}_{\downarrow a^{1}_{2}, \uparrow a^{2}_{2}, a^{3}_{1} +l} \ket{\alpha^1 \downarrow a^{1}_{2}, \alpha^2 \uparrow a^{2}_{2}, \alpha^3 a^{3}_{1} +l} \\
&+  \sum_{a^{2}_{1} \in \mathbb{D}_{1},a^{3}_{2}, a^{1}_{2}  \in \mathbb{D}_{2}} \eta^{3}  \kappa^{\alpha^{1}, \alpha^{2}, \alpha^{3}}_{ \downarrow a^{1}_{2},   a^{2}_{1} +l, \uparrow a^{3}_{2}} \ket{\alpha^1 \downarrow a^{1}_{2}, \alpha^2 a^{2}_{1} +l, \alpha^3 \uparrow a^{3}_{2}} \\
&+ \sum_{a^{1}_{1} \in \mathbb{D}_{1}, a^{2}_{2}, a^{3}_{2}  \in \mathbb{D}_{2}} \eta^{4} \kappa^{\alpha^{1}, \alpha^{2}, \alpha^{3}}_{a^{1}_{1} +l, \uparrow  a^{2}_{2}, \downarrow a^{3}_{2}} \ket{\alpha^1 a^{1}_{1} +l, \alpha^2 \uparrow a^{2}_{2}, \alpha^3 \downarrow a^{3}_{2}} \\
&+  \sum_{a^{1}_{1} \in \mathbb{D}_{1}, a^{2}_{2}, a^{3}_{2}  \in \mathbb{D}_{2}} \eta^{5}  \kappa^{\alpha^{1}, \alpha^{2}, \alpha^{3}}_{a^{1}_{1} +l, \downarrow  a^{2}_{2}, \uparrow a^{3}_{2}} \ket{\alpha^1 a^{1}_{1} +l, \alpha^2 \downarrow a^{2}_{2}, \alpha^3 \uparrow a^{3}_{2}}
\end{aligned}
\end{equation}
where $\alpha^{i} \in \mathbb{S}^{3}$ for $i \in \mathbb{N}_3$.
After projecting the state by the suitable projector so that in each location $s^1$, $s^2$, and $s^3$ have exactly one particle. Finally, we calculate entanglement with $s^1$ and $s^2$ in spin DoF and between $s^1$ and $s^3$ in spin DoF and OAM DoF respectively. Following the above steps, we have
\begin{equation}
\begin{aligned} 
\mathcal{C}^{2}_{s^1 \mid s^2} =& 4 \left( \kappa^{s^{1}, s^{2}, s^{3}}_{\uparrow a^{1}_{2}, \downarrow a^{2}_{2}, a^{3}_{1} +l} \right)^2 \left( \kappa^{s^{1}, s^{2}, s^{3}}_{\downarrow a^{1}_{2}, \uparrow a^{2}_{2}, a^{3}_{1} +l} \right)^2,\\ \mathcal{C}^{2}_{s^1 \mid s^3}=&0, \\ \mathcal{C}^{2}_{s^1 \mid s^2s^3} =& 4 \left( \kappa^{s^{1}, s^{2}, s^{3}}_{\uparrow a^{1}_{2}, \downarrow a^{2}_{2}, a^{3}_{1} +l} \right)^2 \left( \kappa^{s^{1}, s^{2}, s^{3}}_{\downarrow a^{1}_{2}, \uparrow a^{2}_{2}, a^{3}_{1} +l} \right)^2.
\end{aligned}
\end{equation}
So, we get 
\begin{equation}
\mathcal{C}^{2}_{s^1 \mid s^2} + \mathcal{C}^{2}_{s^1 \mid s^3} = \mathcal{C}^{2}_{s^1 \mid s^2s^3}.
\end{equation}

\textbf{Case 5:} Entanglement is calculated between three different DoFs of three particles. 
If three particles are in the eigenstate $\ket{\mathcal{D}}_{j_{k}}$ of the $j$th DoF, $\ket{\mathcal{D}}_{j^{\prime \prime}_{h}}$ of the $j^{\prime \prime}$th DoF, and $\ket{\mathcal{D}}_{j^{\prime}_{l}}$ of the $j^{\prime}$th DoF where $j \neq j^{\prime} \neq j^{\prime \prime} $, then $\mathcal{C}^{2}_{s^{1} \mid s^{2}} = 0$, $\mathcal{C}^{2}_{s^{1} \mid s^{3}} = 0$, and $\mathcal{C}^{2}_{s^{1} \mid s^{2}s^{3}} =  0$. 
Similar result holds for $p$ indistinguishable particles in $\mathbb{S}^p$ locations with each particle having $n$ DoFs where $q$ number of particles are in the the eigenstate $\ket{\mathcal{D}}_{j_{k}}$ of $j$th DoF, $r$ number of particles are in the eigenstate $\ket{\mathcal{D}}_{j^{\prime \prime}_{h}}$ of $j^{\prime \prime}$th DoF and rest of $(p-q-r)$ number of particles are in the eigenstate $\ket{\mathcal{D}}_{j^{\prime}_{l}}$ of the $j^{\prime}$th DoF.

\textbf{Case 6:} Entanglement is calculated in the same DoF of all particles. If two particles are in  $\ket{\mathcal{D}}_{j_{k}}$ and one  particle is in the superpositions of its eigenstate, i.e., $\kappa_{j_{k}} \ket{\mathcal{D}}_{j_{k}} + \kappa_{j_{k^{\prime}}} e^{i \phi} \ket{\mathcal{D}}_{j_{k^{\prime}}} $ where $\kappa^{2}_{j_{k}} + \kappa^{2}_{j_{k^{\prime}}}=1$, then $\mathcal{C}^{2}_{s^{1} \mid s^{2}} \geq 0$, $\mathcal{C}^{2}_{s^{1} \mid s^{3}} \geq 0$, and $\mathcal{C}^{2}_{s^{1} \mid s^{2}s^{3}} \geq  0$. The calculations are similar as for case 2.
Similar result holds for $p$ indistinguishable particles in $\mathbb{S}^p$ locations with each particle having $n$ DoFs where
$(q+r)$ particles are in  $\ket{\mathcal{D}}_{j_{k}}$ and rest of $(p-q-r)$  particles are in the superpositions of its eigenstate, i.e., $\kappa_{j_{k}} \ket{\mathcal{D}}_{j_{k}} + \kappa_{j_{k^{\prime}}} e^{i \phi} \ket{\mathcal{D}}_{j_{k^{\prime}}} $.

\textbf{Case 7:} Entanglement is calculated in the same DoF of all particles.  If two particles are in the eigenstate $\ket{\mathcal{D}}_{j_{k}}$ and  $\ket{\mathcal{D}}_{j_{k^{\prime}}}$ and one  particle is in superpositions of its eigenstate, i.e., $\kappa_{j_{k}} \ket{\mathcal{D}}_{j_{k}} + \kappa_{j_{k^{\prime}}} e^{i \phi} \ket{\mathcal{D}}_{j_{k^{\prime}}} $ where $\kappa^{2}_{j_{k}} + \kappa^{2}_{j_{k^{\prime}}}=1$ of the $j$th DoF, then $\mathcal{C}^{2}_{s^{1} \mid s^{2}} \geq 0$, $\mathcal{C}^{2}_{s^{1} \mid s^{3}} \geq 0$, and $\mathcal{C}^{2}_{s^{1} \mid s^{2}s^{3}} \geq  0$. The calculations are similar as for case 2.
Similar result holds for $p$ indistinguishable particles in $\mathbb{S}^p$ locations with each particle having $n$ DoFs where $q$ number of particles are in  $\ket{\mathcal{D}}_{j_{k}}$, $r$ number of particles are in  $\ket{\mathcal{D}}_{j_{k^{\prime}}}$ and rest of $(p-q-r)$ number of particles are in superpositions of its eigenstate, i.e., $\kappa_{j_{k}} \ket{\mathcal{D}}_{j_{k}} + \kappa_{j_{k^{\prime}}} e^{i \phi} \ket{\mathcal{D}}_{j_{k^{\prime}}} $ where $\kappa^{2}_{j_{k}} + \kappa^{2}_{j_{k^{\prime}}}=1$.

\textbf{Case 8:} Entanglement is calculated in the same DoF of all particles. Each particles are in the superpositions of its eigenstate, i.e., $\kappa_{j_{k}} \ket{\mathcal{D}}_{j_{k}} + \kappa_{j_{k^{\prime}}} e^{i \phi} \ket{\mathcal{D}}_{j_{k^{\prime}}} $ where $\kappa^{2}_{j_{k}} + \kappa^{2}_{j_{k^{\prime}}}=1$.
Now calculations show that $\mathcal{C}^{2}_{s^{1} \mid s^{2}} = 0$, $\mathcal{C}^{2}_{s^{1} \mid s^{3}} = 0$, and $\mathcal{C}^{2}_{s^{1} \mid s^{2}s^{3}} =  0$. This case is similar as case 1 if we take a rotated basis to redefine the eigenstates as $\lbrace \ket{\mathcal{\tilde{D}}}_{j_{k}}, \ket{\mathcal{\tilde{D}}}^{\perp}_{j_{k}} \rbrace$ where $\ket{\mathcal{\tilde{D}}}_{j_{k}} = \kappa_{j_{k}} \ket{\mathcal{D}}_{j_{k}} + \kappa_{j_{k^{\prime}}} e^{i \phi} \ket{\mathcal{D}}_{j_{k^{\prime}}}$.

\textbf{Case 9:} Entanglement is calculated in the same DoF of all particles. 
Each particles are in different superpositions of its eigenstate, i.e., three particles are in the eigenstates $\kappa_{j_{k}} \ket{\mathcal{D}}_{j_{k}} + \kappa_{j_{k^{\prime}}} e^{i \phi_{1}} \ket{\mathcal{D}}_{j_{k^{\prime}}} $, $\kappa^{\prime}_{j_{k}} \ket{\mathcal{D}}_{j_{k}} + \kappa^{\prime}_{j_{k^{\prime}}} e^{i \phi_{2}} \ket{\mathcal{D}}_{j_{k^{\prime}}} $, and  $\kappa^{\prime\prime}_{j_{k}} \ket{\mathcal{D}}_{j_{k}} + \kappa^{\prime\prime}_{j_{k^{\prime}}} e^{i \phi_{3}} \ket{\mathcal{D}}_{j_{k^{\prime}}} $ of the $j$th DoF where $\kappa^{2}_{j_{k}} + \kappa^{2}_{j_{k^{\prime}}}=1$ , $\kappa^{2}_{j^{\prime \prime}_{h}} + \kappa^{2}_{j^{\prime \prime}_{h^{\prime}}} =1$, $\kappa^{2}_{j^{\prime}_{l}} +\kappa^{2}_{j^{\prime}_{l^{\prime}}}=1$ ,  $\phi_1 \neq \phi_2 \neq \phi_3$, $\kappa_{j_{k}} \neq \kappa^{\prime}_{j_{k}} \neq \kappa^{\prime \prime}_{j_{k}}$, and $\kappa_{j_{k^{\prime}}} \neq \kappa^{\prime}_{j_{k^{\prime}}} \neq \kappa^{\prime \prime}_{j_{k^{\prime}}}$. Now calculations show that $\mathcal{C}^{2}_{s^{1} \mid s^{2}} \geq 0$, $\mathcal{C}^{2}_{s^{1} \mid s^{3}} \geq 0$, and $\mathcal{C}^{2}_{s^{1} \mid s^{2}s^{3}} \geq  0$. The calculations are similar as for case 8.
Similar result holds for $p$ indistinguishable particles in $\mathbb{S}^p$ locations with each particle having $n$ DoFs where $q$ number particles are in $\kappa_{j_{k}} \ket{\mathcal{D}}_{j_{k}} + \kappa_{j_{k^{\prime}}} e^{i \phi_{1}} \ket{\mathcal{D}}_{j_{k^{\prime}}} $ eigenstate, $r$ number particles are in $\kappa^{\prime}_{j_{k}} \ket{\mathcal{D}}_{j_{k}} + \kappa^{\prime}_{j_{k^{\prime}}} e^{i \phi_{2}} \ket{\mathcal{D}}_{j_{k^{\prime}}} $   eigenstate and $(p-q-r)$ number of particles are in $\kappa^{\prime\prime}_{j_{k}} \ket{\mathcal{D}}_{j_{k}} + \kappa^{\prime\prime}_{j_{k^{\prime}}} e^{i \phi_{3}} \ket{\mathcal{D}}_{j_{k^{\prime}}} $ eigenstate.

\textbf{Case 10:} Here entanglement is calculated among two different DoFs where two particles are in $\ket{\mathcal{D}}_{j_{k}} $ eigenstate of the $j$th DoF and one particle is in 
$\kappa_{j^{\prime}_{l}} \ket{\mathcal{D}}_{j^{\prime}_{l}} + \kappa_{j^{\prime}_{l^{\prime}}} e^{i \phi} \ket{\mathcal{D}}_{j^{\prime}_{l^{\prime}}} $ eigenstate in the $j^{\prime}$th DoF. Here $j, j^{\prime} \in \mathbb{N}_{n}$ and $\kappa^{2}_{j^{\prime}_{l}} + \kappa^{2}_{j^{\prime}_{l^{\prime}}}=1$. Now calculations show $\mathcal{C}^{2}_{s^{1} \mid s^{2}} = 0$, $\mathcal{C}^{2}_{s^{1} \mid s^{3}} = 0$, and $\mathcal{C}^{2}_{s^{1} \mid s^{2}s^{3}} =  0$. This calculations is easier if we take a rotated basis as shown in case 8.
Similar result holds for $p$ indistinguishable particles in $\mathbb{S}^p$ locations with each particle having $n$ DoFs where $(q+r)$ number of particles are in $\ket{\mathcal{D}}_{j_{k}} $  eigenstate of the $j$th DoF and rest of $(p-q-r)$ number of particles are in 
$\kappa_{j^{\prime}_{l}} \ket{\mathcal{D}}_{j^{\prime}_{l}} + \kappa_{j^{\prime}_{l^{\prime}}} e^{i \phi} \ket{\mathcal{D}}_{j^{\prime}_{l^{\prime}}} $ eigenstate in the $j^{\prime}$th DoF. 

\textbf{Case 11:} Here entanglement is calculated among two different DoFs where 
two particles are in $\ket{\mathcal{D}}_{j_{k}}$ and $\ket{\mathcal{D}}_{j_{k^{\prime}}}$ eigenstate of the $j$th DoF and one particle is in 
$\kappa_{j^{\prime}_{l}} \ket{\mathcal{D}}_{j^{\prime}_{l}} + \kappa_{j^{\prime}_{l^{\prime}}} e^{i \phi} \ket{\mathcal{D}}_{j^{\prime}_{l^{\prime}}} $ eigenstate in the $j^{\prime}$th DoF. Here $j, j^{\prime} \in \mathbb{N}_{n}$ and $\kappa^{2}_{j^{\prime}_{l}} + \kappa^{2}_{j^{\prime}_{l^{\prime}}}=1$. Now calculations show that $\mathcal{C}^{2}_{s^{1} \mid s^{2}} \geq  0$, $\mathcal{C}^{2}_{s^{1} \mid s^{3}} = 0$, and $\mathcal{C}^{2}_{s^{1} \mid s^{2}s^{3}} \geq   0$. If we consider a rotated basis in $j^{\prime}$ DoF as $ \lbrace \ket{\mathcal{\tilde{D}}}_{j^{\prime}_{l}}, \ket{\mathcal{\tilde{D}}}^{\perp}_{j^{\prime}_{l}} \rbrace$ where $\ket{\mathcal{\tilde{D}}}_{j^{\prime}_{l}} =\kappa_{j^{\prime}_{l}} \ket{\mathcal{D}}_{j^{\prime}_{l}} + \kappa_{j^{\prime}_{l^{\prime}}} e^{i \phi} \ket{\mathcal{D}}_{j^{\prime}_{l^{\prime}}}$, then the calculations is similar as case 3.
Similar result holds for $p$ indistinguishable particles in $\mathbb{S}^p$ locations with each particle having $n$ DoFs where
$q$ and $r$ number of particles are in $\ket{\mathcal{D}}_{j_{k}}$ and $\ket{\mathcal{D}}_{j_{k^{\prime}}}$ eigenstate respectively of the $j$th DoF and rest of $(p-q-r)$ number of particles are in 
$\kappa_{j^{\prime}_{l}} \ket{\mathcal{D}}_{j^{\prime}_{l}} + \kappa_{j^{\prime}_{l^{\prime}}} e^{i \phi} \ket{\mathcal{D}}_{j^{\prime}_{l^{\prime}}} $ eigenstate in $j^{\prime}$th DoF. 

\textbf{Case 12:} Here entanglement is calculated among two different DoFs 
two particles are in the $\ket{\mathcal{D}}_{j_{k}}$ eigenstate and $\kappa_{j_{k}} \ket{\mathcal{D}}_{j_{k}} + \kappa_{j_{k^{\prime}}} e^{i \phi} \ket{\mathcal{D}}_{j_{k^{\prime}}} $ eigenstate in the $j$th DoF, one particle is in the superposition, i.e., $\kappa_{j^{\prime}_{l}} \ket{\mathcal{D}_{j^{\prime}_{l}}} + \kappa_{j^{\prime}_{l^{\prime}}} e^{i \phi} \ket{\mathcal{D}}_{j^{\prime}_{l^{\prime}}} $ eigenstate in the $j^{\prime}$th DoF. Now calculations show $\mathcal{C}^{2}_{s^{1} \mid s^{2}} \geq  0$, $\mathcal{C}^{2}_{s^{1} \mid s^{3}} = 0$, and $\mathcal{C}^{2}_{s^{1} \mid s^{2}s^{3}} \geq   0$. Using appropriate rotated basis of the $j$th and $j^{\prime}$th DoF, the calculations is similar as previous case.
Similar result holds for $p$ indistinguishable particles in $\mathbb{S}^p$ locations with each particle having $n$ DoFs where
 $q$ number of particles are in $\ket{\mathcal{D}}_{j_{k}}$ eigenstate of the $j$th DoF, $r$ number of particles are in the superposition, i.e., $\kappa_{j_{k}} \ket{\mathcal{D}}_{j_{k}} + \kappa_{j_{k^{\prime}}} e^{i \phi} \ket{\mathcal{D}}_{j_{k^{\prime}}} $ eigenstate in the $j$th DoF, and rest of $(p-q-r)$ number of particles are in the superposition, i.e., $\kappa_{j^{\prime}_{l}} \ket{\mathcal{D}_{j^{\prime}_{l}}} + \kappa_{j^{\prime}_{l^{\prime}}} e^{i \phi} \ket{\mathcal{D}}_{j^{\prime}_{l^{\prime}}} $ eigenstate in the $j^{\prime}$th DoF.

\textbf{Case 13:} Entanglement is calculated between three different DoFs. 
Here, three particles are in the superpositions of its eigenstate, i.e., $\kappa_{j_{k}} \ket{\mathcal{D}}_{j_{k}} + \kappa_{j_{k^{\prime}}} e^{i \phi} \ket{\mathcal{D}}_{j_{k^{\prime}}} $ where $\kappa^{2}_{j_{k}} + \kappa^{2}_{j_{k^{\prime}}}=1$ of the $j$th DoF;  $\kappa_{j^{\prime \prime}_{h}} \ket{\mathcal{D}}_{j^{\prime \prime}_{h}} + \kappa_{j^{\prime \prime}_{h^{\prime}}} e^{i \phi^{\prime \prime}} \ket{\mathcal{D}}_{j^{\prime \prime}_{h^{\prime}}} $ where $\kappa^{2}_{j^{\prime \prime}_{h}} + \kappa^{2}_{j^{\prime \prime}_{h^{\prime}}} =1$ of the $j^{\prime \prime}$th DoF; and $\kappa_{j^{\prime}_{l}} \ket{\mathcal{D}}_{j^{\prime}_{l}} + \kappa_{j^{\prime}_{l^{\prime}}} e^{i \phi^{\prime}} \ket{\mathcal{D}}_{j^{\prime}_{l^{\prime}}}$ where $\kappa^{2}_{j^{\prime}_{l}} +\kappa^{2}_{j^{\prime}_{l^{\prime}}}=1$ of the $j^{\prime}$th DoF where $j \neq j^{\prime} \neq j^{\prime \prime} $. 
Using appropriate rotated basis of the $j$th, $j^{\prime}$th, and $j^{\prime \prime}$th DoF, the calculations is similar as shown in case 5.
 Now calculations show $\mathcal{C}^{2}_{s^{1} \mid s^{2}} =  0$, $\mathcal{C}^{2}_{s^{1} \mid s^{3}} = 0$, and $\mathcal{C}^{2}_{s^{1} \mid s^{2}s^{3}} =  0$. 
Similar result holds for $p$ indistinguishable particles in $\mathbb{S}^p$ locations with each particle having $n$ DoFs where
 $q$ number of particles are in superpositions of its eigenstate, i.e., $\kappa_{j_{k}} \ket{\mathcal{D}}_{j_{k}} + \kappa_{j_{k^{\prime}}} e^{i \phi} \ket{\mathcal{D}}_{j_{k^{\prime}}} $ of the $j$th DoF, Here, $r$ number of particles are in superpositions of its eigenstate, i.e., $\kappa_{j^{\prime \prime}_{h}} \ket{\mathcal{D}}_{j^{\prime \prime}_{h}} + \kappa_{j^{\prime \prime}_{h^{\prime}}} e^{i \phi^{\prime \prime}} \ket{\mathcal{D}}_{j^{\prime \prime}_{h^{\prime}}} $  of $j^{\prime \prime}$th DoF and rest of $(p-q-r)$ number of particles are in superpositions of its eigenstate, i.e., $\kappa_{j^{\prime}_{l}} \ket{\mathcal{D}}_{j^{\prime}_{l}} + \kappa_{j^{\prime}_{l^{\prime}}} e^{i \phi^{\prime}} \ket{\mathcal{D}}_{j^{\prime}_{l^{\prime}}}$ of $j^{\prime}$th DoF.

One may think that there might be more cases. Upon careful inspection, it can be concluded that all those cases is equivalent with any of the above mentioned cases. 

 From all the above cases we can see that the monogamy holds for pure states with an
equality relation. On the other hand, for mixed states, we use the convexity of the concurrence function to prove the monogamy relation. Since any mixed state can be
expressed as an ensemble of the pure states, we can apply the concurrence function on those
ensembles, i.e., the convex combinations for pure states and minimize it to get the required inequality for any arbitrary mixed states as shown in next section.

\subsection{Monogamy of indistinguishable particles for mixed states} \label{MOE_mixed}
In this section, we generalize the relation for monogamy of entanglement of indistinguishable particles for mixed states. 
We have proved  in Corollary 1.1 the the main text that for all pure states $\rho_{\alpha_{i} \beta_{j} \gamma_{k}}$
\begin{equation} \label{Moe_eq}
\mathcal{C}^{2}_{\alpha_{i} \mid \beta_{j}} \left(  \rho_{\alpha_{i} \beta_{j} }\right) +\mathcal{C}^{2}_{\alpha_{i} \mid \gamma_{k}}\left(  \rho_{\alpha_{i} \mid \gamma_{k} } \right) = \mathcal{C}^{2}_{\alpha_{i} \mid \beta_{j}  \gamma_{k}}   \left( \rho_{\alpha_{i} \beta_{j} \gamma_{k}}\right).
\end{equation}
But this relation is not valid for mixed states as the right hand side is not defined for mixed states. Since all mixed states are convex combination some pure states, we can write $\rho_{\alpha_{i} \beta_{j} \gamma_{k}}$ as a convex combination of pure states, 
as
\begin{equation}
\rho_{\alpha_{i} \beta_{j} \gamma_{k}}= \sum_{m} \text{Pr}_{m} \ket{\psi_{m}}_{\alpha_{i} \beta_{j} \gamma_{k}} \bra{\psi_{m}}_{\alpha_{i} \beta_{j} \gamma_{k}},
\end{equation}
where $\Pr_{m}$ denotes the probability of $\ket{\psi_{m}}_{\alpha_{i} \beta_{j} \gamma_{k}}$.
For each $m$, we can write from Eq.~\eqref{Moe_eq} as
\begin{equation} \label{1st}
\begin{aligned}
 &\mathcal{C}^{2}_{\alpha_{i} \mid \beta_{j}} \left(\ket{\psi_{m}}_{\alpha_{i} \beta_{j}}  \bra{\psi_{m}}_{\alpha_{i} \beta_{j}}    \right)  + \mathcal{C}^{2}_{\alpha_{i} \mid \gamma_{k}} \left(\ket{\psi_{m}}_{\alpha_{i} \gamma_{k}}  \bra{\psi_{m}}_{\alpha_{i} \gamma_{k}}    \right) \\ =& \mathcal{C}^{2}_{\alpha_{i} \mid \beta_{j}\gamma_{k}} \left(\ket{\psi_{m}}_{\alpha_{i} \beta_{j}\gamma_{k}}  \bra{\psi_{m}}_{\alpha_{i} \beta_{j}\gamma_{k}}     \right).
\end{aligned}
\end{equation}

Multiplying both sides with $\text{Pr}_{m}$, we get
\begin{equation}
\begin{aligned}
&\text{Pr}_{m} \mathcal{C}^{2}_{\alpha_{i} \mid \beta_{j}} \left(\ket{\psi_{m}}_{\alpha_{i} \beta_{j}}  \bra{\psi_{m}}_{\alpha_{i} \beta_{j}}    \right)  + \text{Pr}_{m} \mathcal{C}^{2}_{\alpha_{i} \mid \gamma_{k}} \left(\ket{\psi_{m}}_{\alpha_{i} \gamma_{k}}  \bra{\psi_{m}}_{\alpha_{i} \gamma_{k}}    \right) \\ =& \text{Pr}_{m} \mathcal{C}^{2}_{\alpha_{i} \mid \beta_{j}\gamma_{k}} \left(\ket{\psi_{m}}_{\alpha_{i} \beta_{j}\gamma_{k}}  \bra{\psi_{m}}_{\alpha_{i} \beta_{j}\gamma_{k}} \right).
\end{aligned}
\end{equation}
Summing up for all the pure constituents, 
\begin{equation} \label{Sum_lambda_i}
\begin{aligned}
&\sum_{m} \text{Pr}_{m} \mathcal{C}^{2}_{\alpha_{i} \mid \beta_{j}} \left(\ket{\psi_{m}}_{\alpha_{i} \beta_{j}}  \bra{\psi_{m}}_{\alpha_{i} \beta_{j}}    \right)  + \sum_{m}\text{Pr}_{m} \mathcal{C}^{2}_{\alpha_{i} \mid \gamma_{k}} \left(\ket{\psi_{m}}_{\alpha_{i} \gamma_{k}}  \bra{\psi_{m}}_{\alpha_{i} \gamma_{k}}    \right) \\ =& \sum_{m}\text{Pr}_{m} \mathcal{C}^{2}_{\alpha_{i} \mid \beta_{j}\gamma_{k}} \left( \ket{\psi_{m}}_{\alpha_{i} \beta_{j}\gamma_{k}}  \bra{\psi_{m}}_{\alpha_{i} \beta_{j}\gamma_{k}}  \right).
\end{aligned}
\end{equation}
Now consider the decomposition, say $\left\lbrace \left(  \text{Pr}^{*}_{m}, \ket{\psi_{m}}^{*}_{\alpha_{i} \beta_{j} \gamma_{k}} \right)  \right\rbrace$, that minimizes the right hand side of Eq.~\eqref{Sum_lambda_i} and   denote it by
\begin{equation}\label{C_min}
\left( \mathcal{C}^{2}_{\alpha_{i} \mid \beta_{j} \gamma_{k}}\right)^{\text{min}} := \min_{ \left\lbrace \left(  \text{Pr}_{m}, \ket{\psi_{m}}_{\alpha_{i} \beta_{j} \gamma_{k}} \right)  \right\rbrace} \sum_{m}\text{Pr}_{m} \mathcal{C}^{2}_{\alpha_{i} \mid \beta_{j}\gamma_{k}} \left(\ket{\psi_{m}}_{\alpha_{i} \beta_{j}\gamma_{k}}  \bra{\psi_{m}}_{\alpha_{i} \beta_{j}\gamma_{k}} \right).
\end{equation}

Now expressing $\rho_{\alpha_{i} \beta_{j} \gamma_{k}}$ by the above minimizing decomposition as in Eq.~\eqref{C_min}, we have
\begin{equation}
\begin{aligned}
&\mathcal{C}^{2}_{\alpha_{i} \mid \beta_{j}} \left(  \rho_{\alpha_{i} \beta_{j} }\right) +\mathcal{C}^{2}_{\alpha_{i} \mid \gamma_{k}}\left(  \rho_{\alpha_{i} \mid \gamma_{k} } \right)\\
=& \mathcal{C}^{2}_{\alpha_{i} \mid \beta_{j}} \left( \sum_{m} \text{Pr}^{*}_{m}\ket{\psi_{m}}^{*}_{\alpha_{i} \beta_{j}}  \bra{\psi_{m}}^{*}_{\alpha_{i} \beta_{j}}    \right)  + \mathcal{C}^{2}_{\alpha_{i} \mid \gamma_{k}} \left(\sum_{m} \text{Pr}^{*}_{m}\ket{\psi_{m}}^{*}_{\alpha_{i} \gamma_{k}}  \bra{\psi_{m}}^{*}_{\alpha_{i} \gamma_{k}}    \right)\\
\leq & \sum_{m} \text{Pr}^{*}_{m} \mathcal{C}^{2}_{\alpha_{i} \mid \beta_{j}} \left(\ket{\psi_{m}}^{*}_{\alpha_{i} \beta_{j}}  \bra{\psi_{m}}^{*}_{\alpha_{i} \beta_{j}}    \right)  + \sum_{m}\text{Pr}^{*}_{m} \mathcal{C}^{2}_{\alpha_{i} \mid \gamma_{k}} \left(\ket{\psi_{m}}^{*}_{\alpha_{i} \gamma_{k}}  \bra{\psi_{m}}^{*}_{\alpha_{i} \gamma_{k}}    \right) \\
& \hspace{6cm} \text{(by the convexity of  $\mathcal{C}^{2}$~\cite{Wootters98})}\\
=& \sum_{m} \text{Pr}^{*}_{m} \left\lbrace  \mathcal{C}^{2}_{\alpha_{i} \mid \beta_{j}} \left(\ket{\psi_{m}}^{*}_{\alpha_{i} \beta_{j}}  \bra{\psi_{m}}^{*}_{\alpha_{i} \beta_{j}}    \right)  +  \mathcal{C}^{2}_{\alpha_{i} \mid \gamma_{k}} \left(\ket{\psi_{m}}^{*}_{\alpha_{i} \gamma_{k}}  \bra{\psi_{m}}^{*}_{\alpha_{i} \gamma_{k}}    \right) \right\rbrace  \\
=& \sum_{m}\text{Pr}^{*}_{m} \mathcal{C}^{2}_{\alpha_{i} \mid \beta_{j}\gamma_{k}} \left(\ket{\psi_{m}}^{*}_{\alpha_{i} \beta_{j}\gamma_{k}}  \bra{\psi_{m}}^{*}_{\alpha_{i} \beta_{j}\gamma_{k}}     \right) \hspace{1cm} \left( \text{from Eq.~\eqref{1st}} \right) \\
= & \left( \mathcal{C}^{2}_{\alpha_{i} \mid \beta_{j} \gamma_{k}}\right)^{\text{min}} \hspace{1cm} \left( \text{from Eq.~\eqref{C_min}}\right) .
\end{aligned}
\end{equation}

Thus we have for mixed states
\begin{equation} 
\mathcal{C}^{2}_{\alpha_{i} \mid \beta_{j}} \left(  \rho_{\alpha_{i} \beta_{j} }\right) +\mathcal{C}^{2}_{\alpha_{i} \mid \gamma_{k}}\left(  \rho_{\alpha_{i} \mid \gamma_{k} } \right) \leq  \mathcal{C}^{2}_{\alpha_{i} \mid \beta_{j}  \gamma_{k}}   \left( \rho_{\alpha_{i} \beta_{j} \gamma_{k}}\right).
\end{equation}


\section{Physical significance of monogamy of entanglement for indistinguishable particles having multiple DoFs} \label{MoE_phy_sig}
 Existence of indistinguishable particles is a special feature of quantum mechanics. There has been substantial interest in the  community to use indistinguishable particles  as a resource~\cite{Adesso20} for many  quantum information processing tasks such as teleportation~\cite{Ugo15,LFC18}, entanglement swapping~\cite{LFCES19} etc., that are commonly implemented  using distinguishable particles.  
Recently a class of results have been published  which explicitly
demonstrate that certain properties and applications are exclusive to indistinguishable (or distinguishable) particles. We call such results as the separation results  between these two domains. Das \textit{et al.}~\cite{Das20} showed that unit fidelity quantum teleportation can only be performed using distinguishable particles, and on the other hand, hyper-hybrid entangled state can only be produced using indistinguishable particles. Further, if a quantum protocol can be performed using both distinguishable and indistinguishable particles,  one of them may give some advantages over the other. For example entanglement swapping can be performed using minimum two indistinguishable particles~\cite{Das20}, whereas atleast three particles are required for the distinguishable  case~\cite{Pan10,Pan19}.
Another separation result between these two domains is given by Paul \textit{et al.}~\cite{Paul21}. They show that  using two indistinguishable particles each having multiple degrees of freedom, it is possible to violate monogamy of entanglement maximally, which is not feasible for distinguishable particles~\cite{camalet17}.

In continuation of these separation results, this chapter contributes  one property of indistinguishable particles that are different from distinguishable particles.  The inequality of monogamy of entanglement using squared concurrence for three or more distinguishable particles as shown in~\cite{CKW00} becomes equality for pure indistinguishable states, whereas the inequality may hold only for mixed indistinguishable states. The physical significance of this result is that for all pure states, if MoE is calculated for three or more indistinguishable particle, then the residual entanglement in the whole state is zero, i.e., entanglement is distributed among all its bipartitions.  
 Note that, this equality is different from the one proposed in~\cite{Gour18}. This result is extremely helpful to calculate the entanglement in the scenarios where particles are indistinguishable like quantum dots~\cite{Petta05,Tan15}, ultracold atomic gases~\cite{Leibfried03}, Bose-Einstein condensates~\cite{Morsch06,Esteve08}, quantum meteorology~\cite{Giovannetti06,Benatti11}, etc.
\chapter{Applications of entangled distinguishable and indistinguishable particles} \label{Chap9}
In this section, we discuss some applications of distinguishable and indistinguishable particles. First we show an application using extra degrees of freedom as an ancilla qubit to reduce the resource in some cryptographic protocols based on our work in~\cite{Das_SF_21}. Next, we show an how using indistinguishability can create an attack in some cryptographic protocols based on our work in~\cite{MoE}. Finally, we show an entanglement swapping protocol using only two indistinguishable particles and without using Bell state measurement based on our work in~\cite{Das20}.  

This chapter is based on the works in~\cite{Das20,Das_SF_21,MoE}. 

\section{Reducing the resource requirement in cryptographic protocols}  \label{QPQ_app}

\indent Security and efficiency are two major criteria of a cryptographic protocols.
 If two cryptographic systems with  different resource requirements provide the same level of security, the one with less resources becomes the natural choice. Here, we present a generic scheme to reduce the number of particles used in device-independent~\cite{Acin07} (DI) quantum private query~\cite{QPQ07,DIQPQ} (QPQ) without affecting the security that can be used for both distinguishable and indistinguishable particles.
 
In quantum cryptography, DI tests~\cite{Acin07} are required to establish secure key between two parties when the devices are not trustworthy. 
In QPQ, a user queries a database about some specific entry and gets back only the corresponding data without revealing the query to the server. To achieve this, a secure key must be established between the database owner and the user such that the database owner knows the full key but the user knows a fraction of that key. In DI-QPQ~\cite{DIQPQ} using Clauser-Horne-Shimony-Holt test~\cite{CHSH}, the maximum success probability is $\cos^{2}\frac{\pi}{8} \approx 85\%$ using two qubits. 
With an ancilla particle, this success probability can be increased asymptotically to unity using quantum pseudo-telepathy game~\cite{Brassard05,Jyoti18}.

The cost of adding an ancilla particle can be bypassed by using an additional DoF and creating multi-DoF entanglement. In doing so, the success probability remains the same but the generalized SF changes.  


In Section~\ref{Pseudo_tele}, it is shown that the success probability of a quantum private query protocol as described in~\ref{QPQ} can be increased asymptotically to unity using an ancilla particle $X$ as an extra qubit.
Now instead of choosing the ancilla as another particle, if another DoF of $A$ is used as the ancilla, then we can reduce the number of particles used in the quantum pseudo-telepathy test. Then the state in Eq.~\eqref{pseudo_state} can be written as
 \begin{equation} \label{pseudo_state_DoF}
 \ket{\psi}_{BA_1A_2} = \frac{1}{\sqrt{2}} \left( \text{cos}\frac{\theta}{2} \ket{000}_{BA_1A_2}  + \text{sin}\frac{\theta}{2} \ket{010}_{BA_1A_2} + \text{cos}\frac{\theta}{2} \ket{111}_{BA_1A_2} - \text{sin}\frac{\theta}{2} \ket{100}_{BA_1A_2} \right). 
 \end{equation}
 
 Assuming each particle have two DoFs, i.e., $n=2$,  the generalized singlet fraction for the state in Eq.~\eqref{pseudo_state} is
 \begin{equation}
 \begin{aligned}
 F^{(2)}_{g}&=\max \left\{  \max_{i} \left\lbrace  \max_{\psi_{a_{i}b_{{j}}}} \left\lbrace  \sum^{2}_{j=1} \braket{\psi_{a_{i}b_{j}} \mid \rho_{a_{i}b_{j}} |\psi_{a_{i}b_{j}}}  \right\rbrace \right\rbrace ,   \max_{j} \left\lbrace  \max_{\psi_{a_{i}b_{{j}}}} \left\lbrace  \sum^{2}_{i=1} \braket{\psi_{a_{i}b_{j}} \mid \rho_{a_{i}b_{j}} |\psi_{a_{i}b_{j}}}  \right\rbrace  \right\rbrace  \right\}\\
 &=\frac{1}{2}+\cos^{2} \frac{\theta}{2}
 \end{aligned}
 \end{equation}
where $\rho_{a_{i}b_{j}}:= \text{Tr}_{a_{\bar{i}}{b_{\bar{j}}}} ( \rho^{(2)}_{AB} )$ and $\rho^{(2)}_{AB}=\text{Tr}_{X}\left( \ket{\psi}_{BAX} \bra{\psi}_{BAX}\right) $. 

\begin{figure}[t!] 
\centering
\includegraphics[width=\columnwidth]{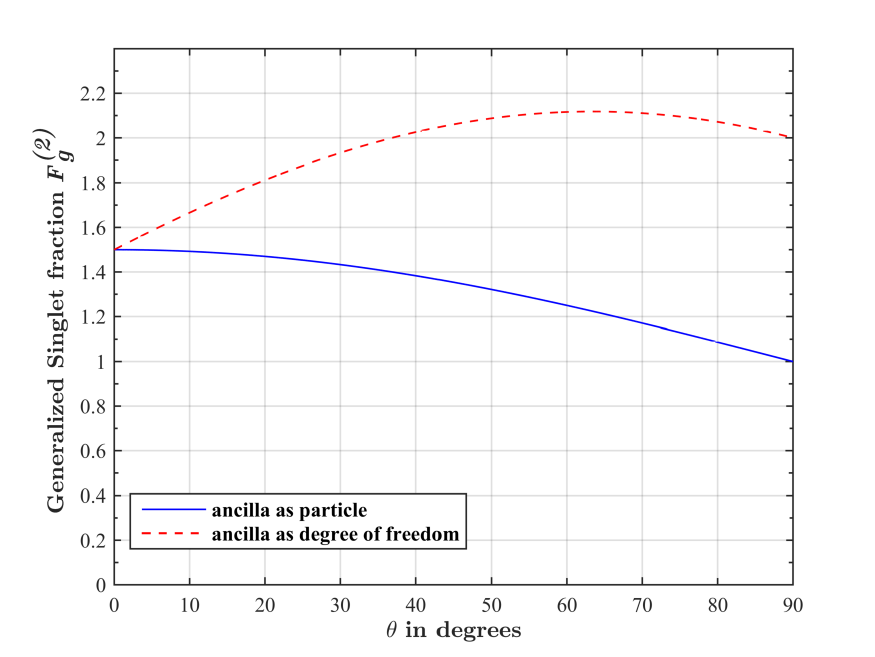} 
\caption{The variation of generalized SF $F^{(n)}_{g}$ for the quantum pseudo-telepathy test using ancilla as particle and using ancilla as degrees of freedom with varying $\theta$ in degrees.}
\label{Appli}
\end{figure}

Similarly if we calculate the generalized singled fraction for the state in Eq.~\eqref{pseudo_state_DoF}, we get $\frac{1}{2}+\cos^{2} \frac{\theta}{2} + 2\cos \frac{\theta}{2}\sin \frac{\theta}{2}$. The values of generalized singlet fraction for the states in Eq.~\eqref{pseudo_state} and Eq.~\eqref{pseudo_state_DoF} are shown in Fig.~\ref{Appli}

Thus in many ancilla assisted  quantum processing protocols like quantum process tomography~\cite{Altepeter03}, entanglement stabilization~\cite{Andersen19}, quantum secret sharing~\cite{Mitra15}, quantum error correction~\cite{Criger12}, quantum measurement~\cite{Brida12}, weak value amplification~\cite{Pang14}, quantum channel discrimination~\cite{Bae19}, coherent-state superposition generation~\cite{Takahashi08}, etc., where the resource can be reduced by using additional DoF as ancilla instead of a particle.

\section{Attacks on the security of cryptographic protocols.} \label{Hardy_appli}
There are scenarios where particles are indistinguishable like  quantum dots~\cite{Petta05,Tan15}, ultracold atomic gases~\cite{Leibfried03}, Bose-Einstein condensates~\cite{Morsch06,Esteve08}, our method will be extremely helpful to calculate the entanglement and monogamy in those situations.
Indistinguishability can be created  manually by particle exchange~\cite{Y&S92PRA,Y&S92PRL}, using which we present a generic one-step indistinguishablity module as a plug-in to many protocols.  This may have an effect in modifying their output.
\begin{figure}[h!] 
\centering
\includegraphics[width=0.6\columnwidth]{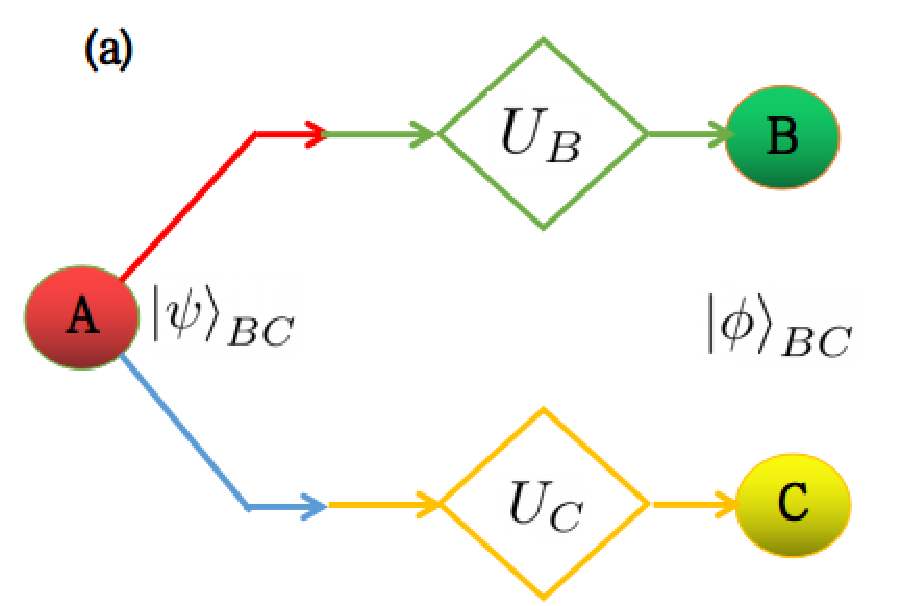} \\
\includegraphics[width=\columnwidth]{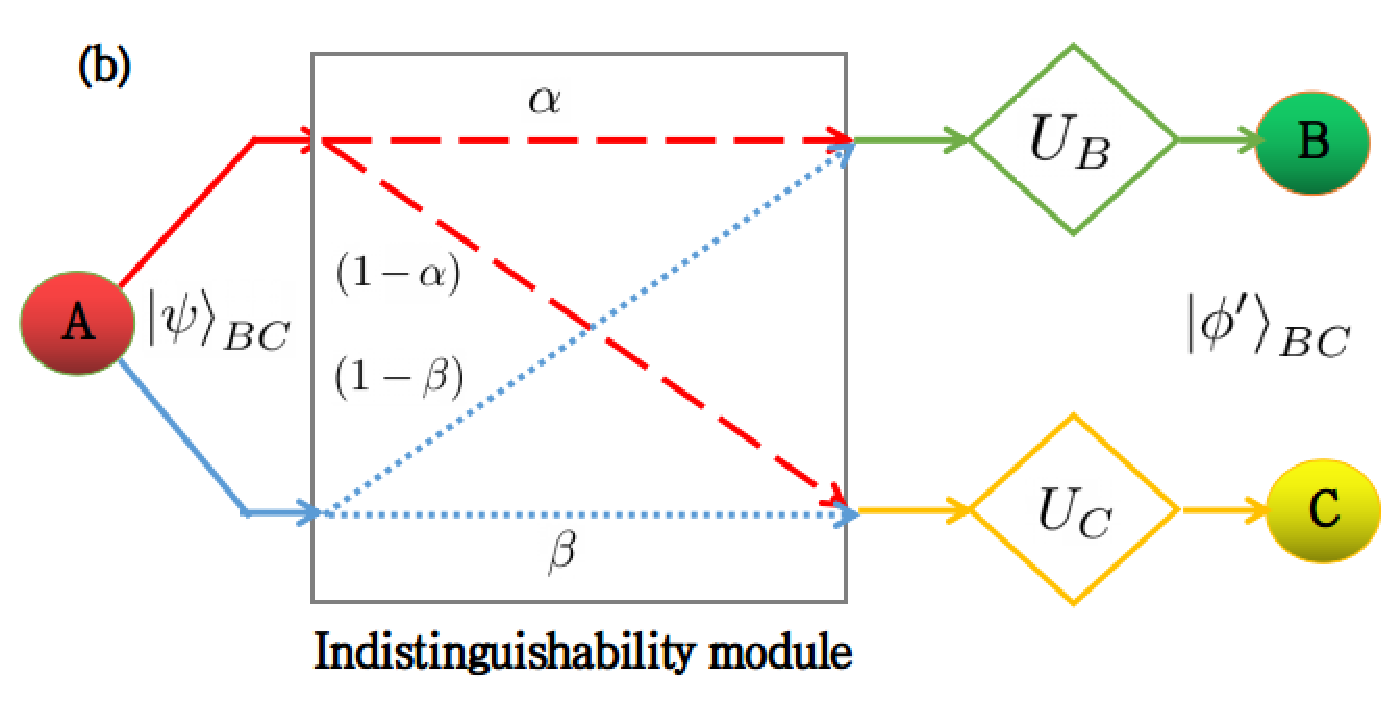} 
\caption{A general protocol with $A$ as the sender and $B$ and $C$ are the receivers.  $\ket{\psi}_{BC}$ is the initial state at $A$ that has to be shared between $B$ and $C$. Here $U_{B}$ and $U_{C}$ are the unitary operations performed by $B$ and $C$ on their respective particles. (a) The final state after the unitary operations performed by $B$ and $C$  is $\ket{\phi}_{BC}$. (b) If an indistinguishability module is present on the path from $A$ to $B$ and $C$, then the final state is $\ket{\phi^{\prime}}_{BC}$.}
\label{ABC_pro}
\end{figure}  

Suppose $A$ is sending a pair of particles to $B$ and $C$ whose joint state is $\ket{\psi}_{BC}$. After $B$ and $C$ perform some local operations $U_{B}$ and $U_{C}$ respectively, the joint state becomes $\ket{\phi}_{BC}$ as shown in Fig.~\ref{ABC_pro}~(a). 

If the particles sent by $A$ become indistinguishable so that the particle intended solely for $B$  now goes to $B$  and $C$  with probability $\alpha$ 	and $\left( 1-\alpha \right)$  respectively. Similarly,  the particle intended solely for $C$ now goes to $C$ and $B$ with probability $\beta$	and $\left( 1-\beta\right)$ respectively.
So the output becomes $\ket{\phi^{\prime}}_{BC}$ as shown in Fig.~\ref{ABC_pro}~(b). If $U_{B}=U_{C}$ and the state $\ket{\psi}_{BC}$ is symmetric with respect to $B$ and $C$, then $\ket{\phi^{\prime}}_{BC}=\ket{\phi}_{BC}$. However, if $U_{B} \neq U_{C}$, or  the state $\ket{\psi}_{BC}$ is asymmetric with respect to $B$ and $C$, or both, then due to indistinguishability the output can be changed, i.e., $\ket{\phi^{\prime}}_{BC} \neq \ket{\phi}_{BC}$.

For example, the output of Hardy's test as shown in Section~\ref{Hardy_gen} which is used in quantum byzantine agreement~\cite{QBA}, random number generation~\cite{ramij_random}, quantum key distribution~\cite{Ramij_QKD} etc., can be different due to indistinguishability as follows.
\begin{figure}[h!]
\centering
\subfloat[]{\includegraphics[width=0.8\columnwidth]{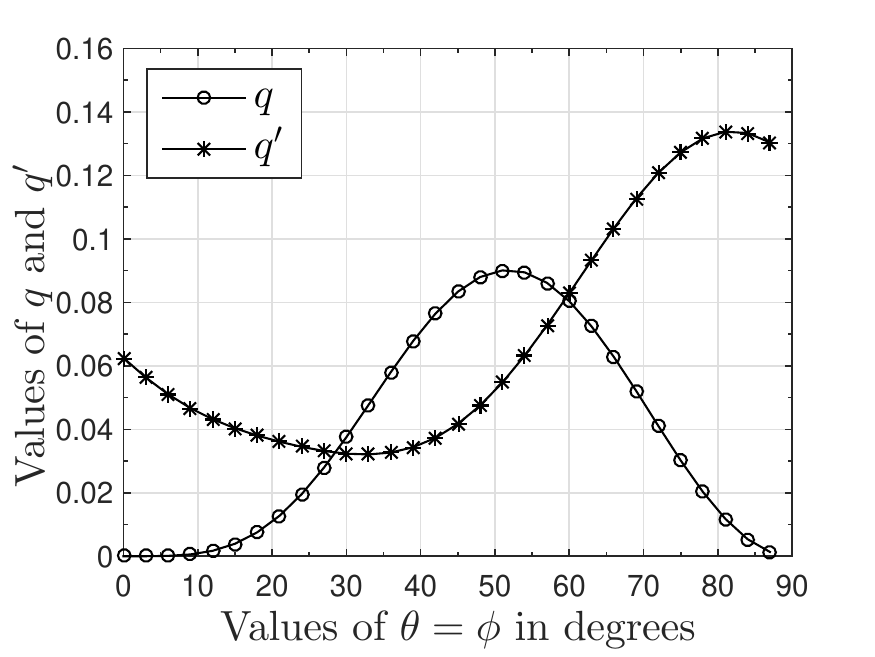}} \\
\subfloat[]{\includegraphics[width=0.8\columnwidth]{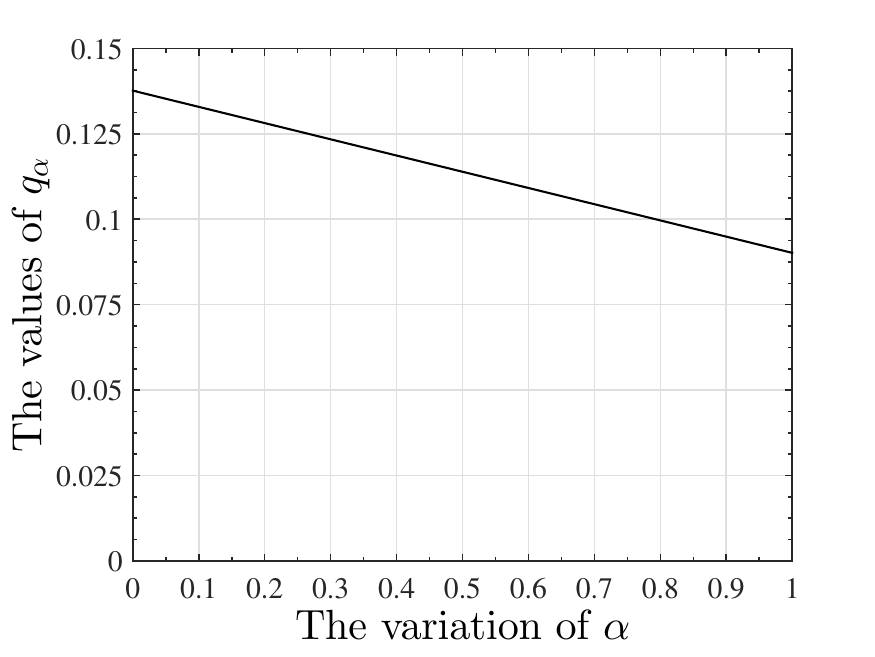} }
        \caption{Effects of indistinguishability on Hardy's probability with $\theta=\phi$ in degrees. (a) The variation of Hardy's probability $q$ and modified Hardy's probability due to indistinguishability $q^{\prime}$  where $\alpha=\beta=\frac{1}{2}$. (b) The variation of modified Hardy's probability $q_{\alpha}$ with $\alpha$ where $\alpha=\beta$ and $\theta=\phi=51.827^{\circ}$.  }\label{Hardy_app}
\end{figure}
Assume Alice and Bob share an entangled state as~\cite{Hardy_soumya}
\begin{eqnarray}
{\small |\psi\rangle_{AB}  = \frac{\text{cos}\theta}{\sqrt{2}}(|00\rangle) + |10\rangle)+ \frac{\text{sin}\theta}{\sqrt{2}}(|01\rangle) + e^{i2\phi}|11\rangle)}, \label{Hardy_state}
\end{eqnarray}
where $0<\theta,\phi< 90^{\circ}$. The Hardy's probability for this state for the measurements stated in~\cite{Hardy_soumya} is 
\begin{equation}
q=\left|\frac{1}{2}\cos \theta \, \cos \chi \left( 1-e^{-2i\phi}\right) \right| ^{2}
\end{equation}
where 
\begin{equation}
\cot\chi =\tan\, \theta \cos \phi
\end{equation}
 The maximum value of Hardy's probability, i.e., $q_{\text{max}} \approx 0.09017$ occurs when $\theta=\phi=51.827^{\circ}$.
If the particle intended for Alice goes to Bob and vice-versa, the  modified probability becomes
\begin{equation}
q^{\prime}=\mid \frac{1}{2} \cos \chi \left( \cos \theta - \sin \theta e^{2i\phi}\right) -\sin \chi e^{-i \phi} \left( \cos  \theta - \sin \theta \right)  \mid^2.
\end{equation}

If $\alpha=\beta$ (as in Fig.~\ref{ABC_pro}), then the final Hardy's probability due to indistinguishability is
\begin{equation}
q_{\alpha}= \alpha^2 q + (1-\alpha)^2 q^{\prime}
\end{equation}

In Fig.~\ref{Hardy_app} (a), we  compare the values  $q$ and $q^{\prime}$. The horizontal axis represents the value of $\theta$ where $0 \leq \theta(=\phi) \leq 90^{\circ}$. We see that the Hardy's probability $q^{\prime}$ increases significantly in the presence of the indistinguishabily module. 
In Fig.~\ref{Hardy_app} (b), we plot the variation of $q_{\text{max}}$ for $\theta=\phi=51.827^{\circ}$ with the value of $\alpha$ vs. $q_{\alpha}$ which shows that the values of $q_{\alpha}$ decreases with increasing $\alpha$. Similar result holds if Hardy's paradox is performed between more than to parties.

\section{Entanglement Swapping using only two indistinguishable particles} \label{2PES}
The seminal work~\cite{ES93} on Entanglement Swapping (ES) required four distinguishable particles as a resource along with Bell state measurement (BSM) and local operations and classical communications (LOCC) as tools. Better versions with only three distinguishable particles were proposed in two subsequent works, one~\cite{Pan10} with BSM and another~\cite{Pan19} without BSM. Recently, Castellini \textit{et al.}~\cite{LFCES19} have shown
that ES for the indistinguishable case is also possible with four particles (with BSM for bosons and without BSM for fermions). This has experimentally realized in~\cite{Wang2021}.
Thus, in terms of resource requirement, the existing best distinguishable versions outperform the indistinguishable one. We turn around this view, by proposing an ES protocol without BSM using only two indistinguishable particles.

 Here we present an ES protocol using two indistinguishable particles, say, $A$ and $B$, without BSM by suitably modifying the circuit of Li \textit{et al.}~\cite{HHNL} as described in Section~\ref{HHNL}. The basic idea is to use any method that destroys the identity of the individual particles, like particle exchange~\cite{Y&S92PRA,Y&S92PRL} or measurement induced entanglement~\cite{Chou05}. Such methods added with suitable unitary operations transfer the intraparticle hybrid entanglement in $A$ (or $B$) to inter-particle hybrid-entanglement between $A$ and $B$. 

\begin{figure}[t!]
\centering
\includegraphics[width=\columnwidth]{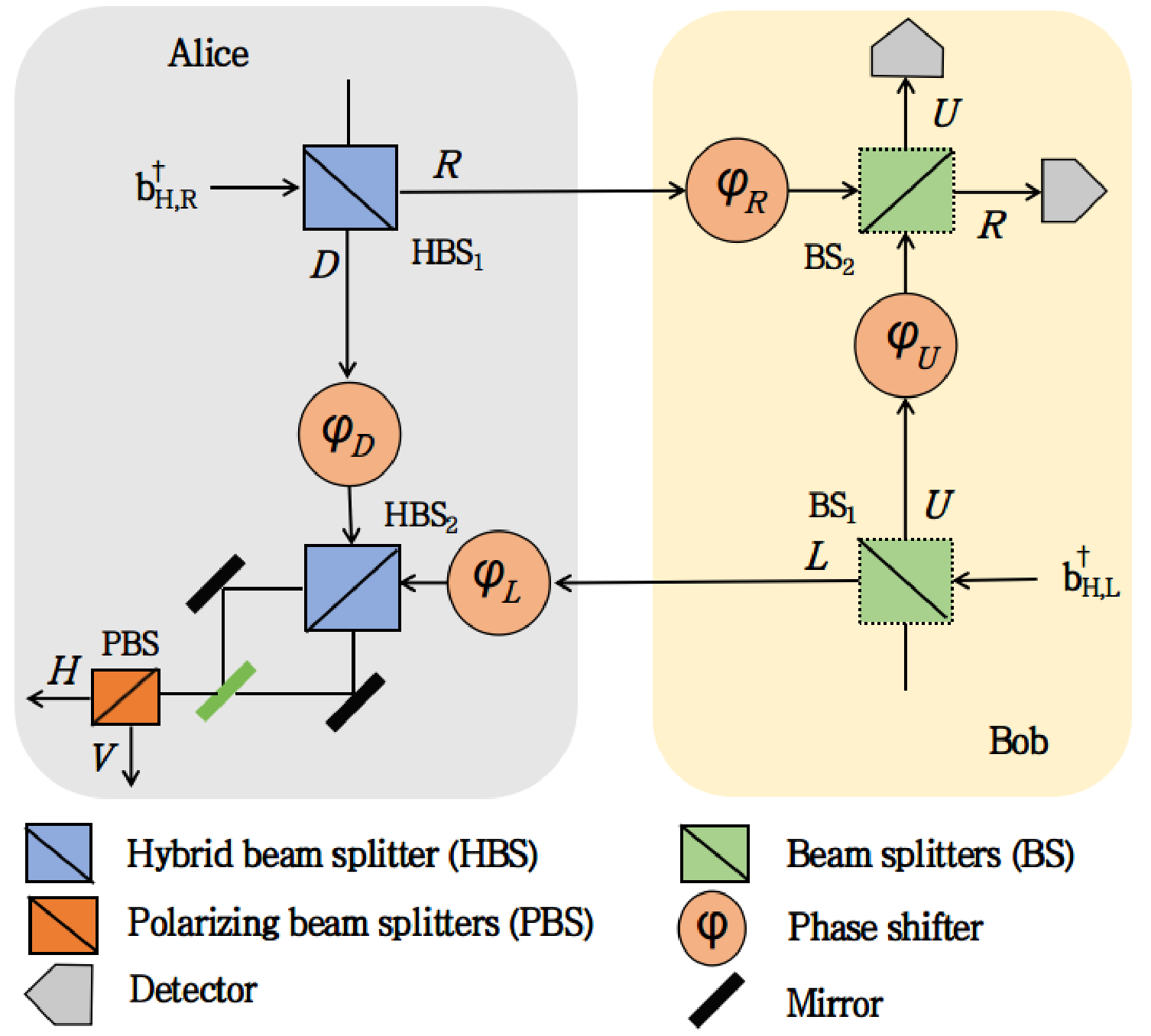} 
\caption{Entanglement swapping with only two indistinguishable particles without Bell state measurements.}
\label{fig:ES}
\end{figure}

Next, we present an optical realization using particle exchange method. Suppose Alice and Bob have two horizontally polarized photons $A$ and $B$, entering into the two modes $R$ and $L$ of a HBS~\cite{HHNL} and a BS, respectively. In second quantization notation, the initial joint state is given by $\ket{\Psi_{i}}=b^{\dagger}_{H,R}b^{\dagger}_{H,L}\ket{0}$, where $\ket{H}$ and $\ket{V}$ denote horizontal and vertical polarization, respectively, and $b_{H,R}$ and $b_{H,L}$ are the corresponding bosonic creation operators satisfying the canonical commutation relations:
\begin{equation}
\left[  b_{i,\textbf{p}_{i}}, b_{j,\textbf{p}_{j}} \right]  = 0,  \left[  b_{i,\textbf{p}_{i}}, b^{\dagger}_{j,\textbf{p}_{j}}\right]  = \delta(\textbf{p}_{i}-\textbf{p}_{j})\delta_{ij}.
\end{equation}

 After passing through HBS$_{1}$, Alice's photon is converted into intraparticle hybrid-entangled state $\frac{1}{\sqrt{2}}\left( b^{\dagger}_{H,R} + i b^{\dagger}_{V,D}\right) $. The particle exchange operation is performed between Alice and Bob, such that the photons coming from $D$ ($U$) and $L$ ($R$) mode go into Alice's (Bob's) side. Next, Alice applies path dependent (or polarization dependent) phase shifts $\varphi_{D}$ and $\varphi_{L}$ on the photons coming from her and Bob's parts, respectively, which go into  HBS$_2$. Similarly, Bob applies path-dependent phase shifts $\varphi_{U}$ and $\varphi_{R}$ on the photons coming from his and Alice's parts, respectively, which go into BS$_2$ as shown in Fig.~\ref{fig:ES}.
The final state is given by
\begin{equation} \label{ESstate}
\begin{aligned}
&\ket{\Psi_{f}}=\frac{1}{4} \left[ e^{i\varphi_{R}}\left( b^{\dagger}_{H,R}+ib^{\dagger}_{H,U}\right) +ie^{i\varphi_{D}} \left( b^{\dagger}_{V,D} + i b^{\dagger}_{H,L}\right)  \right] \\
& \otimes \left[ e^{i\varphi_{L}} \left( b^{\dagger}_{H,L}+ib^{\dagger}_{V,D}\right)  + i e^{i\varphi_{U}}\left( b^{\dagger}_{H,U}+ib^{\dagger}_{H,R}\right) \right] \ket{0}. 
\end{aligned}
\end{equation}
If Alice measures in the polarization DOF and Bob measures in the path DOF, from Eq.~\eqref{ESstate} the probabilities that both of them detect one particle are given by  
\begin{equation}
 \begin{tabular}{c | c c} 
  & B : R & B : U  \\  
 \hline
 A : H \hspace{0.1cm} & $ \frac{1}{4} \text{cos}^{2} \varphi $ \hspace{0.1cm} & $\frac{1}{4} \text{sin}^{2} \varphi$ \\
 A : V \hspace{0.1cm} & $\frac{1}{4} \text{sin}^{2} \varphi$ \hspace{0.1cm} &  $\frac{1}{4} \text{cos}^{2} \varphi$ \\
\end{tabular},
\label{Table:1}
\end{equation}
where $\varphi=\left( \varphi_{D} -\varphi_{L} - \varphi_{R}+ \varphi_{U}\right) / 2$. With suitable values of the phase shifts, one can get the Bell violation in the CHSH~\cite{CHSH} test up to Tsirelson's bound~\cite{Tsirelson}. It can be easily verified that after particle exchange, the particle received at Alice's side (or Bob's side) has no hybrid entanglement, because it is transferred between them. 

Note that the difference between the circuit of Li \textit{et al.}~\cite[Fig. 2]{HHNL} and  Fig.~\ref{fig:ES} is that both the HBSs in Bob's side in the former circuit are replaced by BSs in the latter.
Thus, the intra particle hybrid entanglement of one particle is transferred into the inter-particle hybrid-entanglement of two particles.

\chapter{Conclusion} \label{conclusion}
In this chapter, we discuss the summary, open problems, and future work of each of the previous contributory chapters.

As quantum non-locality is the subset and the strongest form of quantum entanglement, we have performed an experimental verification of Hardy's paradox of non-locality for the first time in superconducting circuits in Chapter~\ref{Chap3}. 
Our initial motivation was to check it for two qubits in the \textit{ibmqx4} five-qubit chip, by choosing any two from the five qubits. 
When $\Sigma_{4} < q \leq q_{max}$, the estimated lower bound $\hat{q}_{lb}$ on Hardy's probability is found to be greater than zero, supporting non-locality. But when $ q \leq \Sigma_{4}$, we get  $\hat{q}_{lb} \leq 0$, because then the errors become of the same order as $q$. 
Interestingly, though $\bar{\epsilon}_{5}^{max}$ decreases with $q$, experimental results show that
$\bar{\epsilon}_{5}^{max}$ does not occur at $q=q_{max}$, rather we get a shift of $\bar{\epsilon}_{5}^{max}$ to the right for the $\left( Q_{3},Q_{4}\right)$ pair of qubits. We also show that the shift direction is not constant, whether it is to the right or left depends on the pair of qubits.
Moreover, we have shown that the above type of shift can occur during the practical implementation of any Hardy's paradox based quantum protocols like quantum Byzantine agreement and we have also discussed possible remedies.
Based on the results of our experiments, we have proposed two performance measures of any quantum computer for two qubits. First, the minimum value of $q$ above which non-locality is established. Second, the amount of shift needed to get the experimental maximum value of $\bar{\epsilon}_{5}^{max}$ of Hardy's probability.
Further, we have performed experiments to show how decreasing the number of gates in the circuits decreases the errors in the circuit for all possible pairs of qubits.
We have also studied the change of errors in IBM quantum computer over time and concluded that errors are decreasing over time.
From the theoretical analysis of the Hardy's experimental set-up, we have found that this test fails for all non-maximally entangled states, where the value of $\phi=90$ and $\theta \neq \left\lbrace 0,45,90\right\rbrace $ degrees. The possibility of a new test for Hardy's paradox for two qubits, so that it does not fail for any non-maximally entangled states is an open problem. Also, the verification of Hardy's test for more than two qubits will be also an interesting work. Further, the comparison between different quantum computers using Hardy's test will be a good research work. We, plan to execute this survey in near future.  

Next, in Chapter~\ref{Chap4} we settle several important open questions that arise due to the recent work on hyper-hybrid entanglement for two indistinguishable bosons by Li \textit{et al.}~\cite{HHNL}.
In particular, we have shown that such entanglement can also exist for two indistinguishable fermions. Further, we have argued that, if in their circuit the particles are made distinguishable, such type of entanglement vanishes. We have also proved the following two no-go results (A) no hyper-hybrid entangled state for distinguishable particles and (B) no unit fidelity quantum teleportation for indistinguishable particles, as in either case the no-signaling principle is violated. 
Our results establish that there exists some quantum correlation or application unique to indistinguishable particles only and yet some unique to distinguishable particles only, giving a separation between the two domains.
The present results can motivate researchers to find more quantum correlations and applications that are either unique to distinguishable or indistinguishable particles or applicable to both. For the latter case, a comparative analysis of the resource requirements and the efficiency or fidelity can also be a potential future work. We plan to find new quantum applications unique for either distinguishable aor indistinguishable particles.

Hereafter, we focus to the most important tool to analysis entangled systems, i.e., the partial trace-out operation in Chapter~\ref{Chap5}. The normal trace-out rule for distinguishable particles was not useful for indistinguishable particles. So, Lo Franco \textit{et al.}~\cite{LFC16,LFC18} have proposed a trace-out rule that is applicable only when each particles has a single degree of freedom. Thus,  we have proposed a  trace-out rule applicable for both distinguishable and indistinguishable particles where each particle have single or multiple degree of freedom. How the new trace-out rule can be applicable to different quantum applications using multiple DoFs is an open area of research. Also, some more tools for indistinguishable particles need to be develop in future to complete the theory of indistinguishability.  In future, we will try to develop new applications of this trace-out rule.

In Chapter~\ref{Chap6}, we found that the relation between teleportation fidelity and singlet fraction is only applicable to distinguishable particles having single degree of freedom. So we established generalized expressions for teleportation fidelity and singlet fraction and derive their relations, applicable for both distinguishable and indistinguishable particles with single or multiple degrees of freedom. We further derive an upper bound for the generalized singlet fraction for distinguishable particles and indistinguishable particles. Our new relation finds application in characterizing different types of composite states in terms of their distinguishability, separability, presence of maximally entangled structure and the number of degrees of freedom. Also we have proposed an an optical circuit to illustrate the non-triviality of our relation. In future, how this relation will be applicable to different quantum information processing protocols will be an interesting research work. Also, one can find new relations for distinguishable particles having single DoF and try to generalized it for both distinguishable and indistinguishable particles having single or multiple DoFs will also be potential future works. We will try to find new applications of the generalized teleportation fidelity and generalized singlet fraction for cryptographic applications.

Monogamy of entanglement, in essence, is a no-go theorem that is a restriction on the shareability of entanglement. We have established the following counter-intuitive result in Chapter~\ref{Chap7}: monogamy of entanglement can be violated maximally for indistinguishable particles by any bipartite entanglement measure that is known to be monogamous for distinguishable particles.  Our result lifts this restriction for indistinguishable particles. It also opens up a new area where researchers can investigate whether the  applications of monogamy of entanglement using distinguishable particles are also applicable using indistinguishable ones and their advantages and disadvantages. Also, what happens to the monogamy relation for indistinguishable particles for more than two-qubits is also an open problem. We try to solve the monogamy relation for more than three qubits for indistinguishable particles having single or multiple DoFs.

Recently, a class of results have been published  which explicitly
demonstrate that certain properties and applications are exclusive to indistinguishable (or distinguishable) particles. In continuation of these separation results, the Chapter~\ref{Chap8} contributes  one property of indistinguishable particles that are different from distinguishable particles.  The inequality of monogamy of entanglement using squared concurrence for three or more distinguishable particles as shown in~\cite{CKW00} becomes equality for pure indistinguishable states, whereas the inequality may hold only for mixed indistinguishable states.
 Note that, this equality is different from the one proposed in~\cite{Gour18}. This result is extremely helpful to calculate the entanglement in the scenarios where particles are indistinguishable like quantum dots~\cite{Petta05,Tan15}, ultracold atomic gases~\cite{Leibfried03}, Bose-Einstein condensates~\cite{Morsch06,Esteve08}, quantum meteorology~\cite{Giovannetti06,Benatti11}, etc.

Finally, we propose two new applications useful for quantum cryptography and one for application useful for quantum networks by entangled distinguishable and indistinguishable particles using the above properties, tools, theorems, relations, and results in Chapter~\ref{Chap9}.
If two cryptographic systems with different resource requirements provide the same level of security, the one with less resources becomes the natural choice. First, we present a generic scheme to reduce the number of particles used in device-independent quantum private query without affecting the security that can be used for both distinguishable and indistinguishable particles. In doing so, the success probability remains the same but the generalized singlet fraction changes. Thus in many ancilla assisted  quantum processing protocols like quantum process tomography~\cite{Altepeter03}, entanglement stabilization~\cite{Andersen19}, quantum secret sharing~\cite{Mitra15}, quantum error correction~\cite{Criger12}, quantum measurement~\cite{Brida12}, weak value amplification~\cite{Pang14}, quantum channel discrimination~\cite{Bae19}, coherent-state superposition generation~\cite{Takahashi08}, etc., where the resource can be reduced by using additional DoF as ancilla instead of a particle. Further, we show that entangled indistinguishable particles may alter certain important parameters in cryptographic protocols.
 In particular, we demonstrate how indistinguishability can change Hardy's probability which is used in quantum byzantine agreement~\cite{QBA}, random number generation~\cite{ramij_random}, quantum key distribution~\cite{Ramij_QKD} etc. Similar attacks can be performed on other quantum cryptographic protocols such as quantum private query~\cite{QPQ07,DIQPQ}, quantum secure direct communication~\cite{Long02,Zhang17}, quantum secret sharing~\cite{Hillery99,Cleve99}, quantum state splitting~\cite{Vitelli13,Passaro13} etc. that use asymmetric entangled states. How these attacks can be neutralized and where our monogamy result can be utilized can be interesting research works. Finally we show a novel entanglement swapping protocol without Bell state measurement using only two indistinguishable particles. 
Thus, these applications will be very useful in the security and reducing the resource requirements of many quantum protocols.

\appendix



\begin{singlespace}

\end{singlespace}

\end{document}